%% file: main.tex
\documentclass[a4paper, twoside]{report}

\usepackage[english]{babel}
\usepackage[utf8]{inputenc}
\usepackage[T1]{fontenc}
\usepackage[%
  backend=bibtex      
 ,style=numeric-comp  
 ,sorting=none        
 ,sortcites=true      
 ,block=none
 ,indexing=false
 ,citereset=none
 ,isbn=true
 ,url=true
 ,doi=true            
 ,natbib=true         
]{biblatex}
\usepackage[a4paper,top=3cm,bottom=3cm,left=2cm,right=2cm]{geometry}
\usepackage{amsmath}
\usepackage{amsfonts}
\usepackage{graphicx}
\usepackage[colorinlistoftodos]{todonotes}
\usepackage[colorlinks=true, allcolors=blue]{hyperref}
\usepackage{csquotes}
\usepackage{epigraph}
\usepackage{tabularx}
    \newcolumntype{L}{>{\raggedright\arraybackslash}X}
    
\usepackage{pdflscape}
\usepackage{afterpage}
\usepackage{capt-of}
\usepackage{siunitx}
\usepackage{subfig}
\usepackage{amsthm}
\usepackage{titlesec}
\usepackage{tikz}
\usepackage{amssymb}
\usepackage{mathtools}
\usetikzlibrary{matrix}
\usepackage[nameinlink]{cleveref}
\usepackage{acronym} 
\usepackage{enumitem}
\usepackage{fancyhdr}
\theoremstyle{definition}
\newtheorem{definition}{Definition}[section]
\usepackage{algorithm}
\usepackage{algpseudocode}
\usepackage{makecell}
\usepackage{multirow}

\newcommand{\etal}{\textit{et al.}}
\newcommand{\cqtOffline}{\textbf{CQT Offline}}
\newcommand{\nsgtOffline}{\textbf{NSGT-CQT Offline}}
\newcommand{\cqtOnline}{\textbf{CQT (Pseudo) Online}}
\newcommand{\nsgtOnline}{\textbf{NSGT-CQT Online}}

\newcommand{\githubFlippy}{\url{https://github.com/flippy-fyp/flippy}}
\newcommand{\githubFlippyQual}{\url{https://github.com/flippy-fyp/flippy-qualitative-testbench}}
\newcommand{\githubFlippyQuant}{\url{https://github.com/flippy-fyp/flippy-quantitative-testbench}}

\begin{document}

\hypersetup{pageanchor=false}
\input{01_front_pages/title/title.tex}
\input{01_front_pages/affirmation/affirmation.tex}
\input{01_front_pages/abstract/abstract.tex}
\input{01_front_pages/acknowledgements/acknowledgements}

\hypersetup{pageanchor=true}
\tableofcontents
\listoffigures
\listoftables
\input{01_front_pages/acronyms/00_acronyms.tex}
\input{01_front_pages/mathsymbols/mathsymbols.tex}

\input{02_parts/00_introduction.tex}
\input{02_parts/01_scofo/01_scofo.tex}

\input{02_parts/02_testbench/01_testbench.tex}

\input{02_parts/03_impl/01_impl.tex}

\input{02_parts/01_conclusion.tex}
\input{03_back_pages/appendix/appendix.tex}

\printbibliography[heading=bibintoc]

\end{document}

%% file: 01_front_pages/title/title.tex
\begin{titlepage}

\newcommand{\HRule}{\rule{\linewidth}{0.5mm}} 

\includegraphics[width=8cm]{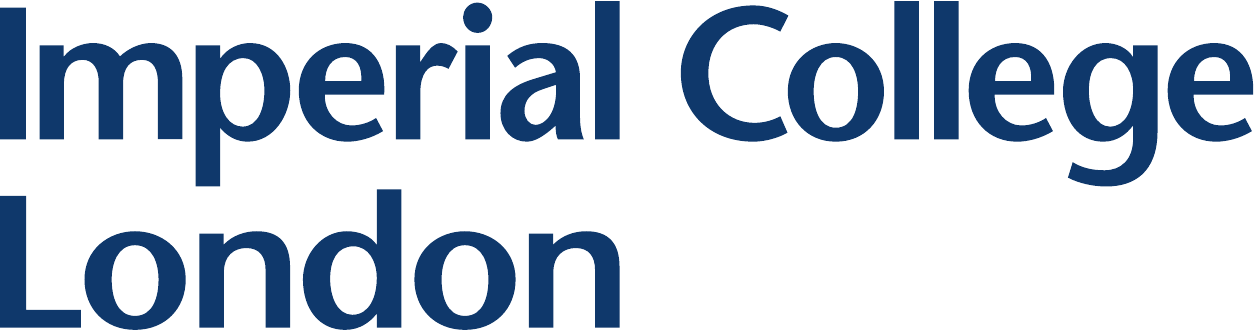}\\[1cm] 
 

\center 


\textsc{\LARGE Final Year Project Report}\\[1.5cm] 
\textsc{\Large Imperial College London}\\[0.5cm] 
\textsc{\large Department of Electrical and Electronic Engineering}\\[3.8cm] 

\makeatletter
\HRule \\[0.4cm]
{ \huge \bfseries Musical Score Following and Audio Alignment}\\[0.11cm] 
\HRule \\[4cm]
 

\begin{minipage}{0.4\textwidth}
\begin{flushleft} \large
\emph{Author:}\\
Lin Hao Lee\\[1.2em]
\emph{CID}:\\
01403154\\[1.2em]
\end{flushleft}
\end{minipage}
~
\begin{minipage}{0.4\textwidth}
\begin{flushright} \large
\emph{Supervisor:} \\
Prof. Patrick A. Naylor \\[1.2em] 
\emph{Second Marker:} \\
Prof. Athanassios Manikas
\end{flushright}
\end{minipage}\\[2cm]
\makeatother



{\large \today}\\[2cm] 
\vfill 

\end{titlepage}

%% file: 01_front_pages/affirmation/affirmation.tex
\renewcommand{\abstractname}{Final Report Plagiarism Statement}
\begin{abstract}
I affirm that I have submitted, or will submit, an electronic copy of my final year project report to the provided EEE link.\\

\noindent
I affirm that I have provided explicit references for all the material in my Final Report that is not authored by me, but is represented as my own work.
\end{abstract}

%% file: 01_front_pages/abstract/abstract.tex
\renewcommand{\abstractname}{Abstract}
\begin{abstract}
Real-time tracking of the position of a musical performance on a musical score, i.e. score following, can be useful in music practice, performance and production. Example applications of such technology include computer-aided accompaniment and automatic page turning. Score following is a challenging task, especially when considering deviations in performance data from the score stemming from mistakes or expressive choices.

In this project, the extensive research present in the field is first explored before two open-source evaluation testbenches for score following--one quantitative and the other qualitative--are introduced. A new way of obtaining quantitative testbench data is proposed, and the \textit{QualScofo} dataset for qualitative benchmarking is introduced. Subsequently, three different score followers, each of a different class, are implemented. First, a beat-based follower for an interactive conductor application--the \textit{TuneApp Conductor}--is created to demonstrate an entertaining application of score following. Then, an Approximate String Matching (ASM) non-real-time follower is implemented to complement the quantitative testbench and provide more technical background details of score following. Finally, a Constant Q-Transform (CQT) Dynamic Time Warping (DTW) score follower robust against major challenges in score following (such as polyphonic music and performance deviations) is outlined and implemented; it is shown that this CQT-based approach consistently and significantly outperforms a commonly used FFT-based approach in extracting audio features for score following.

\end{abstract}

%% file: 01_front_pages/acknowledgements/acknowledgements.tex
\renewcommand{\abstractname}{Acknowledgements}
\begin{abstract}
I would like to thank my supervisor, Professor Patrick A. Naylor, for his invaluable guidance and suggestions. It was Professor Naylor who first suggested this project, and he helped me immensely throughout the project.\\

\noindent
I am extremely grateful to my family and friends who supported and guided me through my academic journey. COVID-19 made this project much tougher--especially when I had to self-isolate in the month this project was due, but these people made it worthwhile and bearable.\\

\noindent
I would like to express my gratitude to my former colleagues from Facebook, Goldman Sachs and Intel for helping develop my 
analytical and engineering skills in industrial contexts.\\

\noindent
A project with user-facing elements cannot be successful without user testing--I am thankful for my patient friends who helped test parts of this project.\\

\noindent
Graphics, figures and diagrams are immensely helpful to illustrate ideas; thus, I thank authors of graphical elements who permitted usage of these elements. See \autoref{appendix_extranotestofigures} for explicit acknowledgements.\\

\noindent
A large part of the motivation for this project stems from my goal to create a cross-platform music practice companion--\textit{TuneApp} \cite{tuneapp}. I am thankful to all who supported my vision and provided feedback to and/or tested the application.

\end{abstract}

%% file: 01_front_pages/acronyms/00_acronyms.tex
\chapter*{List of Acronyms}
\markboth{\MakeUppercase{List of Acronyms}}{}
\begin{acronym}[MIREX] 
    \acro{A-A}{Audio-to-Audio}
    \acro{A-S}{Audio-to-Symbolic}
    \acro{AMT}{Automatic Music Transcription}
    \acro{API}{Application Programming Interface}
    \acro{APT}{Automatic Page Turning}
    \acro{ASM}{Approximate String Matching}
    \acro{CQT}{Constant Q-Transform}
    \acro{CRF}{Conditional Random Field}
    \acro{DCT}{Discrete Cosine Transform}
    \acro{DSP}{Digital Signal Processing}
    \acro{DTW}{Dynamic Time Warping}
    \acro{FFT}{Fast Fourier Transform}
    \acro{GPLv3}{GNU General Public Licence v3.0}
    \acro{GUI}{Graphical User Interface}
    \acro{HMM}{Hidden Markov Model}
    \acro{IR}{Impulse Response}
    \acro{JIT}{Just-in-Time}
    \acro{MFCC}{Mel Frequency Cepstral Coefficient}
    \acro{MIDI}{Musical Instrument Digital Interface}
    \acro{MIR}{Music Information Retrieval}
    \acro{MIREX}{Music Information Retrieval Evaluation eXchange}
    \acro{ML}{Machine Learning}
    \acro{NMF}{Non-negative Matrix Factorisation}
    \acro{NSG}{Nonstationary Gabor}
    \acro{OCR}{Optical Character Recognition}
    \acro{OLTW}{Online Time Warping}
    \acro{OMR}{Optical Music Recognition}
    \acro{PF}{Performance Features}
    \acro{QIP}{Quantum Image Processing}
    \acro{QbH}{Query by Humming}
    \acro{RL}{Reinforcement Learning}
    \acro{RNN}{Recurrent Neural Network}
    \acro{S-S}{Symbolic-to-Symbolic}
    \acro{SF}{Score Features}
    \acro{SMC}{Sequential Monte Carlo}
    \acro{SMF}{Standard MIDI File}
    \acro{STFT}{Short-time Fourier Transform}
    \acro{WTW}{Windowed Time Warping}
\end{acronym}

%% file: 01_front_pages/mathsymbols/mathsymbols.tex
\chapter*{List of Mathematical Symbols}
\markboth{\MakeUppercase{List of Mathematical Symbols}}{}

Mathematical symbols, where relevant, are grouped based on the chapter in which they are first introduced. For a symbol, if a relevant definition is provided in the body of the report, a reference is given.

\section*{General Symbols}

\begin{tabular}{ll}
    $\mathcal{O}$ & Big O notation.   \\[0.2cm]
    $\text{BPM}$  & Beats per minute. \\[0.2cm]
\end{tabular}

\section*{\Cref{tb_quant}}

\begin{tabular}{ll}
    $\delta_s$   & Score note search bound; \Cref{def:bound_ms}.                \\[0.2cm]
    $t_e$        & The estimated note onset time in the performance audio file. \\[0.2cm]
    $t_d$        & The detected time relative to the performance audio file.    \\[0.2cm]
    $t_s$        & The note onset time in the score.                            \\[0.2cm]
    $t_r$        & The true note onset time in the performance audio.           \\[0.2cm]
    $e_i$        & Error; \Cref{def:error}.                                     \\[0.2cm]
    $l_i$        & Latency; \Cref{def:latency}.                                 \\[0.2cm]
    $o_i$        & Offset; \Cref{def:offset}.                                   \\[0.2cm]
    $r_m$        & Miss rate.                                                   \\[0.2cm]
    $r_e$        & Misalign rate.                                               \\[0.2cm]
    $\theta_e$   & Misalignment threshold; \Cref{def:misalign}.                 \\[0.2cm]
    $p_e$        & Piece completion; \Cref{def:pc}.                             \\[0.2cm]
    $\sigma_e$   & Standard deviation of error.                                 \\[0.2cm]
    $\text{MAE}$ & Mean absolute error.                                         \\[0.2cm]
    $\mu_e$      & Mean imprecision.                                            \\[0.2cm]
    $r_p$        & Precision rate.                                              \\[0.2cm]
    $r_{pp}$     & Piecewise precision rate.                                    \\[0.2cm]
    $r_{pt}$     & Total precision rate.                                        \\[0.2cm]
    $\mu_l$      & Mean latency.                                                \\[0.2cm]
    $\sigma_l$   & Standard deviation of latency.                               \\[0.2cm]
    $\text{MAO}$ & Mean absolute offset.                                        \\[0.2cm]
    $\sigma_o$   & Standard deviation of offset.                                \\[0.2cm]
\end{tabular}

\newpage
\section*{\Cref{tb_qual}}

\begin{tabular}{ll}
    $\textsc{Timestamps}$                & List of timestamps; \Cref{def:closesttimestamp}.                             \\[0.2cm]
    $\textsc{Timestamp}_f$               & Timestamp provided by follower; \Cref{def:closesttimestamp}.                 \\[0.2cm]
    $\textsc{Timestamp}_c$               & Timestamp giving the smallest $\textsc{Error}$; \Cref{def:closesttimestamp}. \\[0.2cm]
    $\textsc{Error}$                     & The error between two timestamps; \Cref{def:closesttimestamp}.               \\[0.2cm]
    $\textsc{UpperBoundOrEqual}(x, lst)$ & A function that finds the first element greater than or equals $x$ in $lst$. \\[0.2cm]
    $\textsc{LowerBoundOrEqual}(x, lst)$ & A function that finds the first element less than or equals $x$ in $lst$.    \\[0.2cm]
\end{tabular}

\section*{\Cref{impl_beat}}

\begin{tabular}{ll}
    $\hat{\tau}$   & Instantaneous tempo; \Cref{def:insttempo}.               \\[0.2cm]
    $t_{b}$        & The time of a detected beat; \Cref{def:insttempo}.       \\[0.2cm]
    $a_{d}$        & Acceleration in direction $d$.                           \\[0.2cm]
    $\delta_{|a|}$ & Absolute change in acceleration; \Cref{def:abschgaccel}. \\[0.2cm]
    $T_{delay}$    & Time delay.                                              \\[0.2cm]
    $\tau_a$       & Detected tempo in audio; \Cref{def:targplaybackrate}.    \\[0.2cm]
    $\tau_d$       & Desired audio tempo; \Cref{def:targplaybackrate}.        \\[0.2cm]
\end{tabular}

\newpage
\section*{\Cref{impl_asm}}
\begin{tabular}{ll}
    $S$                     & Score MIDI Data; \Cref{def:score_alignment_asm}.                                               \\[0.2cm]
    $P$                     & Performance MIDI Data; \Cref{def:score_alignment_asm}.                                         \\[0.2cm]
    $L$                     & An alignment between Score and Performance MIDI Data; \Cref{def:score_alignment_asm}.          \\[0.2cm]
    $N$                     & The set of musical notes.                                                                      \\[0.2cm]
    $\Gamma$                & Gap; \Cref{def:score_alignment_asm}.                                                           \\[0.2cm]
    $\alpha$                & Score for a match; \Cref{asm_three_error_def}.                                                 \\[0.2cm]
    $\beta$                 & Score for a mismatch; \Cref{asm_three_error_def}.                                              \\[0.2cm]
    $\gamma$                & Score for an indel; \Cref{asm_three_error_def}.                                                \\[0.2cm]
    $\textsc{Sim}$          & Similarity function; \Cref{sim_func_defn}.                                                     \\[0.2cm]
    $G$                     & A grid system based on $P$ and $S$ to contain alignment scores.                                \\[0.2cm]
    $\textsc{Score}$        & Score function; \Cref{def:scorefunc}.                                                          \\[0.2cm]
    $\textsc{Trace}$        & Backward tracing function to return optimal alignments.                                        \\[0.2cm]
    $\beta_s(c_1, c_2)$     & Score function for a mismatch between two elements $c_1$ and $c_2$.                            \\[0.2cm]
    $\Psi$                  & A list of sets where each set $\psi$ contains parallel notes for a time instance in the score. \\[0.2cm]
    $\textsc{ChordScore}$   & A function returning the score for matching a note in a chord.                                 \\[0.2cm]
    $\text{clamp}$          & The mathematical clamp function that bounds $x$ between $a$ and $b$.                           \\[0.2cm]
    $\textsc{Cost}$         & Cost function.                                                                                 \\[0.2cm]
    $J(a, b)$               & A function that returns the Jaccard index of sets $a$ and $b$.                                 \\[0.2cm]
    $\epsilon$              & Maximum interval for chord execution; \Cref{def:maxintchordexe}.                               \\[0.2cm]
    $\delta$                & Post-alignment parallel voice threshold; \Cref{def:postalignparvoicethres}.                    \\[0.2cm]
    $N_{\Gamma P}$          & Number of gaps in performance.                                                                 \\[0.2cm]
    $N_{\Gamma S}$          & Number of gaps in score.                                                                       \\[0.2cm]
    $N_{\textsc{Mismatch}}$ & Number of mismatches.                                                                          \\[0.2cm]
\end{tabular}

\newpage
\section*{\Cref{impl_dtw}}
\begin{tabular}{ll}
    $\mathcal{S}$                & Feature vector representing the time series of the score; \Cref{def:scofo_dtw}.                    \\[0.2cm]
    $\mathcal{P}$                & Feature vector representing the time series of the performance; \Cref{def:scofo_dtw}.              \\[0.2cm]
    $\mathcal{W}$                & Optimal alignment path between $\mathcal{S}$ and $\mathcal{P}$; \Cref{def:scofo_dtw}.              \\[0.2cm]
    $\mathcal{D}(\mathcal{W})$   & Cost of a path $\mathcal{W}$; \Cref{def:scofo_dtw}.                                                \\[0.2cm]
    $d$                          & Distance function.                                                                                 \\[0.2cm]
    $\sigma_t$                   & Standard deviation of time estimate; \Cref{def:timefreq}.                                          \\[0.2cm]
    $\sigma_\omega$              & Standard deviation of frequency estimate ($\si{\radian}\,\si{\second}^{-1}$); \Cref{def:timefreq}. \\[0.2cm]
    $\sigma_f$                   & Standard deviation of frequency estimate (\si{\hertz}); \Cref{def:timefreq}.                       \\[0.2cm]
    $Q$                          & The ratio of the centre frequency to the bandwidth.                                                \\[0.2cm]
    $f_s$                        & Sampling rate.                                                                                     \\[0.2cm]
    $N_{bin}$                    & Number of bins per octave.                                                                         \\[0.2cm]
    $f_{\min}$                   & Minimum frequency.                                                                                 \\[0.2cm]
    $f_{\max}$                   & Maximum frequency.                                                                                 \\[0.2cm]
    $W$                          & Window function.                                                                                   \\[0.2cm]
    $K$                          & Total number of frequency bins.                                                                    \\[0.2cm]
    $\Omega$                     & Bandwidth.                                                                                         \\[0.2cm]
    $x$                          & A signal of length $L$; \Cref{alg:nsganalysis}.                                                    \\[0.2cm]
    \textbf{$\textbf{(I)FFT}_N$} & (Inverse) FFT of length $N$; \Cref{alg:nsganalysis}.                                               \\[0.2cm]
    $g_k$                        & Real-valued filters centred at $\omega_k$ forming the vector $\textbf{g}$; \Cref{alg:nsganalysis}. \\[0.2cm]
    $I_k$                        & Finite index set where $k \in I_k$; \Cref{alg:nsganalysis}.                                        \\[0.2cm]
    $c$                          & \textbf{CQ-NSGT} coefficients; \Cref{alg:nsganalysis}.                                             \\[0.2cm]
    $h_0$                        & Slicing window centred at $0$.                                                                     \\[0.2cm]
    $s$                          & \textbf{SliCQ} (slice) coefficients.                                                                        \\[0.2cm]
    $C_Q$                        & Constant-Q coefficients; \Cref{def:normenergy}.                                                    \\[0.2cm]
    $E$                          & Energy; \Cref{def:normenergy}.                                                                     \\[0.2cm]
    $w_a$, $w_b$, $w_c$          & Weights used to introduce bias towards a certain step direction in DTW; \autoref{eqn:oltw_dtw}.    \\[0.2cm]
    \texttt{c}                   & Search window.                                                                                     \\[0.2cm]
    \texttt{MaxRunCount}         & A constraint that constrains the slope of the path.                                                \\[0.2cm]
    $M$                          & The cost matrix system containing $\mathcal{P}$, $\mathcal{S}$ and $\mathcal{D}$.                  \\[0.2cm]
    $(\texttt{i}, \texttt{j})$   & Coordinate of the top-rightmost calculated position in $M$.                                        \\[0.2cm]
    \texttt{previous}            & Previous direction(s) incremented.                                                                 \\[0.2cm]
    \texttt{current}             & Current direction(s) to increment.                                                                 \\[0.2cm]
    \texttt{runCount}            & The number of consecutive times the system is incremented in a certain direction.                  \\[0.2cm]
    $(i', j')$                   & The latest coordinate of the lowest cost path.                                                     \\[0.2cm]
\end{tabular}

%% file: 02_parts/00_introduction.tex
\chapter{Introduction}

\epigraph{\itshape I think we will find more and more ways in which technology invades our artistic spaces, so music is something you will need more than ever because it is there in time and in space and for that moment only.}{Simon Rattle}

\section{Motivation}
Portable digital devices, especially tablet computers such as the iPad and the Microsoft Surface, transformed the music education and production industries. In the former industry, the iPad--equipped with interactive touch-enabled music education applications--is now ubiquitous in elementary music classes \cite{ruismaki13, riley15}; in the latter, music composition and performance software running on portable devices are gradually gaining its way into normal use by professional and amateur musicians of many genres \cite{staffpad, musicnotes, tuneapp, forscore}. Moreover, it is worth noting that, these days, professional musicians, whether individual musicians \cite{ravelyuja}, ensembles \cite{debussyborromeo} or even large orchestras \cite{tonkunstlerorch}, can be seen performing on stage with digital sheet music. Further, musical scores are readily available as digital copies through music catalogues such as \textit{IMSLP}\footnote{\url{https://imslp.org/wiki/Main_Page}} and online music stores.

With the exponential growth in popularity and computation power of these devices \cite{tabletgrowth} in mind, there exists even more potential for them to become more indispensable to all musicians--a musician's digital companion should not only be limited to merely a paper sheet music replacement. One such area that is quite needed and useful is the crux of this project: \textbf{score following}, i.e. real-time tracking of the position of a musical performance on a musical score according to the musical performance. Many useful applications of score followers exist, such as automatic page turning (APT) and computer-aided accompaniment. These and other applications are detailed further in \autoref{scofo_apps}.

\section{Challenges}
\label{challenges}
This project recognises three main challenges in score following research and development. The first two relates to research and evaluation problems of score following. The final challenge lies in the actual implementation and development of a robust score follower. The three subsections below detail the challenges.

\subsection{Research}

Research into score following is extensive, ranging from elementary ``string matching'' techniques, first introduced in 1984 \cite{dannenberg84,vercoe84}, to state-of-the-art deep learning algorithms that can follow raw sheet music images \cite{henkel20}. Many works in the literature also develop different types of score followers, such as systems that take in MIDI as opposed to those that take in audio. Hence, it is daunting for new researchers to study existing literature and come up with novel solutions in a specific area or application.

\subsection{Evaluation}

Following from the fact that there are many types of score followers, naturally, evaluation of score followers is not well standardised. While there was effort in 2006 to standardise evaluation \cite{mirex, cont07} as part of MIREX, currently, the evaluation standards are considered out of date given the speed at which score following research is carried out. This had two implications: firstly, many research groups decided to up with their own--often proprietary--evaluation and comparison strategies that unfortunately favours their own approach; secondly, the MIREX evaluation was not popular among researches in the field (there were usually only a few submissions per year and from 2018 onwards there were none). Further, proof of concept score following software that applies score following in practical applications such as computer-aided accompaniment \cite{musicplusone} and APT \cite{arzt08} is often written ad hoc as part of the research project, resulting in software incompatible with standardised evaluations.

\subsection{Score Following}

Finally, the biggest challenge is that score following in itself is a well-known difficult problem \cite{sturm13,mullerdtw,dannerberg03,chen19,heijink00}, significant enough, in fact, that an entire chapter--\autoref{scofo_challenges}--is dedicated to detail the main challenges. In short, these challenges include difficulties in feature extraction from the performance audio, deviations (intentional and unintentional) between human performances and the corresponding score, underspecified sheet music, music improvisation and polyphonic music.






\section{Contributions}
\label{intro:contrib}

The contributions of this project are threefold, each relating to the three respective challenges identified in \autoref{challenges}.

Firstly, the project provides an extensive review of score following's applications, technicalities and challenges before diving into almost 40 years of score following research, providing a detailed and critical overview of research ranging from early pioneering work to recent state-of-the-art solutions.

Secondly, two open-source evaluation testbenches--one quantitative and the other qualitative--for score following are introduced. A novel way of obtaining quantitative testbench data is proposed, and the \textit{QualScofo} dataset is introduced for qualitative benchmarking. The two testbenches have a well-defined API that can be compatible with most score followers. The qualitative testbench also demonstrates practical usages, such as APT, for the score follower in question.

Thirdly, three score followers are created. First, a beat-tracking-based approach is incorporated into an interactive conductor application to showcase an entertaining application for score following. Secondly, an Approximate String Matching (ASM) aligner is implemented as part of a novel way to produce testbench data for the quantitative testbench, as well as to provide more technical background details of score following. Finally, a CQT-DTW score follower is implemented--it is shown that this score follower is robust against major challenges of score following (such as polyphonic music and performance deviations), and that the real-time CQT approach consistently and significantly outperforms a commonly used FFT-based approach in extracting audio features for score following.

\section{Report Structure}

This report is divided into three parts per the threefold contributions given in \autoref{intro:contrib}. The parts are as listed:

\begin{enumerate}
    \item \textbf{\autoref{scofo} — \nameref{scofo}}: \Cref{scofo_apps} first provides applications of score following to give an insight into the motivations of researching and developing such systems. \Cref{scofo_preliminaries} then dives into preliminaries to give sufficient technical and historic background on the many types and details of score following. Further, the challenges of score following are covered in \autoref{scofo_challenges}, before \autoref{scofo_relatedwork} which gives an extensive literature review on related and past work in score following.
    
    \item \textbf{\autoref{testbench} — \nameref{testbench}}: This part deals with formulating systematic and reliable approaches for evaluating score followers. Challenges faced by existing approaches are first detailed in \autoref{tb_preliminaries} before implementation details of a proposed open-source pair of evaluation testbenches are covered in chapters~\ref{tb_quant} and \ref{tb_qual}.
    
    \item \textbf{\autoref{impl} — \nameref{impl}}: A part is dedicated for the completed implementations of three score followers. First, the \textit{TuneApp Conductor} beat-tracking-based score follower is introduced in \autoref{impl_beat}. An Approximate String Matching (ASM) approach is introduced subsequently in \autoref{impl_asm} to produce ground-truth alignments for the quantitative testbench and also to give technical background for the subsequent DTW-based score follower. Finally, a novel CQT-DTW score follower robust against major score following challenges is introduced in \autoref{impl_dtw}.
    
\end{enumerate}

A concluding chapter, \autoref{conclusion}, will discuss findings and contributions of this project, before suggesting areas of future work.

\section{Extra Notes to Readers}

\subsection{Assumed knowledge}

This report assumes elementary musical knowledge. Readers are expected to be able to read simple sheet music (in modern staff 
notation\footnote{A guide can be found at \url{https://en.wikipedia.org/wiki/Musical_notation\#Modern_staff_notation}.}, 
see an example in \autoref{fig:modern-staff-notation}) and understand the basic concepts of sound and music, namely pitch, volume and length. Throughout this report, the British English variant of musical terminology is used (e.g. \textit{quaver} over \textit{eighth note}).

In terms of computer science, readers are assumed to understand basic concepts in algorithms, data structures and software engineering covered by elementary undergraduate-level courses.

Readers are also assumed to understand simple concepts of (audio) signal processing, mainly basic concepts in the time and frequency domains. An undergraduate-level introductory signal processing course is sufficient.

%% file: 02_parts/01_scofo/01_scofo.tex
\part{A Review of Score Following}
\label{scofo}

\input{02_parts/01_scofo/chapters/01_applications.tex}
\input{02_parts/01_scofo/chapters/02_preliminaries.tex}

\input{02_parts/01_scofo/chapters/03_challenges.tex}

\input{02_parts/01_scofo/chapters/04_relatedwork.tex}

%% file: 02_parts/01_scofo/chapters/01_applications.tex
\chapter{Applications}
\label{scofo_apps}
\epigraph{\itshape Creation exists only in the unforeseen made necessary.}{Pierre Boulez}

A review of score following would not be complete without a foray into its practical applications in music practice, performance and production. Further, many score followers--in research or industry--are built for a specific application; thus, knowing the primary goal of a score follower helps give an overview of the problems the authors target. Here, notable applications of score followers are delineated with comprehensive background details for the problems to be solved.

\section{Automatic Page Turning (APT)} 
\label{scofo_apps_apt}

The problem of sheet music page turning for musicians while their hands are preoccupied with playing a musical instrument is well known. In piano performances, a common sighting is that of a human page turner who assists the pianist. However, the vast majority of playing time is actually during practice sessions where the employment of a page turner would prove impractical. 

The appreciation of the problem reflects well on the myriad commercial solutions that exist. While systems that operate on physical paper sheet music exist, these mechanical systems are impractical as they need to balance the speed of turning the page with the relative fragility of the paper to not tear the paper \cite{wolberg12}. Thus, solutions working on digital sheet music are more common. These solutions can be divided into two categories:

\begin{enumerate}
    \item \textbf{Manual page turners} require user input to trigger a page turn. For instance, \textit{AirTurn}\footnote{\url{https://www.airturn.com/}} provides a foot pedal system, requiring the user to use their foot on the Bluetooth-powered pedal to trigger a page turn on a primary (usually tablet) device. A solution requiring no dedicated hardware, \textit{TuneApp}\footnote{\url{https://www.tuneapp.com/}}, operates similarly using any pair of web-enabled devices, with either touch or motion detection available as a trigger on the second device. These solutions, however, are not suitable for musicians playing instruments with foot pedals, such as organs. It is worth noting that \textit{TuneApp}, not limited by the range of Bluetooth, could lead to a solution where a human page turner could be offstage. These solutions do not fully solve the page turning problem; hence, automatic page turners are more desirable.
    \item \textbf{Automatic page turners} such as \textit{ClassicScore}\footnote{\url{https://blog.naver.com/earthcores/}}, \textit{MobileSheets}\footnote{\url{https://www.zubersoft.com/mobilesheets/}} and \textit{Musicnotes}\footnote{\url{https://www.musicnotes.com/apps/}} employ a score-scrolling feature. In \textit{ClassicScore}, the rate of the scrolling is determined from the tempo of a prerecorded playback of the music. \textit{MobileSheets} and \textit{Musicnotes} are similar, with the addition of the ability to adjust the performer's preferred speed. Notably, three score following applications exist: \textit{PhonicScore}\footnote{\url{https://www.phonicscore.com/}}, \textit{Beatik}\footnote{\url{https://www.beatik.com}} and \textit{Tido}\footnote{\url{https://www.tido-music.com/}}; however, real-life tests of these systems show that they are unreliable due to their susceptibility to noise \cite{arzt08}.
\end{enumerate}

Besides commercial solutions, there exists literature in the area that propose better approaches for APT. Eye-gaze tracking systems can actively monitor the musician's on-screen point of regard and turn or scroll pages \cite{tabone17}; however, musicians often look away from the score and toward their hands and elsewhere, leading to the requirement of an eye-gaze prediction model that adds to the computational complexity of the system while still being imperfect \cite{tabone20}. Moreover, eye-gaze systems have no other viable areas of application in the musical context, in contrast to score followers' wide range of applications.

The requirements of an ideal page turning system, mainly supporting operation without additional gestures and being robust to errors or performance improvisations of the musician, strongly motivates the use of a score following system robust to performance deviations.

\begin{figure}[h]
    \centering
    \includegraphics[width=0.5\columnwidth]{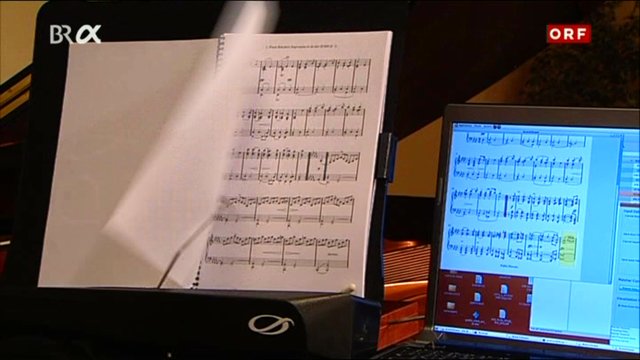}
    \caption{Arzt's page turner shown in a documentary. Source: \cite{arzt16}.}
    \label{fig:arzt-pageturner}
\end{figure}

\autoref{fig:arzt-pageturner} shows an APT system comprising a mechanical page turning device developed by the Viennese company \textit{Qidenus}\footnote{\url{http://qidenus.com/technologies/}} extended by Andreas Arzt to work automatically with a score follower system \cite{arzt16}. The work was demonstrated on a documentary shown on BR Alpha in January 2010.

\section{Computer-aided Accompaniment}
\label{scofo_apps_caa}

\begin{figure}[h]
    \centering
    \includegraphics[width=\columnwidth]{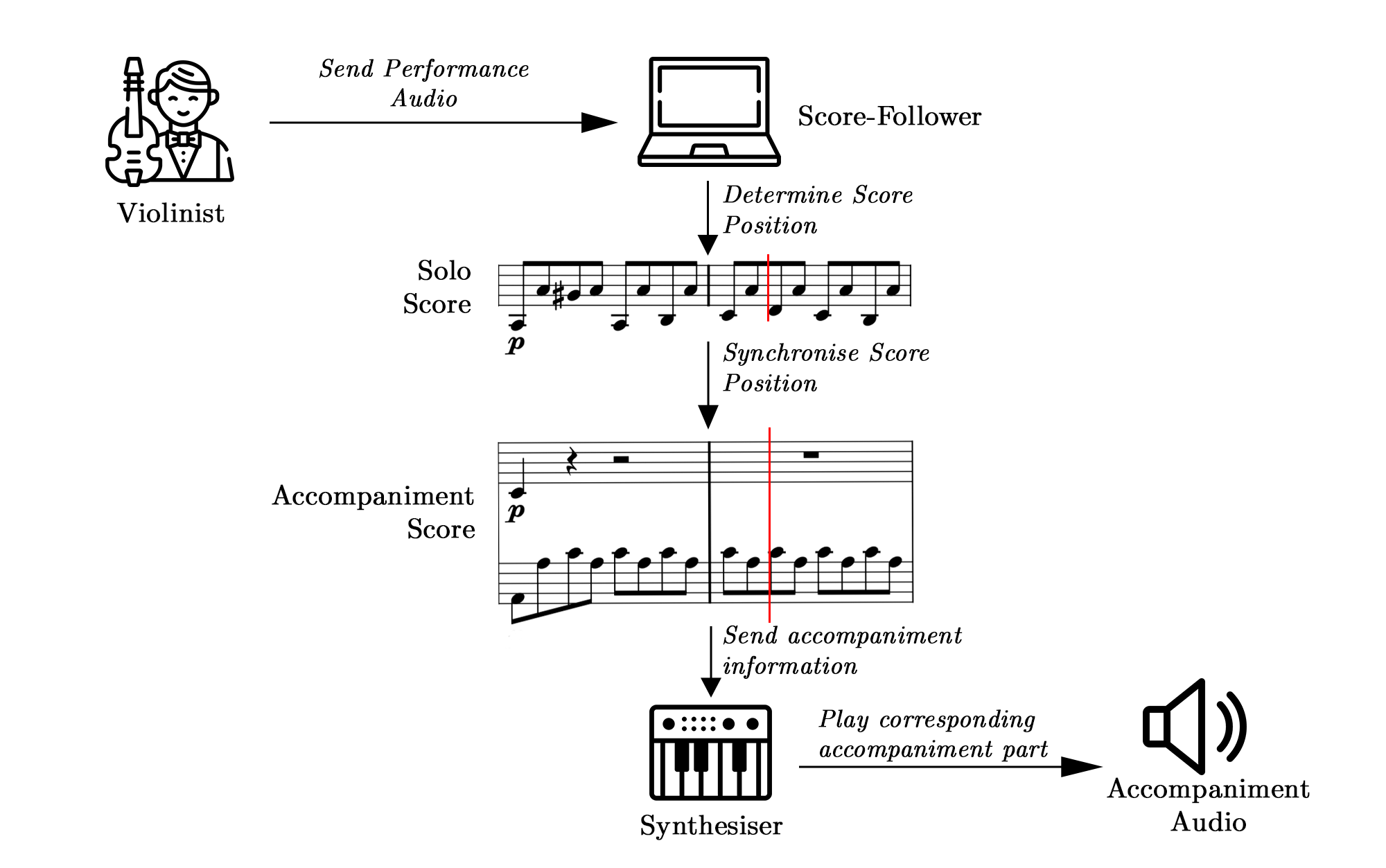}
    \caption{Structure of a computer-aided accompaniment system accompanying a human violinist as a pianist.}
    \label{fig:caa}
\end{figure}

Many live performances are performed accompanied or in ensembles. Musicians are trained to play their instruments in correspondence with others, even if the performance includes uncertainties in and deviations from the musical score. A computer-aided accompaniment system usually comprises an audio-to-symbolic score follower to follow the performance in real time while producing accompaniment music. \autoref{fig:caa} shows a generic structure for a computer-aided-accompaniment system--comprising a score follower--acting as a pianist in a violin-piano duet.

In fact, the earliest score followers in 1984 were built as computer-aided accompaniment tools \cite{vercoe84,dannenberg84}. Music composers such as Manoury and Boulez among others composed music for such systems--these music performances involve electronically generated music synchronised (via basic score followers) with human-generated music \cite{tissot08, cont08}. However, early composers in these experiments are often forced to make compromises so that their music is followed in such a way that the electronic events in the score are correctly triggered \cite{puckette92}. More advanced score following systems would be capable of even richer, accurate and flexible interactions in live performances \cite{musicplusone, cont08, dannenberg06}.

While trained musicians' accompaniment skills--including sensitivity to performance details from basic ones such as rhythm, volume and pitch to advanced ones like amount of breath and bow length left--are difficult for computer-based systems, the latter systems have nearly unlimited technical facility, allowing the coordination of arbitrarily fast notes and complex rhythms. Exploiting this advantage, Jack Beran and Nick Collins composed pieces with complex computer-based accompaniments for the \textit{Music Plus One} music accompaniment system \cite{musicplusone}. 

Commercial solutions providing computer-aided accompaniment exist too, such as \textit{CueTIME}\footnote{\url{https://hub.yamaha.com/cuetime-the-software-that-follows-you/}} and \textit{Metronaut}\footnote{\url{https://www.antescofo.com/}}. \textit{CueTIME} unfortunately works only on certain digital Yamaha keyboards, via aligning keying information to synthesised accompaniment music. \textit{Metronaut} is a relatively new solution (born in IRCAM--see \autoref{scofo_stats_hmm}) capable of following performance audio, but the system is limited to musical pieces (which are preprocessed manually) contained in its library.

Computer-aided accompaniment systems are widely studied in research--in fact it motivated many early approaches to score following detailed in \autoref{scofo_earlywork}. Moreover, literature proposing more advanced systems based on score following for a wide array of instruments and genres exist \cite{xia17, hoffman11, kaliakatsos12, xia15}.

\section{Performance Analysis}
\label{scofo_apps_perfanal}

In music education, training musicians often follow the performances or recordings of other musicians to study a piece of music. While listening to these performances, training musicians usually follow their own copy of the sheet music. To aid this form of training, many videos of aligned performance audio on a scrolling score, often with a marker denoting which note is being played at that point in time, were created. These videos are popular on YouTube; notable examples include \textit{The Scrolling Bach Project}\footnote{\url{https://www.youtube.com/channel/UCNAckPiDYxRWengUlRujs6Q/}}--followed by over 87,000 subscribers as of May 2021--which applies this concept to works by Bach. Usually, these videos are created by dedicated musicians who manually (or sometimes aided by a music alignment system) align notes in the performance audio with notes on a sheet music \cite{gerubach,ashishxiangyi}. 

If a reliable score following solution exists, this idea can be further extended to live performances: users can use this system in a live performance to follow the performance on sheet music. The existence of a reliable score follower also solves a notable problem of projects such as \textit{The Scrolling Bach Project}: users are forced to follow a performance and sheet music combination chosen by the authors. Some users may prefer an interpretation of a music piece by a different performer, or a different sheet music edition published by a different publisher. A score-follower-based system would enable on-the-fly and easy music alignment on any performance and sheet music combination.

A Herculean effort in score following for this application is the work of Arzt \etal{} \cite{arzt16} which introduced a score follower capable of tracking an orchestral performance of the \emph{Alpensinfonie} at the prestigious \emph{Concertgebouw} in Amsterdam \cite{arztconcert}. The system was capable of APT and also provided artistic visualisations and textual information provided by a musicologist to the audience.

\section{Music Practice}

As score following implies that a computer can pinpoint which part of the music is being played by a player, it can also be used in conjunction with musical analysis tools to provide feedback for music practice. This was spotted as an early application of score following; thus, solutions, some commercial, exist.

In 1990, \textit{Piano Tutor} was designed by Dannenberg to model after a human music teacher \cite{dannenberg90}. The system follows the performance of the student via score following, and after the performance, a score alignment system provides feedback on the performance, such as numbers of notes missed or mistimed. \textit{SmartMusic}\footnote{\url{https://www.smartmusic.com/}} is a commercial solution that also offers accompaniment capabilities; notably, studies were carried out with \textit{SmartMusic}, indicating that students who practice with \textit{SmartMusic} accompaniment improve in terms of performance skill, motivation and practice time. A more recently introduced commercial product is \textit{PracticeBird}\footnote{\url{https://www.practicebird.com/}}, which gives real-time intonation and rhythm feedback. It unfortunately suffers from the same problem as the commercial score following APT solutions put forth in \autoref{scofo_apps_apt}: it is not robust enough in noisy environments.

\section{Performance Cues}

In performances where music plays an accompanying or background role such as theatre shows, live TV shows or opera performances, performance cues are often required to synchronise the performance with the music. Notably, in some opera houses and theatre stages, the accompanying orchestra is often seated below stage in an ``orchestra pit'' \cite{tronchin06}. Usually, a conductor is needed to synchronise musicians with the performance happening on stage; the synchronisation often involves cueing performers--sometimes both the musicians and the stage performers. The role of a conductor is often already heavy, and having to divide the conductor's attention between the orchestra and stage performance is not ideal.
    
With a score follower, the job scope of the conductor could be reduced--stage performers could be automatically cued by the system that follows what the orchestra is playing. Further, for small music ensembles, the conductor could even be omitted.

\section{Entertainment}

Score followers can be used for entertainment beyond musical performances--there are many creative applications possible. This project provides an example in \autoref{impl_beat}, in which a beat tracking score follower system is applied to create an interactive conductor application, where users can wave their mobile devices to ``conduct'' music.

Another application of score followers for entertainment was also introduced by the author. This application was timely with regard to the COVID-19 pandemic: 2021's Vienna New Year's Concert\footnote{\url{https://www.wienerphilharmoniker.at/en/newyearsconcert}} was the very first ran without an audience due to the pandemic. The \textit{Radetzky March} is a staple in every year's concert--usually it is the last encore piece played with the audience clapping along. Riccardo Muti, conductor of the 2021 edition, mentioned that 2021 was the very first time the audience can listen to the \textit{March} without the clapping. However, this performance felt empty and weird, prompting the author to create \textit{Radetzky}, a program that can add randomised and volume-sensitive claps to detected beats on any piece of music. An off-the-shelf beat tracking score follower \cite{bock14,krebs15} was used. The source code for \textit{Radetzky} can be found in \url{https://github.com/lhl2617/radetzky}, and 2021's \textit{Radetzky March} with added claps can be watched at \url{https://www.youtube.com/watch?v=dTrGnQCliWg}.

%% file: 02_parts/01_scofo/chapters/02_preliminaries.tex
\chapter{Preliminaries}
\label{scofo_preliminaries}
\epigraph{\itshape I think a lot of people have the impression with classical music because it’s written and it looks very specific on the page that it tells you exactly how to play it, but actually all it tells you is the relative things... nothing is really mathematical.}{Hilary Hahn}

In this chapter, preliminaries of score followers are provided, starting from a definition (\autoref{scofo_definition}) and a generic framework (\autoref{scofo_generic}). Then, the different categories of score followers are covered in \autoref{scofo_types} before a short but important detour into different musical data representations in the context of score followers (\autoref{scofo_mdr}).

\section{Definition}
\label{scofo_definition}

Practical definitions of score following have changed and evolved in the last 40 years, naturally as score following is an intersection of new demands from musicians and new scientific technologies. As such, subjective definitions for score following can lose its account over time. Arshia Cont from IRCAM, a research group active in this area, gave a clear and concise general definition for score following in 2004 \cite{cont04}:

\begin{definition}[Score following]
    Score following serves as a real-time mapping interface from \emph{Audio abstractions} towards \emph{Music symbols} and from performer(s) live performance to the score in question.
\end{definition}

Score following results in a score-performance alignment, which relates to the also heavily researched sibling field of
\emph{music alignment} \cite{chen19, chen14, wang16, sturm13, orio03, hu04, dannerberg03}. Music alignment does not have real-time constraints, which means it can use the entire performance data to perform alignment. Being less flexible, it is more commonly used in research, but practical applications do exist, such as audio querying for MIDI documents \cite{hu04} as well as timestamping of an audio recording according to the desired musical position in a music score \cite{dannerberg03}. These systems are worthy of mention in this context, as music aligners often provide good alignments and benchmarks useful for research and development of score followers.

\section{Generic Framework}
\label{scofo_generic}

The standard approach in most score followers comprises three steps \cite{chen19}:
\begin{enumerate}
    \item \textbf{Feature extraction}: Informative features in the performance data (either audio or other representations such as MIDI) are extracted to characterise the musical content, such as onsets, pitches or chords.
    \item \textbf{Similarity calculation}: Via a defined similarity function, the difference between features extracted in step 1 and the note events in the sheet music is measured.
    \item \textbf{Alignment}: An alignment algorithm is used to determine the closest match between feature sequences and note events.
\end{enumerate}

\begin{figure}[h]
    \centering
    \includegraphics[width=\columnwidth]{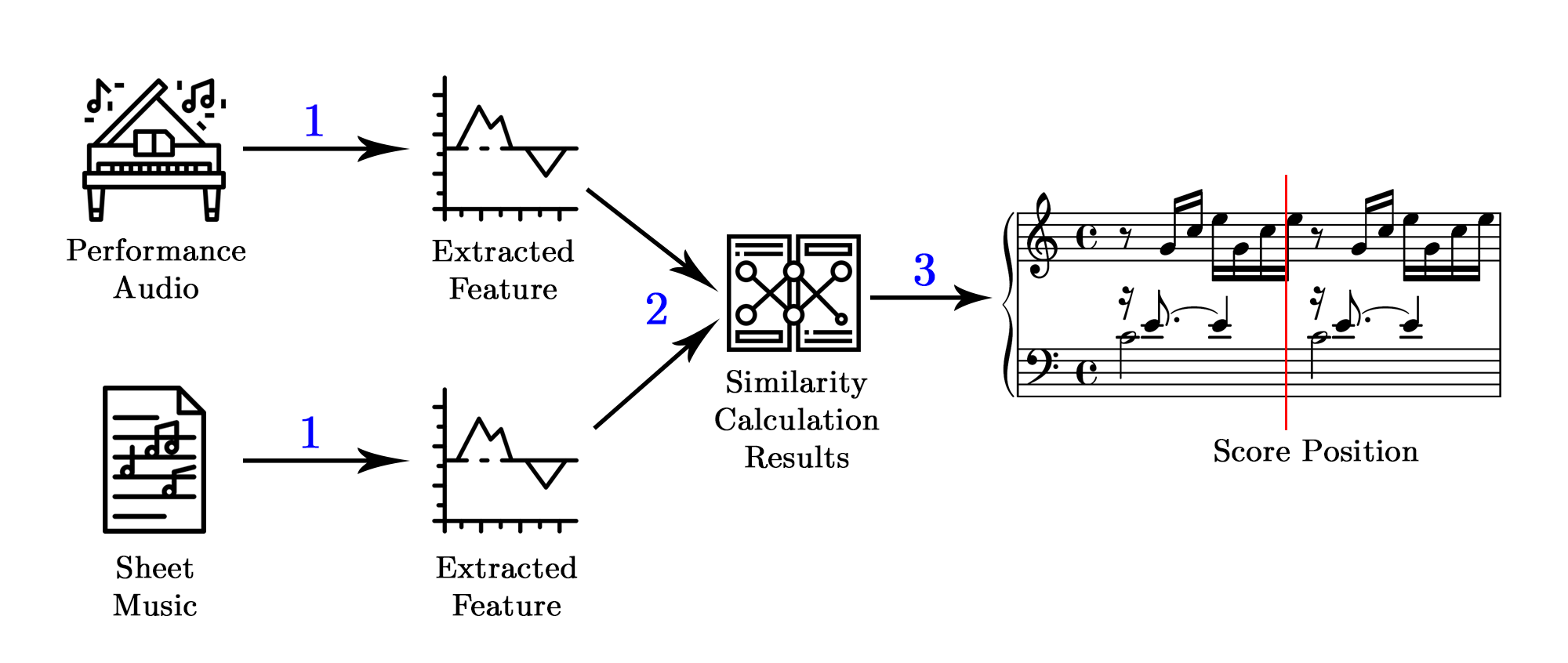}
    \caption{Illustration of a generic framework used in most score followers. The numbers represent each of the three steps delineated in \autoref{scofo_generic}.}
    \label{fig:generic-framework}
\end{figure}

\autoref{fig:generic-framework} illustrates the three steps in a piano performance score following situation.

\section{Types}
\label{scofo_types}

Score followers can usually be categorised into three classes \cite{mullerdtw} defined as such:
\begin{definition}[Audio-to-symbolic score followers]
    \textbf{Audio-to-symbolic} (A-S) systems, which include audio-to-score followers, analyse the content of the input performance audio at a given point in time and maps it to a corresponding time point on the score with a similar musical structure. \autoref{fig:audio-to-symbolic} illustrates this method.
    \begin{figure}[h]
        \centering
        \includegraphics[width=\columnwidth]{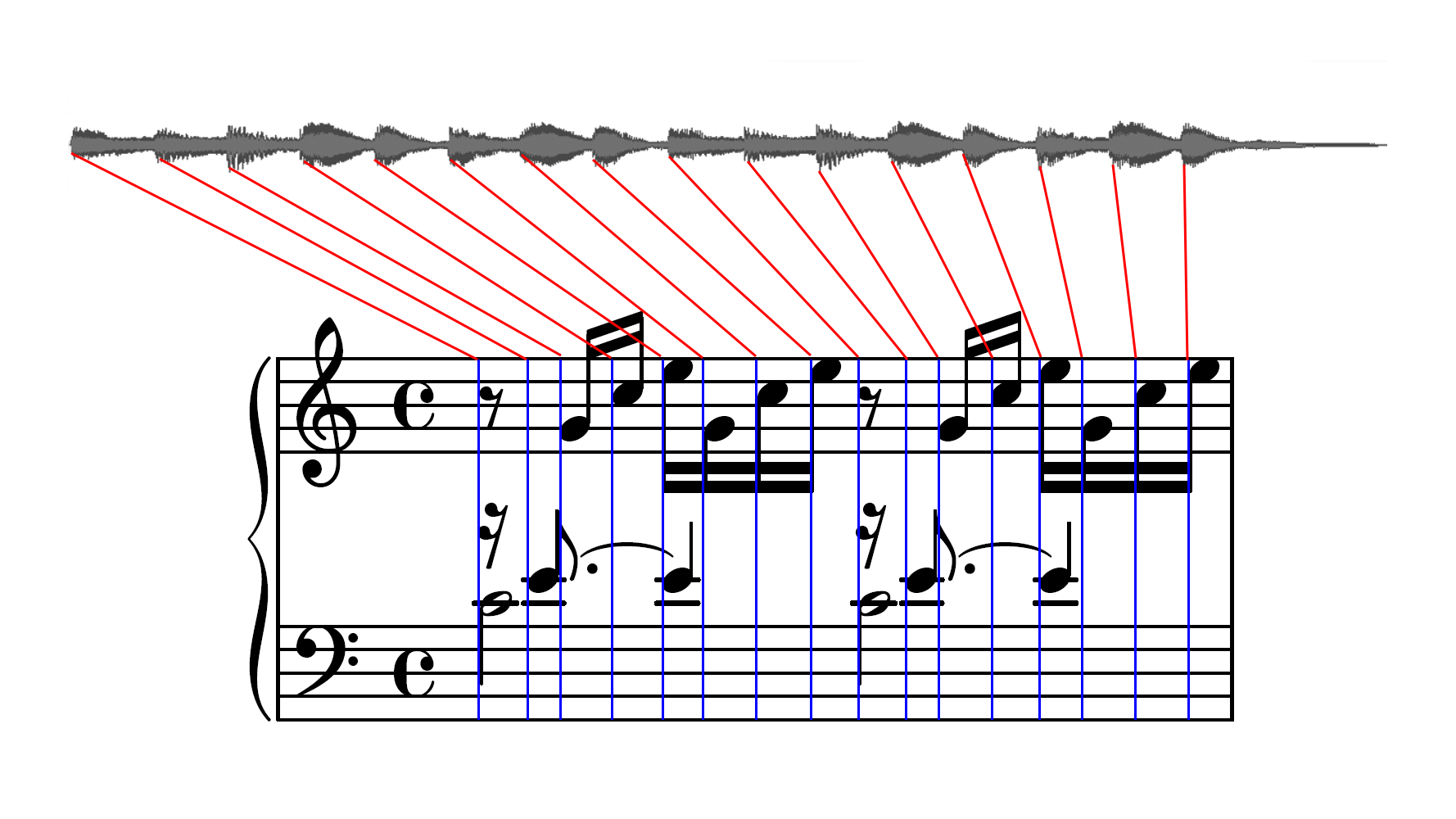}
        \caption{Illustration of audio-to-symbolic matching between an audio waveform and the sheet music of bar 1 of the \textit{Prelude} from Bach's \textit{Prelude and Fugue in C major, BWV 846}. Direct score following from audio to sheet music images is a currently active research topic--see \autoref{scofo_sheetmusicimages}.}
        \label{fig:audio-to-symbolic}
    \end{figure}
\end{definition}

\begin{definition}[Audio-to-audio score followers]
\textbf{Audio-to-audio} (A-A) systems matches a position in the audio of a music performance to the corresponding position in a separate audio recording. \autoref{fig:audio-to-audio} illustrates this method.
\begin{figure}[h]
    \centering
    \includegraphics[width=\columnwidth]{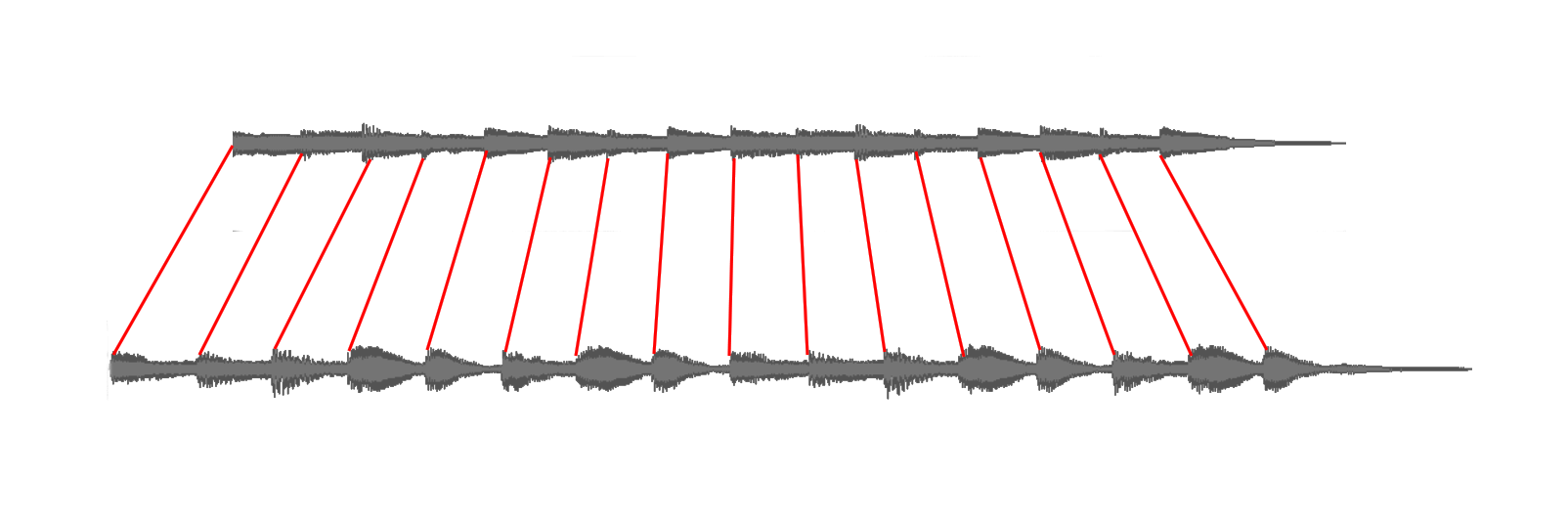}
    \caption{Illustration of audio-to-audio matching between the audio waveforms of two different recordings of bar 1 of the \textit{Prelude} from Bach's \textit{Prelude and Fugue in C major, BWV 846}. This method is common in DTW-based systems, covered in \autoref{scofo_dtw}. This figure also illustrates a problem for score followers: dealing with different performance \textit{tempo}.}
    \label{fig:audio-to-audio}
\end{figure}
\end{definition}

\begin{definition}[Symbolic-to-symbolic score followers]
\textbf{Symbolic-to-symbolic} (S-S) systems work by calculating the correlation between the musical structures of a performed musical note and its corresponding note in the score. \autoref{fig:symbolic-to-symbolic} illustrates this method.
\begin{figure}[h]
    \centering
    \includegraphics[width=\columnwidth]{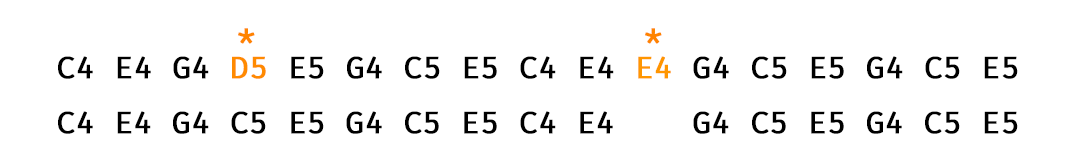}
    \caption{Illustration of symbolic-to-symbolic matching (pitch-based string matching) between two strings representing a sequence of notes (each note is a concatenation of the pitch and octave number) of bar 1 of the \textit{Prelude} from Bach's \textit{Prelude and Fugue in C major, BWV 846}. This method is common in early systems (\autoref{scofo_earlywork}). This figure also features two mistakes a robust score follower should be able to deal with: a wrong note (first asterisk) and a repeated note (second asterisk).}
    \label{fig:symbolic-to-symbolic}
\end{figure}
\end{definition}

It is worth noting that there is overlap for some systems. For example, Orio \etal{} \cite{orio01} showed an approach that synthesises the score's MIDI into an audio file before performing alignment with performance audio data; such a system would be an audio-to-symbolic system comprising an audio-to-audio system. Alonso \etal{} \cite{alonso17} also proposed solutions using a hybrid approach where both symbolic MIDI data and audio of the performance can be used to improve score following.

In this report, systems exhibiting compositional behaviour would be primarily referred by their \emph{inner} system to further emphasise their actual score following alignment method--Orio \etal{}'s system would be classified as an audio-to-audio system. Systems that employ a hybrid approach would be classified using their primary mode of alignment.

\section{Musical Data Representation}
\label{scofo_mdr}

As score followers are software based, computer-based music representations are required for both the performance and the score.

In this section, preliminaries of the musical data representations score followers commonly work with are provided. A focus is placed on open formats popular with score followers, but other lesser-used or proprietary formats are mentioned as an aside.

\begin{figure}[h]
    \centering
    \includegraphics[width=\columnwidth]{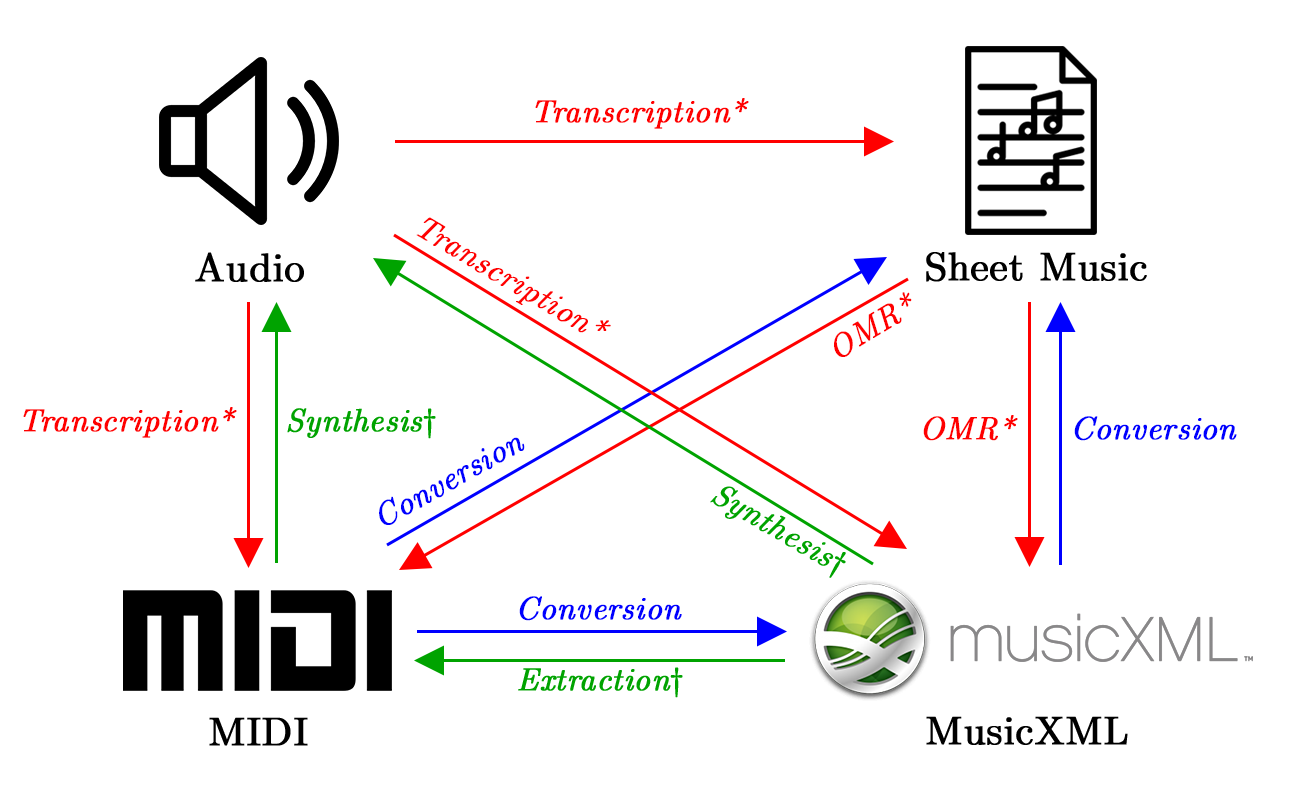}
    \caption{Illustration of common conversion methods between audio, sheet music, MIDI and MusicXML. Red (*-marked) conversions denote lossy and difficult conversions, green ($\dagger$-marked) conversions denote lossless and easy conversions and blue (unmarked) conversions denote lossy and easy conversions.}
    \label{fig:mdr-conv}
\end{figure}

It is worth noting that lossless conversions between some formats listed is impossible. However, conversions between some formats here can be done easier than others. \autoref{fig:mdr-conv} shows the common conversion relationships among four common formats: audio (\autoref{scofo_mdr_audio}), sheet music (\autoref{scofo_mdr_sheetmusic}), MIDI (\autoref{scofo_mdr_midi}) and MusicXML (\autoref{scofo_mdr_musicxml}).

\subsection{Audio}
\label{scofo_mdr_audio}

An obvious medium for score followers to work with is the audio of the performance. Score followers may also convert score data in other forms into its audio form via synthesisers, common for DTW-based score followers covered in \autoref{scofo_dtw}.

In a musical context, training musicians are taught the three attributes of a note: loudness (\emph{piano} vs. \emph{forte}), length (\emph{semiquaver} vs. \emph{semibreve}) and pitch (\textit{A4} vs. \textit{A6}). Beyond that, notes coming together can form chords and, further, polyphonic music. Harmonics may form over notes, and each instrument has a unique \emph{timbre}, giving rise to unique sounds as different instruments' sounds blend together.

It is clear that audio is very complex to work with. Simple notions relating to the music note can be brought over into audio processing rather easily to deal with temporal and pitch features, but more advanced signal processing techniques such as \emph{spectral analysis} are required to process more complex, polyphonic music. \Cref{scofo_speechrecog} contains information on how some score followers preprocess audio.

Another problem of using audio is that microphones may be imperfect, and problems such as reverberation \cite{naylor10} and noise \cite{ali15} increase the difficulty of score following.

Nevertheless, many score followers operate on the \emph{audio-to-audio} \cite{arzt08,Dixon2005MATCHAM} or \emph{audio-to-symbolic}  \cite{henkel19unet,henkel20} paradigms; usually, authors would need to develop more robust audio models or techniques to deal with the imperfections in audio capture and the complexities of multi-instrument polyphonic music. The study of audio processing is of huge interest in the Musical Information Retrieval (MIR) community to deal with other interesting problems besides score following, such as Query by Humming (QbH) \cite{ghias95}.

Audio can be \emph{transcribed} into MusicXML and MIDI via Automatic Music Transcription (AMT) \cite{benetos19} or by human transcription. On the one hand, AMT is a difficult problem, and--so far--no techniques can match the performance of human experts \cite{gowrishankar16}. Human transcription, on the other hand, is tedious.

\subsection{Sheet Music}
\label{scofo_mdr_sheetmusic}

Sheet music may come in analogue (paper) or digital (images, PDF etc.) forms. Generally, analogue sheet music can be easily converted into its digital counterpart easily by scanning the sheet music.

\begin{figure}[h]
    \centering
    \includegraphics[width=\columnwidth]{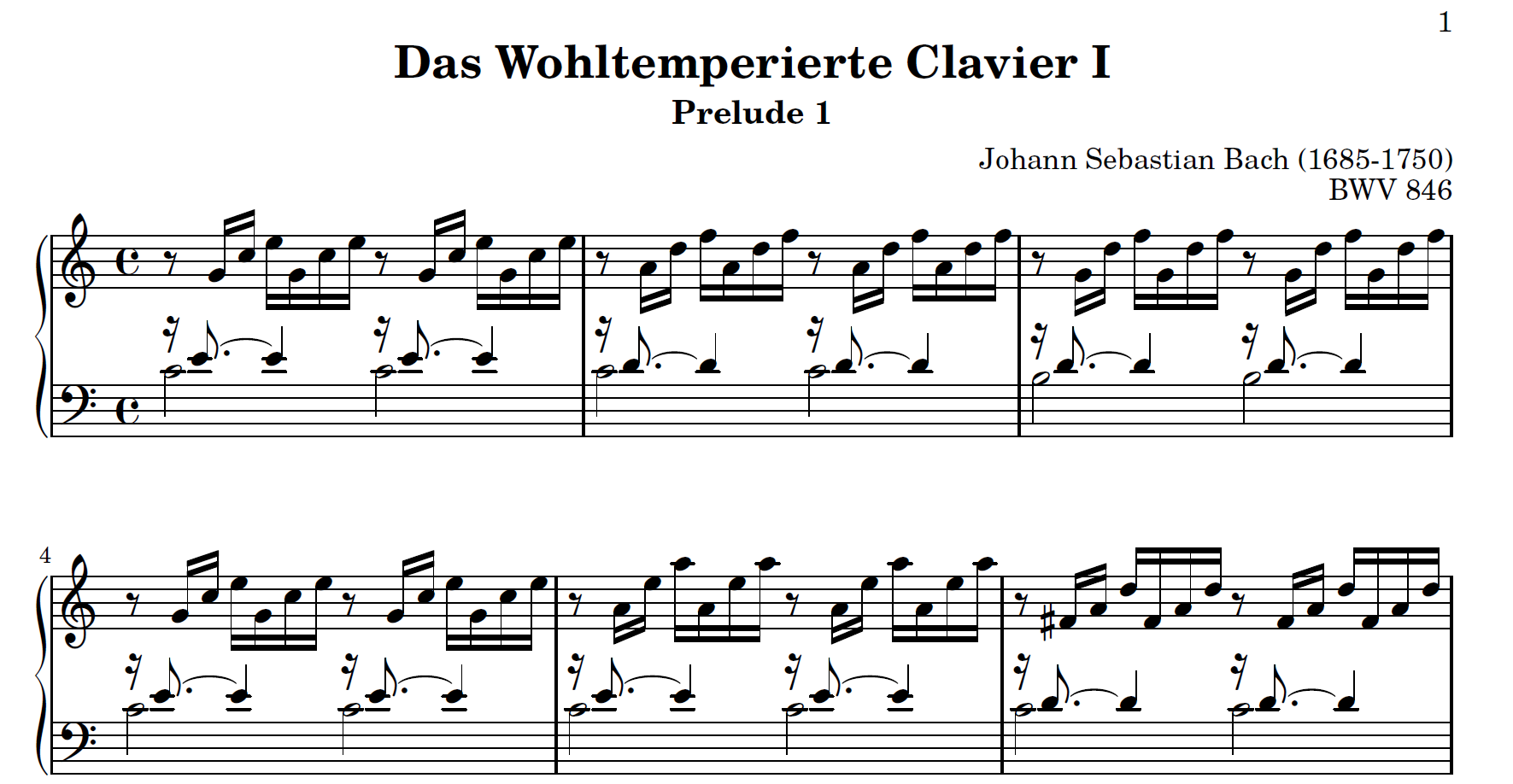}
    \caption{The first six bars of the \textit{Prelude} from Bach's \textit{Prelude and Fugue in C major, BWV 846} in modern staff notation, typeset using \textit{LilyPond}.}
    \label{fig:modern-staff-notation}
\end{figure}

There are many forms of musical notation used in sheet music; however, modern staff notation (also known as common Western music notation) is most commonly used by musicians of numerous genres throughout the world. \autoref{fig:modern-staff-notation} shows an example. Throughout this report, modern staff notation is used extensively. There exists variations on staff notation, e.g. percussion notation, figure bass notation, lead sheets and chord charts, but score following is mostly researched and developed for modern staff notation (rare exceptions include Pardo and Birmingham's 2001 work \cite{pardo01} on following lead sheets and Macrae and Dixon's 2010 \cite{macrae10} paper on following guitar tablature).

Most score followers do not work directly with sheet music images; in fact, it is only in recent years that the challenge of directly working with sheet music images was tackled \cite{dorfer2016}. This is due to the difficulty of processing music notation--score following systems working with sheet music images require the incorporation of computer vision and image processing techniques, which are usually too slow and/or too complex for score following purposes.

Score followers commonly work with symbolic music representations, such as MIDI and MusicXML. Most music creators notate their music in programs capable of outputting MIDI and MusicXML; however, in situations where this is not the case, human transcription from sheet music to symbolic music is possible, but tedious and difficult.  Fortunately, there exists automatic methods to convert sheet music into MIDI, MusicXML and other forms of symbolic music representations. These methods fall under another active sibling field of score following in MIR: Optical Music Recognition (OMR) \cite{zaragoza20}, a superset of its closely named counterpart: Optical Character Recognition (OCR) \cite{awel19}. Nevertheless, even with recent developments incorporating state-of-the-art (and some data-driven) approaches, OMR is still not reliable enough for score following \cite{henkel19}.

\subsection{MIDI}
\label{scofo_mdr_midi}

MIDI (Musical Instrument Digital Interface) is a technical standard, first introduced in 1981 and standardised in 1983, that describes standards for communications, interfaces and connectors that connect electronic instruments, computers and other audio devices \cite{midihistory}.
MIDI is more commonly known as a popular computer music format that can represent both performance and score data; however, it has notable limitations.

The main feature that MIDI brings to MIR is its file format, SMF (Standard MIDI File). SMF data records quantifiable musical data, such as note onsets, offsets, pitches and velocities. However, note that the MIDI protocol does not provide exact representations of musical performances--aspects such as timbre and spectral content are ignored. Further, certain performance directions, notes and auxiliary information (e.g. stem direction, beams, repeats, slurs etc.) usually available in modern sheet music cannot be contained within SMF.

Nevertheless, for MIR research focusing on timing, tempo and articulation, MIDI does convey sufficient information and remains far easier to process than audio recordings \cite{gingras11}. Because a MIDI file contains so little information compared to audio, it also has tiny file sizes. MIDI is, therefore, a sufficiently good format for score followers. In fact, many early score followers (\autoref{scofo_earlywork}) perform string-matching based on score data extracted from MIDI representations of the score and on performance data extracted from MIDI-capable musical instruments. Some DTW-based score followers (\autoref{scofo_dtw}) also work with MIDI score data by first synthesising it into audio.

Conversion of MIDI to audio is possible and common via synthesiser software; in fact, the instrument that synthesises the music can be easily changed, which adds to the flexibility and thus popularity of the format among music creators. However, do note that synthesis of audio from MIDI falls under computer-generated music, and thus will not be able to capture the ``expressiveness'' in performance of a human player. Moreover, music notation software are usually capable of outputting sheet music from MIDI files. Conversion from MIDI to MusicXML, which can be seen as a superset of MIDI (see \autoref{scofo_mdr_musicxml}), is straightforward.

\subsection{MusicXML}
\label{scofo_mdr_musicxml}
As mentioned in \autoref{scofo_mdr_sheetmusic}, MusicXML is a format commonly exported by music notation software. MusicXML is an open digital sheet music interchange format that aims to supersede MIDI as a universal format for modern staff notation. In addition to being a superset of MIDI (i.e. containing data representing how a piece of music should sound), MusicXML also contains data representing how a piece of music should look like, hence capturing all the musical data available on modern sheet music \cite{good03}.

Being a superset of MIDI, MusicXML contains all of MIDI's benefits in terms of representing musical data, and more. Performance direction information in MusicXML files can, in more advanced music synthesis software (such as \textit{Finale}\footnote{\url{https://www.finalemusic.com/}} and \textit{Sibelius}\footnote{\url{https://www.avid.com/sibelius}}), produce music that slightly more resembles what could be produced by a human performer. The fact that MusicXML also contains sheet music information also means it can be used as an alternative to sheet music--notably, MusicXML is the preferred format for music sheets in many applications displaying digital sheet music \cite{tido,phonicscore,musicnotes}. In fact, \textit{PhonicScore}\footnote{\url{https://www.phonicscore.com/}} and \textit{Tido}\footnote{\url{https://www.tido-music.com/}} are two APT (\autoref{scofo_apps_apt}) applications that work by showing a marker on MusicXML-generated sheet music.

MusicXML can be easily synthesised to audio, extracted to MIDI and converted into sheet music. The main problem with MusicXML is that while it is widely adopted, many sheet music publishers do not publish music in that format, instead preferring paper or electronic (such as PDF) sheet music. This means that OMR methods, mentioned in \autoref{scofo_mdr_sheetmusic}, or manual ones, must be used to convert sheet music into MusicXML.

\subsection{Other formats}

Here, other formats that are used by score followers are listed. These formats are not as commonly used as the ones mentioned above. The lower popularity is due to multiple factors--the most prominent ones being the format's proprietary state, limitations in representing musical data and low adoption by the community. Proprietary formats are marked with a $\dagger$.

\begin{itemize}
    \item \textbf{Graphic score editor formats}: \textit{Finale$\dagger$}, \textit{Sibelius$\dagger$}, \textit{NIFF}, \textit{Guido}, \textit{MuseScore}, \textit{Overture$\dagger$}, \textit{Cubase$\dagger$}

    \item \textbf{Typesetting/Mark-up language formats}:
          \textit{LilyPond\footnote{LilyPond \cite{lilypond} is used for typesetting most music in this report (e.g. \autoref{fig:modern-staff-notation}). It is also extensively used to generate sheet music in datasets \cite{dorfer18, seils,tuggener18}.}}, \textit{MusiXTeX}

    \item \textbf{Other machine-readable formats}:
          \textit{Music Encoding Initiative (MEI), IEEE1599, Humdrum}
\end{itemize}

%% file: 02_parts/01_scofo/chapters/03_challenges.tex
\chapter{Challenges}

\label{scofo_challenges}
\epigraph{\itshape Where words fail, music speaks.}{Hans Christian Andersen}

The wide and practical applications of score following detailed in \autoref{scofo_apps} and the ubiquity of powerful portable devices beg the question: 

\begin{displayquote}
    Why are score followers still not common? 
\end{displayquote}

The apparent unpopularity is testament to the difficulty of building a robust score follower. This problem is well known in the research community \cite{sturm13,mullerdtw,dannerberg03,chen19,heijink00}.

The epigraph by Hans Christian Andersen in the header of this chapter suggests that music can be seen as an extension of human expression beyond words. A more scientific insight can be derived: first, note that, on the one hand, music performance can be modelled as speech, and on the other hand, musical scores can be modelled as speech transcripts. This implies that not only there can exist cross-pollination of research and development in score following systems and their speech counterparts (which will be explored in \autoref{scofo_speechrecog}), score following's implementation challenges can be seen as a superset of those in speech recognition and tracking. It is no doubt that research in speech recognition and tracking is highly active \cite{rabiner93}. 
As the difficulties in speech-related research are generally easier to understand, relevant challenges in speech-related research are given to strengthen understanding of some challenges detailed below. 

Further, examples from two musical pieces--one for violin and the other for piano--will be drawn to explain the challenges. 

\begin{figure}[h]
    \centering
    \includegraphics[width=\columnwidth]{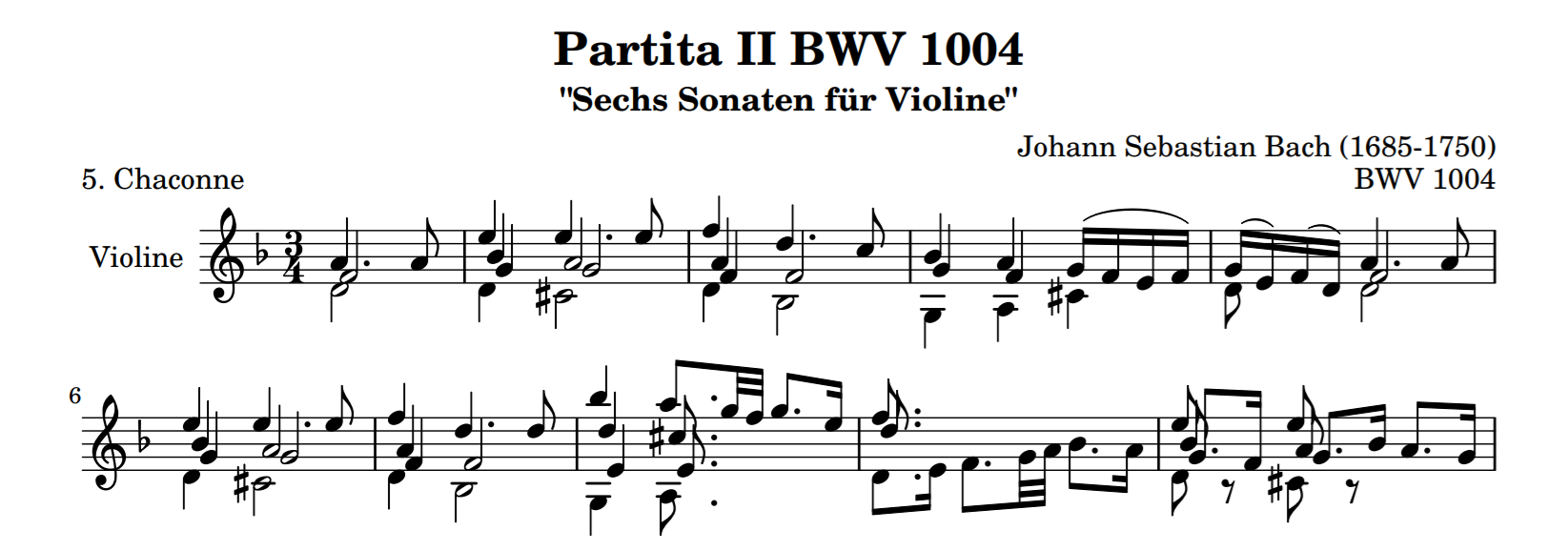}
    \caption{The first 10 bars of Bach's \textit{Chaconne from Partita II, BWV1004}.}
    \label{fig:chaconnefirst10}
\end{figure}

Contrary to popular belief shared by some authors in the field \cite{orio03,nakamuramerged14}, unaccompanied violin pieces \emph{can} be highly polyphonic. Musically informed readers might point to the highly technical works of Niccolo Paganini, but polyphonic violin music existed since Johann Sebastian Bach's time--the baroque period. Examples from Bach's \textit{Chaconne from Partita II, BWV1004} will be drawn--\autoref{fig:chaconnefirst10} shows that this piece is not an ordinary violin piece; it contains chords, rendering it highly polyphonic. Modern players usually play a chord in a technique called ``double-stopping'', i.e. playing two notes at a time with the bow, while baroque-era bows are differently shaped hence allowing the execution of such chords \cite{santos04}.

\begin{figure}[h]
    \centering
    \includegraphics[width=0.5\columnwidth]{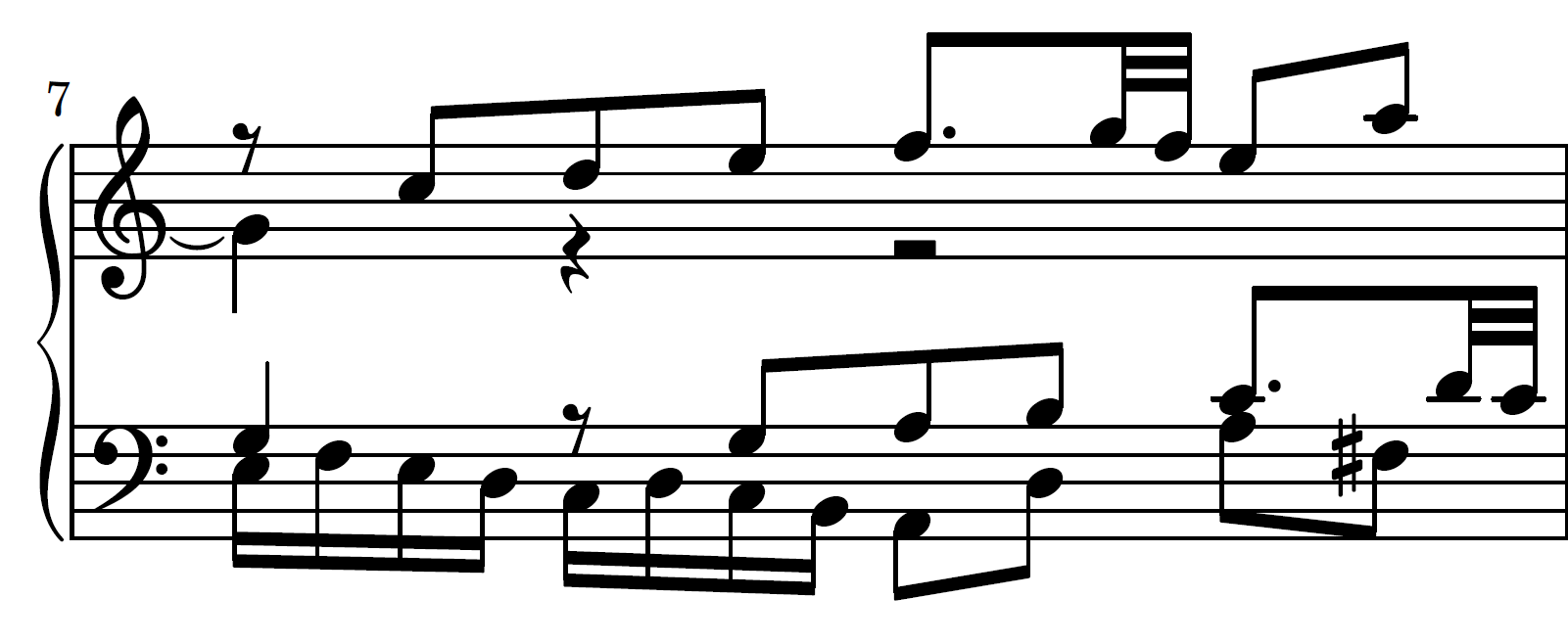}
    \caption{Bar 7 of the \textit{Fugue} from Bach's \textit{Prelude and Fugue in C major, BWV 846}.}
    \label{fig:fuguebar7}
\end{figure}

The accompanying \textit{Fugue} to the previously presented \textit{Prelude} from Bach's \textit{Prelude and Fugue in C major, BWV 846} is also used. While the \textit{Prelude} is low in polyphony (it is a sequence of broken \textit{arpeggios}), the \textit{Fugue} is highly polyphonic--it is written for four voices running parallel to each other. Bar 7 of that piece is shown in \autoref{fig:fuguebar7}, which shows the four voices running in parallel.

\section{Limitations in Feature Extraction}
\label{scofo_challenge_limitfe}

Feature extraction is a core first step of the generic framework for score followers as covered in \autoref{scofo_generic}. Features can be extracted from the audio of the performance or its MIDI representation if the instrument is MIDI capable.

It is clear that if the hardware or software responsible for extracting the feature from the performance data is limited, the subsequent procedures in the score follower would need to work with less complete data and hence come up with a match that is less ideal. Audio as mentioned in \autoref{scofo_mdr_audio} is a challenge to work with, especially if the system needs to be robust to corruption in the audio signal be it by noise, reverberation or hardware limitations. MIDI devices are more robust in this sense, but lower quality instruments producing subpar MIDI data exist.

Relating to speech, speech recognition is also more difficult if the source audio is corrupted.

\section{Performance Deviations}
\label{scofo_challenges_perfdev}

Mismatches between human performances and the corresponding score is possible and sometimes common. Music performance is, after all, a performance art subject to the infinite complexity of human expression and infallibility. Some of these deviations are unintentional, but some are not.

\subsection{Unintentional Mistakes}

Mistakes in performances, even at professional level, exist. This kind of deviation forms the basis of many improvements in score following (see \autoref{scofo_relatedwork}) to make systems more robust to human errors, such as unintentional note repeats, segment skips and pitch mistakes. 

\subsection{Intentional and Artistic Deviations}
\label{scofo_challenges_perfdev_intentional}

\begin{figure}[h]
    \centering
    \includegraphics[width=0.25\columnwidth]{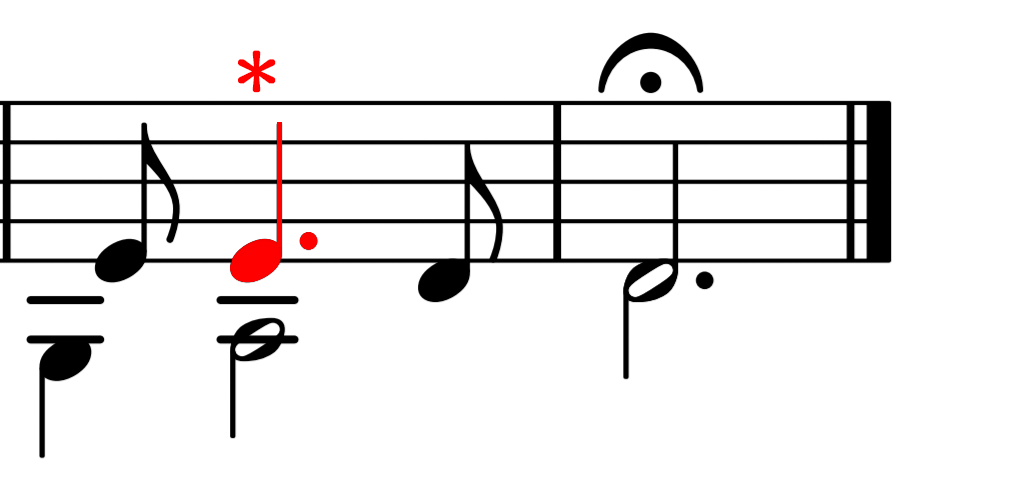}
    \caption{The last two bars of Bach's \textit{Chaconne from Partita II, BWV1004}. The note marked in red (and also an asterisk) is a note where performers frequently add a \textit{trill} to. Also notice the \textit{fermata} symbol on the last note--this denotes ``hold''--violinists can choose however long they wish to hold that note.}
    \label{fig:chaconnelast2}
\end{figure}

Artistic deviations stemming from musicians' expressing themselves exist--musicians could introduce temporal deviations such as variations in note onset, note duration or \textit{tempo}. To enrich the performance, musicians may also add ornaments or variations (examples include a \textit{trill}, \textit{glissando} and \textit{fioritura}) unspecified in the score. An example place where a performer can add an artistic \textit{trill} exists in the second last bar of the \textit{Chaconne}, shown in \autoref{fig:chaconnelast2}. Violinists can hold the second last note (\textit{E}) in the second last bar (marked in red and asterisk) longer and add a \textit{trill} on it to lead in to the final notes.

There are also issues with some pieces that \emph{require} intended artistic or performance-driven deviations--the \textit{Chaconne} in \autoref{fig:chaconnefirst10} serves as a great example. In bar 1, the obvious impracticality of playing three notes is already covered (by ``double-stopping''), but another problem is that the bottom two notes (\textit{D} and \textit{E}) span the length of the two \textit{A} notes. For a bowed instrument like the violin, this is impossible to execute. Some violinists defer to shortening the first chord to the length of the first \textit{A} note, and some violinists do something called a ``rebound''--they replay the lower notes on the onset of the second \textit{A} note, to let the bottom two notes resonate further \cite{santos04}. In fact, this problem recurs frequently in the first 10 bars of the \textit{Chaconne}.

In the context of speech, many professional actors also do not follow written scripts completely, and sometimes are even encouraged to improvise (which is another score following challenge worthy of separate mention in \autoref{scofo_challenge_music_improv}).

A good and robust score follower therefore needs to take these intentional and artistic deviations into account, and as seen in the example above, these deviations can be very hard to track even for an experienced human expert. 

\section{Underspecified Musical Scores}
\label{scofo_challenges_underspec}

\begin{figure}[h]
    \centering
    \includegraphics[width=\columnwidth]{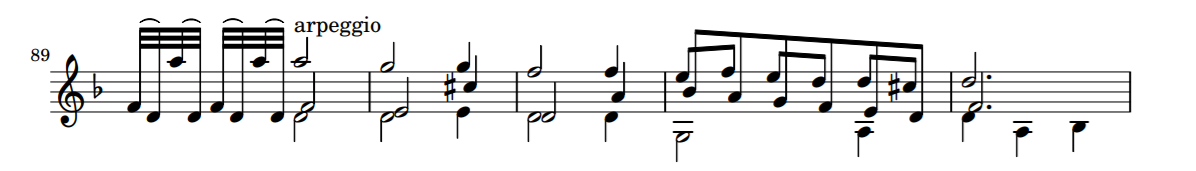}
    \caption{Bars 89-93 of Bach's \textit{Chaconne from Partita II, BWV1004}.}
    \label{fig:chaconnearp}
\end{figure}

Before beginning this section, the reader is referred to the epigraph by Hilary Hahn in \autoref{scofo_preliminaries}.

Musical scores can be seen as a loose set of performance directions to the musician, similar to how some speech scripts can be merely a list of bullet points to be conveyed. Often, musical performance directions are abbreviated, and different performers may execute these directions differently. For instance, there exists different ways to execute a \textit{trill}--the most obvious is its \emph{speed}. 

A more extreme example again can be drawn from the \textit{Chaconne} (in fact, the \textit{fermata} symbol in \autoref{fig:chaconnelast2} is also an example), but now instead in bar 89, see \autoref{fig:chaconnearp}. Notice Bach's direction in the third beat: \textit{arpeggio}. This goes on from bars 89 to 120 and also bars 201 to 208. Bach indicated how to start the first passage by writing out how it should be performed, but it is impossible to follow this pattern for the entire passage. Numerous solutions for these \textit{arpeggios} were given by violinists, and there rarely exists any pair of professional violinists sharing the same treatment to these \textit{arpeggios} \cite{santos04}. In fact, Geminiani in 1751 \cite{geminiani1751} gave 18 systematic ways to execute \textit{arpeggios}--there is no agreed right way to execute them in the context of the \textit{Chaconne}.

Thus, robust score followers must be able to track underspecified music scores and anticipate how an underspecified passage would be undertaken by a performer. There exists ``learning'' methods (see \autoref{scofo_relatedwork}) to get around this, but it is difficult for score followers to generalise to many types of underspecification.

\section{Polyphonic Music}
\label{scofo_challenges_polyphonic}

Polyphonic music can be seen in the speech paradigm as multiple people speaking at the same time. Clearly, this is a difficult problem: recent work is still dealing with speaker identification \cite{tran20}. While it can be argued that polyphonic music is easier to deal with as the parallel voices are usually still structured (lines of melodies or harmonies should still form a lyrical sound with parallel voices), the problem of polyphonic music still remains difficult. In fact, it took over 20 years after the first publishing of the pioneering score following papers before the problem is properly tackled \cite{Dixon2005MATCHAM}, and recent approaches still strive to find new ways to address the challenge \cite{arzt16}.
Polyphonic audio is difficult to deal with as multiple note events occurring at the same time leads to the notes' harmonic series to interfere. This increases the difficulty of identifying the similarity between an audio segment and its corresponding score \cite{chen19}. MIDI as a sequence of notes face a different challenge--notes come in one at a time, therefore making it difficult to determine which note of a set of parallel notes comes first based on the score.

\begin{figure}[h]
    \centering
    \includegraphics[width=\columnwidth]{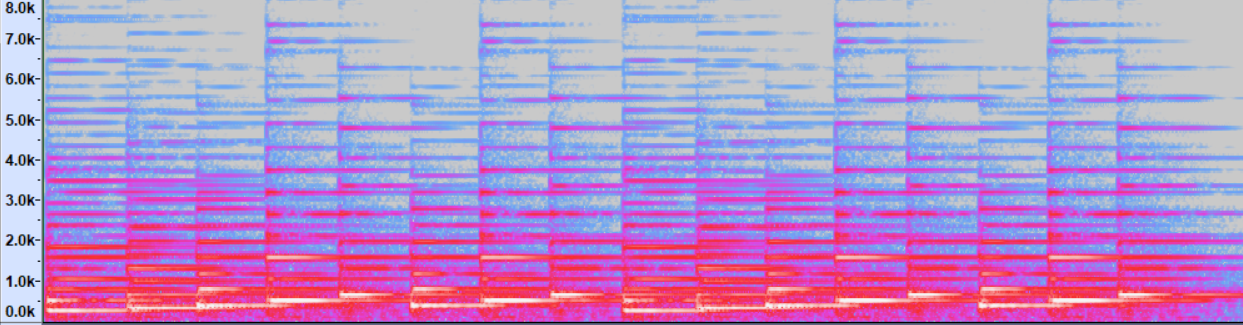}
    \caption{The spectrogram of the audio of Bar 1 of the \textit{Prelude} from Bach's \textit{Prelude and Fugue in C major, BWV 846}.}
    \label{fig:preludespecgram}
\end{figure}

\begin{figure}[h]
    \centering
    \includegraphics[width=\columnwidth]{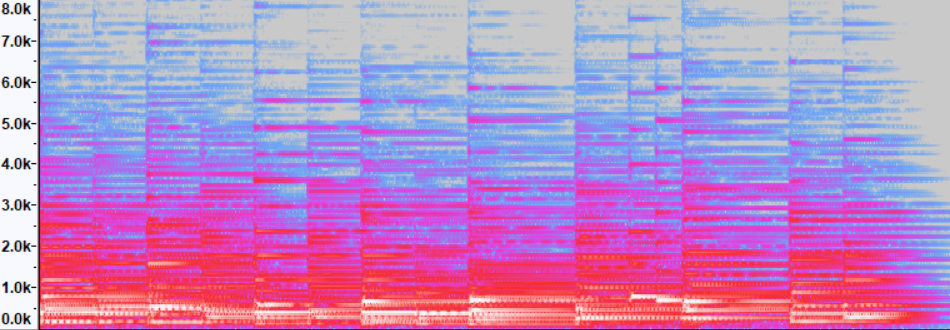}
    \caption{The spectrogram of the audio of Bar 7 of the \textit{Fugue} from Bach's \textit{Prelude and Fugue in C major, BWV 846}.}
    \label{fig:fuguespecgram}
\end{figure}

Examples can again be drawn from the \textit{Chaconne}, a heavily polyphonic piece for the violin whose \textit{timbre} relates to many spectral features via harmonics. However, the \textit{Fugue} is instead used as it is a polyphonic piece that has a monophonic sibling (the \textit{Prelude}). The \emph{spectral analysis} of a sound signal is mentioned in \autoref{scofo_mdr_audio}. This is convenient in this case. Figures~\ref{fig:preludespecgram} and \ref{fig:fuguespecgram} show the spectrograms of the audio in Bar 1 of the \textit{Prelude} and Bar 7 of the \textit{Fugue} respectively\footnotemark.

\footnotetext{These spectrograms are generated using \textit{Audacity}.}

For reference, readers may find Bar 1 of the \textit{Prelude} in \autoref{fig:modern-staff-notation} and Bar 7 of the \textit{Fugue} in \autoref{fig:fuguebar7}.

The spectrogram comprises two axes: the vertical axis represents frequency (in Hertz) and the horizontal axis represents time. The colour of the spectrogram contents is most intense (white) at areas in which the audio frequency at that time is the most intense. Readers familiar with the Fourier Transform could see that the spectrogram is a Short-time Fourier Transform (STFT) of the signal. \autoref{fig:modern-staff-notation} shows that even the monophonic \textit{Prelude} is already complex with many \textit{harmonic overtones} (represented by coloured lines) over the \textit{fundamental} pitches (represented by white lines). This is even more complex in the \textit{Fugue}, where numerous overtones stem from the many parallel fundamental pitches. The overtones interfere with each other, making it difficult to perform pitch-based score following. In fact, pitch detection is another field in MIR still under heavy research and development \cite{kim18}.

So far, only polyphonic music stemming from one instrument was discussed, but note that polyphony from ensemble music is common--this further complicates the problem. Hence, robust score followers should be capable of following polyphonic music--produced by one instrument or many.

\section{Music Improvisation}
\label{scofo_challenge_music_improv}

Improvisation relates to the direct creation or reinterpretation of music on the spot. Improvisation can be used as a way to respond to other people playing in a group (relevant to computer-aided accompaniment (\autoref{scofo_apps_caa})) and usually reflects the music style performed.

The myriad different ways in which the \textit{Chaconne} can be performed can be considered music improvisation--performers often also add extra flair to make each performance ``special''. Interested readers may refer to Santos' 2004 work \cite{santos04} on more issues of the \textit{Chaconne} in practice and performance. 

The fact that even baroque violin pieces offer so much depth and breadth in musical freedom of expression (which is contrary to popular belief) implies that music from other genres and eras are also free and open to improvisation. In fact, Wang's 2017 survey \cite{wang17} demonstrated how Chinese music improvisation varies even across Chinese cities.  

Music improvisation already poses a challenge to human trackers--they need to be trained in the particular domain and performance style. ``Teaching'' or programming this into a score follower is difficult; however, if this is possible, it would open up new doors in computer music. Improvisation in speech is also common (a speaker may take a detour to tell a joke), but this is not a problem of wide interest in the speech community.

%% file: 02_parts/01_scofo/chapters/04_relatedwork.tex
\chapter{Literature Review}
\label{scofo_litreview}

\label{scofo_relatedwork}
\epigraph{\itshape It is sobering to consider that when Mozart was my age he had already been dead for a year.}{Tom Lehrer}

In this chapter, a summary of score following systems proposed in different eras--from pioneering work done in 1984 to recent state-of-the-art systems--is provided.

In general, three eras of score following can be identified; as such, work in these three eras are segmented and presented. The first era (\autoref{scofo_earlywork}) mainly worked with pitch detection and string matching. Later, inspiration from speech recognition and advancements in statistical methods brought about approaches using Dynamic Time Warping (DTW) and Hidden Markov Models (HMMs) (\autoref{scofo_speechrecog}). Exploration in the second era went on for quite long, before researchers introduced recent methods that give new insights into the problem (\autoref{scofo_recent}). Finally, possible future directions of score following are proposed (\autoref{scofo_future}).

Comparisons between all these approaches are very difficult for multiple reasons, which in part motivates testbenches introduced in \autoref{testbench}. Hence, not much focus is placed on comparing systems in this section--more focus is given to new contributions brought by each work.

\textbf{A note on summary tables}: Summary tables covering notable works in the first two eras together with the works' attributes are provided. The attributes used are covered previously in chapters~\ref{scofo_apps} and \ref{scofo_preliminaries}. In these tables, the main author is named, along with the year in which the work is first published (and hence does not account for extensions). The type of each system is presented in initialised form (e.g. A-A for Audio-to-Audio). \textbf{PF} and \textbf{SF} denote the main performance feature and score feature extracted for alignment respectively (approaches clearly using a small part of a feature will not be listed with that feature). The primary alignment technique is provided, along with some notes, such as extensions. As most score followers are built for a particular purpose, its main application is also provided.

\section{Early Work}
\label{scofo_earlywork}
\autoref{table_scofo_earlywork} provides a summary of research in this era.

\afterpage{%
  \clearpage
  \thispagestyle{empty}
  \begin{landscape}

    \begin{table}
      \centering
      \caption{Summary of Early Score Followers (\autoref{scofo_earlywork})}
      \label{table_scofo_earlywork}
      \begin{tabularx}{\linewidth}{|l|l|l|L|L|L|L|L|l|}
        \hline
        \textbf{Author}                                                      &
        \textbf{Year}                                                        &
        \textbf{Type}                                                        &
        \textbf{PF}                                                          &
        \textbf{SF}                                                          &
        \textbf{Alignment}                                                   &
        \textbf{Notes}                                                       &
        \textbf{Application}                                                 &
        \textbf{Papers}
        \\
        \hline
        Dannenberg                                                           &
        1984                                                                 &
        S-S                                                                  &
        Pitch                                                                &
        Pitch                                                                &
        String-matching                                                      &
        First definition, heuristics, heavy extensions                       &
        Computer-aided accompaniment                                         &
        \cite{dannenberg84,bloch85,dannenberg88,allen90}
        \\
        \hline

        Vercoe                                                               &
        1984                                                                 &
        S-S                                                                  &
        Pitch                                                                &
        Pitch                                                                &
        String-matching                                                      &
        First definition, \emph{Synthetic Performer}, ``cost'', ``learning'' &
        Computer-aided accompaniment                                         &
        \cite{vercoe84, vercoe85}
        \\
        \hline

        Baird                                                                &
        1990                                                                 &
        S-S                                                                  &
        Pitch                                                                &
        Pitch                                                                &
        String- and Phrase-matching                                          &
        Segment definition, pre-performance heuristics                       &
        Computer-aided accompaniment                                         &
        \cite{baird90,baird93}
        \\
        \hline

        Puckette                                                             &
        1990                                                                 &
        S-S                                                                  &
        Pitch                                                                &
        Pitch                                                                &
        String-matching                                                      &
        \emph{EXPLODE}, jump heuristics                                      &
        Computer-aided accompaniment                                         &
        \cite{puckette90,puckette92,puckette95}
        \\
        \hline

        Vantomme                                                             &
        1995                                                                 &
        S-S                                                                  &
        Temporal                                                             &
        Temporal                                                             &
        String-matching                                                      &
        Temporal matching via note onset and tempo                           &
        Computer-aided accompaniment                                         &
        \cite{vantomme95}
        \\
        \hline

        Desain                                                               &
        1997                                                                 &
        S-S                                                                  &
        Pitch                                                                &
        Pitch                                                                &
        String-matching                                                      &
        Supported by temporal structures (parallel vs. sequential voices)    &
        -                                                                    &
        \cite{desain97, heijink00}
        \\
        \hline

        Toiviainen                                                           &
        1998                                                                 &
        S-S                                                                  &
        Temporal                                                             &
        Temporal                                                             &
        Beat tracking                                                        &
        Partial score following                                              &
        Computer-aided accompaniment                                         &
        \cite{toiviainen98}
        \\
        \hline

        Pardo                                                                &
        2001                                                                 &
        S-S                                                                  &
        Temporal                                                             &
        Temporal                                                             &
        Beat tracking                                                        &
        Partial score following, Lead sheet following                        &
        Computer-aided accompaniment                                         &
        \cite{pardo01}
        \\
        \hline
      \end{tabularx}
    \end{table}
  \end{landscape}
  \clearpage
}

\subsection{The Pioneers}

Score following was first presented at the 1984 International Computer Music Conference (ICMC) independently by Vercoe \cite{vercoe84} and Dannenberg \cite{dannenberg84}, initially geared towards computer-aided accompaniment.

Dannenberg's initial work \cite{dannenberg84} focused on developing an efficient dynamic programming algorithm for flexible pitch matching. First, the score and audio input (which are MIDI events) are converted to strings and the best match between these strings is computed. This results in a system unsuitable for polyphonic audio produced by, for instance, a piano performance. Later on, Dannenberg's group extended the system to make it capable of handling polyphonic music \cite{bloch85}, correcting extreme mismatching \cite{dannenberg88, allen90} and detecting musical ornaments such as \textit{trills} and \textit{glissandi} \cite{dannenberg88}. Dannenberg also pioneered the idea of multiple matchers running at different locations and picking the matcher with the least error \cite{dannenberg88}.

Vercoe's work in 1984 \cite{vercoe84}, the \emph{Synthetic Performer}, used pitch detection to obtain pitches from flute performances. However, pitch detection was not fast enough; hence, optical sensors are used to obtain fingering information. Vercoe used a matching method based on a theory of ``least cost'' \cite{vercoe85}. The main idea is that a performance deviation adds to the aforementioned cost--the greater the performer's deviation, the higher the cost. For each musical event received, four theories of a predicted position are computed using variations of the cost calculated (e.g. one with a higher weighting on pitch correctness). The theory with the least cost is considered the ``best fit'', i.e. the most likely score position. A notable achievement in Vercoe's system is that it incorporates offline capabilities--Vercoe \etal{} introduced an improvement in 1985 \cite{vercoe85}: a non-real-time statistical learning algorithm that can ``rehearse'' and ``learn'' a performance to incorporate improvements in the next performance--this notion is later more commonly known as ``training''.

\subsection{Continued Work on Pitch Detection and String Matching}
\label{scofo_relatedwork_early_workcontinued}

Early research in score following took heavy inspiration from the pioneers and continued work on extracting pitch features and performing similarity calculations via string matching.

Basing their matching methods on Vercoe's \cite{vercoe84} and Dannenberg's \cite{dannenberg84}, Baird \etal{} \cite{baird90,baird93} devised a new matching algorithm; instead of matching on single events (i.e. individual notes), the matching is performed on segments of a predefined length. The breakthrough for this system is that these segments, as well as a set of heuristics comprising meters, key signatures and a possible tonality, are calculated and analysed pre-performance, meaning that there is an increase of the system's \textit{a priori} knowledge. The use of these data makes it possible to produce cadence points--the system's knowledge of cadential motion means it knows where phrases begin, the system can hence skip to arbitrary phrases in the event of excessive performer error.

The 1990 IRCAM score following system \cite{puckette90, puckette92}, \emph{EXPLODE}, is similar to many others in that time period except for the addition of a backward-skip feature. When a live note is played, a pitched-based system matches it with a note via comparing the pitch of the live note with pitches of skipped notes. Stored backward- and forward-skip times form the boundaries within which skipped notes can be referred to. Notably, compositions with \emph{EXPLODE} in mind were created; for instance, Philippe Manoury composed \emph{En Echo} for soprano and computer. It showed the limits of \emph{EXPLODE}--Puckette \etal{} admitted that often compromises in the composition were forced to be made to ensure the score follower could follow the performance. In 1995, extensions to the system that address some of these problems were introduced \cite{puckette95}.

Though not strictly a score follower but an aligner instead, an Approximate String Matching (ASM) aligner is introduced in \autoref{impl_asm}. Readers interested in the technical background of the problem are highly encouraged to refer to that chapter. The proposed approach is intended to produce high-quality reference data for the use of the quantitative testbench, detailed in \autoref{tb_quant}.

\subsection{Breaking the Mould: Using temporal features}
\label{scofo_relwork_temporal}

After extensive research on pitch methods and realising their limitations, researchers eventually turned to rely more on temporal features.

Vantomme developed in 1995 \cite{vantomme95} a system comprising a main and a backup matching phase--the former is based on note onset and tempo (therefore making it robust when presented with incorrect pitches that were still at their expected onset), and the latter is based on pitch (similar to Dannenberg's 1984 system \cite{dannenberg84}). This two-phased system, in principle, is robust to incorrect pitches with the right rhythm; however, the system fails when many incorrect pitches are received and a call to the backup pitch-based system is required.

While still mainly using pitch information, Desain and Heijink \cite{desain97,heijink00} exploited the temporal structure annotated in the score, such as voices and chords, to predict note orders in the performance. The main idea is that notes in a melodic line are unlikely to be played in a different order, but parallel voices (such as chords) can be played independently of each other. Multiple alternative matches are performed and dynamic programming is used to select the best path and hence match.

An approach based entirely on rhythm was introduced by Toiviainen in 1998 \cite{toiviainen98}. An adaptive oscillator is used to track the beat of a MIDI input. This implies that a specified score can be omitted, meaning that even improvisations can be tracked as well. While a beat tracker can be used as a score follower, it is suboptimal for non-improvisation pieces as scores can be used as a reliable additional source of knowledge.

\subsection{Notable mention: Partial Score Following}
\label{scofo_relwork_partial}

Toiviainen's 1998 approach \cite{toiviainen98} showed that score followers can even be applied on incomplete scores.

Pardo and Birmingham in 2001 \cite{pardo01} took that idea further and showed a system that can follow a lead sheet, common in folk music. Their approach is inspired by gene-sequence analysis, in which gaps between alignments of chords extracted from the MIDI performance and the lead sheet are allowed. Due to limited applications in partial score following, research, while present \cite{stark12}, is less active than that in normal score following.

\section{Speech Recognition's Inspirations}
\label{scofo_speechrecog}

The field of speech recognition is closely related to score following's as mentioned in \autoref{scofo_challenges}. Instead of working with music performance audio and musical scores, speech recognition works with speech audio and speech text. As such, many techniques between these two fields are interchangeable. As speech recognition research--heavily driven by advancements in statistical methods--grew greatly in the 1980s and 1990s, score following researchers borrowed many ideas--the two most successful being Hidden Markov Models (HMMs) and Dynamic Time Warping (DTW).

First, statistical methods which culminated into the use of HMMs (\autoref{scofo_stats}) are explored before diving into DTW methods (\autoref{scofo_dtw}).

See \autoref{table_scofo_speechrecog} for a summary.

\afterpage{%
  \clearpage
  \thispagestyle{empty}
  \newgeometry{top=1.5cm,bottom=1.5cm,left=1.5cm,right=1.5cm}
  \begin{landscape}

    \begin{table}
      \centering
      \caption{Summary of Speech Recognition-Inspired Score Followers (\autoref{scofo_speechrecog})}
      \label{table_scofo_speechrecog}
      \begin{tabularx}{\linewidth}{|L|l|l|L|L|L|L|L|l|}
        \hline
        \textbf{Author}                                                                       &
        \textbf{Year}                                                                         &
        \textbf{Type}                                                                         &
        \textbf{PF}                                                                           &
        \textbf{SF}                                                                           &
        \textbf{Alignment}                                                                    &
        \textbf{Notes}                                                                        &
        \textbf{Application}                                                                  &
        \textbf{Papers}
        \\
        \hline
        Grubb                                                                                 &
        1997                                                                                  &
        S-S                                                                                   &
        Pitch \& Temporal                                                                     &
        Pitch \& Temporal                                                                     &
        Statistical model                                                                     &
        Pioneer in statistical methods                                                        &
        Computer-aided accompaniment                                                          &
        \cite{grubb97}
        \\
        \hline

        Pardo                                                                                 &
        2002                                                                                  &
        S-S                                                                                   &
        Pitch                                                                                 &
        Pitch                                                                                 &
        String-matching                                                                       &
        String-matching extended with statistical methods                                     &
        Computer-aided accompaniment                                                          &
        \cite{grubb97}
        \\
        \hline

        Cano                                                                                  &
        1999                                                                                  &
        A-S                                                                                   &
        Audio (Mostly pitch)                                                                  &
        Note model                                                                            &
        HMM                                                                                   &
        Pioneer in HMM                                                                        &
        -                                                                                     &
        \cite{cano99}
        \\
        \hline

        Raphael                                                                               &
        1999                                                                                  &
        A-S                                                                                   &
        Audio (Spectral)                                                                      &
        Note model                                                                            &
        HMM                                                                                   &
        \emph{Music Plus One}, Spectral features, Viterbi algorithm alternative               &
        Computer-aided accompaniment                                                          &
        \cite{raphael99,musicplusone,raphael01,raphael03}
        \\
        \hline

        IRCAM (Orio, Schwarz, Cont \etal{})                                                   &
        2001                                                                                  &
        A-S \& S-S                                                                            &
        Audio                                                                                 &
        Note model                                                                            &
        HMM                                                                                   &
        \emph{Antescofo}, \emph{Metronaut}
        Extensions for errors, Viterbi algorithm alternative, training, Polyphonic MIDI input &
        Computer-aided accompaniment                                                          &
        \cite{cont08, oriohmm1,schwarz04, cont04, cont06,cuvillier14,donat16}
        \\
        \hline

        Pardo                                                                                 &
        2005                                                                                  &
        S-S                                                                                   &
        Audio                                                                                 &
        Note model                                                                            &
        HMM                                                                                   &
        Model spontaneous changes in performance                                              &
        -                                                                                     &
        \cite{pardo05}
        \\
        \hline

        Dixon                                                                                 &
        2005                                                                                  &
        A-A                                                                                   &
        Audio (Spectral)                                                                      &
        Audio (Spectral, Synthesised)                                                         &
        DTW                                                                                   &
        \emph{MATCH}, Pioneer in DTW score following, improvements in DTW computation         &
        Performance Analysis                                                                  &
        \cite{Dixon2005MATCHAM}
        \\
        \hline

        Arzt                                                                                  &
        2008                                                                                  &
        A-A                                                                                   &
        Audio (Spectral)                                                                      &
        Audio (Spectral, Synthesised)                                                         &
        DTW                                                                                   &
        Many new ideas and improvements (polyphony, DTW computation, orchestra following)     &
        Performance Analysis, APT                                                             &
        \cite{arzt08,arzt10,arzt12,arzt15,arzt16}
        \\

        \hline
      \end{tabularx}
    \end{table}
  \end{landscape}
  \clearpage
}
\restoregeometry

\subsection{Statistical Approaches (including HMMs)}
\label{scofo_stats}

Even with perfect observations in the live performance, performer errors introduce a lot of uncertainty. Therefore, it is natural to consider probabilistic methods for score following. Many early approaches (\autoref{scofo_stats_early}) explored many possibilities brought by advancements in statistics, but HMMs (\autoref{scofo_stats_hmm}) proved seminal. Statistical methods researched did not stop at HMMs--work was done on other statistical models in recent years, which will instead be treated as recent work and covered in \autoref{scofo_recent}.

\subsubsection{Early Approaches}
\label{scofo_stats_early}
In 1997, Grubb and Dannenberg \cite{grubb97} pioneered the use of a statistical approach. The idea in this system is that the position in the score is represented by a probability density function. Observation distributions, specifying the probability of observing any possible value of a detected feature when the performer is playing this event, were defined. New score position probability densities can be calculated using the current score position probability density and the observation densities.

Pardo and Birmingham in 2002 \cite{pardo02} instead used statistical methods to extend string-matching. They defined a statistical model of the transcriber error based on match probabilities; then, an alignment is calculated by dynamic programming.

\subsubsection{Hidden Markov Models (HMMs)}
\label{scofo_stats_hmm}

Hidden Markov Models (HMMs)--ideal models for sequential event recognition (a problem score following can be modelled after)--are immensely popular in speech recognition; interested readers may wish to refer to Rabiner's article on HMMs for speech recognition \cite{rabiner89}. HMMs are also known for their applications in computational finance \cite{petropoulos16} and bioinformatics \cite{li03}. In this context, score following can be formulated as a model containing a sequence of notes in the score, comprising an \emph{observed}
feature sequence extracted from the audio signal and the \emph{hidden} state sequence (which maps to the score's notes) \cite{cano99}. Methods could then be used to compute an estimation for the \emph{hidden} sequence given the \emph{observed} sequences. Further, techniques for training HMMs exist \cite{milks05}.

One of the first HMM-based score followers is Cano \etal{}'s work in 1999 \cite{cano99}. Audio features--namely energy, zero crossing and fundamental frequency--are chosen as observed emissions in the model. The note model architecture is based on three HMMs: a note, a no-note and a silence model. Note lengths are modelled with self-transitions, and the well-known Viterbi algorithm \cite{viterbi67} is used to compute the hidden states which give the alignment on the score.

1999 also saw the early publications in score following of a prominent figure in score following: Christopher Raphael. Raphael's system, first introduced in 1999 \cite{raphael99}, is similar to Cano's except that Raphael's work does not rely on pitch tracking routines--it instead emits spectral features. An alternative decoding technique is also proposed in addition to using the Viterbi algorithm. Raphael \etal{} would go on to create the \emph{Music Plus One} \cite{musicplusone} computer-aided accompaniment system, and his research group would continue to pioneer many solutions to problems of score following and computer-aided accompaniment with and without usage of HMMs \cite{raphael01,raphael03,raphael06,raphael10}.

The HMM-based IRCAM score follower (not to be confused with its predecessor \emph{EXPLODE}), which eventually developed into \textit{Antescofo} \cite{cont08} and a startup named \textit{Metronaut}\footnote{\url{https://www.antescofo.com/}} providing computer-aided accompaniment applications, is based on Raphael's system.
Orio and Dechelle in 2001 \cite{oriohmm1} extended Raphael's system by taking performer's errors into account via the introduction of ``ghost states'' corresponding to local mismatches. Instead of the Viterbi algorithm, another algorithm applied in molecular genetics is used, showing improvements in delay time and robustness in errors. A training method for the HMMs was also proposed.
Further, in 2004, Schwarz \etal{} \cite{schwarz04} showed how the system can be used with polyphonic MIDI data instead of the audio signal of the performance.
Cont \etal{} in 2004 and onwards \cite{cont04,cont06,cont08,cuvillier14,donat16} further extended the system, which is often regarded as the state-of-the-art score follower in its time.

An interesting problem which can be solved by HMMs is the modelling of the possibilities of musicians' spontaneous change in performance, e.g. leaving out a repetition or repeating a part albeit not noted in the score. This idea was proposed by Pardo and Birmingham in 2005 \cite{pardo05}.

Further studies in HMM-based score following up to recent times exist \cite{montecchio08,nakamura14,nakamuramerged14,sagayama14,hori19}, but they mainly dealt with improvements and extensions, such as training methods and better robustness to polyphonic music and mistakes, to pre-existing systems detailed here.

\subsection{Dynamic Time Warping (DTW)}
\label{scofo_dtw}

DTW is a technique for aligning time series--a general introduction to DTW is provided in \cite{mullerdtw}, and a deeper technical background to DTW in the context of score following is given in \autoref{impl_dtw_tech}. DTW is not only used in speech recognition (where it is first introduced) \cite{rabiner93} and score following--it is also used for gesture recognition \cite{gavrila96}, handwriting recognition \cite{niels05} and score following's sibling: music alignment \cite{chen19,muller06,orio01}. In fact, a lot of music alignment approaches use DTW--this is due to the fact that music alignment allows the use of the entire performance audio, meaning that DTW can directly compute the global optimal match of the performance audio to the audio generated from the score \cite{chen19}. Also, note from here that DTW is in principle an audio-to-audio system. Further, DTW can be seen as a special case of HMMs--the cells of the distance matrix correspond to the states and the distances serve as output probabilities for a given state \cite{durbin98}.

The primary computation in DTW is the \emph{minimum cost path} given a \emph{cost matrix}. Often, global path constraints such as the Sakoe-Chiba bound \cite{sakoe78} and Itakura parallelogram \cite{itakura75} are used to reduce the quadratic complexity of this computation. The pioneer of applying DTW in score following, Simon Dixon and Gerhard Widmer, extended the idea of such bounding in their 2005 score following work \emph{MATCH} \cite{Dixon2005MATCHAM}, resulting in a linear-time algorithm. Dixon and Widmer also proposed an interesting approach for extracting audio features--the method results in a low-level spectral representation generated via a windowed Fast Fourier Transform (FFT)--this is in contrast to some approaches in music alignment, such as Dannenberg and Hu's 2003 method \cite{dannerberg03} that uses a chroma representation of 12 elements (each representing the spectral energy corresponding to a pitch class). Following that, the spectral data is mapped into 84 frequency bins, linear at low frequencies and logarithmic at high frequencies to reduce the data and simulate the linear-log frequency sensitivity of the human auditory system.

Arzt \etal{} from 2008 took upon Dixon and Widmer's work and further improved it in many stages \cite{arzt08,arzt10,arzt12,arzt15}. Arzt's thesis in 2016 \cite{arzt16} provides a great summary of these improvements, which involve running multiple trackers simultaneously, introducing backtracking heuristics, using tempo models and computing adaptive distances. All these improvements contribute to a DTW algorithm that is robust--i.e. the minimum cost paths computed at the middle of the piece does not deviate too far from the ideal global minimum--and able to follow polyphonic and even orchestral music.

More advancements in DTW are found in music alignment--examples include a three-dimensional DTW solution proposed by Wang \etal{} \cite{wang15} that extends the 2-D \textit{cost matrix} into a 3-D \textit{cost tensor} capable of incorporating two performance sequences (which could be melody and accompaniment) to make it robust against asynchronies between musical voices in score-performance alignment. Recent papers working on DTW score following mostly involve improvements of Arzt \etal{}'s work \cite{rodriguez16} or its applications in more contexts \cite{brazier20,lin20}. An interesting note is that all high-performers in the MIREX evaluation use advanced approaches involving DTW, see \autoref{scofo_recent_mirex}.

\section{Recent approaches}
\label{scofo_recent}

While some researchers again took inspiration from advancements in existing work to form new approaches (such as combining other methods with existing work, common in top performers in the MIREX evaluation (\autoref{scofo_recent_mirex}) or using other stochastic approaches beyond HMMs (\autoref{scofo_stochastic})), more recent takes of the score following problem take on new paradigms and methods, some fuelled by recent advancements in Machine Learning (\autoref{scofo_sheetmusicimages}).

As these approaches cover more recent and complicated methods that would require more technical background than this report can provide, references to good sources are left for the interested reader.


\subsection{MIREX Evaluation: Top Performers (2010-2020)}
\label{scofo_recent_mirex}

As research interest grew in the area, there came a need for standardised evaluation for different algorithms. Since 2006, as part of the Music Information Retrieval Evaluation eXchange (MIREX) \cite{mirex}, annual evaluations for score following systems were held. The evaluation procedure consists of running score followers on a database of aligned audio to scores where the database contains the score, performance audio and a reference alignment \cite{cont07}. Unfortunately, the evaluation procedure is rather opaque (all links to sample data in the evaluation are now inaccessible), and it has evolved a lot in the early years to render early evaluation results (these were run on very few systems from 2006 to 2009 anyway) incomparable to those of later evaluations. Throughout the running of the evaluation, very few authors submitted their algorithms. From 2018 to 2020 no submissions were received. Further difficulties of running such evaluations (and evaluating score following in general) are covered in \autoref{testbench}--a discussion on the limited success of MIREX is provided in \autoref{tb_mirex_limited}.

Nonetheless, it is interesting to study methods attaining stellar evaluation scores between 2010-2020 when evaluation methods somewhat stabilised. \autoref{table:scofo_recent_mirex} summarises methods scoring $\geq 70\%$ in the total precision score, the main evaluation metric for runs in the period showing precision across all the evaluated musical pieces. Authors who submitted more than one variant of an algorithm throughout the period will have their highest-scoring submission reported.

Note that all high-performing algorithms covered here are DTW-based (\autoref{scofo_dtw}). These systems found improvements mostly in feature extraction, such as using both MIDI and synthesised score data for more score features to operate on \cite{orti15}, or by working on another feature extracted from audio, such as chromagrams \cite{suzuki10}. Worth noting is that Rodriguez-Serrano, Carabias-Orti and their group went on to found \textit{Beatik}\footnote{\url{https://www.beatik.com}}, a company that provides a mobile APT application.

It is perhaps surprising that HMM-based methods were not seen here--that is due to the fact that no HMM-based methods entered. Only one stochastic model was submitted \cite{duan10}, which was an experiment for state-space models (see \autoref{scofo_stochastic}) scoring $49.11\%$--the relevant authors did not resubmit their work after improving it in \cite{duan11,li16}.

Outside the 2010-2020 era, there exists two HMM-based method submissions, in 2008 \cite{montecchio08} and 2006 \cite{cont06}. It is interesting that 2006's running which has only two submissions featured Dannenberg's \cite{dannenberg84} original string-matching algorithm (which was significantly outperformed by Cont's 2006 HMM method \cite{cont06}). Cont \etal{}'s research team continued to improve their HMM-based system (and in fact they also contributed to the running of this evaluation \cite{cont07}), but there is no apparent reason why they did not continue submitting improvements in their work \cite{cont08,cuvillier14,donat16}. On the same note, Arzt \etal{} submitted only one DTW-based algorithm in 2010 (scoring $50.84\%$) \cite{arzt10}, and did not submit their improved work \cite{arzt12,arzt15,arzt16}.

\afterpage{%
  \clearpage
  \thispagestyle{empty}
  \begin{landscape}

    \begin{table}
      \centering
      \caption{Summary of High-Performing (Total Precision $\geq 70\%$) Methods of the 2010-2020 MIREX Score Following Tasks (\autoref{scofo_recent_mirex})}
      \label{table:scofo_recent_mirex}
      \begin{tabularx}{\linewidth}{|L|l|l|L|L|L|L|L|l|l|}
        \hline
        \textbf{Author}                                                                                          &
        \textbf{Year}                                                                                            &
        \textbf{Type}                                                                                            &
        \textbf{PF}                                                                                              &
        \textbf{SF}                                                                                              &
        \textbf{Alignment}                                                                                       &
        \textbf{Notes}                                                                                           &
        \textbf{Papers}
                                                                                                                 &
        \textbf{Score}
        \\
        \hline
        Bris-Peñalver                                                                                            &
        2017                                                                                                     &
        A-A \& A-S                                                                                               &
        Audio (Spectral)                                                                                         &
        Audio (Spectral) \& MIDI (States)                                                                        &
        DTW                                                                                                      &
        Utilises DTW to estimate Real State Sequences                                                            &
        \cite{orti15,alonso16,alonso17}
                                                                                                                 &
        $92.41\%$
        \\
        \hline
        Rodriguez-Serrano                                                                                        &
        2016                                                                                                     &
        A-A \& A-S                                                                                               &
        Audio (Spectral)                                                                                         &
        Audio (Spectral) \& MIDI (Score Information)                                                             &
        DTW                                                                                                      &
        Uses Non-negative Matrix Factorisation (NMF), a low latency signal decomposition method and finally DTW. &
        \cite{orti15}
                                                                                                                 &
        $97.43\%$
        \\

        \hline
        Carabias-Orti                                                                                            &
        2013                                                                                                     &
        A-A \& A-S                                                                                               &
        Audio (Spectral)                                                                                         &
        Audio (Spectral) \& MIDI (States)                                                                        &
        DTW                                                                                                      &
        Introduces usage of Non-negative Matrix Factorisation (NMF)                                              &
        \cite{carabias12}
                                                                                                                 &
        $86.70\%$
        \\

        \hline
        Suzuki                                                                                                   &
        2010                                                                                                     &
        A-A                                                                                                      &
        Audio (Chromagram)                                                                                       &
        Audio (Chromagram)                                                                                       &
        DTW                                                                                                      &
        Extracts sum of chroma and delta chroma vectors from audio                                               &
        \cite{suzuki10}
                                                                                                                 &
        $73.97\%$
        \\
        \hline
      \end{tabularx}
    \end{table}
  \end{landscape}
  \clearpage
}

\subsection{Stochastic approaches beyond HMMs}
\label{scofo_stochastic}

Duan and Pardo in 2011 \cite{duan11} first showed the application of non-HMM state-space models \cite{koller09} in score following, which is improved and further applied in 2016 by Li and Duan \cite{li16}. This approach uses a 2-D state vector to model the underlying score position and tempo of each time frame in the audio performance. They found that a multi-pitch based audio frame model worked well in conjunction with particle filtering to infer the hidden states from observations.

Another class of stochastic modelling methods, conditional random fields (CRFs) \cite{lafferty01}, was employed by Shinji, Yamamoto \etal{}'s group in their 2013 and 2014 papers \cite{shinji14,yamamoto13}. In these works, the music performance is modelled by CRFs, allowing the usage of the delayed-decision Viterbi algorithm which utilises future information to determine past score positions reliably.

Particle filtering (also known as Sequential Monte Carlo (SMC) methods) \cite{doucet09}--a minor role in Li and Duan's method \cite{li16}--is used as a primary method in three methods proposed in 2011-2013 \cite{otsuka11,montecchio11,korzeniowski13}. The main idea of particle filtering is that the hidden state can be continuous instead of discrete like in HMMs. Particle filtering approaches are also capable of modelling both audio-to-audio and audio-to-symbolic score following systems.

\subsection{Paradigm Shift: Sheet Music Images}
\label{scofo_sheetmusicimages}

There exists other unique recent approaches outside the areas covered until this point \cite{jiang19, noto19, macrae10} such as integrating eye-gaze information \cite{noto19} or following different score formats \cite{macrae10}, but none are as ground-breaking as this paradigm shift. So far, all the methods described take in score information via MIDI, MusicXML or another similar format. An issue is that, often, MIDI or MusicXML representations are not available. Optical Music Recognition, OMR, can be used to convert scanned or image-based sheet music into these formats, but the faithfulness of the symbolic score to what is depicted on the sheet image strongly depends on the OMR system \cite{henkel19}.

Gerhard Widmer's group at JKU Linz took another hard look at score following in 2016: this resulted in Matthias Dorfer \etal's 2016 paper on preliminary work for score following using sheet music images \cite{dorfer2016}. The authors proposed a multi-modal deep neural network to predict the position within a sheet snippet based on an audio excerpt. As part of this effort, the Multi-modal Sheet Music Dataset (MSMD) \cite{dorfer17} was created. Moreover, further work by the research group in 2018-2019 culminated in formulating score following as a reinforcement learning (RL) \cite{kaelbling96} problem \cite{dorfer18,henkel19}, where the RL agent's task is to adapt its reading speed in an unrolled sheet image conditioned on an audio excerpt.

Henkel \etal{} in 2019 \cite{henkel19unet} sought to solve a limitation in previous approaches: the scores were required to be represented in an unrolled form. Henkel \etal{}'s 2019 work introduced a system that directly infers positions within full sheet images for monophonic piano music by treating score following as a \emph{referring image segmentation task}, but the system neglects the temporal aspect of score following. The authors' continuation work in 2020 \cite{henkel20} builds upon that foundation and incorporates long term audio context, proposing the first fully capable score following system working on entire sheet images without needing preprocessing steps. This data-driven approach, while exceptional, is still subject to limitations of the dataset used. The dataset unfortunately only contains synthesised, constant-\emph{tempo}, music; demonstrations of Henkel \etal{}'s system on real performances show that the system has generalisation issues. Further, tests on scanned sheet music were not performed.

To alleviate the problem of generalising to real-world audio, Henkel and Widmer in 2021 \cite{henkel21} applied \emph{Impulse Responses} (IRs) as an audio data augmentation technique, allowing the modelling of different recording conditions in the form of microphone and room characteristics. Further, the 2021 work, in contrast to the 2020 work \cite{henkel20}, treats score following as a \emph{bounding box regression task} instead of an image segmentation task. This means that the network predicts the coordinates as well as the width and height of the bounding box that matches the performance audio \emph{query}. While the 2021 work performs reasonably well with some real-world audio, it did not address the problem of generalising in the sheet-image domain. Besides, applying IRs to augment data only aids the network to generalise to different recording characteristics, not to performance deviations as discussed in \autoref{scofo_challenges_perfdev}. A better audio augmentation method would be to use more powerful DSP elements (such as filters, oscillators, reverberation etc.) to augment data, such as those presented in \cite{engel2020ddsp}.

\section{Possible Future Directions}
\label{scofo_future}

The myriad papers published by authors since the pioneering pair in 1984 show how challenging and active the field of score following is. Also, note that the list of challenges in score following in \autoref{scofo_challenges} is extensive but not exhaustive. From the years of work, it is clear that score following is a multi-paradigm problem, requiring expertise from different fields outside of music technology and computer science. Advanced statistical methods, speech recognition ideas, signal processing techniques, machine learning and even computer vision approaches can play a part in solving this problem.

It is important to note that even the best approaches contain future work and discussion sections. For instance, there is growing interest in audio preprocessing to improve the audio quality (e.g. via dereverberation \cite{naylor10} and denoising \cite{ghias95}) before audio is fed into the system. Further, newer approaches that learn from data can benefit from bigger and better datasets encompassing a wide range of music and performances \cite{dorfer17}. Existing systems and algorithms can also be further improved and extended via using new audio feature extraction methods \cite{holighaus13}; this project proposes a new method covered in \autoref{impl_dtw}. Moreover, evaluation and benchmarking of algorithms still pose a huge challenge, in which \autoref{testbench} will attempt to contribute solutions to.

Novel algorithms in this field will no doubt require research teams with expertise in a wide range of techniques; moreover, they will also need to keep up with advancements in related fields. For instance, quantum computing's increased power (quantum image processing (QIP) is already a budding field \cite{qip}) might bring about a paradigm shift in this problem.

As music engraving and performance are fine arts subject to the infinite complexity of humans, score following will likely never have a ``perfect'' solution. This implies that score following systems can always be further improved.





%% file: 02_parts/02_testbench/01_testbench.tex
\part{Testbench}
\label{testbench}

\input{02_parts/02_testbench/chapters/01_preliminaries.tex}
\input{02_parts/02_testbench/chapters/02_quantitative.tex}
\input{02_parts/02_testbench/chapters/03_qualitative.tex}

%% file: 02_parts/02_testbench/chapters/01_preliminaries.tex
\chapter{Preliminaries}
\label{tb_preliminaries}
\epigraph{\itshape The only way to judge art is to wait and see if it becomes evergreen. This takes a bit of time.}{Andrea Bocelli}

As part of the effort to propose fairer and better evaluation testbenches for score followers, the challenges in evaluating followers are identified in \autoref{tb_challenges} before lessons are taken from MIREX's limited success in \autoref{tb_mirex_limited}. Subsequently, other evaluation methods used in the literature are studied in \autoref{tb_literature}. Finally, main requirements sought to be captured in this project's score following evaluation approach is detailed in \autoref{tb_requirements}. 

It is worth noting that music alignment testbenches may be used to evaluate score followers as well: the completed following data can be fed to the alignment testbench. Hence, evaluation work is frequently shared between followers and aligners. 

\section{Challenges}
\label{tb_challenges}

Evaluation to compare score followers is challenging for multiple reasons, including but not limited to:

\begin{enumerate}
    \item \textbf{Generalisation}. Some score followers are optimised for specific kinds of music and therefore address different sets of challenges mentioned in \autoref{scofo_challenges}. Score followers built primarily for monophonic music (most of the early work (\autoref{scofo_earlywork})) are vastly different from those capable of following polyphonic or even multi-instrumental music. Some score followers also primarily deal with certain types of ensembles or instruments, making evaluation datasets biased against trackers not built for following the instruments defined in the datasets. Optimisation areas also span into the periods of the followed music: followers trained to follow baroque music deal with challenges different from those of a follower tracking modern guitar music. 
    \item \textbf{Application-specific Details}. The follower's primary application (among the many detailed in \autoref{scofo_apps}) also plays a role in evaluation. A score follower built for computer-aided accompaniment deals primarily with providing a natural and smooth accompaniment (and hence are better evaluated with experiments with actual performers), whereas a musical-analysis-focused follower would be better suited for quantitative benchmarks dealing with the precision of musical features such as note onsets. Followers for APT have the loosest requirements--these trackers just need to know roughly when to scroll the score or ``flip'' the digital page, implying that the margin of allowed error can be large (seconds in contrast to the milliseconds required in computer-aided accompaniment and musical analysis). 
    \item \textbf{Musical Data Representation}. The many types of musical data representations (see \autoref{scofo_mdr}) score followers work with also pose a challenge. It would be hard to have an evaluation metric comparing score followers working using audio performance data to trackers working with captured MIDI data--transcribing the audio to MIDI for the latter tracker is a solution, but automatic music transcription is a difficult problem \cite{gowrishankar16} and would make the comparison unfair. Recent work on following sheet music images \cite{henkel19} would require a completely new methodology to evaluate as well.
    \item \textbf{Simulating Realistic Performance Deviations}. Many score followers propose algorithms that are robust against deviations in performance, such as skips, repeats and so on. It is hard to formulate a good and fair evaluation strategy to factor these features into the evaluation. Some authors resort to artificially inserting errors into the symbolic data of the score and synthesising the score, which would be limited by the capabilities of the synthesis software to produce performance errors that can be likened to a deviation produced by a human performer. Music datasets also often contain recordings with highly skilled performers who have lower tendencies to make mistakes.
    \item \textbf{Unique Score Following Systems}. Some followers completely detract from notions common in most score followers; for instance, there exists score followers that work only with beat tracking \cite{toiviainen98} or with eye-gaze detection \cite{tabone20}. Vercoe's pioneering system \cite{vercoe85} was also extended to integrate visual information. These score followers may not produce musical data common in score follower evaluation such as note onsets, or may be disadvantaged by the impracticality of running non-automated evaluations.
    \item \textbf{Proprietary Systems}. Most implementations of score following are kept proprietary; further, some are implemented in proprietary frameworks. Public evaluations may merely require the score follower to expose a predefined API, which could be done using a compiled binary targeted at the OS and architecture of the evaluation system (as in MIREX \cite{cont07}). Some authors also keep their evaluation datasets private as production of such datasets is tedious. All these contribute to evaluations being difficult to run and reproduce.
    \item \textbf{Dataset}. Producing a dataset for score following evaluation is difficult. Not only is the production of the ground-truth alignment tedious despite the possibility of using state-of-the-art aligners to help, but the sourcing of good performance data is difficult. Several datasets\footnote{Readers interested in MIR datasets can refer to \url{https://github.com/ismir/mir-datasets}.} were produced fulfilling some score following evaluation requirements \cite{duan11bach10, vienna4x22, MAESTRO}, but not all are of sufficient quality in the score following context.

\end{enumerate}

\section{MIREX's limited success}
\label{tb_mirex_limited}

All the challenges detailed in \autoref{tb_challenges} contributed to the limited success of the MIREX evaluation. MIREX's audio-to-score following evaluation proposal in 2006 \cite{cont07} is a significant step in score following research with the definition of many useful metrics; nevertheless, the primary disadvantage of it lies in the dataset. 

MIREX's initial dataset contains a voice and clarinet work by Mozart, a violin work by Bach and a work composed mainly for computer-aided accompaniment by Boulez. Later editions of the evaluation incorporated more works, but most remain classical music \cite{duan11bach10,miron16}. In addition, MIREX's dataset is not public, this is somewhat understandable--the effort required in creating this dataset means that the datasets are most valuable as unpublished test sets. The disadvantage in making the data completely private means that entrants could not predict how well their algorithms performed prior to evaluation--they would have to wait one year for the next edition. This is in contrast to a popular evaluation in computer vision, the Middlebury evaluation for optical flow \cite{baker07}, that has public and private dataset splits--the publicly available split is useful for initial testing and is often used to compare new state-of-the-art algorithms whose evaluation results are still pending.

In addition to having no public data, the evaluation framework, which could be provided publicly without any apparent issue, is also private. While reading the proposal (\cite{cont07}) and evaluation submission page (which notes deviations from the proposal) could be sufficient to understand the evaluation methodology, having a published software framework, preferably open-source such as that in the Middlebury evaluation, would help authors understand the evaluation further and run some tests themselves to make sure they conform correctly to the evaluation API.

Further, submissions are treated as a black box and the entire performance audio file is supplied \textit{a priori} to the follower. It is up to participants to indicate whether their followers are indeed aligners (and thus not cheat by using the entirety of the performance).

The fact that DTW-based methods dominate recent editions of the evaluation in \autoref{scofo_recent_mirex} also suggests that the evaluation may strongly favour such approaches--this may have contributed to the fact no HMM-based methods entered despite their apparent good performances in real-life tests \cite{metronaut}. Entrants who are also the authors of private datasets submitted to the evaluation also have an unfair advantage; this likely was the reason Cont \etal{} and Arzt \etal{} did not enter their improved solutions--both groups contributed to private datasets used in the evaluation.

The fact that the evaluation is limited to audio-to-score following (the organisers allow symbolic-to-score evaluation via MIDI-to-MIDI, but no entrants used this so far) also automatically disqualified candidates whose solutions did not conform to this paradigm. 

Recent entries by Bris-Peñalver \cite{orti15,alonso16,alonso17} and Rodriguez-Serrano \cite{orti15} also scored extremely high: 92.41\% and 97.43\% in the total precision scores respectively, suggesting that the evaluation may not be sufficient to discriminate top performers in new state-of-the-art approaches. It is also worth noting that these two authors are from the same research group that submitted multiple approaches, sometimes in a single year, and may have inadvertently overfitted their solution on the evaluation dataset.

\section{Solutions in the Literature}
\label{tb_literature}

It was previously mentioned that both music aligners and score followers may share evaluation methodologies--notably the datasets. Here, research done in evaluating aligners is also covered.

Score followers whose methodology detract from the norm (such as eye-gaze-based or beat-tracking-based systems) usually come up with their own evaluation methodology that focuses on the advantages of using their unique approaches \cite{tabone17, tabone20,toiviainen98,henkel19}. In fact, MIREX also runs evaluations for beat tracking systems (although not on real-time systems necessary for score following) \cite{dixon07}. Application-focused systems also usually come up with their own evaluation methods that exercise the approach in the intended area \cite{cont08,arzt16}.

As mentioned, datasets for score following evaluation are difficult to produce. For performance audio data, some authors resort to constructing synthetic datasets using a synthesiser \cite{hu04, remi16, Maezawa15, meron01,orio01}. This method is convenient, but results on such data could mislead: an algorithm can perform vastly differently on human performances; \autoref{scofo_mdr} noted that synthesisers cannot capture the full expressiveness of a human performer, often producing inexpressive constant-tempo performances that can be trivially followed. Some authors appreciate this problem and turn to perturbing onsets and offsets in the score prior to synthesis \cite{ewert12, raffel16}, but these perturbations are still artificial and may not reflect those produced by a human performer.

Further, a way for producing ground-truth alignments is by capturing both performance audio and symbolic data (usually MIDI) using digital instruments such as the Yamaha Disklavier or the Bösendorfer SE/CEUS. Popular and publicly available examples include the MAESTRO \cite{MAESTRO} and MAPS \cite{emiya10} datasets. Nevertheless, these datasets do not contain corresponding scores and are limited to piano music. A subpar solution for the score problem could be to randomly adjust timings of the performance transcripts--Thickstun \etal{} \cite{thickstun20} proposed a better solution: first obtain a subset of performances from the MAESTRO dataset and subsequently pairing it with scores from the KernScores dataset \cite{sapp05}.

\begin{figure}[h]
    \centering
    \includegraphics[width=\columnwidth]{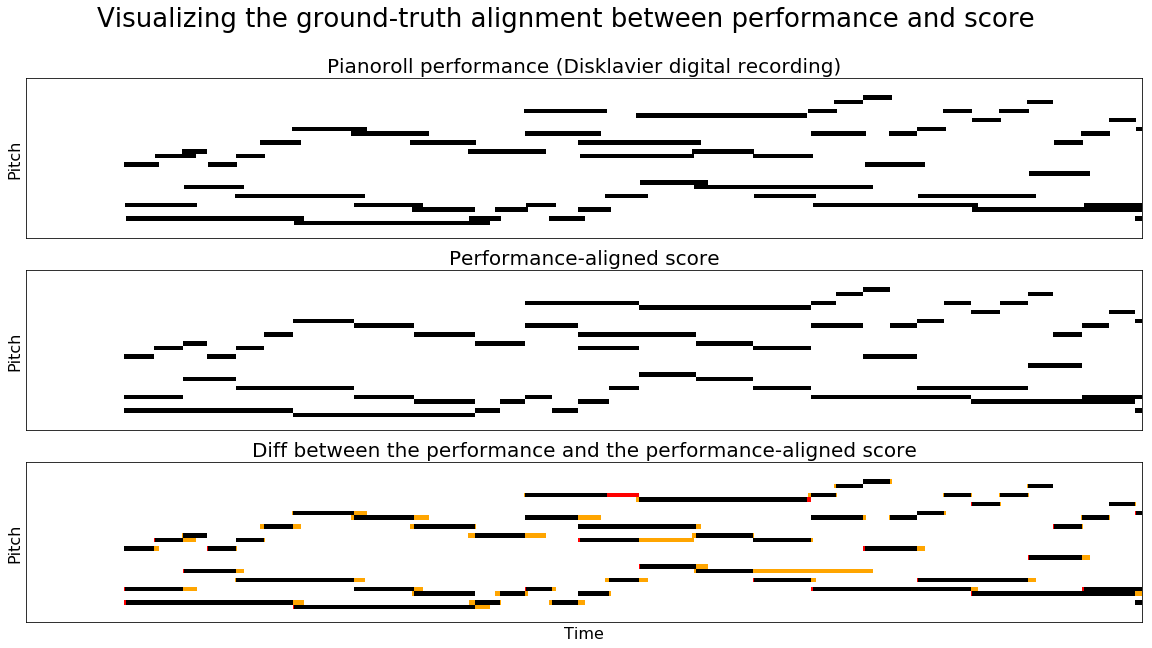}
    \caption{Thickstun \etal{}'s visualisation example to compare the piano-roll performance
    (top) captured by the Yamaha Disklavier to the performance-aligned score created by warping the score according to the ground-truth alignment (middle). In the comparison plot (bottom), red was used to identify missed notes and yellow to identify extra notes. This example visualises the beginning of a performance of Bach’s \textit{Prelude and Fugue in G-sharp minor (BWV 863).} Source: \cite{thickstun20}.}
    \label{fig:thickstunvis}
\end{figure}

As a matter of fact, Thickstun \etal{} in 2020 \cite{thickstun20} proposed a solution to tackle evaluation methodologies for audio-to-score alignment. Their approach culminated in an open-source testbench suite\footnote{Available at \url{https://github.com/jthickstun/alignment-eval}.}, which could be adjusted to work with audio-to-score score followers. Improving upon metrics defined in the MIREX counterpart \cite{cont07}, the authors also introduced a visual component to the evaluation, as seen in \autoref{fig:thickstunvis}--a problem with the visualisation is that it highlights as errors deviations that are not very relevant to score following--most of the errors in \autoref{fig:thickstunvis} identified relate to mistimed note \textit{offsets}; in score following, timings of note \textit{onsets} are more important. In addition, the authors also admit that their testbench is limited to piano music that intersects both the MAESTRO and KernScores datasets.

\section{Testbench Feature Requirements}
\label{tb_requirements}

There can practically exist no one perfect evaluation testbench that can fairly and robustly address challenges identified in \autoref{tb_challenges} to compare score followers of all types from all eras. This project, however, proposes two open-source evaluation solutions--one quantitative and the other qualitative--that both in conjunction will address most of the issues. Most score followers will support both testbenches, with the qualitative one having a slight advantage due to a more flexible API.  The main goals and then the requirements captured by the two proposals are covered.

\subsection{Quantitative Testbench}
\label{tb_quant_req}

The main goal of the quantitative testbench is to provide a wide range of quantitative metrics for evaluating score followers. This testbench shall show a follower's precision, latency, robustness and further.

The quantitative testbench shall also be the automated counterpart in the proposed pair of testbenches.

\begin{enumerate}
    \item \textbf{Open-source}: open-source evaluation testbenches encourage high adoption rates and contribute to reproducible evaluation and research. In addition, community-driven enhancements ensure the continual improvement of the testbench.
    \item \textbf{Flexible Score Follower API}: the testbench should be easily extended to work with different score followers. 
    \item \textbf{Flexible Dataset API}: instead of forcing users to use a predefined dataset as in Thickstun \etal{}'s \cite{thickstun20} approach, users should be free to use datasets they deem suitable for the purpose of evaluating their approaches.
    \item \textbf{Wide array of Evaluation Metrics}: reporting multiple different metrics such as latency, delay in onset detection, misalign percentage etc. as part of a wide set of evaluation metrics gives a more holistic view of the system's performance.
\end{enumerate}

These requirements are covered by the implementation of the quantitative testbench in \autoref{tb_quant}.

\subsection{Qualitative Testbench}
\label{tb_qual_req}

The qualitative testbench's main goal is to provide an interface where human users can evaluate score followers in situations more similar to where these systems will be applied. In this case, the testbench shall show a score being followed in real time.

As opposed to the automated quantitative testbench, the qualitative testbench relies on the manual judgment of a human evaluator, already necessary in score followers aimed at computer-aided accompaniment among other applications.

\begin{enumerate}
    \item \textbf{Open-source}: as proposed in \autoref{tb_quant_req}. 
    \item \textbf{Flexible Score Follower API}: as proposed in \autoref{tb_quant_req}.
    \item \textbf{Flexible Dataset API}: as proposed in \autoref{tb_quant_req}.
    \item \textbf{Clear Visualisation}: the testbench should produce clear indication of a score being followed in real time to be judged by humans.
\end{enumerate}

These requirements are covered in \autoref{tb_qual} which details the project's implementation of the qualitative testbench.

%% file: 02_parts/02_testbench/chapters/02_quantitative.tex
\chapter{Quantitative Testbench}
\label{tb_quant}
\epigraph{\itshape May not music be described as the mathematics of the sense, mathematics as music of the reason? The musician feels mathematics, the mathematician thinks music: music the dream, mathematics the working life.}{James Joseph Sylvester}

In this chapter, the implementation of the quantitative testbench is covered. Firstly, the main goal of the testbench is detailed in \autoref{tb_quant_maingoal}. Then, requirements covered in \autoref{tb_quant_req} are addressed in \autoref{tb_quant_address} before implementation details are covered in \autoref{tb_quant_impl_details}. Last but not least, \autoref{tb_quant_result} details the final product of this testbench.

Actual usage of the testbench will be covered in \autoref{impl} as it will be used to evaluate the project's own score following approaches.

\section{Main Goal}
\label{tb_quant_maingoal}

The main goal of this quantitative testbench is to provide a wide range of quantitative metrics suitable to assess and compare various types of score followers for different applications and music genres. 

\section{Addressing Requirements}
\label{tb_quant_address}

\subsection{Open-source}
The open-source nature of the testbench would encourage adoption and research reproducibility. This testbench is publicly available under the GPLv3 Licence on GitHub\footnote{\githubFlippyQuant}.

\subsection{Flexible Score Follower API}
\label{tb_quant_scofo_api}

The output required from a score follower required by the testbench is the alignment output produced after the follower completes a full following of a piece of music. The only requirement for the score follower is that it must be able to produce \textit{note onset detection data} of the followed piece of music. Note onsets are the primary feature of interest for score followers; hence, most--if not all--score followers produce this feature as an output. Audio-to-symbolic (especially audio-to-score) and symbolic-to-symbolic score followers can be trivially tweaked to produce such an output, whereas audio-to-audio-based systems would require some annotation in the followed audio that maps auditory note onsets to the score, which is almost always made available \cite{arzt16}. 

The project's approach is based on the format proposed by the original MIREX evaluation standards \cite{cont07}. The ID field, however, is not used, nor is the special score format--these are useful only for the private dataset. Instead of using the ID field to match notes in the dataset with notes in the follower's alignment output, the note onset time and the MIDI note number are used--this pair of values uniquely identifies every note sufficiently. Note that this is only \textit{sufficient}--some instruments like the violin are capable of producing two identically pitched notes at the same time, but this poses no problem to score followers: both notes can still be treated as a single note onset. Moreover, as the note onset time in the MIDI is a floating-point value, different programs may output the onset time in varying precision. This is handled by introducing a small search bound (set by default to $\si{1\milli\second}$) defined as such:

\begin{definition}[Score note search bound]
    The score note search bound, $\delta_s$ is a parameter that forms a search window of width $\delta_s$ to match note onset times obtained from score followers to note onset times specified in the score datasets. This search is done efficiently and uses an algorithm similar to the solution to the problem defined in \Cref{def:closesttimestamp}; see \autoref{sec:efficient_marker_position} for more information.
    \label{def:bound_ms}
\end{definition}

In the proposed format, each line in the file denotes a detected note onset and comprises four columns:

\begin{enumerate}
    \item \texttt{est\_time}: $t_e$, the estimated note onset time in the performance audio file ($\si{\milli\second}$, float)
    \item \texttt{det\_time}: $t_d$, the detected time relative to the performance audio file ($\si{\milli\second}$, float)
    \item \texttt{note\_start}: $t_s$, the note onset time in the score ($\si{\milli\second}$, float)
    \item \texttt{midi\_note\_num}: the MIDI note number of the note (integer)
\end{enumerate}

These data adequately capture not only the predicted note onset time which is core to precise score following--latency information, important for evaluating the reactivity and responsiveness of the score follower (crucial for followers targeting time-sensitive applications such as computer-aided accompaniment), are also captured. Further details on how these fields are processed are covered in \autoref{tb_quant_metrics}.

Notice that this data format is also produced by music aligners--the quantitative testbench can evaluate both music aligners and score followers.

Clearly, working on note onset data does not cover all score followers: eye-gaze-based and beat-tracking-based score followers do not produce such data, for instance. This is where the qualitative testbench (\autoref{tb_qual}) fills in the gaps.

\subsection{Flexible Dataset API}
\label{tb_quant_dataset_api}

The quantitative testbench requires reference ground-truth alignments between the audio of the performance and the score. The testbench uses the MIREX reference format \cite{cont07} which is simple and flexible. The reference file format is made up of lines where each line contains data relating to a note onset. Each line contains three columns--the last two are identical to the last two columns in the score follower output required; this means that the note onset and MIDI note number pair address unique note onsets in both formats. The columns are:

\begin{enumerate}
    \item \texttt{tru\_time}: $t_r$, the true note onset time in the performance audio ($\si{\milli\second}$, float)
    \item \texttt{note\_start}: $t_s$, the note onset time in the score ($\si{\milli\second}$, float)
    \item \texttt{midi\_note\_num}: the MIDI note number of the note (integer)
\end{enumerate}

Procuring reference data in this format is non-trivial--the authors of the original format kept their dataset as a privately held test set \cite{cont07}. The only publicly available dataset with such reference data is the Bach10 dataset \cite{duan11bach10}. Other publicly available datasets using different formats for the reference ground-truth alignments exist, such as the Traditional Flute Dataset \cite{tfd}, PHENICX-Anechoic dataset \cite{miron16, PHENICX} and Vienna 4x22 Piano Corpus \cite{vienna4x22}. These formats can be converted to follow the format described.

The production of ground-truth alignments is known to be non-trivial. The naïve approach is to perform manual alignment entirely given a score and performance audio--this was done for the Vienna 4x22 Piano Corpus \cite{vienna4x22}. Authors of the Bach10 dataset \cite{duan11bach10} slightly automated the procedure: software was built to record and modify manually tapped beats. Musicians tapped the beats while listening to the audio file, producing a series of beats. Beat-time alignment is performed, and each note in the score file is then linearly interpolated from the detected beats to produce the reference alignment \cite{duan11bach10}. On another note, the authors of the PHENICX-Anechoic dataset \cite{miron16, PHENICX} instead performed an initial audio-to-score alignment before manually adjusting misalignments under the guidance of a pitch estimator. A more recent approach by Thickstun \etal{} in 2020 \cite{thickstun20} uses a method analogous to DTW that temporally aligns symbolic scores and performances from the KernScores \cite{sapp05} and MAESTRO \cite{MAESTRO} datasets--this method requires minimal human intervention.

To generate high-quality ground-truth alignments flexibly, this project proposes the use of a custom Approximate String Matching (ASM) aligner detailed in \autoref{impl_asm} that requires minimal human intervention. This approach exploits the increasing ubiquity of music performance music datasets (such as the MAESTRO \cite{MAESTRO} and MAPS \cite{emiya10} datasets) comprising performance MIDI and audio data captured by digital instruments. While only common for piano music now, as more research moves into instruments such as the guitar and violin, it is inevitable that such datasets will be produced for instruments besides the piano. As music datasets like the two mentioned do not contain the ground-truth alignments required, reference scores (in the MIDI format, which can be derived from the commonly used MusicXML, LilyPond and Humdrum formats) are first obtained before aligning them to the performance MIDI files to produce high-quality ground-truth alignments. As opposed to Thickstun's \etal{}'s approach \cite{thickstun20} which limits the evaluation on piano music that intersects both the MAESTRO and KernScores datasets, this method is flexible--users can produce ground-truth alignments between any pair of corresponding symbolic performance and score data.

Worse comes to worst, if reference data in this format is unavailable or intractable to produce, users can defer to the qualitative testbench (\autoref{tb_qual}) that does not require such reference data.

\subsection{Wide array of Evaluation Metrics}
\label{tb_quant_metrics}

\Cref{tb_challenges} mentions that different score followers optimise for different applications and music. This therefore means that they also aim for different metrics to optimise. 

The metrics of this testbench is based on--and extended from--the original MIREX score following evaluation proposal \cite{cont07} and Thickstun's 2020 proposal \cite{thickstun20} for evaluating music aligners. It is worth noting that metrics in both papers are in turn derived from earlier audio-to-score alignment works \cite{meron01, orio01, shalevshwatz04}. 

From this point onwards, mathematical notation for note onset timing as defined in subsections~\ref{tb_quant_scofo_api} and \ref{tb_quant_dataset_api} is used. The notation is extended to refer to the $i$-th note onset (according to the reference score): $t_{si}$ refers to the note onset time $t_s$ of the $i$-th note in the score. Three key note-level definitions are presented.

\begin{definition}[Error]
The error, $e_i$, is defined as:
$$
e_i = t_{ei} - t_{ri};
$$
the error is the time between the estimated note onset time and the actual performance note onset time. Score followers that require high accuracy but can sacrifice latency and reactivity, such as those used for music analysis, primarily seek to minimise this metric.
\label{def:error}
\end{definition}

\begin{definition}[Latency]
Latency, $l_i$, is defined as:
$$
l_i = t_{di} - t_{ei} > 0;
$$
latency is a measure of how long the score follower takes to identify the event after it deems it occurred. This metric is important for reactive and low latency systems--it shows how fast the score follower is able to process and report a note onset event.
\label{def:latency}
\end{definition}

\begin{definition}[Offset]
Offset, $o_i$, is defined as such:
$$
o_i = t_{di} - t_{ri};
$$
the offset denotes the lag between the reporting of the detection and the time the event actually occurred. This is particularly important for systems that need to react quickly to note events, such as those in computer-aided accompaniment systems.
\label{def:offset}
\end{definition}

Next, notes that are deemed ``incorrect'' are classified.

\begin{definition}[Missed notes]
Missed notes are defined as events that exist in the reference but are not recognised.
\end{definition}

\begin{definition}[Misaligned notes]
Misaligned notes are recognised notes that are too far (the \emph{misalignment threshold} is usually set to $\theta_e = \si{300\milli\second}$, but is a configurable parameter in this testbench) from the reference onset, i.e. 
$$
|e_i| \ge \theta_e.
$$ 
\label{def:misalign}
\end{definition}
The acceptability of missed or misaligned notes once again depends on the application of the score follower. Let the total number of reference notes, missed notes and misaligned notes be $n$, $n_m$ and $n_e$ respectively. Computer-aided accompaniment followers would be ``lost'' and produce bogus accompaniment if the \textbf{miss rate} $r_m = \frac{n_m}{n}$ (the percentage of missed score onsets) or the \textbf{misalign rate} $r_e = \frac{n_e}{n}$ (the percentage of misaligned score events) are not low enough. Followers for APT could afford a higher miss rate, as long as the notes near the page turning or scrolling position are not missed or too misaligned.

Further, a metric is introduced to measure how far a score follower is capable to follow a piece:

\begin{definition}[Piece completion]
Piece completion, $p_e$, is the percentage of non-misaligned notes followed. All robust score followers should aim to have almost perfect piece completion scores.
\label{def:pc}
\end{definition}

Building from the error, latency and offset metrics defined above, more useful ``global'' metrics are derived for \textit{correct}, i.e. non-misaligned notes, that can be used to compare score followers.

The \textbf{standard deviation of error}, $\sigma_e$, and \textbf{mean absolute error}, $\text{MAE}$, are calculated as
$$
\sigma_e = \sqrt{\frac{\sum e_i^2}{n}}  ~ \forall i ~\text{where}~ |e_i| \leq \theta_e
$$
and 
$$
\text{MAE} = \frac{\sum |e_i| }{n} ~ \forall i ~\text{where}~ |e_i| \leq \theta_e
$$
respectively. These two metrics show the spread and absolute mean of the overall error. The MAE is also known as the \textbf{mean imprecision} $\mu_e$.

The \textbf{precision rate}--the most common metric used to compare score followers in the MIREX evaluation--is simply $$r_p = 1 - r_m - r_e.$$ 

For testbench runs across a suite of pieces, two overall metrics related to the precision rate are the \textbf{piecewise precision rate} $r_{pp}$ and \textbf{total precision rate} $r_{pt}$ which respectively denote the average precision rates across the group of pieces and the percentage of correctly aligned score events across the whole group of pieces. 

Also reported are the \textbf{mean latency} ($\mu_l$), \textbf{standard deviation of latency} ($\sigma_l$), \textbf{mean absolute offset} ($\text{MAO}$) and \textbf{standard deviation of offset} ($\sigma_o$) for non-misaligned notes.

Being an extension of MIREX metrics, results produced by MIREX evaluations can be compared to the metrics proposed here. The reporting of all the proposed metrics in a clear and concise manner encourages a holistic overview on the performance of score followers.

\section{Implementation Details}
\label{tb_quant_impl_details}

\subsection{Choice of Language}

Choosing the language for implementing the quantitative testbench was relatively straightforward. Python\footnote{\url{https://www.python.org/}} was chosen--similar to Thickstun \etal{}'s \cite{thickstun20} approach. The heavy mathematical nature of the quantitative testbench benefits from Python's wide library of numeric packages, such as \texttt{numpy}. In addition, in recent years, Python is widely used in the scientific community--recent score followers also started to move away from proprietary frameworks in favour of Python \cite{henkel19,dorfer18}, in part due to the ease of applying Machine Learning on the score following problem via the many Machine Learning tools built for Python.

Python is also simple and easy to learn, leading to wide usage inside and outside the scientific community as users need not worry about nits in software development--they can focus more on the core logic and intent of the program. This wide acceptance adds to the benefit that the source code of the testbench can be easily understood--and even extended--by users.

While Python lacks in speed \cite{benchmarksgame}, note that the running of the testbench is not time sensitive and that the runtime of the testbench is linear time with respect to the number of notes in the score. Further, in practice, testbench runs are multitudes faster than real time.

More of Python's benefits are covered in \autoref{impl}--Python is extensively used to prototype this project's score followers.

\subsection{Engineering Practices}
\label{tb_quant_bettereng}

Python is a dynamic and weakly typed interpreted language. This implies that Python programs may be hard to maintain especially as projects age and grow; for instance, not knowing what \textit{type} is valid for a variable makes mistakes (e.g. mistakenly assigning strings to \texttt{est\_time} which is supposed to take an integer) hard to avoid in the long run. To counter Python's weaknesses, a three-part strategy is used: static analysis, code testing and automated testing.

Firstly, \texttt{mypy}\footnote{\url{http://mypy-lang.org/}}, a static type checker, is used to statically analyse the source code, catching statically analysable errors before they bubble into runtime errors that are more difficult to fix. The code is also generously annotated with types in Python's \texttt{typing} module, which not only assists \texttt{mypy} in producing better static checks--these annotations also make it easier for users to understand the code, helping users produce better code changes.

Further, the code is extensively tested (via Python's \texttt{unittest} feature), which includes mocking testbench runs, to ensure that results the testbench deliver are reliable. Many edge cases (such as empty scores) are covered to make sure they behave properly, saving valuable debugging time by preventing these errors from arriving at runtime.

Finally, GitHub Actions\footnote{\url{https://github.com/features/actions}} is used to run automated tests, including unit tests and mock testbench runs. This adds another layer of protection--contributors may sometimes forget to run tests after making code changes, and this layer would prevent bugs and regressions from arriving in the code. In addition, developers develop on different types of machines--having a central workflow running tests on a baseline environment (Ubuntu 20.04) ensures clear signal of test results in code changes (it is, however, worth noting that the solution is designed to be runnable on all Python 3-capable environments).

\section{Resultant Product}
\label{tb_quant_result}

\subsection{Open-source Repository}

The testbench is released publicly under the GPLv3 Licence on GitHub\footnote{\githubFlippyQuant}.

\subsection{Usage Guide}

The README\footnote{\url{https://github.com/flippy-fyp/flippy-quantitative-testbench/blob/main/README.md}} of the repository shows a detailed usage guide of all features in this testbench. The repository for the score follower to be introduced in \autoref{impl_dtw}\footnote{\url{https://github.com/flippy-fyp/flippy}} also demonstrates usage of this testbench.

\subsection{Tools}

Besides offering the features discussed in this chapter, convenience tools are provided--these include:

\begin{enumerate}
    \item \textbf{ASM Score Aligner}. This aligner produces testbench reference data from performance and reference scores, as described in \autoref{tb_quant_dataset_api}. The algorithm is fully detailed in \autoref{impl_asm}.
    \item \textbf{MIDI to Score Converter}. This converter takes in MIDI files and outputs a score format with each line comprising two columns representing each note's start time (\si{\milli\second}, float) and MIDI note number respectively. This converter makes it convenient to extract note onset details from MIDI files.
    \item \textbf{MusicXML to MIDI/Score Converter}. This converter takes in MusicXML files and is capable of extracting MIDI data out of the file (described in \autoref{scofo_mdr_musicxml}). This MIDI data can also be converted into the score format described above (effectively, the MIDI output is piped into the MIDI to Score Converter).
    \item \textbf{Score to MIDI Converter}. This converter is the opposite of the MIDI to Score Converter.
    \item \textbf{Reference Score to MIDI Converter}. This converter takes a file containing a score of the reference format described in \autoref{tb_quant_dataset_api} and outputs a MIDI file. The first column (\texttt{tru\_time}) is used as the note onset time in the MIDI.
\end{enumerate}

\subsection{Reproduction Suite}
\label{tb_quant_repro_suite}

A reproduction suite is provided to not only demonstrate actual usage of this testbench, but also to reproduce results of the ASM score aligner to be introduced in \autoref{impl_asm}.

%% file: 02_parts/02_testbench/chapters/03_qualitative.tex
\chapter{Qualitative Testbench}
\label{tb_qual}
\epigraph{\itshape If it sounds right, then it is.}{Eddie Van Halen}

The implementation of the qualitative testbench is detailed in this chapter. First, the main goal to be achieved by this testbench is presented in \autoref{tb_qual_maingoal} before requirements captured in \autoref{tb_qual_req} are addressed in \autoref{tb_qual_address}. Also detailed in \autoref{tb_qual_impl_details} are the implementation details of the testbench. Last but not least, \autoref{tb_qual_results} details the final product of this testbench.

Actual usage of the testbench will be covered in \autoref{impl} as it will be used to evaluate the project's score following approaches.

\section{Main Goal}
\label{tb_qual_maingoal}

The main goal of the qualitative testbench is to provide a user interface showing score following in action: the score follower shall follow the score provided in real time. The position of the tracking shall be showed live to the user for evaluation purposes.

\section{Addressing Requirements}
\label{tb_qual_address}

\subsection{Open-source}
Similar to this testbench's quantitative counterpart (\autoref{tb_quant}), this testbench is publicly available under the GPLv3 licence on GitHub\footnote{\url{https://github.com/flippy-fyp/flippy-qualitative-testbench/}}.

\subsection{Flexible Score Follower API}
In \autoref{tb_quant_scofo_api} it was noted that the quantitative testbench only works for score followers capable of producing tracked note onset data. This automatically disqualifies approaches that do not produce such data (such as eye-gaze and beat tracking systems). This testbench resolves this issue--the only output required from score followers is the \textbf{instantaneous time} of the currently followed position in the score.

An alignment algorithm shall take the timing information provided by the score follower to overlay a marker on a provided sheet music to indicate the estimated position.


\subsection{Flexible Dataset API}
\label{tb_qual_flexdatasetapi}

This approach requires no tedious creation of valuable reference datasets. Since a human user is the primary agent evaluating the followers, the only dataset required here is sheet music and its corresponding performance audio (streamed live or playback). Currently, only the \nameref{scofo_mdr_musicxml} format is accepted. It is however possible to perform staff detection for sheet music images which would make the approach viable for recent approaches working on sheet music images (this was experimented in \cite{vreefollow})--this feature can be added when the need arises.

For the purpose of qualitatively evaluating the score following system proposed subsequently in \autoref{impl_dtw}, the \textit{QualScofo} dataset comprising a wide variety of musical pieces, was compiled--see \autoref{tb_qual_prod_dataset} for more information.

\subsection{Clear Visualisations}

Readers may note that the visualisation provided by Thickstun \etal{} \cite{thickstun20} (shown in \autoref{fig:thickstunvis}) may be somewhat helpful despite its shortcomings. In fact, the adding of visualisations immune to the offset problem in the quantitative testbench was considered, but it quickly became clear that the proposed quantitative metrics adequately capture such information. Users can also quickly compare the aligned files with their respective reference files to obtain more information. That said, the benefits of clear visualisations are appreciated. Thus, a friendly user-interface for score following visualisation is introduced in this qualitative testbench.

All the user needs to do is to load the MusicXML sheet music, then play the performance audio after initiating the testbench and follower. The user can then evaluate the actual following of the score follower by the visual marker provided by this testbench. This is useful for evaluating systems prior to their user acceptance testing stage in real-life applications. Users can judge subjectively how well a score follower is performing; they can also look for specific features targeted by the score follower to optimise.

\section{Implementation Details}
\label{tb_qual_impl_details}

\subsection{Choice of Framework and Language}
\label{tb_qual_impl_details_choice}

The decision of which framework and language to implement this testbench is, in contrast to the quantitative counterpart, not as straightforward. The fact that the quantitative testbench works on artifacts in the form of completed alignment text files means that the quantitative testbench can in fact use any language capable of reading in text files and performing mathematical analyses.

The qualitative testbench, on the other hand, may be required to work with score followers written in different languages, tools and frameworks. Rewriting the testbench ad hoc in the target score follower's language defeats the intention of an open and widely adopted test suite. In fact, the motivation behind the flexible APIs of the qualitative testbench stems from part of Ken Thompson's well-known Unix philosophy \cite{unixphilo}:

\begin{displayquote}
    Make each program do one thing well. To do a new job, build afresh rather than complicate old programs by adding new ``features''.
\end{displayquote}

Therefore, the qualitative testbench is intended to be an easy plug-and-play testbench solution highly capable of doing just its job. Hence, it should be written in a language and/or framework that facilitates such a requirement. Also note that the user interface requirement for visualisations means that a framework and language supporting these visualisations are required.

Python remains a viable alternative with the benefits detailed in \autoref{tb_quant_bettereng}. However, Python's graphical user interface (GUI) options (e.g. \texttt{Tkinter}, \texttt{PyGtk} etc.) are not as strong as those in an area that has seen huge strides in GUI development: the Web. The \texttt{Electron} framework\footnote{\url{https://www.electronjs.org/}} was thus chosen. The \texttt{Electron} framework allows easy development of cross-platform desktop apps with the classic Web stack of JavaScript, HTML and CSS. Well-known applications built with \texttt{Electron} include Visual Studio Code, WhatsApp Desktop and Microsoft Teams.

The Web support in \texttt{Electron} comes from its inclusion of \texttt{Chromium}\footnote{\url{https://www.chromium.org/}}, an open-source browser that is the core of the well-known Google Chrome browser. JavaScript execution in \texttt{Electron} is powered by the performant V8 JavaScript engine that just-in-time (JIT) compiles JavaScript code to run quickly on machines \cite{v8}--experimental results show that even though JavaScript is widely regarded as a slow interpreted language, JavaScript on V8 proves to be very performant (much more than Python) in benchmarks \cite{benchmarksgame} for a language of its class. Cross-platform desktop apps may also require OS-specific functions--\texttt{Electron} offers this in its vast, well-documented, suite of APIs.

JavaScript being the \textit{lingua franca} of the Web also means that the community of developers is active and strong. There hence exists many open-source libraries. In fact, the user interface for rendering MusicXML is built upon an existing library\footnote{\url{https://opensheetmusicdisplay.org/}}. Migration of the codebase into a Web-based application (or a Web-based mobile application framework such as React Native\footnote{\url{https://reactnative.dev/}}) is also straightforward for \texttt{Electron} applications.

The downside of \texttt{Electron} is clearly its bundle size after building--the application has to effectively bundle a whole web browser (\texttt{Chromium}). The qualitative testbench can be written for the Web to be used in a browser, but that would disqualify the vast majority of score followers not written in a Web-supported framework or language. In addition, the capability of working with native OS calls would prove more convenient for this testbench's purposes, effectively facilitating an interface capable of working with many score followers.

\subsection{Follower-Testbench Interface}
\label{tb_qual_follower_testbench_interface}

A significant challenge in building the testbench is in the implementation of the follower-testbench interface--a requirement for a flexible testbench is that it should support followers written in any framework or language. While there are user-friendly abstractions (a strong contender is ZeroMQ\footnote{\url{https://zeromq.org/}}) that facilitate communication between programs written in different languages, these are not as easily set up as plain low-level sockets, which are simple to implement and available in most programming languages and systems.

Using sockets not only allow the score follower to run in a separate and independent process--it also means that the score follower can run on a \emph{separate device}. This is beneficial when used in ensemble settings--the follower could be run on one system equipped with a well-placed microphone to take in audio from the performance, which is then fed into the score follower. The resultant timestamps can be broadcasted to the devices running this testbench, which can, in effect, replace the players' sheet music. Further, the computation-heavy score following algorithm could be off-loaded to a high-performance server, and the devices displaying the sheet music and following result do not need to be highly performant systems.

To initialise the testbench, users would indicate the host's port in the testbench to which the follower would send UDP packets--UDP was chosen for its lower overhead and the fact that it supports individual, separable message packets containing timestamps from the follower. Despite UDP's nature of dropping packets and possible out-of-order message receiving which are not present in TCP, in practice the UDP interface worked seamlessly given the low bandwidth required; further, using TCP would require extra programming to parse the required timestamps as TCP is a stream-oriented protocol--this would incur extra overhead in addition to the acknowledgement overhead present in TCP. As the responsiveness of the testbench is paramount, UDP is ideal.

\begin{figure}[h]
    \centering
    \includegraphics[width=\columnwidth]{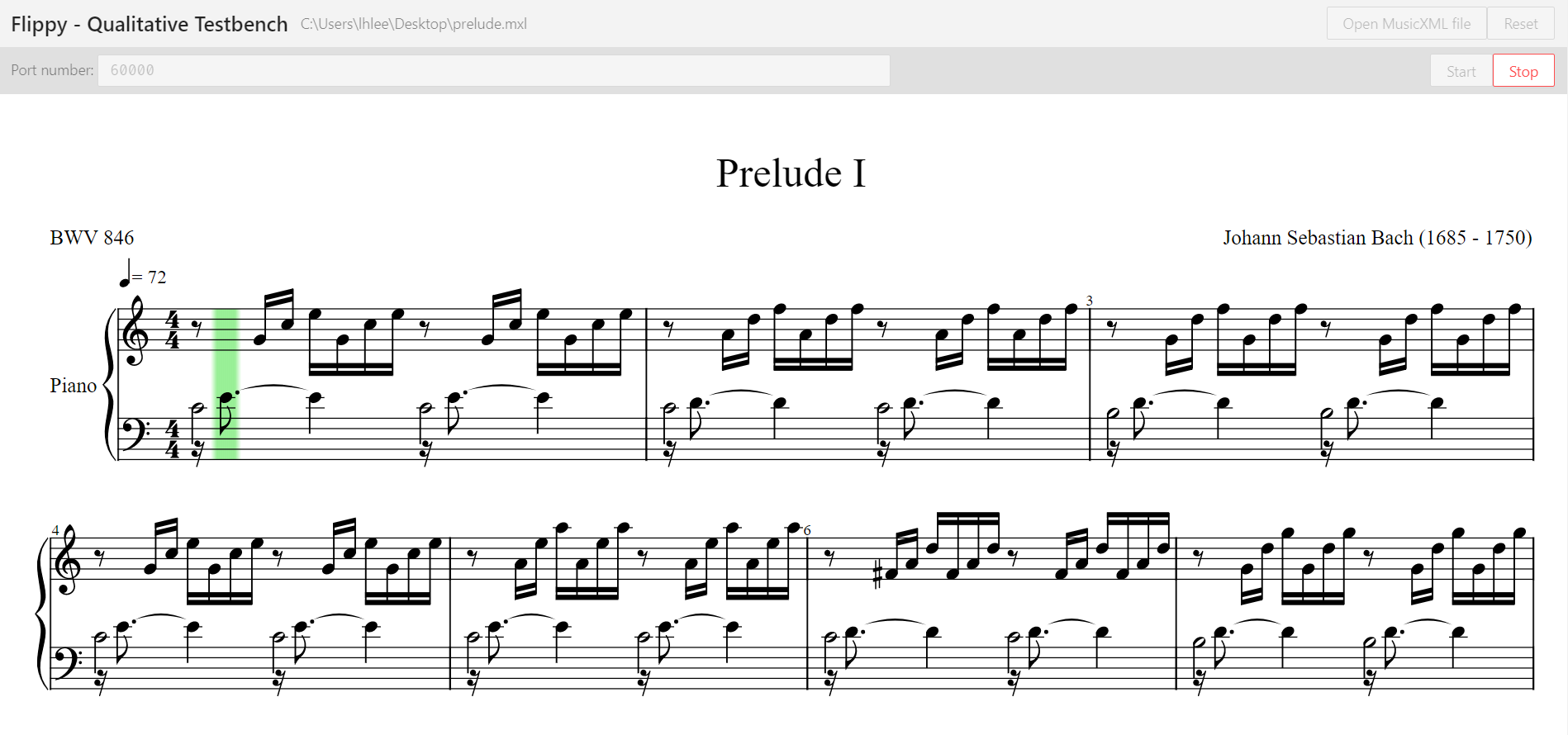}
    \caption{A screenshot of the qualitative testbench software at work. The piece being followed is Bach's \textit{Prelude in C major, BWV 846}. The user interface is simple and straightforward: the first line of the header shows the currently open (if any) MusicXML's file path, together with file controls. The second line contains a textbox for users to key in the local host's port number to which the follower would send timestamp output (via UDP messages), along with controls to start and stop the following.}
    \label{fig:qualbwv846}
\end{figure}

During score following, the testbench receives and parses timestamps in UDP packets received from the follower, which is then used to calculate and indicate the position on the displayed sheet music. \autoref{fig:qualbwv846} shows the qualitative testbench at work.

\subsection{Efficient Marker Positioning}
\label{sec:efficient_marker_position}

A requirement to pursue the qualitative testbench's goal of giving clear visualisations on the performance of a score follower is the testbench's responsiveness. The most computationally intensive part of this testbench is the positioning of the visual marker on the music sheet.

Score followers may output arbitrary timestamps that are not necessarily close to and/or larger than the last timestamp produced--this means that the testbench needs to responsively handle arbitrary jumps. Some score followers may also produce timestamps that do not correspond directly to a note onset timestamp in the score.

The library used for MusicXML visualisation\footnote{\url{https://opensheetmusicdisplay.github.io/}} offers an interface to position the marker on note onsets of the MusicXML document. In the testbench's preprocessing step, each note onset is tied to its timestamp in seconds. Thus, the testbench needs to performantly position the visual marker to the (closest) note onset based on the timestamp given by the score follower.

\subsubsection{Finding the Closest Timestamp}
\label{tb_qual_finding_closest}

The problem is first formulated:

\begin{definition}[The problem of Finding the Closest Timestamp]
    Given a list of timestamps $\textsc{Timestamps}$ and a timestamp from the follower $\textsc{Timestamp}_f$, find $\textsc{Timestamp}_c$, the timestamp within $\textsc{Timestamps}$ that minimises $\textsc{Error} = |\textsc{Timestamp}_c - \textsc{Timestamp}_f|$. Where there exists ties, the preceding timestamp is returned.
    \label{def:closesttimestamp}
\end{definition}

A naïve approach is to calculate $\textsc{Error}$ for every single element in $\textsc{Timestamps}$. Immediately the calculation can be optimised--note that the values of $\textsc{Error}$ form a parabola with a minimum; therefore, the calculation can be stopped when $\textsc{Error}$ first increases--the preceding timestamp is the required result. An example evaluation of this algorithm is as such: given $\textsc{Timestamps} = [0, 2, 4, 6, 8]$ and $\textsc{Timestamp}_f = 3.3$, the $\textsc{Error}$s calculated are $[3.3, 1.3, 0.7, 2.7]$--the calculation stops at the fourth element and the result is thus $\textsc{Timestamp}_c = 4$.

The above approach takes $\mathcal{O}(n)$ time and space, where $n$ is the number of elements in $\textsc{Timestamps}$. As mentioned, the responsiveness of the visual marker is of utmost importance, and any optimisation is welcome--the number of note onsets in large complex scores can be on the order of tens of thousands to hundreds of thousands, and each score following session requires this function to be called as many times as there are note onsets in the score (and therefore in the context of the whole score following the time complexity is quadratic--$\mathcal{O}(n^2)$). All the latency caused by this calculation adds up and results in an overall drop in the testbench's responsiveness.

A more efficient approach is used--$\textsc{Timestamps}$ can be preprocessed and stored as a binary search tree. Two binary search tree operations that help solve the problem are
$$\textsc{UpperBoundOrEqual}(\textsc{Timestamp}_f, \textsc{Timestamps})$$ and
$$\textsc{LowerBoundOrEqual}(\textsc{Timestamp}_f, \textsc{Timestamps})$$ that find the first element greater than or equals and lesser than or equals $\textsc{Timestamp}_f$ respectively, both in $\mathcal{O}(\log n)$ time. For the example given above, the former function returns $4$, the latter $2$--the former results in a smaller $\textsc{Error}$ and therefore is the required result. The optimisation from $\mathcal{O}(n)$ to $\mathcal{O}(\log n)$ is significant, especially when the complexity is instead calculated in the context of the whole score following procedure: the resulting complexity of calculating required timestamps is quasilinear: $\mathcal{O}(n \log n)$.

\subsubsection{Placing the Visual Marker Efficiently}

Because the score follower may produce arbitrary timestamps at any point in time, the procedure of calculating and/or placing the visual marker of the preceding timestamp may be incomplete when a new timestamp is received. Instead of finishing this operation and therefore causing \emph{compounded} drops in responsiveness when multiple successive big jumps are required, the procedure for the preceding note can be preempted.

Procedure preemption varies across different languages, frameworks and systems. As the main programming language chosen, of which the reasons of choice are detailed in \autoref{tb_qual_impl_details_choice}, JavaScript offers first-class support for \emph{asynchronous programming} via Promises\footnote{\url{https://web.dev/promises/}}. In most multi-threaded languages, asynchronous programming is a means of \emph{parallel programming} in which a unit of work runs separately from the main application thread and notifies the calling thread of its completion, failure or progress \cite{asyncdotnet}. However, JavaScript, as the informed reader may point out, is single threaded. JavaScript's Promises allow the emulation of such asynchronous behaviour--Promise evaluation does not block the main JavaScript thread. The benefit of using the asynchronous programming paradigm is that the idea and execution of preemption become simpler: via cancelling the in-progress asynchronous job that is independent of the main thread.

The calculation of the closest timestamp as well as the placement of the visual marker are thus performed asynchronously via Promises. By current design, however, preemption in JavaScript Promises is not well supported; fortunately, wrappers for Promises that add preemption support to JavaScript Promises exist. This testbench uses one such wrapper: Bluebird\footnote{\url{http://bluebirdjs.com/docs/getting-started.html}}.

As an aside, another possible way of dealing with preempting asynchronous code in JavaScript is via Observables\footnote{\url{https://rxjs-dev.firebaseapp.com/guide/observable}}. Observables offer a different and easier interface, but currently require the inclusion of a significantly large and complex library. Note however that Observables are currently a proposed feature for JavaScript\footnote{\url{https://github.com/tc39/proposal-observable}}.

With preemptive asynchronous programming in JavaScript, the placing of the visual marker on the sheet music is made more efficient--the latest timestamp takes priority and may cause the preemption of any calculations and placement operations of preceding timestamps.

\subsection{Engineering Practices}

The good engineering practices employed in the development of the quantitative testbench as per \autoref{tb_quant_bettereng} are replicated. Fortunately, while the choice of language and framework for this testbench was tougher, better practices are more easily maintained as there are many existing tools created and supported by the community. Again, the maintainability and reliability of the source code are tackled via the three core steps: static analysis, code testing and automated testing.

The first two steps are conveniently covered by using React\footnote{\url{https://reactjs.org/}} and TypeScript\footnote{\url{https://www.typescriptlang.org/}} on top of \texttt{Electron}. React is a JavaScript library framework that abstracts away the view layer for Web applications besides allowing the creation of reusable components. Further, React applications are fast, scalable and simple--data on the interface can be changed without reloading the page. This results in less code being written to enable the interface features of this testbench, and therefore there are fewer areas to make mistakes in. TypeScript, on the other hand, is a \textit{typed} superset of JavaScript. TypeScript has a type system vastly stronger than that of \texttt{mypy} for Python, and \textit{transpiles} down to JavaScript for runtime execution.

Static analysis and testing are covered by these two tools--React abstracts away user interface rendering code; TypeScript provides powerful static analysis. These features do not mean there is no need for tests--tests are still necessary; however, the number of tests required is fewer due to React's abstractions and TypeScript's strong static analysis.

GitHub Actions\footnote{\url{https://github.com/features/actions}} is once again used to run automated tests--the setup is similar to that in \autoref{tb_quant_bettereng}.

\subsection{Known Limitations}

\subsubsection{MusicXML Library Issues}

The library used for MusicXML visualisation, \textit{OpenSheetMusicDisplay}\footnote{\url{https://opensheetmusicdisplay.github.io/}}, is currently a limiting factor for the performance of this qualitative testbench--it only reached version \texttt{1.0.0} on 6 May 2021, and still has unresolved issues, such as failing to handle score repeats and inaccurately rendering some score components. Nevertheless, this project is actively being developed and updating this dependency for this testbench is trivial when a new version becomes available. The inability to handle repeats can be worked around by unrolling repeats using a MusicXML editor, and the score rendering inaccuracies are not significant enough to cause major issues in using this testbench.

\section{Resultant Product}
\label{tb_qual_results}

\subsection{Open-source Repository}

The testbench is released publicly under the GPLv3 Licence on GitHub\footnote{\githubFlippyQual}.

\subsection{Usage Guide}

The README\footnote{\url{https://github.com/flippy-fyp/flippy-qualitative-testbench/blob/main/README.md}} of the repository shows a detailed usage guide of all features in this testbench. The README in the repository for the score follower to be introduced in \autoref{impl_dtw}\footnote{\githubFlippy} also details a usage guide for this testbench.

Further, the README shows a development guide, including details on how to install dependencies, set up a development environment, run a development instance of the application that will refresh upon live code changes, lint code and build the project.

\subsection{Pre-built Installers}

Pre-built installers, automatically built via GitHub Actions, are available for Mac, Windows and Linux systems. These installers give most users a simple way to quickly set up the testbench. Should these pre-built installers not work (perhaps due to the usage of an esoteric system), users can also follow the development guide to build the testbench for their system.

\subsection{Mock Follower}

A mock follower program is written in Python to demonstrate usage of the testbench and also serve as an example on how to implement a score follower that complies with the follower-testbench interface described in \autoref{tb_qual_follower_testbench_interface}. Effectively, this mock follower serves as a simple toy example that helps users understand the simple interface, which is core for the adoption of this testbench.

\subsection{QualScofo: A Qualitative Score Following Dataset}
\label{tb_qual_prod_dataset}

For the purpose of qualitative evaluations using this testbench, a dataset--the \textit{QualScofo} (\textbf{Qual}itative \textbf{Sco}re \textbf{fo}llowing) dataset--comprising a variety of music pieces with different instrument configurations is compiled\footnote{The dataset can be found at \url{https://github.com/flippy-fyp/QualScofo}.}.

The dataset comprises five \textit{groups} each made up of several \textit{pieces} with similar instrument configurations. There are 16 pieces in total. \textit{Pieces} are usually short excerpts (up to 2 minutes). Each piece may be uniquely identified with its Group ID and Piece ID. Tables~\ref{table:qual_dataset_1} and \ref{table:qual_dataset_2} give detailed information on each piece in the dataset. The tables are sorted by \textbf{Group ID}, \textbf{Piece ID} and \textbf{Year} (of composition). More details pertinent in the context of score following for each piece are provided in the dataset's usage in \autoref{impl_dtw_eval_qual}.

As can be seen, the \textit{QualScofo} dataset covers a wide range of classical music encompassing music composed between 1717 and 1905. Further, the instrumentation configurations range from unaccompanied violin and cello to romantic-era symphony orchestras. Some performances are recorded in a studio, while others are live performance recordings. The most notable limitation of this dataset is that it does not span a large range of performer skill levels--this is due to the difficulty of procuring recordings by less-skilled players.

The scores are obtained from \textit{MuseScore.com}\footnote{Not to be confused with its sister organisation--\textit{MuseScore.org}--that produces the score-editing program \textit{MuseScore}.} and converted into the MIDI, MusicXML and PDF formats; the corresponding performance audio files are obtained via \textit{YouTube}\footnote{\url{https://www.youtube.com/}. Downloadasaur (\url{https://www.downloadasaur.com}) was used to obtain the audio representations of the performance videos.}. As this dataset is made for research purposes, it falls under the fair use allowance\footnote{Under section 107 of the Copyright Act of 1976, allowance is made for ``fair use'' for purposes such as criticism, comment, news reporting, teaching, scholarship, education and research. Fair use is a use permitted by copyright statute that might otherwise be infringing.}.
The source \textit{MuseScore.com} and \textit{YouTube} URLs for each piece can be formed via the following format: \texttt{\url{https://bit.ly/qualscofo-(musescore|youtube)-<GROUP_ID>-<PIECE_ID_PREDASH>}}. \texttt{<GROUP\_ID>} denotes the Group ID, and \texttt{<PIECE\_ID\_PREDASH>} denotes the Piece ID characters prior to the first occurring dash (\texttt{-}), if any. For instance, the \textit{YouTube} URL for the \texttt{chaconne-arp} piece in the \texttt{violin} group is \url{https://bit.ly/qualscofo-youtube-violin-chaconne}. Note that the performance audio of the pieces \texttt{prelude} and \texttt{fugue} in the \texttt{piano} group are not obtained from \textit{YouTube} but from the MAESTRO dataset \cite{MAESTRO}.

\afterpage{%
    \clearpage
    \thispagestyle{empty}
    \begin{landscape}

        \begin{table}
            \centering
            \caption{Details of the qualitative dataset (Part 1 of 2).}
            \label{table:qual_dataset_1}
            \begin{tabularx}{\linewidth}{|l|l|l|L|L|L|L|L|L|L|}
                \hline
                \textbf{Group ID}                                      &
                \textbf{Piece ID}                                      &
                \textbf{Year}                                          &
                \textbf{Piece Name}                                    &
                \textbf{Composer}                               &
                \textbf{Instrumentation}                               &
                \textbf{Excerpt Range}                                 &
                \textbf{Notes}
                \\
                \hline
                \texttt{cello}                                         &
                \texttt{suite1}                                        &
                1717-23                                                &
                Cello Suite No. 1 in G Major (Prélude) BWV1007         &
                Johann Sebastian Bach                                  &
                Solo Cello                                             &
                Bars 1-22                                          &
                Studio recording by professional cellist Yo-Yo Ma.
                \\
                \hline
                \texttt{octet}                                         &
                \texttt{mendelssohn}                                   &
                1825                                                   &
                String Octet in E-flat major, Op. 20                   &
                Felix Mendelssohn                                      &
                String Octet                                           &
                Bars 1-21                                          &
                Live performance recording by professional musicians.
                \\
                \hline
                \texttt{orchestra}                                     &
                \texttt{eine}                                          &
                1787                                                   &
                Eine kleine Nachtmusik, K. 525                         &
                Wolfgang Amadeus Mozart                                &
                Chamber String Orchestra                               &
                Bars 1-23                                          &
                Live performance recording by professional musicians.
                \\
                \hline
                \texttt{orchestra}                                     &
                \texttt{peer}                                          &
                1875                                                   &
                Peer Gynt, Op. 23                                      &
                Edvard Grieg                                           &
                Romantic-era Symphony Orchestra                        &
                Bars 1-29                                          &
                Studio recording by Berliner Philharmoniker Orchestra under Herbert von Karajan.
                \\
                \hline
                \texttt{orchestra}                                     &
                \texttt{1812-1}                                        &
                1880                                                   &
                1812 Overture, Op. 49                                  &
                Pyotr Ilyich Tchaikovsky                               &
                Romantic-era Symphony Orchestra, Brass Band, Artillery &
                Bars 358-379                                       &
                Studio recording, unknown orchestra, but most likely Morton Gould Orchestra and Band.
                \\
                \hline
                \texttt{orchestra}                                     &
                \texttt{1812-2}                                        &
                1880                                                   &
                1812 Overture, Op. 49                                  &
                Pyotr Ilyich Tchaikovsky                               &
                Romantic-era Symphony Orchestra, Brass Band, Artillery &
                Bars 380-422                                       &
                Studio recording, unknown orchestra, but most likely Morton Gould Orchestra and Band.
                \\
                \hline
                \texttt{piano}                                         &
                \texttt{fugue}                                         &
                1722                                                   &
                Fugue in C major, BWV 846                              &
                Johann Sebastian Bach                                  &
                Piano                                                  &
                Full piece                                             &
                Obtained from the MAESTRO dataset \cite{MAESTRO}.
                \\
                \hline
                \texttt{piano}                                         &
                \texttt{prelude}                                       &
                1722                                                   &
                Prelude in C major, BWV 846                            &
                Johann Sebastian Bach                                  &
                Piano                                                  &
                Full piece                                             &
                Obtained from the MAESTRO dataset \cite{MAESTRO}.
                \\
                \hline
            \end{tabularx}
        \end{table}
    \end{landscape}
    \clearpage
}

\afterpage{%
    \clearpage
    \thispagestyle{empty}
    \begin{landscape}

        \begin{table}
            \centering
            \caption{Details of the qualitative dataset (Part 2 of 2).}
            \label{table:qual_dataset_2}
            \begin{tabularx}{\linewidth}{|l|l|l|L|L|L|L|L|L|L|}
                \hline
                \textbf{Group ID}                                                  &
                \textbf{Piece ID}                                                  &
                \textbf{Year}                                                      &
                \textbf{Piece Name}                                                &
                \textbf{Composer}                                           &
                \textbf{Instrumentation}                                           &
                \textbf{Excerpt Range}                                             &
                \textbf{Notes}
                \\
                \hline
                \texttt{piano}                                                     &
                \texttt{turkish}                                                   &
                1783                                                               &
                ``Turkish March'' from Piano Sonata No. 11 in A major, K. 331/300i &
                Wolfgang Amadeus Mozart                                            &
                Piano                                                              &
                Bars 1-25 (Movement III)                                       &
                Studio recording by professional pianist Rousseau.
                \\
                \hline
                \texttt{piano}                                                     &
                \texttt{moonlight}                                                 &
                1801                                                               &
                Piano Sonata No. 14 ``Moonlight'', Op. 27                          &
                Ludwig van Beethoven                                               &
                Piano                                                              &
                Bars 1-16 (Movement I)                                         &
                Live performance recording by Claudio Arrau.
                \\
                \hline
                \texttt{piano}                                                     &
                \texttt{unsospiro}                                                 &
                1845-49                                                            &
                ``Un sospiro'' from Three Concert Études, S.144                    &
                Franz Liszt                                                        &
                Piano                                                              &
                Bars 1-10 (Étude No. 3)                                        &
                Live performance recording by Marc-André Hamelin
                \\
                \hline
                \texttt{piano}                                                     &
                \texttt{gnossienne}                                                &
                1890                                                               &
                Gnossienne No. 1                                                   &
                Erik Satie                                                         &
                Piano                                                              &
                Bars 1-35 (Assuming each bar comprises 4 beats)                &
                Studio recording by professional pianist Alessio Nanni.
                \\
                \hline
                \texttt{piano}                                                     &
                \texttt{entertainer}                                               &
                1902                                                               &
                The Entertainer                                                    &
                Scott Joplin                                                       &
                Piano                                                              &
                Bars 1-16                                                      &
                Audio recorded from a piano roll (pianola).
                \\
                \hline
                \texttt{piano}                                                     &
                \texttt{clair}                                                     &
                1905                                                               &
                ``Clair de Lune'' from Suite bergamasque, L. 75                    &
                Claude Debussy                                                     &
                Piano                                                              &
                Bars 1-17                                                      &
                Studio recording by professional pianist Lang Lang.
                \\
                \hline
                \texttt{violin}                                                    &
                \texttt{chaconne-arp}                                              &
                1717-1720                                                          &
                ``Chaconne'' from Partita in D minor for solo violin, BWV1004      &
                Johann Sebastian Bach                                              &
                Solo Violin                                                        &
                Bars 84-123 (Chaconne)                                         &
                Studio recording by professional violinist Hilary Hahn.
                \\
                \hline
                \texttt{violin}                                                    &
                \texttt{chaconne-front}                                            &
                1717-1720                                                          &
                ``Chaconne'' from Partita in D minor for solo violin, BWV1004      &
                Johann Sebastian Bach                                              &
                Solo Violin                                                        &
                Bars 1-24 (Chaconne)                                           &
                Studio recording by professional violinist Hilary Hahn.
                \\
                \hline
            \end{tabularx}
        \end{table}
    \end{landscape}
    \clearpage
}

%% file: 02_parts/03_impl/01_impl.tex
\part{Implementations}
\label{impl}

\input{02_parts/03_impl/chapters/01_beattracking.tex}
\input{02_parts/03_impl/chapters/02_asm.tex}
\input{02_parts/03_impl/chapters/04_dtw.tex}

%% file: 02_parts/03_impl/chapters/01_beattracking.tex
\chapter{A Beat Tracking Approach: The TuneApp Conductor}
\label{impl_beat}
\epigraph{\itshape If the difference between 1911 and 2011 is electricity and computation, then Max Mathews is one of the five most important musicians of the 20th Century.}{Miller Puckette}

Beat-tracking-based score followers were briefly mentioned in subsections~\ref{scofo_relwork_temporal} and \ref{scofo_relwork_partial}. These type of trackers are, of course, suboptimal as they do not use reliably available musical information available in the sheet music. Nevertheless, beat tracking itself is a widely studied field in MIR \cite{matthewdavies19}, and there exists \textit{two} beat-tracking-related MIREX evaluations \cite{mirexaudiobeattracking,mirexaudiodownbeatestimation}. While not directly related to work that will be presented in the subsequent chapters, there exists motivation in creating a simple beat-tracking-based score follower.

\section{Motivation}
\subsection{Author's Backstory}
\label{impl_beattrack_backstory}

\textit{In this subsection, ``I'' shall refer to the author.}

I am not a musician, but I used to play the violin and some piano. Growing up with electrical engineering influences (thanks to my father, who worked at \textit{AT\&T Bell Labs}) eventually led me to read Jon Gertner's \textit{The Idea Factory: Bell Labs and the Great Age of American Innovation} \cite{gertner2012idea} (a key factor in my acceptance at Imperial College London--I discussed contents of this book with my interviewer). Bell Labs had no shortage of legends, but one that struck me the most was Max Mathews: undoubtedly the father of computer music. Max Mathews gave the music community so much--from little toys like his \textit{Radio-Baton} to now-ubiquitous synthesisers. He showed that it is possible to fuse my musical and engineering interests to create fun and interactive engineering projects.

\begin{figure}[h]
    \centering
    \includegraphics[width=0.5\columnwidth]{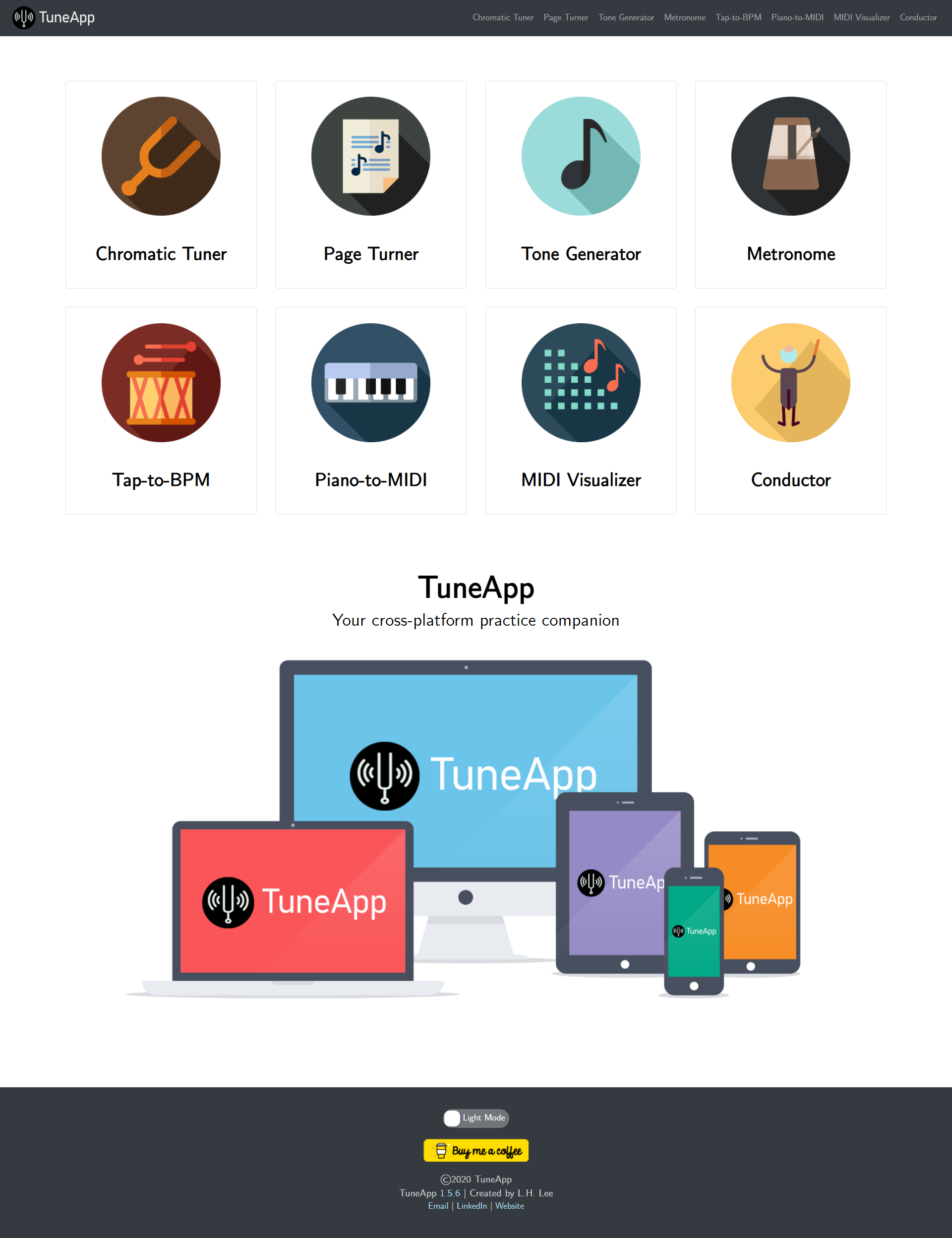}
    \caption{Desktop home page of \textit{TuneApp 1.5.6 (November 2020)}}
    \label{fig:tuneapp156}
\end{figure}

Fast-forward some years, \textit{TuneApp}\footnote{\url{https://www.tuneapp.co/}}--a widely available, cross-platform, music practice companion with core music practice features like a chromatic tuner, metronome, page turner, beat detector etc. was born. The home page of \textit{TuneApp} is shown in \autoref{fig:tuneapp156}. In fact, I plan to incorporate score follower systems introduced in the work in the page turner (manual as of time of writing). 

\begin{figure}[h]
    \centering
    \includegraphics[width=0.3\columnwidth]{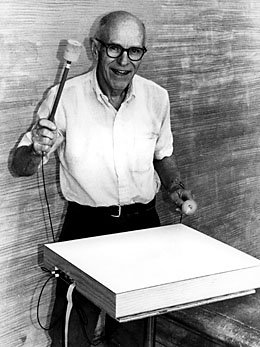}
    \caption{Max Mathews waving the \textit{Radio-Baton}.}
    \label{fig:mathewsradiobaton}
\end{figure}

One of Max Mathews' work that I particularly liked was the \textit{Radio-Baton} \cite{radiobaton}--an instrument that allows the user to ``conduct'' music. \autoref{fig:mathewsradiobaton} shows Mathews waving the \textit{Radio-Baton}. 

The high engineering sophistication required for the \textit{Radio-Baton} begs the question:
\begin{displayquote}
Now that sensors that could deliver the same information required for the functionality of the \textit{Radio-Baton} are more common (and even embedded into mobile devices), can a similar contraption be easily made in today's widely available technology? 
\end{displayquote}

The answer, spoiler alert, is \textit{yes}. This chapter of the report details one weekend I spent for this score following-related work.

\subsection{A different Score Following Model}

Work in this section via a simpler beat tracking model serves to show how the technology stack of a score follower should look like, in addition to giving a practical example of how the generic framework of score following (\autoref{scofo_generic}) is fulfilled.

\section{Requirements}
\label{impl_beattrack_req}

Requirements that are partly fulfilled by the \textit{Radio-Baton} are captured to create a ``conductor'' experience. In this case, no specialised hardware is required. The beats are tracked via common sensors widely available on user devices--be it PCs or mobile phones.

\subsection{Beat Tracking on User Input}
A core part of conducting is to provide the fundamental \textit{beats} of the music to be followed by the player. This application shall be able to track beats via user input. 

Users on a PC/Mac shall be able to use a button to tap, whereas users on accelerometer-enabled mobile phones (most modern smartphones) shall be able to wave their phones to indicate the beats of the music.

\subsection{Real-time Beat-sensitive Music Playback}

After tracking beats input by the user, the application shall have a music player that can play back--in real time--the music at the tempo derived from the user input. The music is provided by the user; three forms of input are supported: \nameref{scofo_mdr_midi}, \nameref{scofo_mdr_audio} and \nameref{scofo_mdr_musicxml}, each in a separate interface.

\subsection{Music Playback Visualisation}

In addition to auditory feedback on user input, a visualisation, core to the \textit{score following} aspect of the application, is provided. This is similar to the visual feedback provided in the qualitative testbench (\autoref{tb_qual}).

\section{Technical Background}

The technical background on how the requirements defined in \autoref{impl_beattrack_req} are fulfilled is first provided.

\subsection{Beat Tracking on User Input}

\subsubsection{Algorithm to calculate instantaneous tempo}

The metric of interest to be measured from user input is the \textbf{instantaneous tempo}. 

    \begin{definition}[Instantaneous tempo]
        The instantaneous tempo, $\hat{\tau}$, in $\text{BPM}$ (beats per minute) is defined as
        $$
        \hat{\tau} = \frac{60}{t_{b,n} - t_{b,n-1}},
        $$
        where $t_{b,n}$ and $t_{b,n-1}$ are the times (measured in $\si{\second}$) 
        of the last two detected beats.
        \label{def:insttempo}
    \end{definition}

$\text{BPM}$ is a common unit for tempo widely used in music. As this equation requires two beats, the tempo prior to receiving two beats is set to $\hat{\tau}_0$: the music's original tempo (detected from MIDI, Audio or MusicXML).

There exists possibilities of creating a tempo model based on information recorded prior to the two most recent beats, but experimentally this did not bring about good results--real conductors and musicians respond most sensitively to the most recent beats. Further, when the next beat is not detected, the music is usually played at the last indicated tempo--thus motivating the use of the instantaneous tempo here. Another possibility is to incorporate a temporal model that slows down the music exponentially over time when no further beats are detected, but again this is not natural: musicians usually continue playing at the set tempo unless an indication to stop is provided by the conductor.

\subsubsection{Button Input: All Web devices}

A button will be provided for users to ``tap'' on--this can be via a mouse click on a PC or Mac, or a touch on a touchscreen device. This registers the time of the beats ($t_{b,n}$ and $t_{b,n-1}$).

\subsubsection{Accelerometer: Most modern mobile devices}

In addition to a button input, an accelerometer--common in modern mobile devices--can be used to measure acceleration in three axes. This can thus be used to calculate whether a strong and sudden change in acceleration is detected in any direction (which is the primary sign of a beat used by conductors).

Taking the two most recent sets of acceleration in three directions as $\{a_{d,n}\}$ and $\{a_{d,n-1}\}$ for $d \in \{x,y,z\}$, the \textbf{absolute change in acceleration} is derived:

\begin{definition}[Absolute change in acceleration]
The absolute change of acceleration, $\delta_{|a|}$, is calculated as follows:
$$
\delta_{|a|} = \sum_{d=x,y,z} |a_{d,n} - a_{d,n-1}|.
$$
\label{def:abschgaccel}
\end{definition}

Moving on, what constitutes a \textbf{detected beat} is defined:

\begin{definition}[Detected beat]
A beat shall be treated as detected if $\delta_{|a|}$ is larger than a user-adjustable threshold $\theta_a$ (important due to hardware differences and user preferences). 
\end{definition}

When a beat is detected, the system is suspended for a set time delay: $T_{delay} = \si{200\milli\second}$ (the gap translates to $300\text{BPM}$). This is used to cap the tempo at a sensible maximum of $300\text{BPM}$, and also serves to stop erroneous series of beats being detected in a single intended beat.

During beat tracking, the accelerometer sensors are constantly polled to detect changes in acceleration to update $\delta_{|a|}$, save for the period when the system is suspended for $T_{delay}$.

\subsection{Real-time Beat-sensitive Music Playback}

Players for MIDI, Audio and MusicXML that are able to play music at the indicated instantaneous tempo $\hat{\tau}$ are required. 

MIDI and MusicXML contain score data that indicates beats, so implementing change of speed is trivial--the synthesiser just needs to play the indicated notes at a different tempo.

The Audio component is difficult for two reasons:
\begin{enumerate}
    \item \textbf{Non-pitch-shifting speed change}: A naïve and fast time-stretching method to change the speed of audio would cause a pitch shift. Transposition, or pitch scaling, is required to counter the pitch-shifting effects.
    \item \textbf{Beat detection}: The playback speed is determined by \textit{aligning} beats detected from the user to beats on the audio. A beat detection algorithm is therefore required to extract beat data from the audio recording.
\end{enumerate}

First, the more trivial non-pitch-shifting speed change problem is tackled. Firstly, the \textbf{target playback rate} is defined:

\begin{definition}[Target playback rate]
First, take $\tau_a$ as the audio's detected tempo (for simplicity's sake, assume that the audio has constant tempo, but note that this system is capable of processing audio with varying tempo). $\tau_d$ is defined as the desired audio tempo (derived from user input). The target playback rate is therefore 
$$\frac{\hat{\tau}_d}{\tau_a}.$$ 
\label{def:targplaybackrate}
\end{definition}

A playback rate of $2$ without pitch adjustments means that the frequency of the audio will be \textit{doubled} (the wavelengths of the audio frequencies are halved): the resulting audio is said to be an \textit{octave} higher than the original. Therefore, it is required to \textit{transpose} (pitch adjust/scale) \cite{rai19} the resulting audio, effectively halving the frequencies, to compensate for the pitch shift resulting from the change of playback rate. 

Moving on, the harder problem of audio beat detection is resolved. A beat detector is required for this. A real-time beat detector can be used, but is not paramount to this application; the beats in the audio can be detected more reliably using an offline algorithm prior to the user's conducting.

\subsection{Music Playback Visualisation}

A visual interface showing the tracking of the beats and the response of the application greatly improves user interaction and experience. The type of music visualisation depends on the available musical information. For MIDI, note information can be generated in a scrolling piano roll with the currently played note(s) highlighted. Audio, on the other hand, does not contain note information--the waveform or the spectrogram can be instead shown with a playback marker placed on the current audio position. MusicXML is the most straightforward--the interface is similar to the qualitative testbench's described in \autoref{tb_qual}: the rendered score is shown with a marker denoting the score position.

\section{Score Follower Framework}

The application fulfils the generic score follower framework described in \autoref{scofo_generic}. The \textbf{features extracted} from the score and ``performance'' (in this case conducting) are \textit{beats}. The \textbf{similarity calculation} step is fused with the \textbf{alignment} step, in which the music playback is synchronised by aligning the \textit{instantaneous tempo} in the score and the user's conducting.

It follows that this tracker is of the \textbf{symbolic-to-symbolic} type--the follower works by matching symbolic temporal information extracted from the conducting and the score.

Notice also that the system could be modified to use actual musical performance audio as the input--this would require a real-time beat tracking solution to detect beats in the performance audio.

\section{Implementation}
\label{impl_beattrack_impl}

As this application is an integration to the existing \textit{TuneApp} Web stack, the technologies used are therefore the same, i.e. React\footnote{\url{https://reactjs.org/}} and TypeScript\footnote{\url{https://www.typescriptlang.org/}}. While \textit{TuneApp} is kept proprietary, the source code is available for non-commercial purposes upon request.

To access device acceleration data, the well-defined \texttt{DeviceMotionEvent}\footnote{\url{https://developer.mozilla.org/en-US/docs/Web/API/DeviceMotionEvent}} Web API that offers applications access to the data is used. As for non-pitch-shifting playback rate changing audio playback, the Web Audio API\footnote{\url{https://developer.mozilla.org/en-US/docs/Web/HTML/Element/audio}} is used. This API gives access to audio processing implementations, often native and efficient, by different browsers.

To detect the beats in an audio file, a Web implementation of a fast and reliable non-real-time method proposed by Simon Dixon in 2001, \textit{BeatRoot} \cite{beatroot}, is used. \textit{BeatRoot} is capable of detecting beats from a wide range of musical styles, including classical, jazz and modern. \textit{BeatRoot} works by detecting and analysing rhythmic events to generate tempo hypotheses at various metrical levels--the best fit is chosen from these hypotheses. As \textit{BeatRoot} works on only a single channel of audio, audio data from all channels are averaged before being fed into the system.

The MIDI visualisation is built upon prior work in \textit{TuneApp}'s \textit{MIDI Visualizer}\footnote{\url{https://www.tuneapp.co/visualizer}}, which is in turn based on the \texttt{Magenta.js}\footnote{\url{https://magenta.github.io/magenta-js/music/index.html}} library. The \texttt{SGM} soundfont\footnote{\url{https://www.polyphone-soundfonts.com/documents/27-instrument-sets/256-sgm-v2-01}} is used for MIDI synthesisation. The interface is a piano roll that scrolls with the music, see the implementation in \autoref{fig:tuneappconductordeployed}(a): currently active notes are \textit{magenta}, and other notes are \textit{grey}. The Audio visualisation, on the other hand, is a waveform of the average audio over all channels, together with grey vertical lines denoting instances of the detected beats. The waveform scrolls in line with the playback, with a red vertical line showing the current playback position, see \autoref{fig:tuneappconductordeployed}(b). A MusicXML display library is used for MusicXML visualisation\footnote{\url{https://opensheetmusicdisplay.github.io/}}--the playback shows the sheet music in addition to a green marker showing the current position in the score--see \autoref{fig:tuneappmxmlconductor}.

For each section, users can upload their own files or use a sample file provided. 

\section{Results}

Readers interested in trying the \textit{TuneApp Conductor} can do so by visiting \url{https://www.tuneapp.co/conductor}. 

\subsection{Fulfilment of Requirements}

Following the implementation details in \autoref{impl_beattrack_impl}, all requirements listed in \autoref{impl_beattrack_req} are met. A small exception is that the MusicXML-based interface, while complete and working, is not released to the public--\autoref{impl_beattrack_appuse} contains more information on this.

\subsection{Application Usage and Interface}
\label{impl_beattrack_appuse}

\begin{figure}[h]
    \centering
    \subfloat[\centering The MIDI-based conductor]{{\includegraphics[width=5cm]{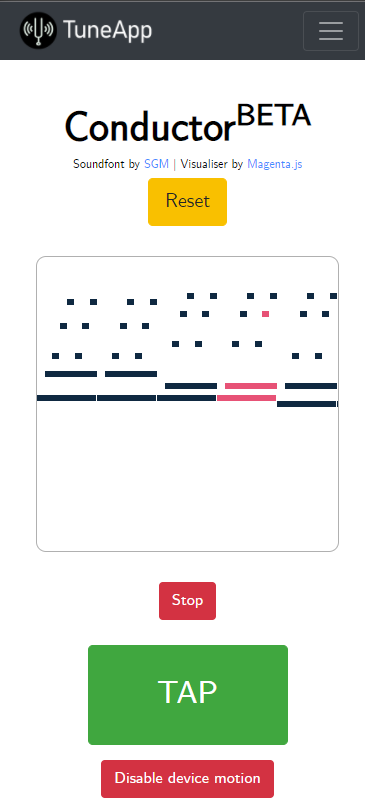} }}
    \qquad
    \subfloat[\centering The audio-based conductor]{{\includegraphics[width=5cm]{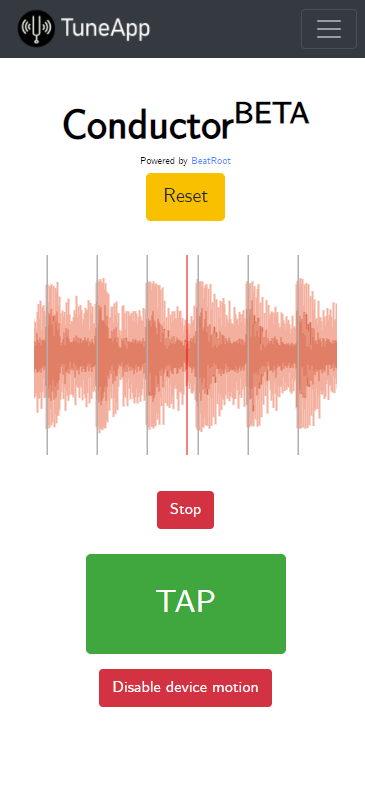} }}
    \caption{The two released variants of the \textit{TuneApp Conductor}.}
    \label{fig:tuneappconductordeployed}
\end{figure}

\autoref{fig:tuneappconductordeployed} shows the two variants released for users: the MIDI-based conductor (a) and the audio-based one (b). Users on their mobile devices can either tap the green \textbf{TAP} button, or start conducting by treating their devices as a conducting baton. Users on a PC or Mac can only use the button.

\begin{figure}[h]
    \centering
    \includegraphics[width=0.5\columnwidth]{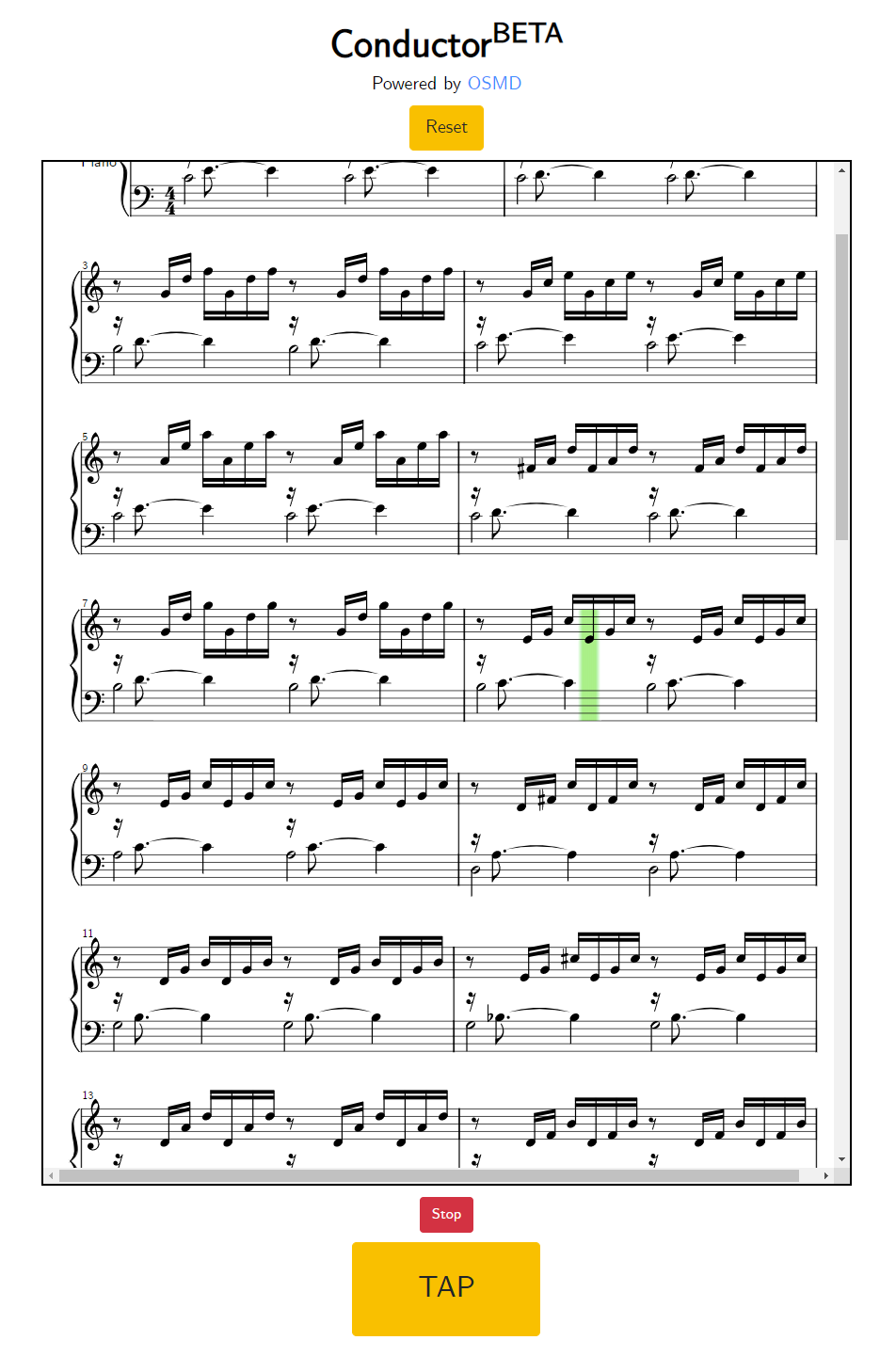}
    \caption{The MusicXML-based \textit{TuneApp Conductor}, not released to users.}
    \label{fig:tuneappmxmlconductor}
\end{figure}

The MusicXML-based conductor was not deployed, as it was hard to optimise for mobile: even the most modern devices that were tested on showed significant sluggishness in the application, likely due to the large MusicXML rendering power required on top of the resources needed for beat-aligned music playback. Nevertheless, it works well on a modern laptop--see \autoref{fig:tuneappmxmlconductor}, but it was not released to the public: \textit{TuneApp}'s core goal is to be as cross-platform as possible.

\subsection{User Feedback}
\label{impl_beattrack_feedback}

User feedback collected by informal survey show that the \textit{TuneApp Conductor} is well received. Users liked the idea in general, and the performant loading and functioning of the application is often mentioned. Some comments and reviews received include:
\begin{itemize}
    \item ``This is such a good idea! It's so fun to use and works very fast!'' 
    \item ``I am surprised that this works so well on my phone!''
    \item ``This is great for people who are `closet conductors' (like me), but most notably it has potential to be used for training musicians' rhythmic abilities.''
    \item ``Such a cool use for common sensors we easily take for granted these days.''
\end{itemize}

Users surveyed enjoyed both the MIDI- and Audio-based conductors, indicating no preference of one to the other. A select few testers allowed to test the MusicXML-based conductor echoed the sentiments mentioned in \autoref{impl_beattrack_appuse}: the MusicXML rendering and playback was not responsive and performant enough. Hence, the feature was dropped entirely, but may come back in the future--either by optimisation of the rendering and playback of the interface, or simply via the increase in computation power allowing more powerful devices to run it better.

\subsection{Evaluation}

As this score follower does not produce note onset information, the quantitative testbench (\autoref{tb_quant}) cannot be used. Note, however, that the qualitative testbench (\autoref{tb_qual}) is in fact already built into this approach via the MusicXML-based interface. In addition, qualitative feedback is obtained from the two other interfaces. Extensive user testing is used and informal surveying is performed to collate qualitative feedback. User feedback is as reported in \autoref{impl_beattrack_feedback}.

Interestingly, one user compared the application to one of Google's experiments: the \textit{Semi-Conductor}\footnote{\url{https://semiconductor.withgoogle.com/}}. While the \textit{Semi-Conductor} features more nuances in conducting as it uses pose detection to detect ``conducting information'' via the users' cameras, the user did not find Google's approach responsive enough--it is after all an application that is intended to demonstrate \textit{PoseNet}--a Web-based pose detection model \cite{posenet}--running in-browser rather than to provide a musically focused application, which is the \textit{TuneApp Conductor}'s core goal.

%% file: 02_parts/03_impl/chapters/02_asm.tex
\chapter{An Approximate String Matching (ASM) Approach}
\label{impl_asm}
\epigraph{\itshape
    Music helps immensely with math skills, and math skills help immensely with music skills.}{Michael Giacchino}

This chapter goes back to basics to create a basic music alignment system via Approximate String Matching, a notion popular in early score followers. Performance of such score followers and aligners are inferior to the many score followers introduced after the first ten years of score following, but for this project there are motivations to introduce a simple alignment system.

\section{Motivation}

The creation of the ground-truth alignment dataset for the quantitative testbench (via widely available performance datasets such as the MAESTRO \cite{MAESTRO} and MAPS \cite{emiya10} datasets) introduced in \autoref{tb_quant} requires the alignment of a performance MIDI sequence to MIDI data extracted from a score. This alignment can either be done by tedious manual alignment, or by the use of an Approximate String Matching (ASM) aligner. This latter method would allow straightforward generation of reference ground-truth alignment datasets, with minimal manual adjustments required post-alignment. Thus, this ASM aligner is a core part of the new data procurement method explained in \autoref{tb_quant_dataset_api}.

In addition, such an aligner can be converted into a score follower, but most importantly, an ASM aligner--being simple and easy to explain--is important in the context of this report to help develop complex concepts used in more advanced approaches proposed in the next chapter.

\section{Requirements}
\label{impl_asm_req}

\subsection{Offline MIDI Data Alignment}
The primary goal of this system remains to complement the quantitative testbench introduced in \autoref{tb_quant}. The system shall, given MIDI data obtained from performance recordings on MIDI-capable instruments as well as reference MIDI data from scores, provide offline alignments.

\subsection{Robust Score Alignment}
\label{impl_asm_robust}
As offline score aligners share many of their online score follower counterparts' challenges, which are covered in \autoref{scofo_challenges}, the aligner shall produce alignments that are high quality and robust to performance deviations. Note also that this aligner is intended to process performance in datasets, which usually have limited performance deviations \cite{MAESTRO,emiya10}.

\subsection{Misalignment Indication}
\label{impl_asm_indicate}
This system shall produce indication of misalignments that it could not resolve automatically to aid manual adjustments. 


\subsection{Input Data Size}
\label{impl_asm_input_size}
The aligner shall work reasonably efficiently with a pair of sequences, each of length up to $10000$ elements. This number is arrived at after computing the average number of notes per performance in the MAESTRO dataset ($5517$) \cite{MAESTRO}.

\subsection{Output Format}
This system shall produce alignments in the reference format required by the quantitative testbench, described in \autoref{tb_quant_dataset_api}. In short, each aligned note is described in a line with three columns denoting the true note onset time in the performance, the note onset time in the score and the MIDI note number of the note respectively.

\section{Technical Background}
\label{impl_asm_techback}
\subsection{Approximate String Matching (ASM)}
\label{impl_asm_techback_asm}


The crux of robust symbolic-to-symbolic score alignment and score following systems is \textbf{Approximate String Matching (ASM)}. ASM is more commonly known as \textit{sequence alignment} in the bioinformatics field where the establishment of homologies between genomic sequences is of large interest \cite{marcosola17}. In fact, the third chapter in \cite{marcosola17} as well as Navarro's 2001 review \cite{navarro01} of ASM provide excellent overviews on ASM, useful for the interested reader. ASM is also colloquially referred to in the computer science community as \textit{fuzzy string searching}.

Applied in both score following and score alignment, both online and offline ASM algorithms exist. This aligner deals with the \textit{offline} counterpart. Algorithms may also be classified as \textit{global} or \textit{local}--the former means that the ends of the sequences are not \textit{free}: the final shape of the alignment spans throughout both sequences attempting to match them completely; the latter is the opposite: the ends of the sequences are \textit{free}. \textit{Semi-global} (or ``glocal'') aligners is another classification of interest: it works with aligning a smaller sequence end-to-end to a local region of the larger sequence, in which the larger sequence has free ends.

ASM score followers may use a special form of an online semi-global aligner--this method can make it robust to skips and repeats. ASM score aligners on the other hand use offline global ASM algorithms--score and performance data are usually sequences that are similar and of roughly equal size.

Clearly, identical sequences are trivial to align. When sequences differ, as in the case of performance data mismatching with score data, it is necessary to understand how these errors are dealt with. Before that, score alignment is first formulated as an ASM problem.

\subsubsection{Problem Formulation}

\begin{definition}[Score alignment as an ASM problem]
    The score's MIDI data is first represented as a string $S = s_1 s_2 \ldots s_n$ and the performance MIDI data as a string $P = p_1 p_2 \ldots p_m$. The goal is to find the best scoring \textit{global alignment} between $S$ and $P$, giving the respective resulting strings $S' = \sigma_1 \ldots \sigma_\nu$ and $P' = \pi_1 \pi_2 \ldots \pi_\nu$, where $S'$ and $P'$ are \textit{substrings} of $S$ and $P$ respectively with \textit{gaps} allowed in them.
    The aligned result can therefore be represented as a list of tuples
    $$L = [(\sigma_1,\pi_1), \ldots, (\sigma_\nu,\pi_\nu)].$$
    For any arbitrary tuple in $L$, the non-gap elements may \textit{match}, \textit{mismatch} or be matched to a \textit{gap} $\Gamma$. The final case is referred to as an \textit{indel} (INsertion or DELetion).
    \label{def:score_alignment_asm}
\end{definition}

Defining the set of musical notes $N \in \{A, B, C, D, E\}$, an example of aligning score and performance data is given. Say that $P = ABCDABE$ and $S = ACDDCBC$, an alignment may be $P' = ABCDA \Gamma BE$ and $S' = A \Gamma CDDCBC$, giving
$$L = [(A,A), (B, \Gamma), (C, C), (D, D), (A, D), (\Gamma, C), (B, B), (E, C)].$$ $(A,A)$ denotes a \textit{match}, meaning that the right note is played. $(B, \Gamma)$ and $(\Gamma, C)$ are cases of \textit{indel}; the former indicating an extra note not in the score being played, the latter a missed note in the score not being played. $(A, D)$ indicates a \textit{mismatch}: a wrong note $A$ played instead of the correct note $D$.



\subsubsection{Scoring System}




There are three ways in which non-gap elements in tuples of $L$ may be matched. Given an arbitrary tuple $(\sigma,\pi)$ where $\sigma \neq \Gamma$ (i.e. $\sigma$ is not a gap element), there are three cases:
\begin{enumerate}
    \item $\sigma = \pi$, \textbf{Match}: the two elements are the same.
    \item $\sigma \neq \pi$, \textbf{Mismatch}: the two elements are different.
    \item $\pi = \Gamma$, \textbf{Indel}: the best alignment involves aligning $\sigma$ to a gap.
\end{enumerate}

Notice that a tuple may never be both gaps, as deleting the tuple will yield a better alignment. The cases above therefore generalise for tuples where $\pi \neq \Gamma$.

Moving on, how these three cases affect the \textit{scoring function} to be maximised is detailed. A \textbf{match} is not an error and therefore should be assigned a zero or positive score. \textbf{Mismatch} and \textbf{indel} errors are assigned negative values. These values can be tuned according to the data or use case; for example, for alignments where gaps are undesirable, the indel score (which is also known as the \textit{gap penalty}) can be significantly negative. Needleman-Wunsch assigns 1 to match and -1 for mismatch and indel \cite{needleman70}.

\subsubsection{Algorithm Presentation}

The alignment problem set forth can be brute-forced. While the alignment here is not time sensitive, the problem quickly becomes intractable for large sequences (which is in this case, see \autoref{impl_asm_input_size}). The ways of aligning two sequences of lengths $m, n$ is given by
$$
    \binom{n+m}{m}\frac{(m+n)!}{(m!)^2}.
$$
For $n = m = 100$, this evaluates to approximately $9 \times 10^{58}$. Assuming each enumeration takes one millisecond, the algorithm would take up to $3 \times 10^{48}$ \textit{years} to complete!

Thus, the algorithm's core part is based on the Needleman-Wunsch algorithm \cite{needleman70}, an efficient algorithm that performs ASM besides being a popular example of dynamic programming. For now, the Needleman-Wunsch algorithm's scoring system is used. An example of the use of the proposed algorithm is given: the goal is to produce a score alignment of the sequences $P = ABCDABE$ and $S = ACDDCBC$ where the set of musical notes is $N \in \{A, B, C, D, E\}$. The steps required for alignment are outlined:

\begin{enumerate}
    \item \textbf{Grid Construction}. First a grid $G$ based on $P$ and $S$ is constructed. For convenience, the grid is indexed using a coordinate system $(x, y)$, where the last row of the first column is the origin $(0, 0)$--this cell is referred to as $G(0, 0)$. $x^+$ is in the rightward direction and $y^+$ is in the upward direction. $P$ is placed on the last row starting from $G(2, 0)$ to the right and $S$ is placed on the first column starting from $G(0, 2)$ to the top. Spaces are left for scores.

          \begin{tikzpicture}
              \matrix[matrix of nodes,nodes={draw=gray, anchor=center, minimum size=.6cm}, column sep=-\pgflinewidth, row sep=-\pgflinewidth] (A) {
                  $C$                        & \phantom{} & \phantom{} & \phantom{} & \phantom{} & \phantom{} & \phantom{} & \phantom{} & \phantom{}   \\
                  $B$                        & \phantom{} & \phantom{} & \phantom{} & \phantom{} & \phantom{} & \phantom{} & \phantom{} & \phantom{}   \\
                  $C$                        & \phantom{} & \phantom{} & \phantom{} & \phantom{} & \phantom{} & \phantom{} & \phantom{} & \phantom{}   \\
                  $D$                        & \phantom{} & \phantom{} & \phantom{} & \phantom{} & \phantom{} & \phantom{} & \phantom{} & \phantom{}   \\
                  $D$                        & \phantom{} & \phantom{} & \phantom{} & \phantom{} & \phantom{} & \phantom{} & \phantom{} & \phantom{}   \\
                  $C$                        & \phantom{} & \phantom{} & \phantom{} & \phantom{} & \phantom{} & \phantom{} & \phantom{} & \phantom{}   \\
                  $A$                        & \phantom{} & \phantom{} & \phantom{} & \phantom{} & \phantom{} & \phantom{} & \phantom{} & \phantom{}   \\
                  \phantom{} & \phantom{} & \phantom{} & \phantom{} & \phantom{} & \phantom{} & \phantom{} & \phantom{} & \phantom{}   \\
                  \phantom{} & \phantom{} & $A$                        & $B$                        & $C$                        & $D$                        & $A$                        & $B$                        & $E$ \\};
          \end{tikzpicture}

    \item \textbf{Table Filling}. First, the score for each cell is calculated row-by-row. Following from the scoring system set forth, parameters for the scores for a match, a mismatch and an indel are defined:

          \begin{definition}[Scores for the three error cases]
              \label{asm_three_error_def}
              $\alpha$, $\beta$ and $\gamma$ denote the scores for a match, mismatch and an indel respectively.
          \end{definition}

          In this example, $\alpha = 1$ and $\beta = \gamma = -1$.

          The \textit{Similarity Function}, $\textsc{Sim}$, is defined as:
          \begin{definition}[Similarity Function]
              \label{sim_func_defn}
              $$
                  \textsc{Sim}(c_1, c_2) = \begin{cases}
                      \alpha, & \text{if }c_1 = c_2 \\
                      \beta,  & \text{otherwise}    \\
                  \end{cases}
              $$
              where $c_1$ and $c_2$ are elements in the strings $P$ and $S$.
          \end{definition}

          The score function is defined as
          \begin{definition}[Score function] The score for a cell $G(x, y)$ is given by
              $$
                  \textsc{Score}(x, y) = \max
                  \begin{cases}
                      \textsc{Score}(x-1, y-1) + \textsc{Sim}(G(x, 0), G(0, y)) \\
                      \textsc{Score}(x-1, y) + \gamma                           \\
                      \textsc{Score}(x, y-1) + \gamma                           \\
                  \end{cases}
              $$
              where $\textsc{Score}(0, y) = \textsc{Score}(x, 0) = -\infty$ and $\textsc{Score}(1, 1) = 0$. The first line constitutes moving diagonally upwards and rightwards and matching (or mismatching) $G(x, 0)$ and $G(0, y)$, whereas the second line denotes moving rightwards by matching $G(0, y)$ to a gap. The final line indicates moving upwards by matching $G(x, 0)$ to a gap.
            \label{def:scorefunc}
          \end{definition}

          First, the simplest row $G(x, 1)$ and column $G(1, y)$ are filled:

          \begin{tikzpicture}
              \matrix[matrix of nodes,nodes={draw=gray, anchor=center, minimum size=.6cm}, column sep=-\pgflinewidth, row sep=-\pgflinewidth] (A) {
                  $C$                         & -7                          & \phantom{}  & \phantom{}  & \phantom{}  & \phantom{}  & \phantom{}  & \phantom{}  & \phantom{}   \\
                  $B$                         & -6                          & \phantom{}  & \phantom{}  & \phantom{}  & \phantom{}  & \phantom{}  & \phantom{}  & \phantom{}   \\
                  $C$                         & -5                          & \phantom{}  & \phantom{} & \phantom{} & \phantom{} & \phantom{} & \phantom{} & \phantom{}  \\
                  $D$                         & -4                          & \phantom{} & \phantom{} & \phantom{} & \phantom{} & \phantom{} & \phantom{} & \phantom{}  \\
                  $D$                         & -3                          & \phantom{} & \phantom{} & \phantom{} & \phantom{} & \phantom{} & \phantom{} & \phantom{}  \\
                  $C$                         & -2                          & \phantom{} & \phantom{} & \phantom{} & \phantom{} & \phantom{} & \phantom{} & \phantom{}  \\
                  $A$                         & -1                          & \phantom{} & \phantom{} & \phantom{} & \phantom{} & \phantom{} & \phantom{} & \phantom{}  \\
                  \phantom{} & 0                           & -1                          & -2                          & -3                          & -4                          & -5                          & -6                          & -7                           \\
                  \phantom{} & \phantom{} & $A$                         & $B$                         & $C$                         & $D$                         & $A$                         & $B$                         & $E$ \\};
          \end{tikzpicture}

          The necessary subgrid required to calculate $\textsc{Score}(2,2)$ is then extracted for simplicity:

          \begin{tikzpicture}
              \matrix[matrix of nodes,nodes={draw=gray, anchor=center, minimum size=.6cm}, column sep=-\pgflinewidth, row sep=-\pgflinewidth] (A) {
                  $A$                         & -1                          & \phantom{}  \\
                  \phantom{} & 0                           & -1                           \\
                  \phantom{} & \phantom{} & $A$ \\};
          \end{tikzpicture}

          Here,

          \begin{equation*}
              \begin{split}
                  \textsc{Score}(2, 2) &= \max
                  \begin{cases}
                      \textsc{Score}(1, 1) + \textsc{Sim}(G(2, 0), G(0, 2)) = 0 + \textsc{Sim}(A, A) & = 1  \\
                      \textsc{Score}(1, 2) + \gamma              = -1 + -1                           & = -2 \\
                      \textsc{Score}(2, 1) + \gamma              = -1 + -1                           & = -2 \\
                  \end{cases}\\
                  &= 1.
              \end{split}
          \end{equation*}
          This calculation is straightforward as it is clear that the best score arises from aligning $A$ to $A$ without introducing any gaps. The \emph{backwards} movement direction that gives the best score is also recorded. In this case, this is represented as an \textbf{arrow} from $G(2,2)$ to $G(1, 1)$.

          Further, \textsc{Score}(3, 2), \textsc{Score}(2, 3) and \textsc{Score}(3, 3) are calculated.

          \begin{tikzpicture}
              \matrix[matrix of nodes,nodes={draw=gray, anchor=center, minimum size=.6cm}, column sep=-\pgflinewidth, row sep=-\pgflinewidth] (A) {
                  $C$                         & -2                          & \phantom{} & \phantom{}  \\
                  $A$                         & -1                          & 1                           & \phantom{}  \\
                  \phantom{} & 0                           & -1                          & -2                           \\
                  \phantom{} & \phantom{} & $A$                         & $B$ \\};
          \end{tikzpicture}

          Here,

          \begin{equation*}
              \begin{split}
                  \textsc{Score}(3, 2) &= \max
                  \begin{cases}
                      \textsc{Score}(2, 1) + \textsc{Sim}(G(3, 0), G(0, 2)) = 0 + \textsc{Sim}(B, A) & = -1 \\
                      \textsc{Score}(2, 2) + \gamma              = 1 + -1                            & = 0  \\
                      \textsc{Score}(3, 1) + \gamma              = -2 + -1                           & = -3 \\
                  \end{cases}\\
                  &= 0,
              \end{split}
          \end{equation*}

          \begin{equation*}
              \begin{split}
                  \textsc{Score}(2, 3) &= \max
                  \begin{cases}
                      \textsc{Score}(1, 2) + \textsc{Sim}(G(2, 0), G(0, 3)) = 0 + \textsc{Sim}(A, C) & = -1 \\
                      \textsc{Score}(1, 3) + \gamma              = -2 + -1                           & = -3 \\
                      \textsc{Score}(2, 2) + \gamma              = 1 + -1                            & = 0  \\
                  \end{cases}\\
                  &= 0,
              \end{split}
          \end{equation*}
          and

          \begin{equation*}
              \begin{split}
                  \textsc{Score}(3, 3) &= \max
                  \begin{cases}
                      \textsc{Score}(2, 2) + \textsc{Sim}(G(3, 0), G(0, 3)) = 1 + \textsc{Sim}(B, C) & = 0  \\
                      \textsc{Score}(2, 3) + \gamma              = 0 + -1                            & = -1 \\
                      \textsc{Score}(3, 2) + \gamma              = 0 + -1                            & = -1 \\
                  \end{cases}\\
                  &= 0.
              \end{split}
          \end{equation*}

          For $\textsc{Score}(3, 2)$ and $\textsc{Score}(2, 3)$, the cases in which gaps are introduced are more optimal than moving diagonally.

          This is repeated until the whole grid is filled, giving

          \begin{tikzpicture}
              \matrix[matrix of nodes,nodes={draw=gray, anchor=center, minimum size=.6cm}, column sep=-\pgflinewidth, row sep=-\pgflinewidth] (A) {
                  $C$                         & -7                          & -5  & -3  & -1  & -2  & -2  & 0   & 0                            \\
                  $B$                         & -6                          & -4  & -2  & -2  & -1  & -1  & 1   & 0                            \\
                  $C$                         & -5                          & -3  & -3  & -1  & 0   & 0   & 0   & -1                           \\
                  $D$                         & -4                          & -2  & -2  & -1  & 1   & 1   & 0   & -1                           \\
                  $D$                         & -3                          & -1  & -1  & 0   & 2   & 1   & 0   & -1                           \\
                  $C$                         & -2                          & 0   & 0   & 1   & 0   & -1  & -2  & -3                           \\
                  $A$                         & -1                          & 1   & 0   & -1  & -2  & -3  & -4  & -5                           \\
                  \phantom{} & 0                           & -1  & -2  & -3  & -4  & -5  & -6  & -7                           \\
                  \phantom{} & \phantom{} & $A$ & $B$ & $C$ & $D$ & $A$ & $B$ & $E$ \\};
          \end{tikzpicture}

          As an aside, note that the score for a cell can be given from two or all of the three cases. An example is in the calculation of $\textsc{Score}(6,6)$, in which the calculation is

          \begin{equation*}
              \begin{split}
                  \textsc{Score}(6, 6) &= \max
                  \begin{cases}
                      \textsc{Score}(5, 5) + \textsc{Sim}(G(6, 0), G(0, 6)) = 1 + \textsc{Sim}(A, C) & = 0  \\
                      \textsc{Score}(5, 6) + \gamma              = 0 + -1                            & = -1 \\
                      \textsc{Score}(6, 5) + \gamma              = 1 + -1                            & = 0  \\
                  \end{cases}\\
                  &= 0.
              \end{split}
          \end{equation*}

          This implies that two optimal alignments can be derived, either by mismatching $(G(6,0), G(0,6)) = (A, C)$ or by adding a gap to give the tuple $(G(6, 0), \Gamma) = (A, \Gamma)$. It also means that there will be two branches of the optimal path: one from $G(6,6)$ to $G(5,5)$, another from $G(6,6)$ to $G(6,5)$.

    \item \textbf{Tracing the optimal path}.
          Once final grid is produced, the optimal path(s) is(are) traced to give the global alignment(s). The path is traced by starting from the top right corner $G(8, 8)$ and moving according to the available arrows from the cells. The path ends at $G(1, 1)$.

          The path-tracing algorithm can be written as a function $\textsc{Trace}$ where a call to $\textsc{Trace}(1+|P|, 1+|S|, [\,])$ gives all the optimal solutions.
          $$
              \textsc{Trace}(x, y, a) = \begin{cases}
                  \textsc{Trace}(x-1, y-1, [(G(x, 0), G(0, y))] + a), & \text{if diagonal arrow present} \\
                  \textsc{Trace}(x-1, y, [(G(x, 0), \Gamma)] + a),    & \text{if left arrow present}     \\
                  \textsc{Trace}(x, y-1, [(\Gamma, G(0, y))] + a),    & \text{if down arrow present}     \\
              \end{cases}
          $$

          where $\textsc{Trace}(1, 1, a) = a$ and the $+$ operator for two lists is defined as concatenation.

          This approach does not enumerate over all possible optimal paths but instead returns the first optimal alignment.

          An example alignment result is given by the path traced by

          \begin{tikzpicture}
              \matrix[matrix of nodes,nodes={draw=gray, anchor=center, minimum size=.6cm}, column sep=-\pgflinewidth, row sep=-\pgflinewidth] (A) {
                  $C$                         & -7                          & -5                          & -3                          & -1                          & -2                          & -2                          & 0                           & \textbf{0}  \\
                  $B$                         & -6                          & -4                          & -2                          & -2                          & -1                          & -1                          & \textbf{1} & 0                            \\
                  $C$                         & -5                          & -3                          & -3                          & -1                          & 0                           & \textbf{0} & 0                           & -1                           \\
                  $D$                         & -4                          & -2                          & -2                          & -1                          & 1                           & \textbf{1} & 0                           & -1                           \\
                  $D$                         & -3                          & -1                          & -1                          & 0                           & \textbf{2} & 1                           & 0                           & -1                           \\
                  $C$                         & -2                          & 0                           & 0                           & \textbf{1} & 0                           & -1                          & -2                          & -3                           \\
                  $A$                         & -1                          & \textbf{1} & \textbf{0} & -1                          & -2                          & -3                          & -4                          & -5                           \\
                  \phantom{} & \textbf{0} & -1                          & -2                          & -3                          & -4                          & -5                          & -6                          & -7                           \\
                  \phantom{} & \phantom{} & $A$                         & $B$                         & $C$                         & $D$                         & $A$                         & $B$                         & $E$ \\};
          \end{tikzpicture}

          giving $P'_1 = ABCDA \Gamma BE$, $S'_1 = A \Gamma CDDCBC$ and
          $$L_1 = [(A,A), (B, \Gamma), (C, C), (D, D), (A, D), (\Gamma, C), (B, B), (E, C)].$$

          Other optimal alignments are possible by tracing alternative paths. Another optimal alignment is given by

          \begin{tikzpicture}
              \matrix[matrix of nodes,nodes={draw=gray, anchor=center, minimum size=.6cm}, column sep=-\pgflinewidth, row sep=-\pgflinewidth] (A) {
                  $C$                         & -7                          & -5                          & -3                          & -1                          & -2                          & -2                          & 0                           & \textbf{0}  \\
                  $B$                         & -6                          & -4                          & -2                          & -2                          & -1                          & -1                          & \textbf{1} & 0                            \\
                  $C$                         & -5                          & -3                          & -3                          & -1                          & 0                           & \textbf{0} & 0                           & -1                           \\
                  $D$                         & -4                          & -2                          & -2                          & -1                          & \textbf{1} & 1                           & 0                           & -1                           \\
                  $D$                         & -3                          & -1                          & -1                          & \textbf{0} & 2                           & 1                           & 0                           & -1                           \\
                  $C$                         & -2                          & 0                           & 0                           & \textbf{1} & 0                           & -1                          & -2                          & -3                           \\
                  $A$                         & -1                          & \textbf{1} & \textbf{0} & -1                          & -2                          & -3                          & -4                          & -5                           \\
                  \phantom{} & \textbf{0} & -1                          & -2                          & -3                          & -4                          & -5                          & -6                          & -7                           \\
                  \phantom{} & \phantom{} & $A$                         & $B$                         & $C$                         & $D$                         & $A$                         & $B$                         & $E$ \\};
          \end{tikzpicture}

          giving $P'_2 = ABC \Gamma DABE$, $S'_2 = A \Gamma CDDCBC$ and
          $$L_2 = [(A,A), (B, \Gamma), (C, C), (\Gamma, D), (D, D), (A, C), (B, B), (E, C)].$$

\end{enumerate}

\subsection{Complexity Improvements}
\label{impl_asm_complexity}

Taking that the \textsc{Score} function takes $\mathcal{O}(1)$ time to compute, the algorithm requires $\mathcal{O}(nm)$ time and space.

It is worth noting that for the majority of global alignments, the optimal path lies within a certain region in the diagonal. If optimality can be given up for the sake of time and space complexity, a common optimisation is to introduce a \textit{bounded} grid, i.e., scores are only computed for a maximum distance $d$ from the main diagonal. For $d = 2$ applied on the above example, the calculations are limited to the scores shown on the following grid:

\begin{tikzpicture}
    \matrix[matrix of nodes,nodes={draw=gray, anchor=center, minimum size=.6cm}, column sep=-\pgflinewidth, row sep=-\pgflinewidth] (A) {
        $C$                         & \phantom{} & \phantom{} & \phantom{} & \phantom{} & \phantom{} & -2                          & 0                           & \textbf{0}  \\
        $B$                         & \phantom{} & \phantom{} & \phantom{} & \phantom{} & -1                          & -1                          & \textbf{1} & 0                            \\
        $C$                         & \phantom{} & \phantom{} & \phantom{} & -1                          & 0                           & \textbf{0} & 0                           & -1                           \\
        $D$                         & \phantom{} & \phantom{} & -2                          & -1                          & \textbf{1} & 1                           & 0                           & \phantom{}  \\
        $D$                         & \phantom{} & -1                          & -1                          & \textbf{0} & 2                           & 1                           & \phantom{} & \phantom{}  \\
        $C$                         & -2                          & 0                           & 0                           & \textbf{1} & 0                           & \phantom{} & \phantom{} & \phantom{}  \\
        $A$                         & -1                          & \textbf{1} & \textbf{0} & -1                          & \phantom{} & \phantom{} & \phantom{} & \phantom{}  \\
        \phantom{} & \textbf{0} & -1                          & -2                          & \phantom{} & \phantom{} & \phantom{} & \phantom{} & \phantom{}  \\
        \phantom{} & \phantom{} & $A$                         & $B$                         & $C$                         & $D$                         & $A$                         & $B$                         & $E$ \\};
\end{tikzpicture}

It is clear that less space and time are required in this setting. The shape of the bounds can be a parallelogram as well, which may give better results for certain cases. The grid, with a parallelogram bound applied, would look roughly like:

\begin{tikzpicture}
    \matrix[matrix of nodes,nodes={draw=gray, anchor=center, minimum size=.6cm}, column sep=-\pgflinewidth, row sep=-\pgflinewidth] (A) {
        $C$                         & \phantom{} & \phantom{} & \phantom{} & \phantom{} & \phantom{} & \phantom{} & \phantom{} & \textbf{0}  \\
        $B$                         & \phantom{} & \phantom{} & \phantom{} & \phantom{} & -1                          & -1                          & \textbf{1} & \phantom{}  \\
        $C$                         & \phantom{} & \phantom{} & -3                          & -1                          & 0                           & \textbf{0} & 0                           & \phantom{}  \\
        $D$                         & \phantom{} & \phantom{} & -2                          & -1                          & \textbf{1} & 1                           & 0                           & \phantom{}  \\
        $D$                         & \phantom{} & -1                          & -1                          & \textbf{0} & 2                           & 1                           & \phantom{} & \phantom{}  \\
        $C$                         & \phantom{} & 0                           & 0                           & \textbf{1} & 0                           & -1                          & \phantom{} & \phantom{}  \\
        $A$                         & \phantom{} & \textbf{1} & \textbf{0} & -1                          & \phantom{} & \phantom{} & \phantom{} & \phantom{}  \\
        \phantom{} & \textbf{0} & \phantom{} & \phantom{} & \phantom{} & \phantom{} & \phantom{} & \phantom{} & \phantom{}  \\
        \phantom{} & \phantom{} & $A$                         & $B$                         & $C$                         & $D$                         & $A$                         & $B$                         & $E$ \\};
\end{tikzpicture}

However, bounding may give suboptimal results--there will always be a bound strict enough to give suboptimal matches save for the trivial case where $P = S$.

As optimality is important, ways to improving the footprint of the algorithm without compromising global optimality are focussed on. A space optimisation can be achieved using Hirschberg's algorithm \cite{hirschberg75} which uses divide and conquer to reduce the space complexity to $\mathcal{O}(\min(m,n))$. With regard to improvements in running time, the Method of Four Russians achieves an $\mathcal{O}(mn / \log{n})$ time complexity for the algorithm \cite{sung2010algorithms}.

This therefore means that the alignment algorithm can be optimised to run in \textbf{linear memory} and \textbf{subquadratic time}.

Also notice that while the \textit{bottom-up} approach was presented, the algorithm can also be computed via a \textit{top-down} approach in which the algorithm starts by calculating $\textsc{Score}(8, 8)$. Notice however that the top-down approach may include repeat score calculations; this can be averted using \textit{memoisation}, i.e. storing the results of calls to the $\textsc{Score}$ function and returning the cached result when the function is called with the same inputs again.

\subsection{Similarity Matrix}

The similarity function defined in \Cref{sim_func_defn} can be further developed to produce more favourable alignments with regard to score alignment. In principle, mistakes that are more likely to happen should be penalised less.

Instead of penalising a mismatch naïvely by assigning a score of $-1$, note that mistakes constituting accidentally playing a neighbouring note may be more probable than one playing a distant note; that is, a mismatch of $(A, B)$ is more likely than a mismatch of $(A, E)$ in actual music performances. This is especially probable for keyboard instruments as opposed to multi-string instruments like the lute and violin, where mistakes involving the accidental playing of neighbouring strings that are not close in pitch are more common. The likeliness of pitch mismatch mistakes also varies dynamically: at higher notes on a string for stringed instruments, neighbouring pitch mistakes are more likely--the notes are \textit{spatially} closer together.

Thus, different instruments may use different penalising systems for mismatches; however, it still remains that neighbouring note mistakes are significantly likely and this system would generalise well to most instruments. Therefore, a \textbf{similarity matrix} that penalises mismatches between distant notes more is introduced. This matrix for the set of musical notes $N \in \{A, B, C, D, E\}$ is given by

\begin{tikzpicture}
    \matrix[matrix of nodes,nodes={draw=gray, anchor=center, minimum size=.6cm}, column sep=-\pgflinewidth, row sep=-\pgflinewidth] (A) {
        \phantom{} & $A$                         & $B$                         & $C$                         & $D$                         & $E$                                                  \\
        $A$                         & \textbf{1} & -1                          & -2                          & -3                          & -4                                                   \\
        $B$                         & -1                          & \textbf{1} & -1                          & -2                          & -3                                                   \\
        $C$                         & -2                          & -1                          & \textbf{1} & -1                          & -2                                                   \\
        $D$                         & -3                          & -2                          & -1                          & \textbf{1} & -1                                                   \\
        $E$                         & -4                          & -3                          & -2                          & -1                          & \textbf{1} \\};
\end{tikzpicture}

The parameter $\alpha$ in this case is equal to $1$ and the parameter $\beta$ is a function given by

$$
    \beta_s(c_1, c_2) = -|c_1 - c_2| ~\text{where}~ c_1 \neq c_2 ~\text{and}~ c_1, c_2 \in N
$$
taking the definition of subtraction of two elements in $N$ as simply their distance of their indices in $N$.

Given that there are 128 MIDI pitches, this system may overpenalise mistakes that are too far away; hence, a lower bound, $\hat{\beta}$, for the most negative error can be set, where mismatches for pairs of pitches that have scores lower than the bound would instead have an error equal to the prescribed lower bound. $\beta_s$ is therefore updated to be

$$
    \beta_s'(c_1, c_2) = \max(-|c_1 - c_2|, \hat{\beta}) ~\text{where}~ c_1 \neq c_2 ~\text{and}~  c_1, c_2 \in N.
$$

To penalise values increasingly up to an octave interval, $\hat{\beta} = -12$ is chosen--there are 12 semitones (and hence MIDI pitches) in an octave.

\subsection{A sensible gap (indel) penalty}
Gaps may be required to optimally align score and performance sequences due to extra notes being played or notes being missed. This is not uncommon in performance and may also be caused by inferior MIDI devices unable to capture all the performance data. This is a parameter subject to tuning according to the dataset, but a sensible value is $\gamma = -1$, similar to a small mistake made by accidentally playing a neighbouring note.

\subsection{Dealing with Chords and Polyphony}
\label{impl_asm_polyphony}

Parallel voices, such as chords on the piano, cause two problems for the current setup. Firstly, there is no \textit{correct} way to order parallel voices in the chord defined in the score to be aligned. Secondly, chords in the performance may not be played and/or recorded at the same time instance--a chord $\{A,B,C\}$ comprising the three notes $A$, $B$ and $C$ may be played in any of the $3! = 6$ orders. This is further complicated if pieces with multiple instruments are aligned. Three different approaches are explored for this system--the third approach is chosen.

\begin{enumerate}
    \item \textbf{Double Alignment}.
    
This solution is inspired by Dannenberg and Mukaino's work \cite{dannenberg88}. Chords identified in the score are grouped into a set--a chord with the parallel notes $A$, $B$, $C$ would hence be denoted as $\{A,B,C\}$. The problem now comes at the alignment stage: the algorithm needs to be extended to work with these sets. This can be done by allowing notes for the chord in $P$ to come in any of the $3! = 6$ orders.
    
A naïve approach would be to generate multiple $S$ each denoting a permutation of the chords in the score and pick the $S$ giving the best alignment, but this quickly becomes intractable. A better technique is to modify the algorithm to allow accepting of multiple notes in $P$ for the alignment. This again requires a redefinition of the similarity and $\beta$ function. As a reminder, the similarity function is currently

$$
\textsc{Sim}'(c_1, c_2) = \begin{cases}
    \alpha, & \text{if }c_1 = c_2, \\
    \beta_s'(c_1, c_2),  & \text{otherwise}  \\
\end{cases}
$$
where
$$
\beta_s'(c_1, c_2) = \max(-|c_1 - c_2|, \hat{\beta}) ~\text{where}~ c_1 \neq c_2 ~\text{and}~  c_1, c_2 \in N.
$$

First, the score sequence $S = s_1 s_2 \ldots s_n$ is converted into a list of sets $\Psi = \psi_1 \psi_2 \ldots \psi_\nu$. Each set $\psi$ contains parallel notes at one time instance (non-parallel notes are therefore singletons).  The alignment problem is redefined to find an alignment between $P$ and $\Psi$, giving $P'$ and $\Psi'$ as the optimal alignments and a list of tuples representing the alignment result.

The similarity function required is 
$$
\textsc{Sim}''(c, \psi) = \begin{cases}
    \alpha, & \text{if } c \in \psi \\
    \beta_s''(c, \psi),  & \text{otherwise}  \\
\end{cases}
$$
where $c \in P$ and $c \in N$.

$\beta_s''(c, \psi)$ is defined as
$$
\beta_s''(c, \psi) = \max(-\textsc{MinAbsDist}(c, \psi), \hat{\beta}) ~ ~\text{where}~ c \notin \psi
$$
where $\hat{\beta} = -12$ as noted.

The $\textsc{MinAbsDist}(c, \psi)$ function returns the minimum absolute distance (calculated by the absolute difference in pitch number) of $c$ with any element in $\psi$. 

The $\textsc{Score}$ function also needs to be changed for the second case: this is because playing the next note $G(x, 0)$ in a possible chord $G(0, y)$ should not be considered an indel. In addition, a match or mismatch should still be more favourable and hence the value for $\textsc{Sim}$ is capped at 0. The redefined function is therefore
$$
\textsc{Score}''(x, y) = \max
\begin{cases}
    \textsc{Score}''(x-1, y-1) + \textsc{Sim}''(G(x, 0), G(0, y)) \\
    \textsc{Score}''(x-1, y) + \textsc{ChordScore}(x, y)\\
    \textsc{Score}''(x, y-1) + \gamma                          \\
\end{cases}
$$

where
$$
\textsc{ChordScore}(x, y) = \begin{cases}
    \text{clamp}(\gamma,\textsc{Sim}''(G(x, 0), G(0, y)),\alpha) & ~\text{if} ~x \neq 1, y \neq 1 ~\text{and}~ |G(0, y)| > 1,\\ 
    \gamma & ~\text{otherwise}. \\ 
\end{cases}\\
$$
($\text{clamp}(a, x, b)$ is the mathematical clamp function that bounds $x$ between $a$ and $b$: $\text{clamp}(a, x, b) = max(a, min(x, b))$.)

As an example, consider $\Psi = \{A\}\{A,B,C\}\{A,C\}\{D\}$ and $P = ACBACAD$. For convenience, say $\psi_1 = \{A\}$, $\psi_2 = \{A,B,C\}$ and so on. The completed grid which gives an optimal solution is



\begin{tikzpicture}
    \matrix[matrix of nodes,nodes={draw=gray, anchor=center, minimum size=.6cm}, column sep=-\pgflinewidth, row sep=-\pgflinewidth] (A) {
      $\psi_4$ & -4 & -2 & 0 & 1 & 3 & 4 & 5 & \textbf{7} \\
      $\psi_3$ & -3 & -1 & 1 & 2 & 4 &  \textbf{5} &  \textbf{6} & 5 \\
      $\psi_2$ & -2 & 0 &  \textbf{2} &  \textbf{3} &  \textbf{4} & 5 & 6 & 5 \\
      $\psi_1$ & -1 & \textbf{1} & 0 & -1 & -2 & -3 & -4 & -5 \\
        \phantom{} & \textbf{0} & -1 & -2 & -3 & -4 & -5 & -6 & -7 \\
        \phantom{} & \phantom{} & $A$ & $C$ & $B$ & $A$ & $B$ & $C$ & $D$ \\};
\end{tikzpicture}

where the alignment is 
$$
L =  [(A, \{A\}), (C, \{A, B, C\}), (B, \{\Gamma\}), (A, \{\Gamma\}), (C, \{A, C\}), (A, \{\Gamma\}), (D, \{D\})].
$$

An extra alignment step is added in the end to give every $p \in P$ a time, or in this case a pairing to $\psi \in \Psi$. The gaps are hence removed--a tuple $(p, {\Gamma})$ will have $p$ matched to the closest $\psi$ that precedes it. For the case of $(\Gamma, \psi)$ where $\psi \neq \Gamma$, otherwise known as a set of notes in the score with no mapping to a performance note, the set of notes is dropped. The final alignment is given by a list of tuples $L' = [(\rho_1, \psi_1), \ldots, (\rho_k, \psi_k)]$, 
$$
L' = [(\{A\}, \{A\}), (\{C, B, A\}, \{A, B, C\}), (\{C, A\}, \{A, C\}), (\{D\}, \{D\})].
$$

As seen, even though the chords are played in the wrong order in $P$, they are still aligned in the end.
    
A problem in the current setup quickly arises, however, for repeat chords. Say $P = ABAB$ and $\Psi = \{A, B\}\{A, B\}$. The three alignments 
\begin{equation*} 
    \begin{split}
        [(\{A, B\}, \{A, B\}), (\{A, B\}, \{A, B\})],& \\
        [(\{A, B\}, \{A, B\}), (\{A, B\}, \{B\})],& ~\text{and}~ \\
        [(\{A, B\}, \{A\}), (\{A, B\}, \{A, B\})]& \\
    \end{split}
\end{equation*}

are all ``optimal''.  It is clear that the more ideal solution is the first alignment--two notes are played for each chord comprising two notes.

This problem arises because a chord set $\psi$ is not penalised to accept more notes than it contains. This can be done in the calculation of $\textsc{Score}$, but it further complicates the already complex function.

Note that using timing information may give rise to a better solution, but that unnecessarily complicates the problem--it is already assumed in \autoref{impl_asm_robust} that the performances used to create reference datasets are likely of high quality and adding temporal information to the algorithm may complicate the problem more than help it.

Notice that the sequence is now aligned \emph{twice} after the algorithm is modified to work with polyphonic music--$L$ is the result of the first alignment and $L'$ the second. Therefore, the definition of optimality needs to be extended to include the minimisation of a \emph{cost} function in the final alignment. This cost function, $\textsc{Cost}$, is defined as

$$
\textsc{Cost}(L') = \sum_{\forall (\rho, \psi) \in L'}  J(\rho, \psi) 
$$

where $J(\rho, \psi)$ is the \textbf{Jaccard index}, a statistic used to gauge the similarity of sets,
$$
J(\rho, \psi) =\begin{cases}
    1 & ~\text{if}~ \rho = \psi = \varnothing,  \\
    \frac{\rho \cap \psi}{\rho \cup \psi} & ~\text{otherwise.} \\
\end{cases}
$$

However, note that this now comes back to the original unoptimised problem--there may be up to 
$$
\binom{n+\nu}{\nu}\frac{(\nu+n)!}{(\nu!)^2}
$$
``optimal'' paths to consider! 

Methods like these are better in score following (as in the source of inspiration of this solution \cite{dannenberg88}) where the path needs not to be optimal--it just needs to be ``good enough''--heuristics can therefore be used to speed up calculations. For instance, Desain and Heijink's approach \cite{desain97,heijink00} uses a window to limit the alignment size, making the calculations significantly more tractable.

    \item \textbf{Sorting Parallel Performance and Score Notes}.

    
    This solution, simpler than the first method, makes use of performance temporal information to preprocess performance data. Another parameter, $\epsilon$, is defined the maximum interval for chord execution.
    
        \begin{definition}[Maximum interval for chord execution]
            The \textbf{maximum interval for chord execution}, $\epsilon$, is the maximum allowed time for a chord to be fully executed, i.e. for all its notes onsets to be recorded.
            \label{def:maxintchordexe}
        \end{definition}
        
        The determination of $\epsilon$ is rather trivial for piano music. A data-driven solution is to base it on the time interval between the two closest notes in the score. However, a sensible estimate of $\epsilon = 50 \si{\milli\second}$ for piano music should work well--this is a touch lower than the time interval between two demisemiquaver notes at $120\text{BPM}$.
        
        Before alignment is performed, a preprocessing step will first run through $P$ and sort (in ascending order) note onsets that can be grouped within the time interval $\epsilon$. The grouping will be \textit{greedy} with regard to the number of notes in a chord--if there exists multiple overlapping solutions to neighbouring notes, the group with the most notes will be selected. For instance, taking 
        $$
        A_{500} \quad D_{549}\quad  C_{551}\quad  B_{552},
        $$
        
        where the substrings of the note denote the note onset time in $\si{\milli\second}$, the resulting sequence should be 
        $$
        A_{500}\quad B_{552} C_{551} D_{549}
        $$
        instead of
        $$
        A_{500} D_{549} \quad C_{551} B_{552}.
        $$
    
        Additionally, parallel voices in the score data will also be sorted in ascending order.
        
        More musically informed readers may point to technically demanding works of Rachmaninoff where pianists with limited finger span may need to \textit{break} large chords. However, notice that the notes in the chord are sorted in ascending order--the chords in $P$ and $S$ will remain in the correct order as pianists usually break the chord and play the notes in ascending order. 
        
        Where this may cause a problem is in strummed string instruments like the guitar--a chord may be played from top to bottom slowly, causing the chord to take more than $\epsilon$ to completely execute--\textit{in the ``wrong'' direction}. More examples can be drawn from \autoref{scofo_challenges_polyphonic}--chords in the \textit{Chaconne}, see \autoref{fig:chaconnefirst10}, may be executed in either direction at varying speeds. Moreover, the requirement to determine the parameter $\epsilon$ also makes the algorithm harder to generalise to more instruments and musical styles.
    
        Further, this procedure mangles performance data and may cause more mistakes in the alignment than try to fix them.

    \item \textbf{Sorting Parallel Score Notes and Running a Post-alignment Step}.
    
This solution uses part of the second approach; first, parallel voices in the score are sorted (which is likely the way a chord is executed), then, gaps and mismatches are corrected after automatic alignment. These gaps and mismatches will be indicated to the user as per \autoref{impl_asm_indicate}. There is already an assumption made in \autoref{impl_asm_robust} that the performances used to create reference datasets likely have few mistakes and this should still give the best possible alignment prior to more corrections--it does not overcomplicate the problem as in the first solution, and it does not perturb performance data as in the second method. The preprocessing of the score data--i.e. sorting parallel voices, still gives correct score information.

Besides sorting parallel voices in the score, an optional post-alignment step is proposed. Note that each misalignment may have a corresponding misalignment that both cancel each other out. An example for misaligned gaps is given by the (mis)alignment

$$[(A,\Gamma), (B, B), (\Gamma, A)].$$

If all the elements in the above are part of a set of parallel voices, the two gaps with $A$ should be aligned together to give $$[(A, A), (B, B)].$$ A more complex case involving also a mismatch is the (mis)alignment

$$[(A,\Gamma), (B, B), (\Gamma, C), (C, A)].$$

Similarly, if all the elements are part of a set of parallel voices, the mismatches and gaps can be fixed to give the alignment $$[(A,A), (B,B), (C,C)].$$

Thus, a post-alignment step can iterate through the alignment to find trivially fixable misalignments that correspond to false negatives involving parallel notes. As such misalignments may also occur with fast passages in which the sequence of notes may be misplayed, an adjustable threshold is used to determine how far apart between two misalignments to consider them to be ``close enough'' to be fixed by this post-alignment step. The threshold parameter is defined as follows:

\begin{definition}[Post-alignment Parallel Voice Threshold] The \textbf{Post-alignment Parallel Voice Threshold}, $\delta$, is the maximal absolute difference in score time between two notes to treat the two notes as ``close enough'' to be considered eligible for post-alignment parallel voice fixes.
    \label{def:postalignparvoicethres}
\end{definition}

The choice of $\delta$ should vary from piece to piece depending on how ``polyphonic'' the performance is--monophonic performances may not need post-alignment at all, and pieces with low polyphony can sensibly have $\delta$ set anywhere from $0$ to the time taken for a beat in the score.

This post-alignment step introduces a temporal aspect in the alignment--the original ASM algorithm does not use any temporal information in source data.

\end{enumerate}

It is thus clear that the third approach--i.e. sorting parallel notes in the score and then running a post-alignment step is superior when compared to the two other methods. The \textbf{third method} is therefore chosen.

\subsection{Dealing with Ornaments}
\label{impl_asm_ornaments}

The \textit{trill}, a common music ornament, consists of a rapid alternation between two adjacent notes, usually a semitone or tone apart. In some MIDI files, ornaments are not reflected and therefore a \textit{trill} is reduced to its base note onset in the data. Following this, Dannenberg and Mukaino \cite{dannenberg88} dealt with \textit{trills} and \textit{glissandi} by introducing a special mode in their score follower that allows the mapping of many performance notes (the actual many notes making up the \textit{trill} or the \textit{glissando}) to one score note (the base note of the \textit{trill} or \textit{glissando}).

However, recent MIDI-producing musical programs \cite{lilypond} are able to produce MIDI data that reflects the ornament. Nevertheless, notice that performers may interpret ornaments in different ways--see sections~\ref{scofo_challenges_perfdev} and \ref{scofo_challenges_underspec}. This means that manual alignment will occasionally be required after automatic alignment. However, the advantage this has over Dannenberg and Mukaino's approach is that the score has more onsets--albeit artificial--to work with, giving better temporal resolution: in Dannenberg and Mukaino's solution, if an event is required to be triggered in the time between a \textit{trill} event and the subsequent note onset, the system would have to estimate when to trigger that event.

Ornaments are not dealt in a special way in this system, giving the users the choice to either use MIDI data that reflects ornaments or otherwise. Thus, further manual adjustment may be required for MIDI files not reflecting ornaments, but as ornaments generally do not span the majority of pieces, these adjustments are projected to be minimal.

\section{Score Follower Framework}
As this system does not run in real time, it is a score aligner (some authors use the term \textit{offline score follower}). However, the generic framework of score followers described in \autoref{scofo_generic} can be used to model this system. The \textbf{features extracted} from the performance and the score are \textit{pitch}, \textit{note order} and \textit{note onset time} information in the form of a sequence of MIDI notes. The \textbf{similarity calculation} is via the minimisation of an offline ASM cost function defined in \autoref{impl_asm_techback}, which provides a globally optimal \textbf{alignment}.

Clearly, this system works on strings and hence is of the \textbf{symbolic-to-symbolic} type.

\section{Implementation}

\subsection{Algorithm Information}

The final algorithm used is as presented in \autoref{impl_asm_techback_asm}. In summary, the used definitions are as follows:
$$
    \textsc{Score}^{*}(x, y) = \max
    \begin{cases}
        \textsc{Score}^{*}(x-1, y-1) + \textsc{Sim}^{*}(G(x, 0), G(0, y)) \\
        \textsc{Score}^{*}(x-1, y) + \gamma^{*}                           \\
        \textsc{Score}^{*}(x, y-1) + \gamma^{*}                           \\
    \end{cases}
$$
where $\textsc{Score}^{*}(0, y) = \textsc{Score}^{*}(x, 0) = -\infty$ and $\textsc{Score}^{*}(1, 1) = 0$,

$$
    \textsc{Sim}^{*}(c_1, c_2) = \begin{cases}
        \alpha^{*},          & \text{if }c_1 = c_2, \\
        \beta^{*}(c_1, c_2), & \text{otherwise}     \\
    \end{cases}
$$
where
$$
    \beta^{*}(c_1, c_2) = \max(-|c_1 - c_2|, \hat{\beta}^{*}) ~\text{where}~ c_1 \neq c_2 ~\text{and}~  c_1, c_2 \in N^{*}.
$$

($N^{*}$ is the set of all 128 MIDI note representations.)

The final solution is the first complete optimal path traced by the $\textsc{Trace}^{*}(1+|P|, 1+|S|, [\,])$ function:

$$
    \textsc{Trace}^{*}(x, y, a) = \begin{cases}
        \textsc{Trace}^{*}(x-1, y-1, [(G(x, 0), G(0, y))] + a),  & \text{if diagonal arrow present} \\
        \textsc{Trace}^{*}(x-1, y, [(G(x, 0), \Gamma^{*})] + a), & \text{if left arrow present}     \\
        \textsc{Trace}^{*}(x, y-1, [(\Gamma^{*}, G(0, y))] + a), & \text{if down arrow present}     \\
    \end{cases}
$$
where $\textsc{Trace}^{*}(1, 1, a) = a$ and the $+$ operator for two lists is defined as concatenation.

The parameters $\alpha^{*}$, $\hat{\beta}^{*}$ and $\gamma^{*}$ are $1$, $-12$ and $-1$ respectively as discussed. To deal with chords and polyphony, the third method detailed in \autoref{impl_asm_polyphony} is used; the score sequence $S$ is preprocessed: parallel notes are sorted in ascending order. The post-alignment step that fixes close and trivial misalignments is used as required with its corresponding threshold, $\delta$, adjusted as necessary.


\subsection{Output Data}
Data is output in two streams: \texttt{stdout} and \texttt{stderr}.
\begin{enumerate}
    \item \texttt{stdout}: In this stream, alignment data that fulfil the reference data format as per \autoref{tb_quant_dataset_api} are output. In the event of a mismatch an indication is provided in the format

          \begin{verbatim}
// MISMATCH: <PERFORMANCE_NOTE> - <SCORE_NOTE>
\end{verbatim}
          where note information comprises the note onset time of the note and the MIDI number of the note.

          A gap, on the other hand, is reported in the following format
          \begin{verbatim}
// GAP: <GAP|PERFORMANCE_NOTE> - <GAP|SCORE_NOTE>
\end{verbatim}

    \item \texttt{stderr}: Auxiliary information such as the number of mismatches and gaps are output. The first line prints if the post-alignment step is enabled--the other lines are always printed. The metrics introduced each line from the second line onwards correspond to $|L|$, $N_{\Gamma P}= |[(a, b) ~ \forall ~ (a, b)\in L ~\text{where}~ a = \Gamma]|$,  $N_{\Gamma S} = |[(a, b) ~ \forall ~ (a, b) \in L ~\text{where}~ b = \Gamma]|$ and $N_{\textsc{Mismatch}} = |[(a, b) ~ \forall ~ (a, b) \in L ~\text{where}~ a \neq b]|$ respectively. The output format is given by
          \begin{verbatim}
Running PostAlign with threshold <POST_ALIGNMENT_THRESHOLD>
Length of alignment: <...>
Total number of gaps in performance: <...>
Total number of gaps in score: <...>
Total number of mismatches: <...>
\end{verbatim}

\end{enumerate}

\subsection{Implementation Details}
\label{impl_asm_impl_details}

This algorithm is introduced in Python, as it is meant to be used in the quantitative testbench (\autoref{tb_quant}). The rationale for using Python is straightforward as the code is to be added into the testbench\footnote{\githubFlippyQuant} which is written in Python. It is easier to maintain a codebase in one language. The other benefits and best practices of Python mentioned in \autoref{tb_quant_bettereng} apply here as well.


\section{Evaluation}
\label{impl_asm_eval}

As the ASM aligner is a tool that needs to be reliable enough to produce ground-truth alignments, it is first evaluated against a widely available dataset that contains manually aligned data in \autoref{impl_asm_eval_bach10}. The ASM aligner is then used for its purpose in \autoref{impl_asm_eval_bwv846}--it is used to produce reference alignments for a selected suite in the MAESTRO dataset \cite{MAESTRO}. 

The results for both evaluations in subsections~\ref{impl_asm_eval_bach10} and \ref{impl_asm_eval_bwv846} can be reproduced in the qualitative testbench repository's \nameref{tb_quant_repro_suite}.

\subsection{Performance on Bach10}
\label{impl_asm_eval_bach10}

The Bach10 dataset \cite{duan11bach10} contains ground-truth alignments in the required format stipulated in \autoref{tb_quant_dataset_api} and the output of this aligner. Moreover, it contains the MIDI of the reference score. This means that the Bach10 dataset is an ideal candidate to evaluate the aligner and also the quantitative testbench proposed in \autoref{tb_quant}: the output of the aligner can be directly compared against the reference ground-truth alignments via the quantitative testbench. The misalignment threshold $\theta_e$ (refer to \Cref{def:misalign}) is set to $\si{300\milli\second}$, the standard as per MIREX evaluations \cite{cont07}. The score note search bound $\delta_s$ (described in \Cref{def:bound_ms}) is set to the default $\si{1\milli\second}$.

Evaluation is run by first converting the MIDI of the score and the performance part (extracted from the ground-truth alignments) into a pair of strings which are then fed into the ASM aligner algorithm. While the Bach10 dataset is highly polyphonic--it comprises 10 short chorales, each with four voices each played in four instruments--the parallel voices in the performance audio are well quantized and -aligned (as the performance audio is strictly beat-aligned resulting in parallel voices having the exact same timestamp \cite{duan11bach10}). As a result, the post-alignment step designed in \autoref{impl_asm_polyphony} to deal with misaligned parallel voices is not required.

The output of the aligner is then converted into the score follower output defined in \autoref{tb_quant_scofo_api}--as this is a non-real-time alignment, $t_d$, the time the note is detected, is simply set to $t_e$, the estimated note onset time in the performance audio file.
This means all metrics dealing with latency $l_i$ and offset $o_i$ are not valid--the metrics thus ignored are mean latency $\mu_l$, standard deviation of latency $\sigma_l$,
mean absolute offset $\text{MAO}$ and standard deviation of offset $\sigma_o$. Additionally, piece completion $p_e$ is also irrelevant as this is a global alignment that matches both ends of both strings of data.

Added to the evaluation metrics are the metrics from the ASM aligner: the length of alignment $|L|$, the number of gaps in performance $N_{\Gamma P}$, the number of gaps in score $N_{\Gamma S}$ and the number of mismatches $N_{\textsc{Mismatch}}$.

All the metrics considered are given in \autoref{table:asmbach10}. Readers may wish to consult \autoref{tb_quant_metrics} for more information about metrics related to the quantitative benchmark.

\begin{table}[ht]
    \caption{Metrics used to evaluate the ASM Score Aligner against the Bach10 dataset.}
    \centering 
    \begin{tabular}{c c} 
        \hline\hline 
        Metric                        & Symbol                  \\ [0.5ex] 
        \hline 
        Miss Rate                     & $r_m$                   \\
        Misalign Rate                 & $r_e$                   \\
        Mean Absolute Error           & $\text{MAE}$            \\
        Standard Deviation of Error   & $\sigma_e$              \\
        Precision Rate                & $r_p$                   \\
        Length of Alignment           & $|L|$                   \\
        Number of Gaps in Performance & $N_{\Gamma P}$          \\
        Number of Gaps in Score       & $N_{\Gamma S}$          \\
        Number of Mismatches          & $N_{\textsc{Mismatch}}$ \\
        \hline 
    \end{tabular}
    \label{table:asmbach10} 
\end{table}

The results are shown in \autoref{table:asmbach10scores}. The repository of the testbench\footnote{\githubFlippyQuant} contains instructions on how to reproduce these results. As the runs were almost instantaneous, their runtimes were not recorded.

\begin{table}[ht]
    \caption{Results of the ASM Score Aligner against the Bach10 dataset. The arrows, where present, denote whether a metric should be minimised ($\downarrow$) or maximised ($\uparrow$).} 
    \centering 
    \begin{tabular}{l c c c c c c c c c} 
        \hline\hline 
        Piece                                & $r_m\downarrow$ & $r_e\downarrow$ & $\text{MAE}\downarrow$ & $\sigma_e\downarrow$ & $r_p\uparrow$ & $|L|$ & $N_{\Gamma P}\downarrow$ & $N_{\Gamma S}\downarrow$ & $N_{\textsc{Mismatch}}\downarrow$ \\ [0.5ex] 
        \hline 
        \textit{AchGottundHerr}              & 0.00 & 0.00 & 0.00        & 0.00 & 1.00     & 142   & 3              & 0              & 0                       \\
        \textit{AchLiebenChristen}           & 0.00 & 0.00 & 0.00        & 0.00 & 1.00     & 283   & 1              & 0              & 0                       \\
        \textit{ChristederdubistTagundLicht} & 0.00 & 0.00 & 0.00        & 0.00 & 1.00     & 182   & 1              & 0              & 0                       \\
        \textit{ChristeDuBeistand}           & 0.00 & 0.00 & 0.00        & 0.00 & 1.00     & 233   & 9              & 0              & 0                       \\
        \textit{DieNacht}                    & 0.00 & 0.00 & 0.00        & 0.00 & 1.00     & 197   & 7              & 0              & 0                       \\
        \textit{DieSonne}                    & 0.00 & 0.00 & 0.00        & 0.00 & 1.00     & 212   & 6              & 0              & 0                       \\
        \textit{HerrGott}                    & 0.00 & 0.00 & 0.00        & 0.00 & 1.00     & 189   & 0              & 0              & 0                       \\
        \textit{FuerDeinenThron}             & 0.00 & 0.00 & 0.00        & 0.00 & 1.00     & 149   & 0              & 0              & 0                       \\
        \textit{Jesus}                       & 0.00 & 0.00 & 0.00        & 0.00 & 1.00     & 170   & 6              & 0              & 0                       \\
        \textit{NunBitten}                   & 0.00 & 0.00 & 0.00        & 0.00 & 1.00     & 234   & 1              & 0              & 0                       \\
        \hline 
    \end{tabular}
    \label{table:asmbach10scores} 
\end{table}

As it can be seen in \autoref{table:asmbach10scores}, the ASM aligner performed exceptionally well on the Bach10 dataset. Upon closer inspection, it can be seen that most of the very minimal errors where $N_{\Gamma P} \neq 0$ are due to the fact that the performance data does not contain all the notes in the score--this means that the performances has \emph{missed notes} when compared to the full score data. These can be trivially fixed--authors of the Bach10 dataset resorted to removing these missed notes entirely from their ground-truth alignments, which explains results with $r_m = 0$ but $N_{\Gamma P} \neq 0$.


From these results it can be shown that this ASM alignment method is indeed a reliable method to produce ground-truth alignments between symbolic score and performance data. A caveat, however, is that the symbolic performance data used in this evaluation were obtained from the ground-truth alignment as it was not captured directly from the musicians--the instruments recorded in the Bach10 dataset were not MIDI capable. Further, the annotation of the ground-truth alignment was done via manual beat alignment, so parallel notes on different voices are aligned on the same annotated note onset time. It can, nevertheless, be expected that if MIDI-capable instruments were used, similar results can be obtained: the issue with parallel notes can be resolved with the post-alignment step with a reasonable setting for the post-alignment parallel voice threshold $\delta$. The next evaluation in \autoref{impl_asm_eval_bwv846} uses data that do not produce evaluation results with such caveats.

\subsection{Performance on Bach's BWV 846}
\label{impl_asm_eval_bwv846}

Bach's BWV 846 is a recurring theme for this project--its value is that it is a pair of pieces that showcases two important contrasts: the \textit{Prelude} is primarily monophonic and lacks ornaments; the \textit{Fugue} is highly polyphonic and contains some ornaments. Thus, it is apt to use both pieces to evaluate the aligner. The performance MIDI of both pieces are obtained from the MAESTRO dataset \cite{MAESTRO}. The sheet music is obtained from \textit{MuseScore.com}\footnote{\url{https://musescore.com/}}, which contains high-quality sheet files for both performances.

\begin{figure}[h]
    \centering
    \includegraphics[width=0.5\columnwidth]{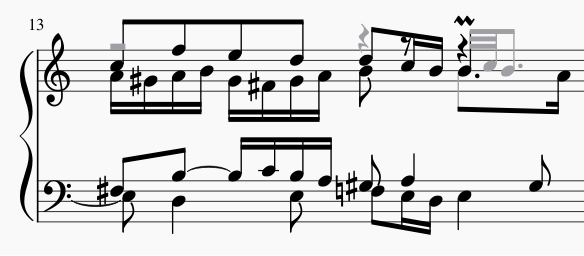}
    \caption{Bar 13 of Bach's \textit{Fugue in C major, BWV 846} in \textit{MuseScore}. The semi-opaque notes are manual indications for the execution of the \textit{mordent} (the ornament shown above the fourth beat of the top staff).}
    \label{fig:846fugue13}
\end{figure}

The \textit{MuseScore}\footnote{\url{https://musescore.org/}} score-editing program provides features to annotate how ornaments should be played. The reference sheet file for the \textit{Fugue} contains the standard interpretation of the written ornaments. For instance, \autoref{fig:846fugue13} shows how the interpretation of a \textit{mordent} is specified in the 13th bar of the \textit{Fugue}.

These indications represent the standard--and most widely accepted--way of executing these ornaments and are left as-is--the player however is free to interpret the ornament however they wish; in fact, the performer in the MAESTRO dataset played this \textit{mordent} with a few more alternating notes, making it more akin to a short \textit{trill}. Misalignments are thus expected, and manual adjustments after the initial alignment can be done to fix these deviations.

As the MAESTRO dataset does not contain ground-truth alignments like the Bach10 dataset used in \autoref{impl_asm_eval_bach10}, only metrics reported by the aligner are evaluated--these are summarised in \autoref{table:asmbwv846}. Manual inspection is further performed to check the alignment between the performance and the score.

\begin{table}[ht]
    \caption{Metrics used to evaluate the ASM Score Aligner against BWV 846.}
    \centering 
    \begin{tabular}{c c} 
        \hline\hline 
        Metric                        & Symbol                  \\ [0.5ex] 
        \hline 
        Length of Alignment           & $|L|$                   \\
        Number of Gaps in Performance & $N_{\Gamma P}$          \\
        Number of Gaps in Score       & $N_{\Gamma S}$          \\
        Number of Mismatches          & $N_{\textsc{Mismatch}}$ \\
        \hline 
    \end{tabular}
    \label{table:asmbwv846} 
\end{table}

Because the performance data are obtained from a MIDI-capable device, the problems related to polyphonic notes mentioned in \autoref{impl_asm_polyphony} arise. It is expected that the \textit{Prelude} requires minimal fixes, as opposed to the \textit{Fugue}. Therefore, the post-alignment step is run with $\delta$ adjusted as necessary.

It is worth mentioning that the MAESTRO dataset comprises performances by performers with a range of skill with some performances containing substantial errors--it is after all collected from an international junior piano competition\footnote{\url{https://piano-e-competition.com/}}. The \textit{Fugue} is a much more challenging piece than the \textit{Prelude}, and more deviations were found in the performance.

\subsubsection{Prelude}

\begin{figure}[h]
    \centering
    \includegraphics[width=\columnwidth]{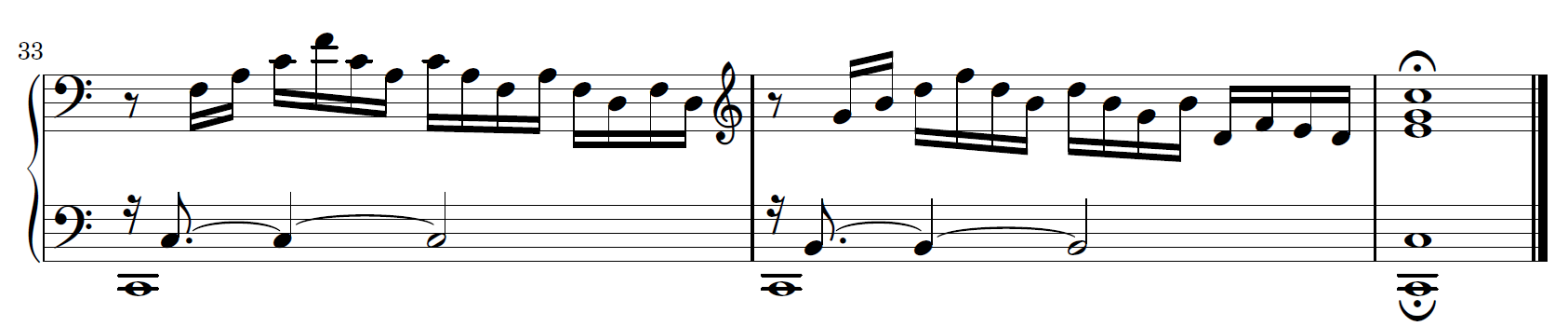}
    \caption{Bars 33-35 of Bach's \textit{Prelude in C major, BWV 846}. The final bar contains a chord in this otherwise highly monophonic piece.}
    \label{fig:prelude33}
\end{figure}

While this project treats the \textit{Prelude} as a highly monophonic piece, it is worth noting that the final bar contains a chord. This can be seen in \autoref{fig:prelude33}. Throughout the piece the two base notes of each broken \textit{arpeggio} are also held--but this property can be ignored since only note onsets are dealt with.

It is thus expected that the post-alignment step is needed to handle that last chord. Alignment was first run with $\delta = \text{N/A}$, i.e. the post-alignment step is not run. It was then found that setting $\delta = 0$ is sufficient to handle the last chord. Results are summarised in \autoref{table:asmpreludescores}. The repository of the testbench\footnote{\githubFlippyQuant} contains instructions on how to reproduce these results.

\begin{table}[ht]
    \caption{Results of the ASM Score Aligner against the Prelude, BWV 846. The arrows, where present, denote whether a metric should be minimised ($\downarrow$) or maximised ($\uparrow$).} 
    \centering 
    \begin{tabular}{c c c c c c c c} 
        \hline\hline 
        $\delta$ & $|L|$ & $N_{\Gamma P}\downarrow$ & $N_{\Gamma S}\downarrow$ & $N_{\textsc{Mismatch}}\downarrow$ \\ [0.5ex] 
        \hline 
        N/A      & 551   & 3              & 2              & 0                       \\
        0        & 549   & 1              & 0              & 0                       \\
        \hline 
    \end{tabular}
    \label{table:asmpreludescores} 
\end{table}

As expected, the ASM aligner was able to align the piece well even without the post-alignment step. Note that when the post-alignment step is not run, gaps arise in the final bar--the chord is not executed in parallel perfectly. The last 7 lines of the alignment output are as following:
\begin{verbatim}
// GAP: GAP - 113333.288 36
// GAP: 134664.1666666667 72 - GAP
134670.83333333334 113333.288 48
// GAP: 134674.1666666667 36 - GAP
134675.83333333334 113333.288 64
// GAP: GAP - 113333.288 67
// GAP: GAP - 113333.288 72
\end{verbatim}

When post-alignment was run with $\delta = 0$, most of the gaps are fixed, save for a gap in the performance--this is due to performer mistake: in the final chord, the performer omitted MIDI note number 64 which corresponds to the \textit{G} above middle-\textit{C}. This gap can therefore be removed for the ground-truth alignment, and thus this shows that the ASM approach handles the \textit{Prelude} exceptionally well.

\subsubsection{Fugue}

The highly polyphonic \textit{Fugue} is expected to introduce more gaps and mismatches without the post-alignment step. Further, the \textit{Fugue} is more technically demanding than the \textit{Prelude}--this is reflected in
the higher number of mistakes by the performer in the piece.

With regard to polyphony, $\delta$ is adjusted further higher. For reference, the time taken for a beat in the score is $1000\si{\milli\second}$. Key results with $\delta \in \{\text{N/A}, 0\si{\milli\second}, 500\si{\milli\second}, 1000\si{\milli\second}\}$ are shown in \autoref{table:asmfuguescores}.

\begin{table}[ht]
    \caption{Results of the ASM Score Aligner against the Fugue, BWV 846. The arrows, where present, denote whether a metric should be minimised ($\downarrow$) or maximised ($\uparrow$).} 
    \centering 
    \begin{tabular}{c c c c c c c c} 
        \hline\hline 
        $\delta (\si{\milli\second})$ & $|L|$ & $N_{\Gamma P}\downarrow$ & $N_{\Gamma S}\downarrow$ & $N_{\textsc{Mismatch}}\downarrow$ \\ [0.5ex] 
        \hline 
        N/A                           & 950   & 196            & 197            & 29                      \\
        0                             & 932   & 178            & 179            & 26                      \\
        500                           & 787   & 33             & 34             & 0                       \\
        1000                          & 549   & 23             & 24             & 0                       \\
        \hline 
    \end{tabular}
    \label{table:asmfuguescores} 
\end{table}

The number of deviations, as projected, are higher in the \textit{Fugue} than that in the \textit{Prelude}. It is clear that a higher $\delta$ is required, however it must be noted that higher $\delta$ values introduce the possibility of \emph{false positives}--the post-alignment algorithm may be too aggressive in resolving gaps and mismatches to the extent that it mistakenly aligns some notes. Upon closer inspection of the results, $\delta = 1000\si{\milli\second}$ contains some false positives--some performance notes are matched to wrong notes in the score. The alignment results for $\delta = 500\si{\milli\second}$, which corresponds to half a beat in score time, had a close-to-zero false positive rate--the gaps were later identified as missed or extra notes in the score. The extra notes were mostly in areas with \textit{mordent} ornaments (an example is shown in \autoref{fig:846fugue13})--in which, as mentioned, the player plays the \textit{mordent} as a short \textit{trill} instead, meaning that there were extra notes in the performance. Gaps caused by missed notes were mainly mistakes, and were easily fixable.

For the purpose of creating the ground-truth alignments, these minor mistakes can actually be omitted entirely should the benefit of the doubt be given to the follower for missed or extra notes. Overall, the performance of the ASM aligner on the \textit{Fugue} is exceptional, and the fact that the alignment only required trivial manual adjustments to be of ground-truth quality is testament to the robustness and reliableness of the ASM aligner.

\section{Discussion}

\subsection{Testbench Complement: Ground-Truth Aligner}

As can be seen from the results in \autoref{impl_asm_eval}, this ASM aligner proved capable and reliable of quickly producing high-quality ground-truth alignment datasets from music performance datasets containing performance MIDI data, namely the MAESTRO dataset. This result means that researchers of score followers and score aligners may exploit the increasing ubiquity of such datasets to produce high-quality reference datasets for evaluation of score followers. Moreover, all the requirements listed in \autoref{impl_asm_req} are met.

This aligner therefore serves as a good complement to the quantitative testbench introduced in \autoref{tb_quant}, and will resolve the challenge of sourcing high-quality ground-truth alignments for evaluation. The proposed approach also is less tedious than the approaches taken by authors of the Bach10 dataset \cite{duan11bach10}, Traditional Flute Dataset \cite{tfd}, PHENICX-Anechoic dataset \cite{miron16, PHENICX} and Vienna 4x22 Piano Corpus \cite{vienna4x22}, as covered in \autoref{tb_quant_dataset_api}. Further, while Thickstun \etal{} \cite{thickstun20} proposed a beat-alignment algorithm that--like this approach--requires minimal manual work, this approach is more flexible: Thickstun \etal{}'s approach was limited only for music that intersects both the MAESTRO \cite{MAESTRO} and KernScores \cite{sapp05} datasets; this proposed approach can work with any arbitrary pair of performance and score data.

\subsection{Conversion to Score Following}

Most early work in score following, as covered in \autoref{scofo_earlywork}, involve ASM-based methods. It is therefore interesting to study how this aligner can be converted to its real-time counterpart.

First, note that this aligner produces \emph{offline optimal global} alignments given two complete performance and score strings. Score followers do not have access to the whole performance string, and hence have to settle for an \emph{online suboptimal local} alignment. Upholding optimality is difficult mainly due to the time complexity of alignment algorithms. The \emph{local} counterpart of the Needleman-Wunsch algorithm, the Smith-Waterman algorithm \cite{smithwaterman}, has a time complexity of $\mathcal{O}(mn)$ to align two sequences of lengths $m$ and $n$, which is not sufficiently fast enough for real-time purposes. Most of the early ASM-based followers use certain heuristics to reduce the time complexity to linear or quasilinear \cite{dannenberg84, vercoe84}, resulting in systems that can run \emph{online}.

Further, ASM-based methods are plagued by limitations in performance feature extraction. ASM-based methods require the string representation of performances, and a lot of work was put into making string-based feature extraction from performances robust--see \autoref{scofo_relatedwork_early_workcontinued}. MIDI-capable devices prove trivial as the data can be directly captured, but work with performance audio proved difficult for ASM-based methods, especially with ones that work with pitch data.

Moreover, much effort was put into dealing with ornaments and parallel voices. These issues are difficult even for non-real-time systems as shown in subsections~\ref{impl_asm_ornaments} and \ref{impl_asm_polyphony}. \Cref{scofo_challenges_polyphonic} details more of the challenges in dealing with polyphony. Even with myriad extensions for ASM-based followers to handle polyphony \cite{dannenberg88}, many researchers in the field do not consider the problem of polyphony properly handled until an audio-to-audio DTW-based approach was introduced roughly 20 years since the first score following papers \cite{Dixon2005MATCHAM}.

The complications stated here therefore explain why ASM-based methods were quickly made obsolete by approaches involving statistical models (\autoref{scofo_stats_hmm}) and DTW (\autoref{scofo_dtw}) that can work directly on performance audio. Nevertheless, it is worth noting that even recent advancements in score following build upon the work and concepts from basic ASM-based score aligners; for instance, the common path constraints used in DTW algorithms are similar to those presented in \autoref{impl_asm_complexity}. Therefore, the detailing of this ASM-based aligner serves as a good technical introduction for core concepts used in more advanced score following approaches.

%% file: 02_parts/03_impl/chapters/04_dtw.tex
\chapter{A CQT-DTW Score Follower}
\label{impl_dtw}
\epigraph{\itshape The notes are just signs. You have to go behind them.}{Mariss Jansons}

\section{Motivation}

The previous score followers introduced in chapters~\ref{impl_beat} and \ref{impl_asm} do not sufficiently address the challenges in score following this project aims to tackle. This CQT-DTW approach therefore is set out to resolve as many challenges listed in \autoref{scofo_challenges} as possible. The specific challenges targeted are listed in \autoref{impl_dtw_req}.

\section{Requirements}
\label{impl_dtw_req}

This score follower shall address the following challenges previously listed in \autoref{scofo_challenges}:
\begin{enumerate}
    \item \textbf{\Cref{scofo_challenge_limitfe} -- \nameref{scofo_challenge_limitfe}}.

          This DTW-based system shall work with reliable features extracted from the audio of both the performance and score. No extra processing will be required from the users' side, and digital score data is optional.

    \item \textbf{\Cref{scofo_challenges_perfdev} -- \nameref{scofo_challenges_perfdev}}.

          This system shall aim to provide robust score following even with the presence of performance deviations, either intentional or unintentional.
    \item \textbf{\Cref{scofo_challenges_underspec} -- \nameref{scofo_challenges_underspec}}.

          Music ornaments, such as the \textit{trill}, \textit{mordent} and \textit{glissandi}, shall be properly tracked by this follower.

    \item \textbf{\Cref{scofo_challenges_polyphonic} -- \nameref{scofo_challenges_polyphonic}}.

          Polyphonic music, either mono-instrumental or multi-instrumental, shall be handled reliably by this system.
\end{enumerate}

While the final challenge--\nameref{scofo_challenge_music_improv} (\autoref{scofo_challenge_music_improv})--is omitted, further work can be done to extend this system to address the challenge. As trackers are rarely used for improvised music, this is not currently targeted by the implementation.

\section{Technical Background}

Technicalities that build up this score following system are provided here, beginning with the motivation backing the choice of CQT in \autoref{impl_dtw_cqt_compared} before a deep dive into its technicalities given in \autoref{impl_dtw_cqt_tech}; \autoref{impl_dtw_cqt_to_note} is dedicated to detail how the resultant features vectors for alignment are derived. Similarity calculation via DTW is then covered in subsections~\ref{impl_dtw_compared} (where a comparison of DTW with other approaches is given) and \ref{impl_dtw_tech} (where DTW's technicalities are covered in detail). \Cref{impl_dtw_dist_comp} deals with the distance computation required for DTW. Finally, \autoref{impl_dtw_baseline_worked_example} covers OLTW--the online DTW implementation used in this approach; a worked example is also included.

\subsection{CQT: Compared to other Features}
\label{impl_dtw_cqt_compared}

As mentioned in \autoref{scofo_mdr_audio}, the audio from the score and performance can be very difficult to work with. Therefore, application-specific lower-level features are often extracted from the audio. Numerous differing examples of such features exist, and many were previously applied in score following. The choice of CQT over other methods is defended by comparing CQT directly against other methods; more in-depth analyses of these features can be found in
\cite{grachten13, joder10}.

It is also worth noting that CQT was used in Chen and Jang's DTW approach \cite{chen19} recently, which showed that CQT was indeed robust for score following; however, their approach mainly focuses on optimising offline onset detection methods for alignment--the real-time following relies on an off-the-shelf online onset detector that works via Recurrent Neural Networks (RNNs) \cite{eyben10, bock12}. Chen and Jang also used score information to determine the frequency range of CQT to use--its dependency on score information is seen as an inflexibility when compared to this system.

\subsubsection{Versus Short-time Fourier Transform (STFT)}

The spectrogram of an audio signal can be produced via computing the STFT--an example of a spectrogram is shown in \autoref{fig:preludespecgram}. Background for STFT related to CQT can be found in \autoref{impl_dtw_cqt_tech}. The STFT is well researched and applied in many areas of signal processing due to its simplicity, but there are two major issues in the context of score following.

Firstly, the complexity required to compare two spectrograms is intensive. Early DTW score aligners therefore use more sparse mid-level features \cite{muller04, orio01}, which were not reliable due to the next issue: the standard FFT uses a constant bin size throughout all frequencies, leading to a consistent and fully continuous transform that unfortunately does not reflect human auditory perception--humans perceive frequency on a logarithmic scale. Mapping these constant-sized bins into a logarithmic scale results in peaks on the lower end that are incredibly wide, lacking much required detail. The pioneering DTW score follower by Dixon and Widmer in 2005 \cite{Dixon2005MATCHAM} went around this limitation by mapping the windowed FFT of the signal into 84 frequency bins that are linear at low frequencies and logarithmic at high frequencies; however, the resolution problem at low frequency persists.

CQT neatly solves the second issue by increasing the buffer size at lower frequencies and reducing the buffer size at larger frequencies--\autoref{impl_dtw_cqt_tech} contains more technical background for the CQT and develops concepts of CQT based on the Fourier transform.  The first issue is actually not addressed by the naïve implementation of the CQT, and in fact the computation of CQT is more intensive. Real-time CQT is however made possible in advancements by Dörfler, Holighaus, Grill and Velasco \cite{holighaus13,angelo11}--more details are given in \autoref{impl_dtw_cqt_tech}.

\subsubsection{Versus Chromagrams}

Discrete chromagrams, introduced by Wakefield in 1999 \cite{wakefieldchroma}, are sequences of \emph{chroma vectors}, where each chroma vector representation is a 12-element vector. Each element in the vector represents the spectral energy corresponding to one pitch class in an octave (e.g. C, C\#, D, ...). Chromagrams were in fact used by numerous approaches in both alignment and following, and the computation of chromagrams can be done via a number of different ways, including via the STFT and CQT \cite{dannerberg03, ellis07, mauch10, muller11, peeters06, suzuki10}. Systems using features derived from the chroma representation were also developed \cite{ewert09, muller05}.

A latent problem with using chromagrams exists: the pitches of a note class across all octaves are grouped into one element in the chroma vector. This means that the representation is unable to determine from which octave a note is from--while this may be seen as an advantage in terms of handling deviations from the score in the event that the performer plays the piece with the correct notes in the wrong octave; nevertheless, such a system would fail when there are parts in the score that comprise distinct parts with the same notes transposed to different octaves.

The CQT is capable of representing the range required for multi-octave music, and hence is immune to this problem.

\subsubsection{Versus Mel Frequency Cepstral Coefficients (MFCC)}

Mel Frequency Cepstral Coefficients (MFCCs) are more commonly used in speech methods \cite{mermelstein76}, but are useful features in MIR \cite{logan00} and were previously applied in music alignment \cite{grachten13}. MFCCs are in fact based on the STFT, and are commonly derived as follows \cite{SAHIDULLAH2012543}:

\begin{enumerate}
    \item Take the STFT of a signal.
    \item Using triangular or cosine overlapping windows, map the powers of the spectrum obtained onto the \emph{mel scale} (a perceptual scale of pitches judged by listeners to be equidistant from one another) \cite{stevens37}.
    \item Take the logs of the powers at each of the mel frequencies.
    \item Take the DCT (Discrete Cosine Transform) of the list of mel log powers.
\end{enumerate}

These amplitudes of the resulting spectrum are finally the MFCCs. Other approaches to compute MFCCs can be found in \cite{SAHIDULLAH2012543}.

MFCCs, however, are less useful for capturing pitch features (a task in which the CQT excels at) essential for score following \cite{hu04}. MFCCs are more useful for capturing the formant structure and timbre of speech and music signals \cite{grachten13}.

\subsubsection{Versus Other Features}

Features other than the ones listed above, such as chroma onsets \cite{ewert09} and spectral factorisation \cite{orti15}, were also used in the literature, but thus far no known methods reliably outperform the flexibility, reliability and robustness of the CQT to capture comprehensive audio features required for score following.

\subsection{CQT: Technicalities}
\label{impl_dtw_cqt_tech}

Before diving into the technicalities behind CQT, related background of the Fourier Transform and STFT is first given. Note that, for simplicity, the \emph{continuous} variants are discussed--their \emph{discrete} analogues \cite{nagathil13} are used in digital signal processing, as in the case of this score following implementation.

\subsubsection{Relevant background in Fourier Transform}

First, a definition of the Fourier Transform is given:

\begin{definition}[Fourier Transform]
    The Fourier transform of $f(t)$ is given by
    $$
        F(\omega) = \int_{-\infty}^{\infty} f(t) e^{-j\omega t} \, dt.
    $$
\end{definition}

If the frequency of $f(t)$ changes over time, $F(\omega)$ does not capture \emph{what} these changes are and \emph{when} these changes occur. This is because $e^{j\omega t}$ does not have \emph{compact support}, defined as:

\begin{definition}[Compact Support \& Set]
    A function has compact support if it is zero outside a \emph{compact set}. A compact set $\mathfrak{S}$ of a topological space $\mathfrak{X}$ is compact if for every open cover of $\mathfrak{S}$ there exists a finite subcover of $\mathfrak{S}$.
\end{definition}

\begin{figure}[h]
    \centering
    \includegraphics[width=0.5\columnwidth]{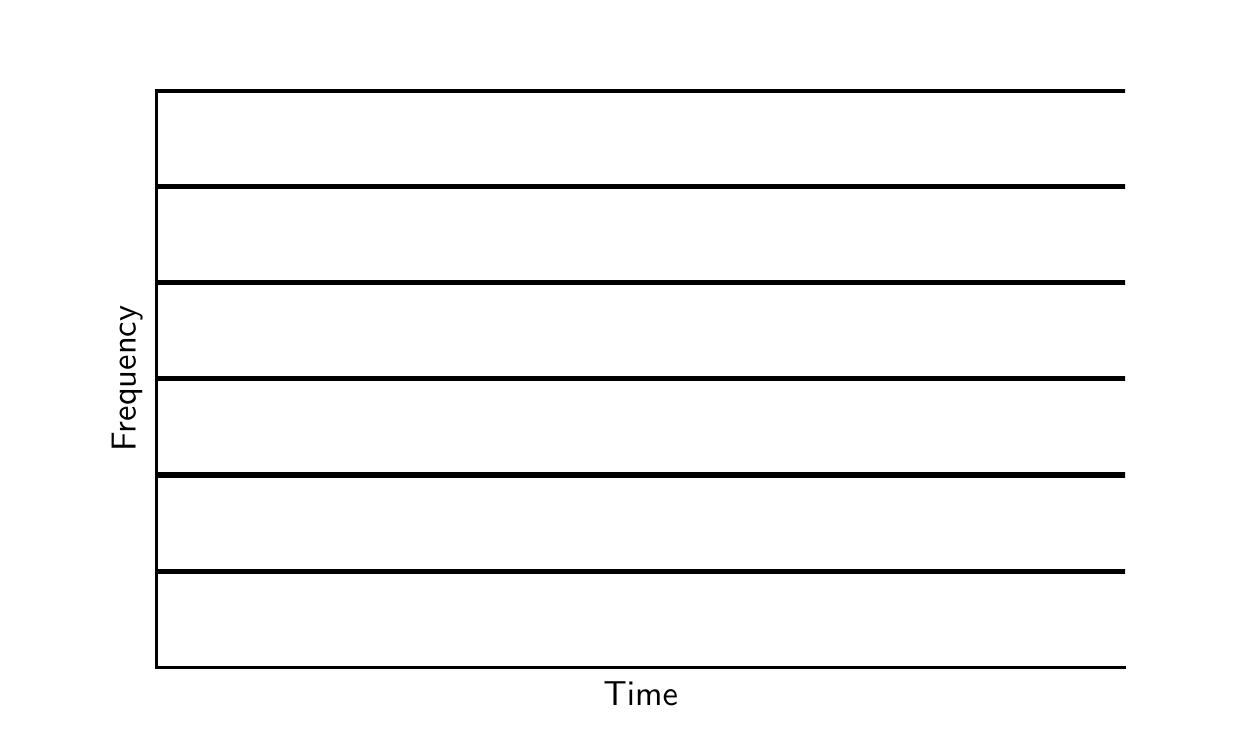}
    \caption{The partition of the time-frequency plane representing Fourier Transform behaviour.}
    \label{fig:ft-partitions}
\end{figure}

Thus, the Fourier transform is not ideal for studying \emph{non-stationary} process signals, which includes musical audio as it has non-constant frequencies and spectral contents. \autoref{fig:ft-partitions} shows the behaviour of the Fourier Transform--it is not able to capture changes in frequency over time.

\subsubsection{Relevant background in STFT}

STFT is a \emph{Fourier-related transform} capable of performing time-frequency analysis of non-stationary process signals. The definition is most commonly given as:

\begin{definition}[Short-time Fourier Transform]
    The Short-time Fourier Transform of $f(t)$ first multiplies $f(t)$ by a real-valued even \emph{windowing function} $g(t)$ (which is nonzero over only a short, finite interval and has a norm of $1$). The Fourier transform of the resulting signal is taken as the window is slid along the time axis \emph{in constant intervals}, resulting in a 2D representation of the signal. Mathematically,

    $$
        STFT\{f(t)\}(u, \omega) = \int_{-\infty}^{\infty} f(t) g(t-u) e^{-j \omega t} \, dt.
    $$
    \label{eqn:stft1}
\end{definition}

The ``sliding along the time axis'' involves the studying of contents of $f(t)$ around the time $u$ by varying $u$. What can be decided is the size around $u$ to consider at any point in time--this is determined by the \emph{support} of the windowing function $g(t)$.

\begin{figure}[h]
    \centering
    \includegraphics[width=0.5\columnwidth]{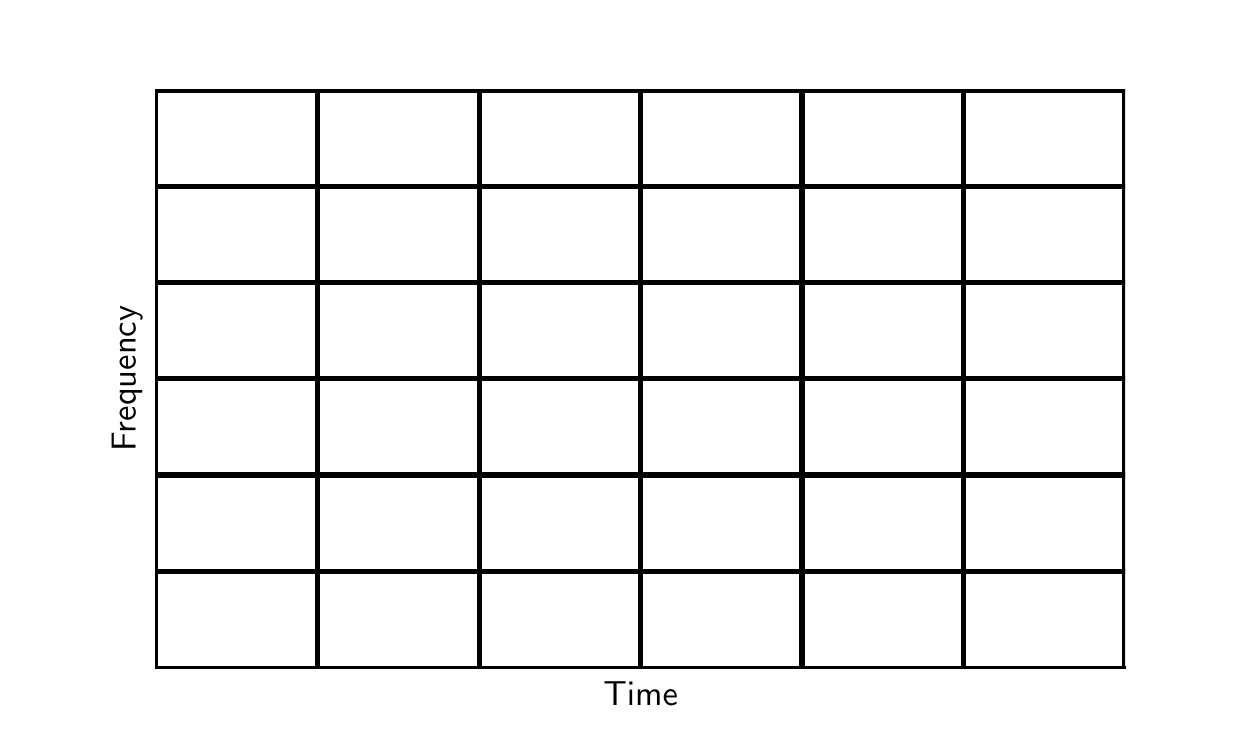}
    \caption{The partition of the time-frequency plane representing STFT behaviour.}
    \label{fig:stft-partitions}
\end{figure}

Relating to \autoref{fig:ft-partitions}, the STFT tiling of the time-frequency plane can be represented as a grid of boxes each centred at $(u, \omega)$, shown in \autoref{fig:stft-partitions}.

One pitfall for STFT is that it has a fixed resolution--the width of the windowing function $g(t)$ directly affects how the signal is represented in terms of resolutions in time and frequency. This is known as the \emph{time-frequency uncertainty principle}, or the \emph{Gabor limit}, which is related to Heisenberg's uncertainty principle.

\begin{definition}[Time-frequency Uncertainty Principle]
    Given that the standard deviations of the time and frequency estimates (in $\si{\radian}\,\si{\second}^{-1}$) are $\sigma_t$ and $\sigma_\omega$ respectively,
    $$
        \sigma_t \sigma_\omega \geq \frac{1}{2},
    $$
    which implies that a function cannot be both time- and band-limited; alternatively from $\omega = 2\pi f$ this can be rewritten in terms of the standard deviation of frequency estimate $\sigma_f$ (in \si{\hertz}),
    $$
        \sigma_t \sigma_f \geq \frac{1}{4\pi}.
    $$
    \label{def:timefreq}
\end{definition}

\begin{figure}[h]
    \centering
    \subfloat[]{{\includegraphics[width=0.45\linewidth]{assets/plots/stft.pdf} }}
    \qquad
    \subfloat[]{{\includegraphics[width=0.45\linewidth]{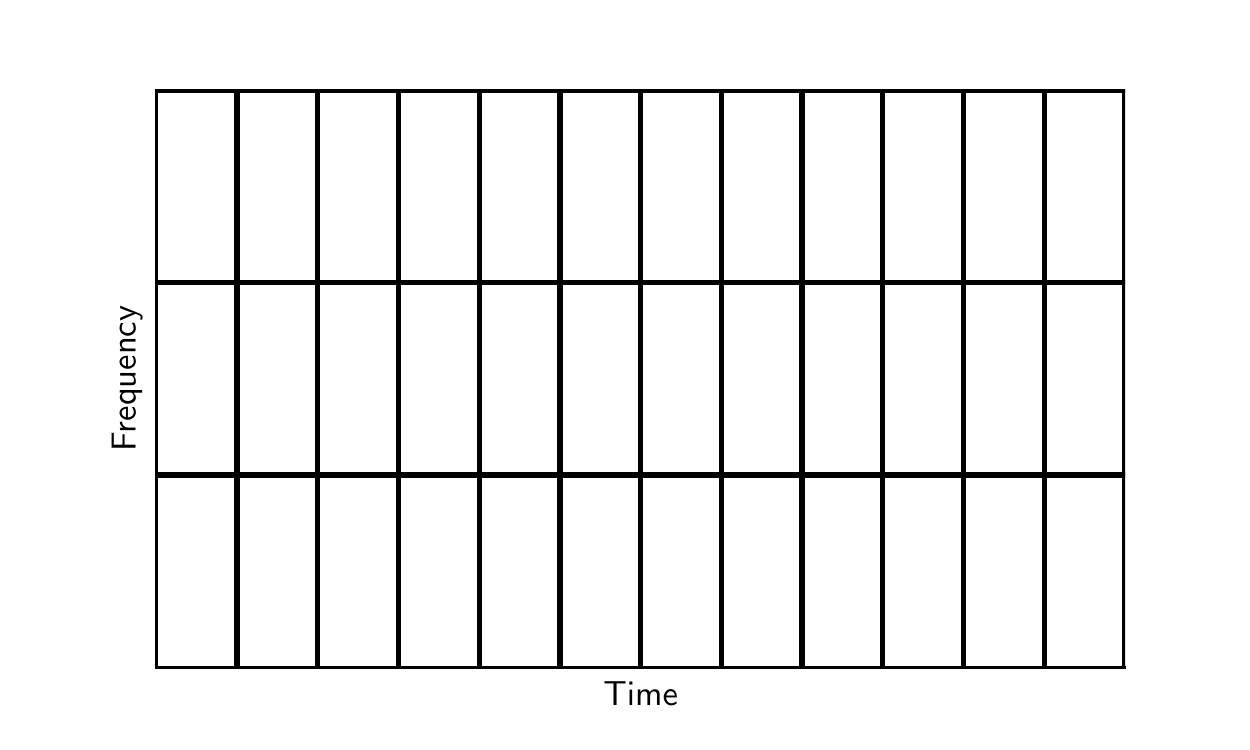} }}
    \caption{Illustration of the time-frequency uncertainty principle. The left figure has a wider window, giving it better frequency resolution but worse time resolution. The right figure gets better time resolution with a thinner window, but loses out in frequency resolution.}
    \label{fig:stft2}
\end{figure}

\autoref{fig:stft2} shows a comparison of using two window widths.

Another problem of STFT is that the frequency bins are constant sized and cannot be changed--humans perceive pitch on a \emph{logarithmic} scale, and the constant frequency bins in STFT when mapped to a logarithmic scale leads to poor resolution in the lower frequency bins.

\subsubsection{CQT: Background}

The CQT is another Fourier-related transform that was first proposed in 1978 \cite{youngbergcqt} and subsequently introduced to musical applications in 1991 \cite{browncqt}. The $Q$ in CQT denotes the ratio of the centre frequency to the bandwidth--thus Constant-Q implies that the aforementioned ratio is kept constant.

CQT therefore is capable of capturing auditory features that relate to the resolution of the human auditory system. This is opposed to the STFT that has constant bandwidths and window sizes. However, it is worth noting that the CQT is still subject to the time-frequency uncertainty principle defined in \Cref{def:timefreq}.

CQT's constant-Q ratio is achieved by varying the time shifts for each window, giving rise to two key observations:

\begin{itemize}
    \item At high frequencies, resolution in \emph{time} improves.
    \item At low frequencies, resolution in \emph{frequency} improves.
\end{itemize}

The CQT depends on a few key parameters:

\begin{itemize}
    \item $f_s$, the sampling rate.
    \item $N_{bin}$, the number of bins per octave.
    \item $f_{\min}$ and $f_{\max}$, the minimum and maximum frequencies respectively.
\end{itemize}

$f_s$ is typically set to $44100\si{\hertz}$, the common sampling rate used by music on Audio CD, which captures the typical range of human hearing ($20-20000\si{\hertz}$) as a consequence of the Nyquist-Shannon sampling theorem \cite{nyquist}. It is worth noting, however, that the highest pitch on a piano--$C8$--corresponds to a frequency of approximately $4186.0\si{\hertz}$, which implies that lower $f_s$ (e.g. $22050\si{\hertz}$) may still capture adequate information.

Naturally as in musical applications there are 12 notes in an octave, $N_{bin}$ is set to $12$. $f_{\min}$ should logically be set to the lowest pitch on the piano $A0$ ($27.5\si{\hertz}$), however this is usually set much higher, e.g. at $C3$ ($130.8\si{\hertz}$), and the reason can be traced back to \Cref{def:timefreq}--the bin size required for the gap between $A0$ and $A\#0$ is around $1.6\si{\hertz}$; setting $\sigma_f$ to half of the smallest bin size gives $\sigma_f \approx 0.8 \si{\hertz}$ and $\sigma_t \approx 100\si{\milli\second}$. Notably, $\sigma_t \approx 100\si{\milli\second}$ leads to relatively unresponsive applications due to low time resolution. On the other hand, setting $f_{\min} = C3$ gives $\sigma_f \approx 65.4 \si{\hertz}$ and $\sigma_t \approx 1200\si{\micro\second}$ in a similar configuration. While there certainly is information lost when $f_{\min}$ is set higher, this is countered by the fact that the sounds made by an instrument produces \emph{overtones} above the note (thus part of these overtones are captured within the range). Worth noting is that Chen and Jang \cite{chen19} instead used score information to determine both $f_{\min}$ and $f_{\max}$, but there is a key optimisation used by them that will be detailed later on.

Clearly, $f_{\max}$ is bounded by $f_s$ by the Nyquist-Shannon sampling theorem as such:
$$
    f_{\max} \leq \frac{1}{2} f_s;
$$
however, again for optimisation reasons to result in fewer bins to process, $f_{\max}$ is usually set at--or slightly above--$4186.0\si{\hertz}$, the highest pitch of the piano.

Once $N_{bin}$, $f_{\min}$ and $f_{\max}$ are chosen, a sequence of geometrically spaced centre frequencies $f_0, f_1, ..., f_K$ can be formed:

$$
    f_k = f_{\min} \times 2^{\frac{k}{N_{bin}}}
$$
where $0 \leq k \leq K, k \in \mathbb{Z}$
and $K \in \mathbb{Z}$ is the largest frequency such that $f_K \leq f_{\max}$ implying $f_{K+1} \geq f_{\max}$. (This further implies that $K$ is the total number of frequency bins.)

From here, the bandwidth at the $k$-th frequency is therefore
$$
    \Omega_k = f_{k+1} - f_{k-1}.
$$

Thus, as required by CQT, $Q$, the ratio of $f_k$ (the centre frequency) to $\Omega_k$ (its corresponding bandwidth) is \textbf{constant} and independent of $k$:

\begin{align*}
    Q & = \frac{f_k}{\Omega_k}                                                                                                          \\
      & = \frac{f_k}{f_{k+1} - f_{k-1}}                                                                                                 \\
      & = \frac{f_{\min} \times 2^{\frac{k}{N_{bin}}}}{f_{\min} \times 2^{\frac{k+1}{N_{bin}}}-f_{\min} \times 2^{\frac{k-1}{N_{bin}}}} \\
      & = \frac{2^{\frac{k}{N_{bin}}}}{2^{\frac{k+1}{N_{bin}}}-2^{\frac{k-1}{N_{bin}}}}                                                 \\
      & = \frac{1}{2^\frac{1}{N_{bin}} - 2^{-\frac{1}{N_{bin}}}}.
\end{align*}

It is quite clear that the calculation of the STFT involves moving a window along the \emph{time} axis. The CQT requires a different paradigm--the window function is instead the real-valued, zero-centred and even continuous function $W(f)$, which is moved along the \emph{frequency} axis. $W(f)$ is positive in the interval $(-0.5, 0.5)$ and zero elsewhere. To achieve varying bandwidths, $W(f)$ is scaled accordingly as it is translated, the window required for $f_k$ is given by

$$
    W_k(f) = \frac{f - f_k}{\Omega_k}.
$$


\subsubsection{CQT: Calculation}
\label{sec:cqt_calc}
The CQT, as can be seen, is much more complex than the STFT. This most likely motivated Dixon and Widmer in 2005 \cite{Dixon2005MATCHAM} to instead bin spectral features calculated via the more efficient STFT into linear-log bins. This technique was in fact used even in 2021 by Henkel and Widmer to extract features for their object-detection-based sheet music score follower \cite{henkel21}. Chen and Jang in their 2019 score following system \cite{chen19} instead limited the number of times the CQT is calculated--they only calculated the CQT \emph{twice} for each detected onset (before and after). CQT's inefficiency strongly demotivated its use in real-time approaches--CQT (and its derivatives) is more commonly used in non-real-time applications \cite{mauch10, muller11, peeters06, suzuki10,thickstun20}.

In 1992 a method employing the FFT to efficiently calculate the CQT (and its inverse) was proposed \cite{brown92}; this was improved further via a method that efficiently calculates CQT octave-by-octave using low-pass filtered and downsampled results for consecutively lower pitches \cite{christiancqt}. However, neither approach was capable in real-time processing of an audio stream.

Nevertheless, in \cite{holighaus13} a bounded-delay CQT implementation was proposed; it was shown to be real-time capable. This approach was referred to as the \emph{sliced CQT} (sliCQ), as it involved \emph{slicing} the signal into overlapping time slices via Tukey windows--this is in contrast to prior approaches that rely on a Fourier transform of the entire signal.

A detailed overview of sliCQ can be found in \cite{holighaus13}; here, a high-level overview on the analysis part is given to highlight key features of the algorithm and detail modifications required to enable real-time processing of a streamed signal. First, the \textbf{$\textbf{CQ-NSGT}_L$} NSG (Nonstationary Gabor) analysis algorithm for a signal $x$ of length $L$ is given in \Cref{alg:nsganalysis}. Key functions and parameters are as follows:
\begin{itemize}
    \item \textbf{$\textbf{(I)FFT}_N$} denotes a (inverse) FFT of length $N$ (including the necessary preprocessing--periodisation or zero-padding--to ensure that the input vector is of length $N$).
    \item $a_k$ denotes selected frequency-dependent time-shift parameters (hop sizes) forming the vector $\textbf{a}$.
    \item $g_k$ are real-valued filters centred at $\omega_k$ representing windows constructed in the frequency domain forming the vector $\textbf{g}$.
    \item $I_K$ is the finite index set where $k \in I_K$.
\end{itemize}

\begin{algorithm}[h]
    \caption{NSG analysis: $c = \textbf{CQ-NSGT}_L(x,\textbf{g},\textbf{a})$}\label{alg:nsganalysis}
    \begin{algorithmic}[1]
        \State $y\gets \textbf{FFT}_L(x)$
        \For{$k\in I_K,~n=0,\ldots,\frac{L}{a_k}-1$}
        \State $c \gets \sqrt{\frac{L}{a_k}}\cdot\textbf{IFFT}_{\frac{L}{a_k}}(y\bar{g_k})$
        \EndFor
    \end{algorithmic}
\end{algorithm}

NSG analysis in \Cref{alg:nsganalysis} results in $c$, \textbf{CQ-NSGT} coefficients that mimic CQT coefficients calculated by classical methods \cite{brown92, christiancqt}. Further, \cite{holighaus13} also proposes the inverse algorithm: NSG synthesis, which can transform $c$ back into $x$--readers interested in the details as well as proofs of the propositions may consult \cite{holighaus13}.



The computational complexity of NSG analysis, shown in \Cref{alg:nsganalysis} is
$$\mathcal{O}(L \log L).$$

\begin{figure}[h]
    \centering
    \includegraphics[width=0.5\columnwidth]{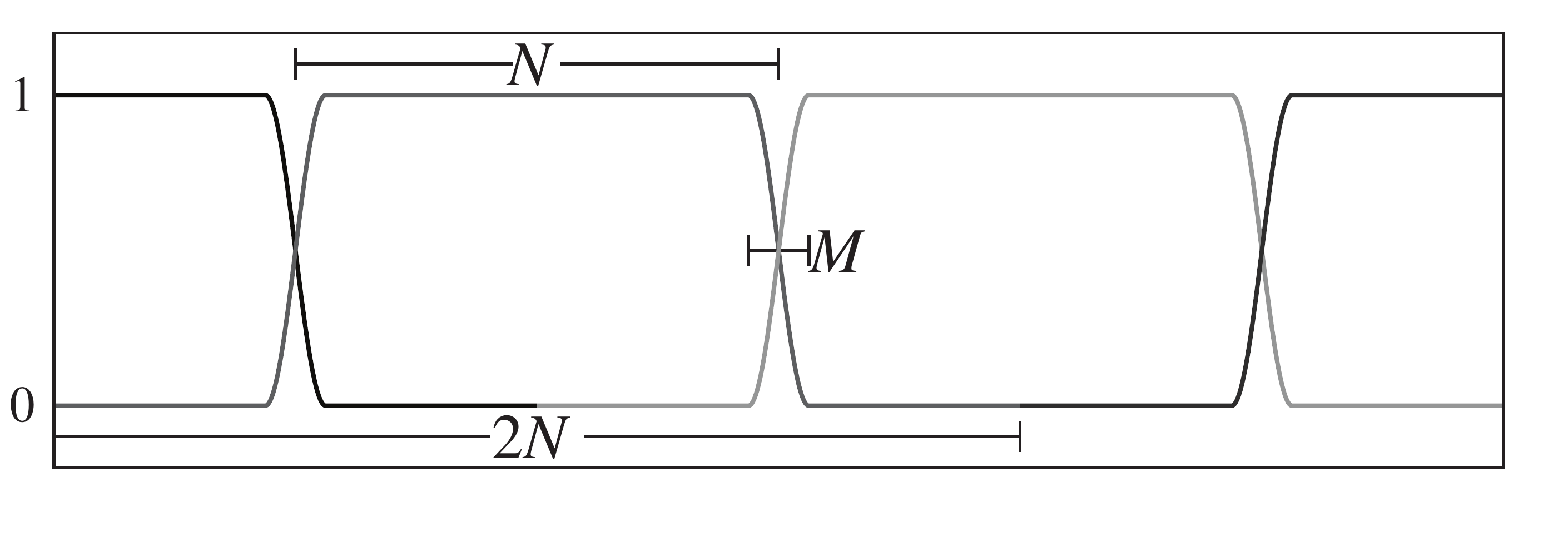}
    \caption{Tukey windows used for slicing with essential length $N$ and transition areas
        of length $M$, for some $N,M\in \mathbb{N}$ with $M < N$.
        Note that the chosen amount of zero-padding leads to a half-overlap situation.
        Source: \cite{holighaus13}.}
    \label{fig:tukeywins}
\end{figure}

\begin{figure}[h]
    \centering
    \includegraphics[width=0.5\columnwidth]{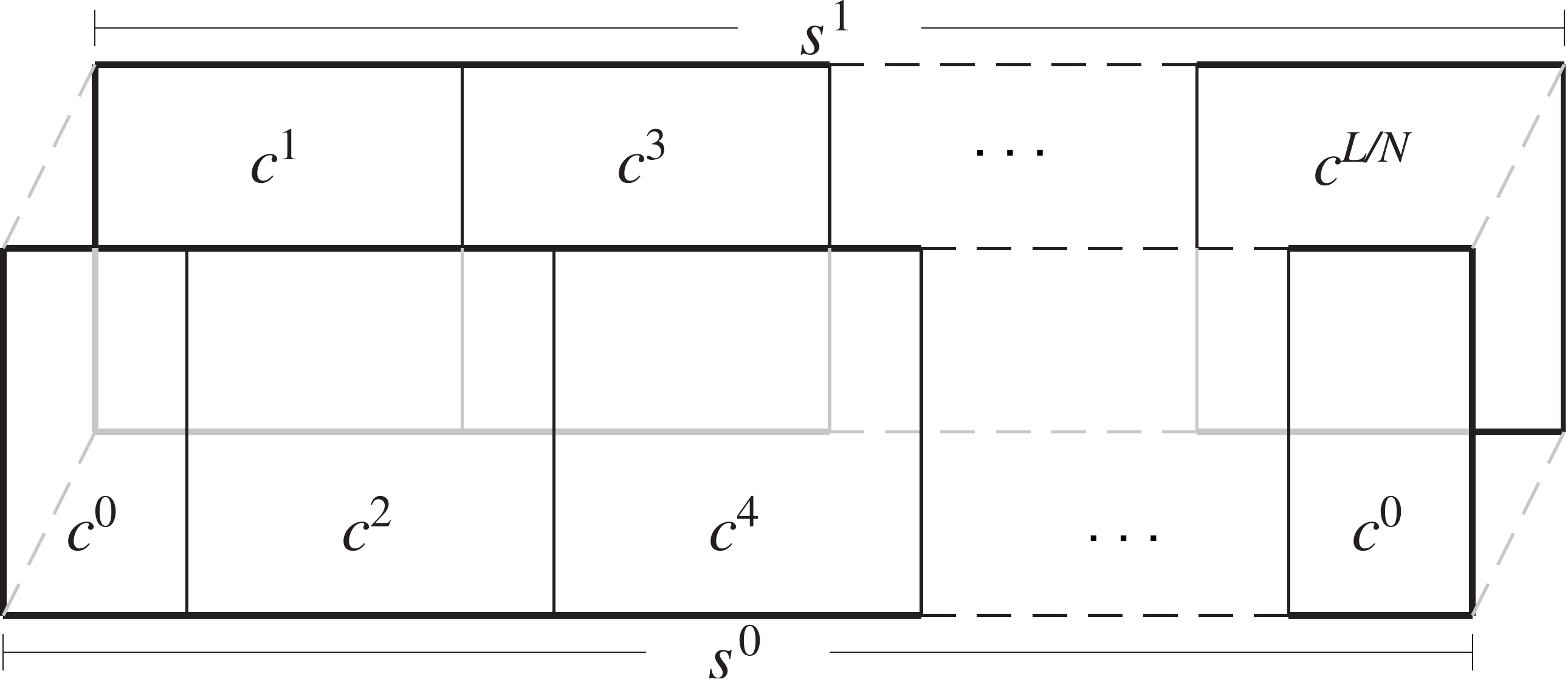}
    \caption{Schematic illustration of the sliCQ coefficients' structure. Source: \cite{holighaus13}.}\label{fig:slicqarray}
\end{figure}

While \Cref{alg:nsganalysis} deals with a full signal (analogous to the FFT), sliCQ analysis (analogous to the STFT) cuts the signal into slices along the time axis before computing the \textbf{sliCQ coefficients} $s$, which mimic CQ coefficients computed for each time slice. The process is as follows:

\begin{enumerate}
    \item The signal $x \in \mathbb{C}^{L}$ of length $L$ is cut into overlapping slices $x_m$ of length $2N$ via multiplying with uniform translations of $h_0$ (a $0$-centred slicing window)--the slicing process is illustrated in \autoref{fig:tukeywins}.
    \item For each $x_m$, the coefficients
          $$c^m \in\mathbb{C}^{\frac{2N}{a_k} \times |I_K|}$$
          are obtained by applying  $\textbf{CQ-NSGT}_{2N}(x,\textbf{g},\textbf{a})$ (\Cref{alg:nsganalysis}).
    \item  Because the slicing windows overlap as per \autoref{fig:tukeywins}, each time index corresponds to two consecutive slices. Thus, the slice coefficients $c^m$ are rearranged into a $2$-layer array $s = \{s^l\}_{l\in\{0,1\}} \in \mathbb{C}^{2\times \frac{L}{a_k}\times |I_K|}$; see \autoref{fig:slicqarray}.
\end{enumerate}

Like for \Cref{alg:nsganalysis}, \cite{holighaus13} proposed sliCQ synthesis which complements sliCQ analysis. SliCQ synthesis reconstructs the original signal $x$ from the sliCQ coefficients $s$.

Assuming that the slice length $2N$ is independent of $L$, the resulting computational complexity of sliCQ analysis is
$$
    \mathcal{O} (L).
$$

SliCQ analysis' properties bring about several observations important for feature extraction in this system's context:

\begin{itemize}
    \item The \textbf{efficiency} and \textbf{bounded computational complexity} of the algorithm allows for \emph{real-time processing} of a signal.
    \item The \textbf{slicing} process implies that NSG analysis resulting in high-quality CQ coefficients can be done with individual slices of the signal, contrary to earlier CQT calculation approaches that require the full signal \cite{brown92, christiancqt}. This means that sliCQ analysis can be performed on \emph{streamed} slices of an input signal.
\end{itemize}

\begin{figure}[h]
    \centering
    \includegraphics[width=0.75\columnwidth]{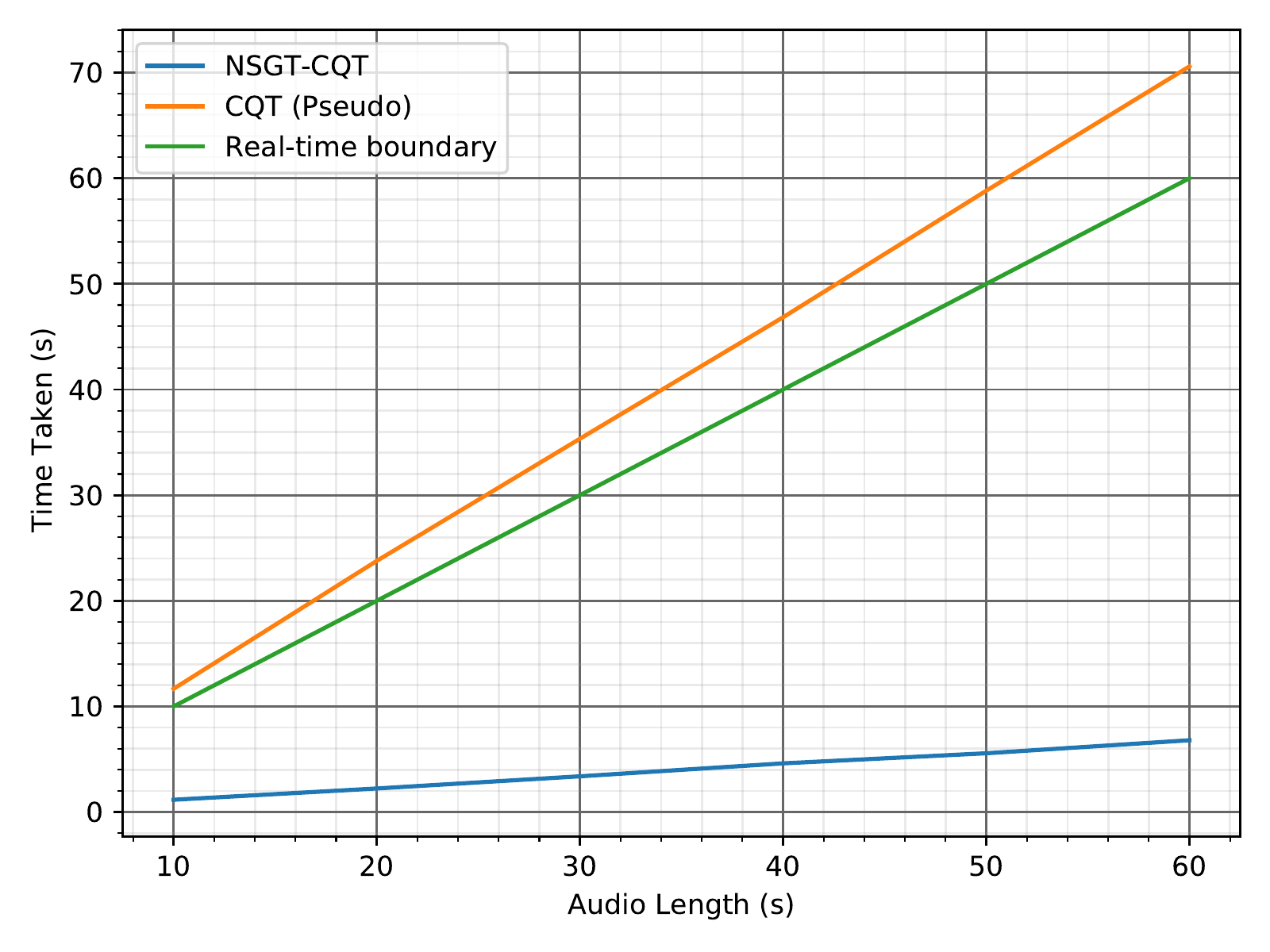}
    \caption{Plot of time taken for feature extraction versus length of audio.
        NSGT-CQT denotes sliCQ analysis with transition and slice lengths of 2048 and 8192 respectively, and CQT (Pseudo) denotes pseudo CQT computed via FFT with hop and frame lengths of 2048 and 8192 respectively.
        The audio used is a 44100Hz recording of \textit{AchGottundHerr} from the Bach10 dataset \cite{duan11bach10}, and the results are obtained on the system described in \autoref{appendix_benchmark_system}.}
    \label{fig:cqt_time}
\end{figure}

Thus, sliCQ analysis is suitable for a score following system as it can extract features from a \emph{streamed} performance audio signal in \emph{real time}. \autoref{fig:cqt_time} contrasts the time taken by sliCQ and pseudo CQT (computed via FFT)\footnote{\url{https://librosa.org/doc/main/generated/librosa.pseudo_cqt.html}} to extract features from varying lengths of audio. As expected, the time taken for both approaches are linear with regard to the length of the audio (in other words, the time complexity is $\mathcal{O}(L)$ where $L$ is the length of the signal). Notably, pseudo CQT is not able to perform within real-time constraints. SliCQ on the other hand comfortably meets the real-time boundary, requiring only a little over 10\% of the audio's length. \Cref{impl_dtw_prod_repro} contains instructions on how to reproduce these results.

Nevertheless, a challenge in using sliCQ for a \textit{streamed} sequence persists: careful choices of slicing and window properties must be made to ensure the high quality of the computed coefficients, e.g. a slice length that is too long in length implies low response times, and a slice length too short is unable to capture adequate features required. Further implementation details of this algorithm can be found in \autoref{cqt_impl}; for this system, windowing and slicing properties are exposed as configurable parameters.

\subsection{From CQT Features to Feature Vectors}
\label{impl_dtw_cqt_to_note}




In this system, CQT is run in real time continuously on slices of the audio signals.
The extracted CQT coefficients for each time slice are converted to their absolute values then $\ell_1$-normalised to sum up to $1$. Formally,

\begin{definition}[Normalised Energy for a Time Slice]
    Given that the extracted CQT coefficients for a time slice $t$ is represented by $C_Q(t)$ (these in fact correspond to sliCQ coefficients $s$ as computed in \nameref{sec:cqt_calc}), the normalised energy for each time slice, $E(t)$ is given by
    $$
        E(t) = \frac{|C_Q(t)|}{\|C_Q(t)\|_1}.
    $$
    $E(t, f_k)$ represents an element in $E(t)$ that comprises the energy for the frequency bin centred at $f_k$.
    \label{def:normenergy}
\end{definition}

Thus, the \textbf{Feature Vector} for aligning performance and score audio signals (see \Cref{def:scofo_dtw}) is $E(t)$.

\begin{figure}[h]
    \centering
    \subfloat[\centering \textit{Prelude}.]{{\includegraphics[width=0.45\linewidth]{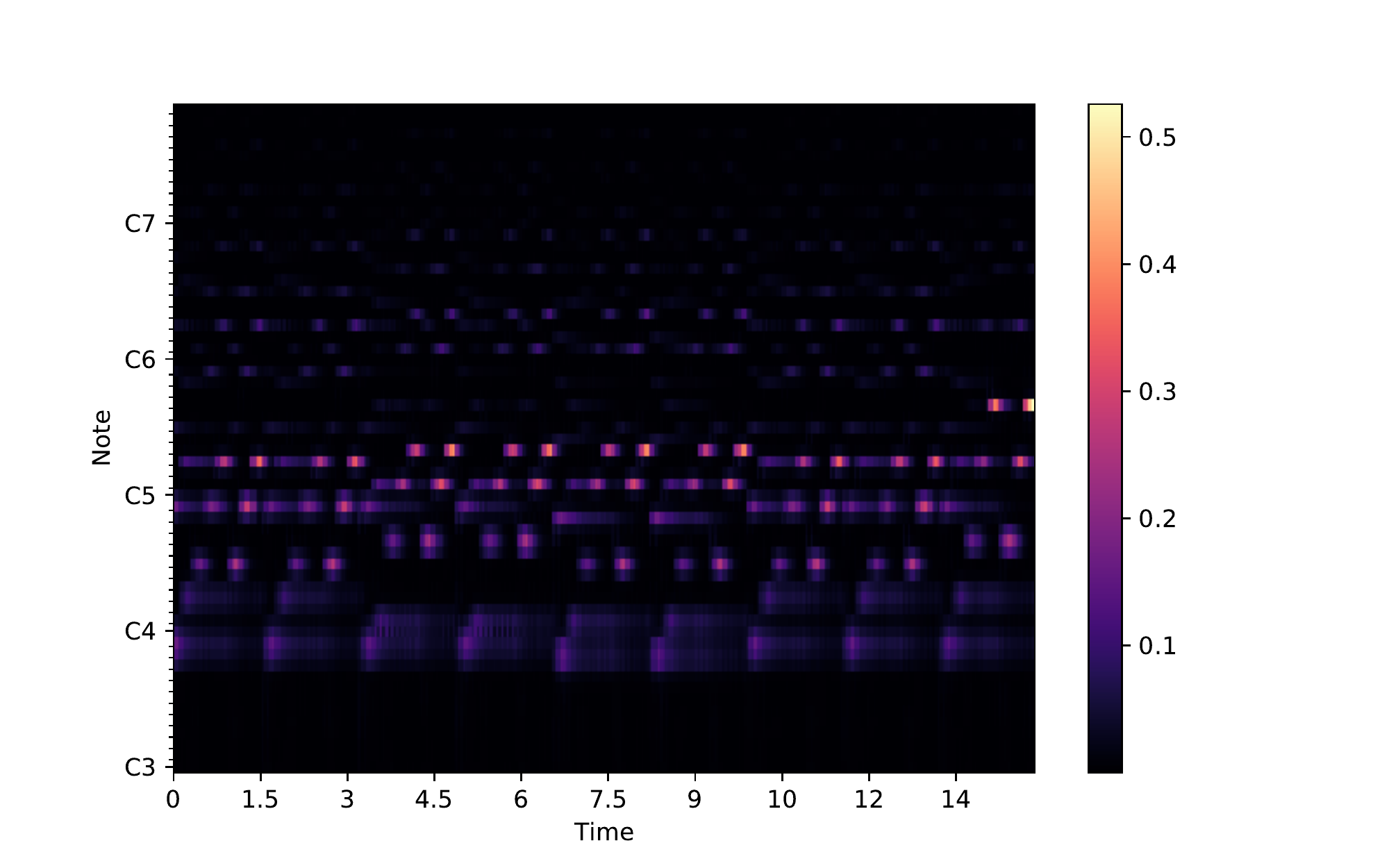} }}
    \qquad
    \subfloat[\centering \textit{Fugue}.]{{\includegraphics[width=0.45\linewidth]{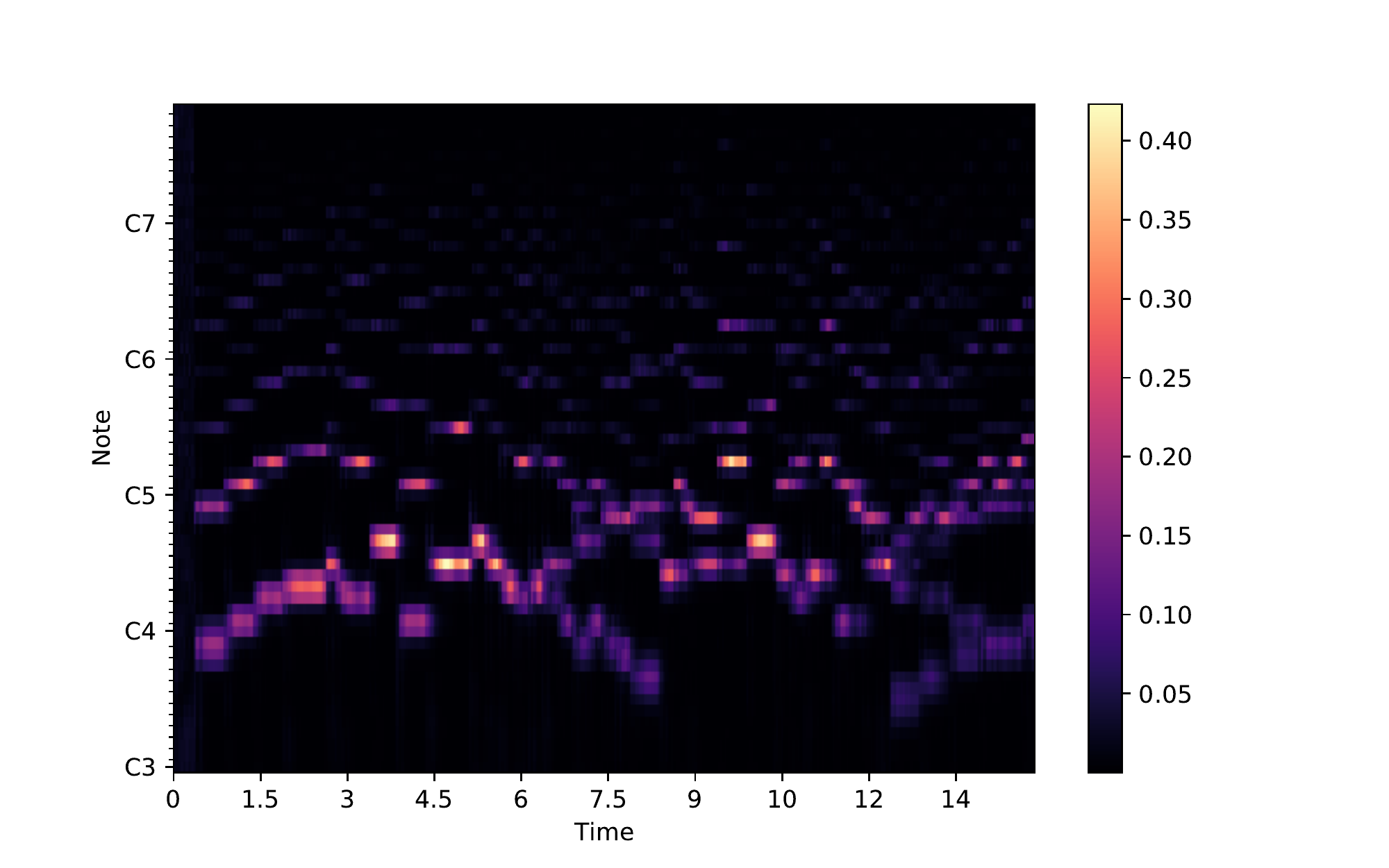} }}
    \caption{Visualisation of the feature vectors extracted from the first 15 seconds of an audio performance (obtained from the MAESTRO dataset \cite{MAESTRO}) of Bach's \textit{Prelude and Fugue in C major, BWV 846}.}
    \label{fig:featuresbwv846}
\end{figure}

\autoref{fig:featuresbwv846} shows example visualisations of extracted feature vectors from performance audio following the system detailed in subsections~\ref{impl_dtw_cqt_tech} and \ref{impl_dtw_cqt_to_note}. \Cref{impl_dtw_prod_repro} contains instructions on how to reproduce these results. For reference, the sheet music for the music pieces can be found on \textit{IMSLP}\footnote{\url{https://imslp.org/wiki/Prelude\_and\_Fugue\_in\_C\_major\%2C\_BWV\_846\_(Bach\%2C\_Johann\_Sebastian)}}. It is first worth noting that the clarity obtained here is far superior when compared to the spectrograms (obtained via STFT) shown in Figures~\ref{fig:preludespecgram} and \ref{fig:fuguespecgram}. While harmonics--shown primarily by the fainter notes an octave and a compound major fifth above the real notes in the score--are still present, the actual notes in the score clearly dominate the visualised features. Another observation is related to the time-frequency uncertainty principle (see \Cref{def:timefreq}): the frequency resolution at lower frequencies is poorer than that of higher frequencies; observe the blurrier features extracted from the lower notes.

\subsection{DTW: Compared to other Approaches}
\label{impl_dtw_compared}

A review of major work in DTW-based followers is provided in \autoref{scofo_dtw}. Indeed, \autoref{scofo_litreview} outlines myriad possible approaches for score following; therefore, firstly the motivations for choosing DTW is outlined by comparing DTW directly to other major approaches.

\subsubsection{Versus Approximate String Matching (ASM)}
\label{dtw_vs_asm}

DTW is closely related to ASM, and many concepts from the ASM aligner implemented in \autoref{impl_asm} are shared with DTW. In short, the major difference between DTW and ASM is that in the former, the notion of \emph{gaps} and \emph{arrows} does not exist--one sequence in the DTW alignment is treated as a \emph{time-warped} version of the other. The concept of \emph{time-warped} sequence alignment in DTW renders DTW highly capable of aligning audio-to-audio, which in turn motivated the myriad uses of DTW in speech recognition \cite{rabiner89}.

The primary advantage of DTW over ASM is the fact that the DTW does not require complex feature extraction techniques to extract string data from performance or score data. The effort in converting the ASM aligner featured in \autoref{impl_asm} to a real-time method is futile considering the work required to render it capable of following audio performances. ASM-based methods were also quickly made obsolete since the introduction of DTW and HMM approaches--the performance of ASM-based methods is significantly far behind its successors.

\subsubsection{Versus Hidden Markov Models (HMMs)}

\Cref{scofo_dtw} mentioned how DTW is a special case of HMMs. What is more pertinent is the fact that DTW-based followers are simpler and more flexible. HMM-based methods usually require the full detailed score of the performance to produce \emph{hidden states} required for score following; DTW-based methods, on the other hand, do not strictly require score data. Further, the technicalities and work required to develop a HMM model is overly complex for measly returns.

Further, DTW's audio-to-audio following nature means that the digital score representation can be either synthesised from any form of digital score, or obtained from performance recordings. This key advantage was exploited by Arzt \etal{} in their demonstration of a score follower in the \textit{Concertgebouw}. The researchers were not able to find sufficiently good digital scores of the \textit{Alpensinfonie}, and therefore resorted to manually annotate downbeats on a recording of a separate performance. Score following is then performed based on this annotated recording, which maps to corresponding sections on scanned sheet music shown on tablets to the audience \cite{arzt16, arztconcert}. This workaround would not have been possible for a HMM (and ASM) method.

\subsubsection{Versus Sheet Music Image Following}

The paradigm-shifting methods documented in \autoref{scofo_sheetmusicimages} are capable of following sheet music images. However, one major disadvantage identified is that these approaches are data driven, and the major implementations are trained on the MSMD dataset \cite{dorfer17} which contains sheet music in the LilyPond format. The ground-truth information used to map performance audio to pixels on a sheet music image produced from the LilyPond files is not available, and thus the researchers resorted to synthesising constant-\textit{tempo} performance audio from the MIDI data extracted from the LilyPond scores.
While Henkel and Widmer's most recent work in 2021 \cite{henkel21} shows some improvements in generalising to performance audio, the problem of generalising to sheet-music images and more diverse human-produced performances is yet to be solved. It is however hopeful to be able to train the networks on datasets with human performance audio mapped to scanned sheet music images; nevertheless, such datasets do not currently exist, and the creation of such a dataset would be of really high effort and budget, though a feasible--but imperfect--way is to perform audio-to-audio alignment between human-generated performance audio and the synthesised audio, then map the human performance audio to the score data in the MSMD dataset. Further, a method could be devised to align score data in the MSMD dataset to data in corresponding scanned sheet music images.

DTW is therefore chosen over sheet music image following despite the latter's capability in providing an end-to-end solution able to follow performance audio on sheet music images.

\subsubsection{Other Key Advantages}

In \autoref{scofo_recent_mirex}, DTW is shown to be dominant in the MIREX evaluations. DTW shines in its flexibility to be extended or modified to work with different aspects and features of audio, such as the STFT and chromagram representations. Further, score data can also be used to augment audio features extracted from the score. These advantages were exploited and experimented by the top performers of the MIREX evaluation \cite{orti15,alonso16,alonso17,carabias12,suzuki10}.

\subsection{DTW: Technicalities}
\label{impl_dtw_tech}

Here, a deeper dive into the technicalities of DTW in the score following context is given. Readers preferring a more general introduction may refer to \cite{rabiner93}.

\subsubsection{Introduction}

DTW, as mentioned in \autoref{dtw_vs_asm}, is similar to ASM. First, score following is defined as a DTW problem, based on the score alignment ASM definition in \Cref{def:score_alignment_asm}. The following definition shows how closely related DTW and ASM are, with the exception that the concept of \emph{gaps} is absent in DTW.

\begin{definition}[Score following as a DTW problem]
    Let the time series of the score and the performance to be the \emph{feature vectors} $\mathcal{S}=s_1,s_2,...s_m$ and $\mathcal{P}=p_1,p_2,...,p_n$ respectively. Note that $\mathcal{P}$ is \emph{streamed} and \emph{partial} (with regard to $\mathcal{S}$). An alignment between $\mathcal{S}$ and $\mathcal{P}$ is a path $\mathcal{W} = \mathcal{W}_1, \mathcal{W}_2, ..., \mathcal{W}_i$ through an $m \times n$ cost matrix where each $\mathcal{W}_k$ is an ordered pair $(i_k, j_k)$ such that $(i, j) \in \mathcal{W}$ means that $s_i$ and $p_i$ are aligned. $\mathcal{W}$ is constrained to be monotonous, continuous and bounded by the ends of both sequences $\mathcal{S}$ and $\mathcal{P}$. The alignment is done with respect to the \emph {local} cost matrix, and the cost of a path $\mathcal{D(W)}$ is the sum of the local match costs of the path. The goal is to minimise $\mathcal{D(W)}$.
    \label{def:scofo_dtw}
\end{definition}

\subsubsection{Local Step Constraint Calculation}

Rabiner \etal{} in 1993 \cite{rabiner93} proposed several local step constraints which can be used for the computation for optimal $\mathcal{D}$. The simplest and most common one is as follows:
\begin{equation}
    \mathcal{D}(i, j) = d(i, j) + \min
    \begin{cases}
        \mathcal{D}(i, j-1)   \\
        \mathcal{D}(i-1, j)   \\
        \mathcal{D}(i-1, j-1) \\
    \end{cases}
    \label{eqn:orig_dtw}
\end{equation}
given that $d(i, j)$ is the distance between $s_i$ and $p_i$, $\mathcal{D}(i, j)$ is the cost of the minimum cost path from $(1,1)$ to $(i, j)$ and $\mathcal{D}(1, 1) = d(1, 1)$.

An optimal alignment path can be extracted by tracing the recursion backwards from $\mathcal{D}(m, n)$.

\subsubsection{Optimisations in Path Computation}

To speed up DTW path calculations, global path constraints are often introduced--this is similar to the bounds added in \autoref{impl_asm_complexity} to optimise the ASM aligner; in fact, the constraint in which the path lies within a fixed distance from the diagonal is known in DTW as the Sakoe-Chiba bound \cite{sakoe78}. On the other hand, the constraint for the path that lies within a parallelogram around the diagonal of the matrix is known as the Itakura parallelogram \cite{itakura75}. A deeper study into these two constraints can be found in \cite{dtwitakuravssakoe}. It is again worth noting that there always exists a constraint strict enough to give suboptimal alignments save for the trivial case where $\mathcal{P} = \mathcal{S}$.

Nevertheless, even with these constraints, the naïve implementation DTW recursion in \autoref{eqn:orig_dtw} has quadratic time and space complexity with regard to the length of the sequences to align (assuming $n \leq m$, the time complexity can be said to be $\mathcal{O}(n^2)$; the space complexity is simply $\mathcal{O}(mn)$). SparseDTW \cite{sparsedtw} and PrunedDTW \cite{pruneddtw} both retain the original complexities, but offer decreased computation times. However, there exist recent developments that break these bounds: Gold and Sharir in 2018 \cite{gold18} broke the almost-50-year-old quadratic time bound, giving a method that runs in $\mathcal{O}(n^2 \log \log \log n / \log \log n)$; further, Tralie and Dempsey in 2020 \cite{tralie2020} proposed an optimised DTW method that runs in $\mathcal{O}(m+n)$ space complexity.


To further improve runtimes, optimality is often sacrificed--the optimal path is \emph{approximated}. MultiscaleDTW \cite{muller06} is one such example, but the most popular algorithm is perhaps FastDTW \cite{fastdtw}, which has linear time and space complexity. Nevertheless, Wu and Keogh in 2020 \cite{wu20} showed that FastDTW is approximate and generally slower than the exact DTW in application. Further, SparseDTW \cite{sparsedtw}, while being an exact calculation of DTW (and has its original complexity), was shown to be comparable in speed to FastDTW. The caveat, thus, is that complexity notation may mislead when the algorithm is deployed into production--the $\mathcal{O}(1)$ step could be vastly different among different algorithms.

\subsubsection{Online Time Warping (OLTW)}

Whilst many recent optimised (and sometimes approximate) DTW algorithms may calculate the required path in a duration less than the length of the aligned time sequences (therefore making them technically capable of real-time applications), the real-time constraint usually does not involve a \emph{streamed} sequence required for score following as per \Cref{def:scofo_dtw}.  Dixon in 2005 \cite{dixonOLTW} presented a DTW variant--known as on-line time warping (OLTW)--that is capable of incrementally aligning sequences of arbitrary lengths in real time.\footnote{Some authors refer to OLTW as ODTW (Online Dynamic Time Warping) \cite{henkel21}.}

Dixon's algorithm calculates the path in a \emph{forward manner} as such: in the main loop, a partial row or column of the path cost matrix is calculated (using the standard DTW recursion as per \autoref{eqn:orig_dtw}, restricted to use only matrix entries which are already calculated). The decision if a row or a column should be calculated depends on where the current minimum cost path on the current row and column lies--the tending ``direction'' of the current path is used as a constraint to reduce the total amount of computation required.

\begin{figure}[h]
    \centering
    \includegraphics[width=0.75\columnwidth]{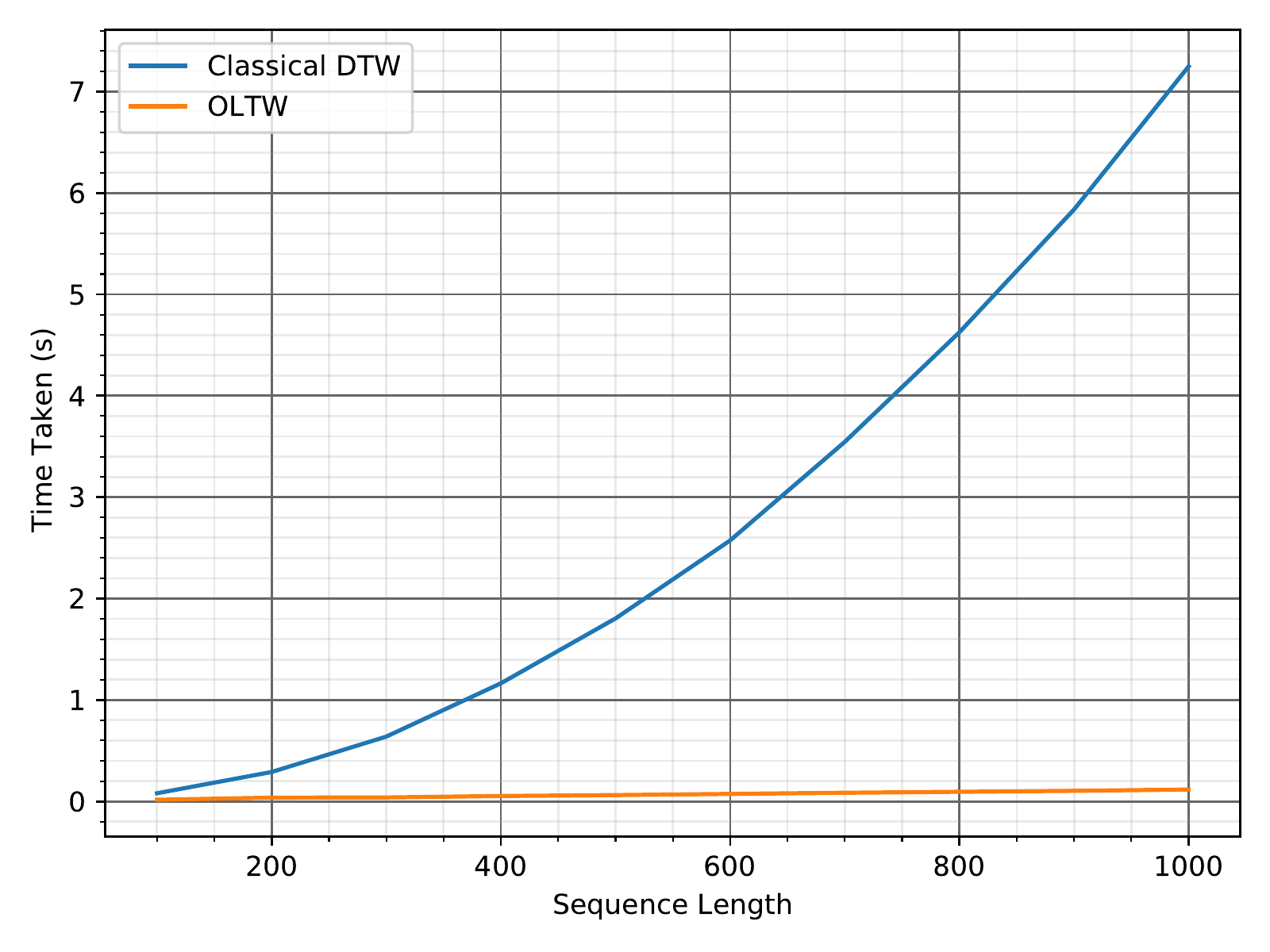}
    \caption{Plot of time taken for alignment versus sequence length.
    Classical DTW denotes an unoptimised and recursive version of DTW, and OLTW denotes Dixon's OLTW algorithm with the parameters $\texttt{MaxRunCount} = 3$ and $\texttt{c} = 500$ (these parameters are detailed subsequently in \autoref{impl_oltw_param}). The sequences are made up of 50-unit elements, populated with random samples of floating-point numbers from a uniform distribution over $[0, 1)$. To reduce noise, the average of three runs is taken. The results are obtained on the system described in \autoref{appendix_benchmark_system}.}
    \label{fig:dtw_time}
\end{figure}

Dixon's complete algorithm runs in linear time with regard to the length of the series (thus meaning that the incremental step is bounded by a constant). \autoref{fig:dtw_time} compares the time required to align sequences of varying length between classical DTW and OLTW. The quadratic time complexity of the classical DTW algorithm is apparent, which is in contrast to OLTW's linear time complexity. \Cref{impl_dtw_prod_repro} contains instructions on how to reproduce these results.

Further, OLTW has been applied and extended successfully in many score following applications, including musical analysis \cite{Dixon2005MATCHAM}, APT \cite{arzt08} and computer-aided accompaniment \cite{schauer17}. Dixon's OLTW algorithm would go on to form the basis of Arzt's works \cite{arzt08,arzt10,arzt12,arzt15, arzt16}. Macrae and Dixon in 2010 \cite{macrae10dtw} further developed OLTW to incorporate ideas of the A* algorithm \cite{hart68}. Other derivations of OLTW were also introduced by Suzuki \etal{} \cite{suzuki10}, Carabias-Orti \etal{} \cite{carabias12} and Chen and Jang \cite{chen19} among many others; numerous approaches involve adding more weights and terms (usually to introduce temporal elements) to the classic DTW recursion shown in \autoref{eqn:orig_dtw}.

Dixon's OLTW algorithm was chosen to be part of this follower's system, and its technicalities and a detailed worked example is presented in \autoref{impl_dtw_baseline_worked_example}.


The feature vector used for DTW in this system is the \textbf{Normalised Energy for a Time Slice} as defined in \Cref{def:normenergy}.

While Dixon and Widmer \cite{Dixon2005MATCHAM} used the half-wave rectified first-order difference computed from every consecutive pair of energy vectors computed to simulate note onsets, this configuration performed worse when trialled in this score follower. In further experiments in DTW aligners it was also found that not performing rectification offered superior performance \cite{thickstun20}.




\subsection{Distance Computation}
\label{impl_dtw_dist_comp}

To compute the distance $d(i, j)$ between the feature vectors $E_{i}(t)$ and $E_{j}(t)$, the $\ell_1$ (Manhattan) distance is used:

\begin{definition}[Distance Function used for DTW]
    $$
        d(i, j) = \|  E_{i}(t) - E_{j}(t) \|_1
    $$
    \label{def:distfunc}
\end{definition}

An example calculation is given in \autoref{impl_dtw_baseline_worked_example}.

As an aside, Arzt and Dixon's work prior to 2016 in \cite{Dixon2005MATCHAM,arzt08} incorrectly stated that the Euclidean (i.e. $\ell_2$-norm) was used--this was corrected in 2016 in \cite{arzt16}: the Manhattan ($\ell_2$) distance between the two vectors normalised by a logarithmically weighted sum of their norms was actually used. In practice, it was found that the simple $\ell_1$ (Manhattan) distance works sufficiently well, and using more complex methods like the aforementioned would incur unnecessary computation complexity while not offering measurable improvements.


%




\subsection{OLTW: Technicalities and Worked Example}
\label{impl_dtw_baseline_worked_example}
\input{02_parts/03_impl/chapters/04a_dtw_workedexample.tex}

\section{Score Follower Framework}

This score follower conforms to the generic framework given in \autoref{scofo_generic}.

The \textbf{features extracted} from the score and performance is the \emph{normalised energy in frequency bins captured by CQT} as described in subsections~\ref{impl_dtw_cqt_tech} and \ref{impl_dtw_cqt_to_note}. \textbf{Similarity calculation} is performed via taking the $\ell_1$ distance between two feature vectors as detailed in \autoref{impl_dtw_dist_comp}. Finally, the \textbf{alignment} step is done via OLTW as delineated in \autoref{impl_dtw_baseline_worked_example}.

Further, this score follower is of the \textbf{audio-to-audio} class--the score can either be a notated audio file (or stream), or synthesised from MIDI/MusicXML to audio.

\input{02_parts/03_impl/chapters/04b_dtw_impl.tex}

\input{02_parts/03_impl/chapters/04c_dtw_eval.tex}
\input{02_parts/03_impl/chapters/04d_dtw_discussion.tex}

\input{02_parts/03_impl/chapters/04e_dtw_product.tex}
\input{02_parts/03_impl/chapters/04f_future.tex}

%% file: 02_parts/03_impl/chapters/04a_dtw_workedexample.tex
\subsubsection{Parameters}
\label{impl_oltw_param}
Dixon's OLTW algorithm \cite{Dixon2005MATCHAM} is used as the baseline algorithm for this project's system, and here technicalities of the algorithm are presented before a worked step-by-step example.

The parameters for the algorithm are listed:

\begin{itemize}
    \item \texttt{c}: The search window.
    \item \texttt{MaxRunCount}: A parameter which constrains the slope of the path to be between $\frac{1}{\texttt{MaxRunCount}}$ and \texttt{MaxRunCount}.
\end{itemize}


Extending the original local path constraint algorithm given in \autoref{eqn:orig_dtw}, weights are introduced to introduce bias towards a certain step direction. The resultant formula is shown in \autoref{eqn:oltw_dtw}; $w_a$, $w_b$ and $w_c$ are configurable parameters, which can be positive or negative. Similar weighting systems were also used in multiple approaches in the literature \cite{arzt08, arzt16, chen19, orti15,wang16}.

\begin{equation}
    \mathcal{D}(i, j) = \min
    \begin{cases}
        w_a \times d(i, j) + \mathcal{D}(i, j-1)            \\
        w_b \times d(i, j) + \mathcal{D}(i-1, j)            \\
        w_c \times d(i, j) + \mathcal{D}(i-1, j-1) \\
    \end{cases}
    \label{eqn:oltw_dtw}
\end{equation}

In fact, the original approach in OLTW by Dixon \cite{dixonOLTW} is a case of \autoref{eqn:oltw_dtw} with $w_a = w_b = 1$ and $w_c = 2$--this is to make the comparison of different path lengths in the algorithm possible. Note further that \cite{arzt16} presented an incorrect version of \autoref{eqn:oltw_dtw}.

Another constraint is added to limit the runtime of the algorithm: if the calculation of a certain clause (out of the three in \autoref{eqn:oltw_dtw}) in $\mathcal{D}$ depends on an uncalculated entry, that entry is set to $\infty$ (i.e. ignored). The algorithm is set up so that at least one clause is not $\infty$.

\subsubsection{Algorithm Presentation}
\label{impl_oltw_presentation}

\begin{itemize}
    \item \textbf{Preprocessing}

    \begin{enumerate}
        \item \textbf{Compute $\mathcal{S}$}.

            $\mathcal{S}$ is first computed. This can either be from the synthesised score (MIDI or MusicXML), or from a preprocessed piece of audio (the preprocessing maps relevant points in the audio to its position in a score). For simplicity, the former case is assumed from this point onwards.



        \item \textbf{Construct Cost Matrix}.

            A cost matrix $M$ containing the costs $\mathcal{D}$ is constructed. This is similar to the grid constructed in \autoref{impl_asm_techback_asm}.

            $M$ is indexed using a coordinate system $(x, y)$ where the last row of the first column is the origin $(0, 0)$--this cell is referred to as $M(0, 0)$. $x^+$ is in the rightward direction and $y^+$ is in the upward direction. $\mathcal{S}$ is placed on the first column starting from $M(0, 1)$ in the $y^+$ direction:

            \begin{tikzpicture}
                \matrix[matrix of nodes,nodes={draw=gray, anchor=center, minimum size=.6cm}, column sep=-\pgflinewidth, row sep=-\pgflinewidth] (A) {

                    $s_m$  \\
                    \vdots \\
                    $s_2$  \\
                    $s_1$  \\
                    \phantom{} \\};
            \end{tikzpicture}

            $M$ will expand in the score following process in the $x^+$ direction as $\mathcal{P}$ is streamed in--$\mathcal{P}$ is placed starting from $M(1, 0)$ tending towards the $x^+$ direction.

            $M(0, 0)$ is left empty, $M(x, y)$ for $x > 0$ and $y > 0$ are spaces left for the costs $\mathcal{D}$--in fact, for $i \neq 0, j \neq 0$,
            $$
                \mathcal{D}(i, j) = M(i, j).
            $$

    \end{enumerate}

    \item \textbf{Following Algorithm}

    Here, the generic following algorithm is presented.

    Step 4-7 forms the main \emph{loop} of the algorithm.

    \begin{enumerate}
        \item \textbf{Initialise variables}

            Let \texttt{i} and \texttt{j} be the latest (top-rightmost) calculated position in $M$ representing $M(i, j)$. \texttt{i} and \texttt{j} are both set to $1$.

            The \texttt{current} and \texttt{previous} variables track the direction(s) of the current and previous direction(s) of increment and have values that can be \texttt{i}, \texttt{j}, \texttt{ij} or \texttt{None}. Both variables are initialised to \texttt{None}.

            The \texttt{runCount} variable which is an integer is initialised to 1. This variable keeps track the consecutive number of times the system is incremented in a certain direction.

        \item \textbf{Obtain $p_\texttt{i}$}.
        \item \textbf{Calculate $\mathcal{D}(\texttt{i}, \texttt{j})$}.
        \item \textbf{Determine whether to increment \texttt{i}, \texttt{j} or both (\texttt{ij})}.
            \begin{itemize}
                \item If $\texttt{i} < \texttt{c}$, increment both (\texttt{ij}).

                        (This is always the case for the beginning of the system, so that a square block of length $\texttt{c}$ is formed.)
                \item Else if $\texttt{runCount} > \texttt{MaxRunCount} $, increment $j$ if \texttt{previous = i}, increment $i$ otherwise.

                        (This prevents the algorithm from ``running away'' too far in one direction. The effect of this is that the slope of the path is bound between $\frac{1}{\texttt{MaxRunCount}}$ and $\texttt{MaxRunCount}$.)
                \item Else obtain the latest coordinate of the lowest cost path $(i', j')$ fulfilling $\min(\mathcal{D}(i', j'))$ for $i' = \texttt{i}$ or $j' = \texttt{j}$.
                        If $i' < \texttt{i}$ increment $\texttt{j}$; else if $j' < \texttt{j}$ increment $\texttt{i}$; else increment \texttt{ij}.

                        (This limits the search path--only a row or column in the tending direction of the path is calculated, save for the case where $(i', j') = (\texttt{i}, \texttt{j})$ where both \texttt{i} and \texttt{j} are incremented.)
            \end{itemize}

            The \texttt{current} variable holds this decision, and can be \texttt{i}, \texttt{j} or \texttt{ij}.
        \item \textbf{Increment \texttt{i}, \texttt{j} depending on \texttt{current}}.
            \begin{itemize}
                \item $\texttt{i} \in \texttt{current}$:
                        \begin{enumerate}
                            \item Increment \texttt{i}.
                            \item Exit if performance is complete; otherwise, obtain $p_\texttt{i}$.
                            \item Compute $\mathcal{D}(\texttt{i}, \texttt{J})$ for $\texttt{J} \in \mathbb{Z}, \max(0, \texttt{j}-\texttt{c} + 1) < \texttt{J} \leq \texttt{j}$.
                        \end{enumerate}
                \item $\texttt{j} \in \texttt{current}$:
                        \begin{enumerate}
                            \item Increment \texttt{j}.
                            \item Compute $\mathcal{D}(\texttt{I}, \texttt{j})$ for $\texttt{I} \in \mathbb{Z}, \max(0, \texttt{i}-\texttt{c} + 1) < \texttt{I} \leq \texttt{i}$.
                        \end{enumerate}

            \end{itemize}

        \item \textbf{Update \texttt{runCount}}.

            Increment \texttt{runCount} if $\texttt{current} = \texttt{previous} \neq \texttt{ij}$; otherwise, reset \texttt{runCount} to $1$.

            Assign $\texttt{previous}$ to $\texttt{current}$.

        \item \textbf{Go to Step 4}.

    \end{enumerate}
\end{itemize}

\subsubsection{Algorithm Analysis}
As can be seen, for every step of the loop a maximum of $2 \times \texttt{c}$ calculations of $\mathcal{D}$ are required. These calculations are also memoised and kept in $M$, meaning that bounded time is required for each computation, giving a time complexity of $\mathcal{O}(1)$. From this, every step of the loop also has a time complexity of $\mathcal{O}(1)$, and finally it can be said the whole following procedure is \emph{linear} in time complexity with regard to the longest of the sequences $\mathcal{P}$ and $\mathcal{S}$: $\mathcal{O}(\max(|\mathcal{P}|, |\mathcal{S}|))$. On the other hand, the space complexity is simply $\mathcal{O}(|\mathcal{P}||\mathcal{S}|)$.

As was discussed in \autoref{impl_dtw_tech}, \autoref{fig:dtw_time} demonstrates the linear time complexity ($\mathcal{O}(\max(|\mathcal{P}|, |\mathcal{S}|))$) of OLTW, which is in contrast to the quadratic time complexity ($\mathcal{O}(|\mathcal{P}||\mathcal{S}|)$) of DTW.

\subsubsection{Worked Example}
\label{impl_dtw_baseline_example_sequences}

The OLTW algorithm is worked out in a step-by-step example here. For this example, \texttt{c} is set to 3, \texttt{MaxRunCount} is ignored, and $w_a = w_b = w_c = 1$.

Referring to \Cref{def:scofo_dtw}, assume the example feature vectors $\mathcal{P}$ and $\mathcal{S}$ in \autoref{eqn:exampleps} are extracted from the audio following steps detailed in subsections \ref{impl_dtw_cqt_tech} and \ref{impl_dtw_cqt_to_note}. Note that the complete $\mathcal{P}$ is given here--during the example $\mathcal{P}$ will be partially fed into the system. Also, assume that $K = 2$, i.e. there are $2$ frequency bins; further, for simplicity, the normalisation step is skipped.

\begin{align}
    \begin{split}
        \mathcal{P} & =
        \begin{bmatrix}
            1 \\
            2
        \end{bmatrix}
        \begin{bmatrix}
            3 \\
            3
        \end{bmatrix}
        \begin{bmatrix}
            2 \\
            2
        \end{bmatrix}
        \begin{bmatrix}
            2 \\
            3
        \end{bmatrix}
        \begin{bmatrix}
            6 \\
            6
        \end{bmatrix}       \\
        \mathcal{S} & =
        \begin{bmatrix}
            1 \\
            2
        \end{bmatrix}
        \begin{bmatrix}
            3 \\
            3
        \end{bmatrix}
        \begin{bmatrix}
            2 \\
            2
        \end{bmatrix}
        \begin{bmatrix}
            4 \\
            3
        \end{bmatrix}
        \begin{bmatrix}
            2 \\
            2
        \end{bmatrix}
    \end{split}
    \label{eqn:exampleps}
\end{align}

For convenience, the distances $d(i, j)$ between every pair of vectors in $\mathcal{P}$ and $\mathcal{S}$ (calculated as per \autoref{impl_dtw_dist_comp}) are given in the grid below. $i$ and $j$ respectively index $\mathcal{P}$ and $\mathcal{S}$. Note also that not all distances are required should the algorithm not need to calculate the distance in question. An example calculation is
$$
    d(2, 1) = \| p_2 - s_1 \|_1 = |3-1| + |3-2| = 3.
$$

\begin{tikzpicture}
    \matrix[matrix of nodes,nodes={draw=gray, anchor=center, minimum size=.6cm}, column sep=-\pgflinewidth, row sep=-\pgflinewidth] (A) {
        $d$   & $p_1$ & $p_2$ & $p_3$ & $p_4$ & $p_5$                     \\
        $s_1$ & 0     & 3     & 1     & 2     & 9                         \\
        $s_2$ & 3     & 0     & 2     & 1     & 6                         \\
        $s_3$ & 1     & 2     & 0     & 1     & 8                         \\
        $s_4$ & 4     & 1     & 3     & 2     & 5                         \\
        $s_5$ & 1     & 2     & 0     & 1     & 8\\};
\end{tikzpicture}

\begin{enumerate}
    \item First, the preprocessing steps are run. Only $s_1$, $s_2$ and $s_3$ are shown now for simplicity.

          \begin{tikzpicture}
              \matrix[matrix of nodes,nodes={draw=gray, anchor=center, minimum size=.6cm}, column sep=-\pgflinewidth, row sep=-\pgflinewidth] (A) {
                  $s_3$ \\
                  $s_2$ \\
                  $s_1$ \\
                  \phantom{} \\};
          \end{tikzpicture}
    \item The algorithm is run for $\texttt{i}, \texttt{j} < \texttt{c}$. For each step the minimum cost path is \textbf{bolded}. $(i', j')$ is the top-rightmost bolded cell within $M$--this represents the current position in the score of the performance.

          \begin{tikzpicture}
              \matrix[matrix of nodes,nodes={draw=gray, anchor=center, minimum size=.6cm}, column sep=-\pgflinewidth, row sep=-\pgflinewidth] (A) {
                  $s_3$                  & \phantom{}         \\
                  $s_2$                  & \phantom{}         \\
                  $s_1$                  & \textbf{0}         \\
                  \phantom{} & $p_1$ \\};
          \end{tikzpicture}
          \begin{tikzpicture}
              \matrix[matrix of nodes,nodes={draw=gray, anchor=center, minimum size=.6cm}, column sep=-\pgflinewidth, row sep=-\pgflinewidth] (A) {
                  $s_3$                  & \phantom{} & \phantom{}         \\
                  $s_2$                  & 3                      & \textbf{0}         \\
                  $s_1$                  & \textbf{0} & 3                              \\
                  \phantom{} & $p_1$                  & $p_2$ \\};
          \end{tikzpicture}
          \begin{tikzpicture}
              \matrix[matrix of nodes,nodes={draw=gray, anchor=center, minimum size=.6cm}, column sep=-\pgflinewidth, row sep=-\pgflinewidth] (A) {
                  $s_3$                  & 4                      & 2                      & \textbf{0}         \\
                  $s_2$                  & 3                      & \textbf{0} & 2                              \\
                  $s_1$                  & \textbf{0} & 3                      & 4                              \\
                  \phantom{} & $p_1$                  & $p_2$                  & $p_3$ \\};
          \end{tikzpicture}

          Thus far, the calculations are trivial as $p_1 p_2 p_3 = s_1 s_2 s_3$. The optimal diagonal path is found trivially. Here, $(i', j') = (\texttt{i}, \texttt{j}) = (3, 3)$.

    \item The next direction to increment is both \texttt{i} and \texttt{j} as $(i', j') = (\texttt{i}, \texttt{j})$.

          First, \texttt{i} is incremented and relevant costs are computed; then, \texttt{j} is incremented with its relevant costs computed as well.

          \begin{tikzpicture}
              \matrix[matrix of nodes,nodes={draw=gray, anchor=center, minimum size=.6cm}, column sep=-\pgflinewidth, row sep=-\pgflinewidth] (A) {
                  $s_4$                  & \phantom{} & \phantom{} & \phantom{} & \phantom{}         \\
                  $s_3$                  & 4                      & 2                      & \textbf{0} & 1                              \\
                  $s_2$                  & 3                      & \textbf{0} & 2                      & 3                              \\
                  $s_1$                  & \textbf{0} & 3                      & 4                      & 6                              \\
                  \phantom{} & $p_1$                  & $p_2$                  & $p_3$                  & $p_4$ \\};
          \end{tikzpicture}
          \begin{tikzpicture}
              \matrix[matrix of nodes,nodes={draw=gray, anchor=center, minimum size=.6cm}, column sep=-\pgflinewidth, row sep=-\pgflinewidth] (A) {
                  $s_4$                  & \phantom{} & 3                      & 3                      & 2                              \\
                  $s_3$                  & 4                      & 2                      & \textbf{0} & \textbf{1}         \\
                  $s_2$                  & 3                      & \textbf{0} & 2                      & 3                              \\
                  $s_1$                  & \textbf{0} & 3                      & 4                      & 6                              \\
                  \phantom{} & $p_1$                  & $p_2$                  & $p_3$                  & $p_4$ \\};
          \end{tikzpicture}

          Notice that now $\texttt{i} = i'$ but $\texttt{j} \neq j'$.

    \item As $\texttt{i} = i'$ but $\texttt{j} \neq j'$, the next step is to increment \texttt{i}:

          \begin{tikzpicture}
              \matrix[matrix of nodes,nodes={draw=gray, anchor=center, minimum size=.6cm}, column sep=-\pgflinewidth, row sep=-\pgflinewidth] (A) {
                  $s_4$                  & \phantom{} & 3                      & 3                      & \textbf{2} & 7                              \\
                  $s_3$                  & 4                      & 2                      & \textbf{0} & \textbf{1} & 9                              \\
                  $s_2$                  & 3                      & \textbf{0} & 2                      & 4                      & 9                              \\
                  $s_1$                  & \textbf{0} & 3                      & 4                      & 6                      & \phantom{}         \\
                  \phantom{} & $p_1$                  & $p_2$                  & $p_3$                  & $p_4$                  & $p_5$ \\};
          \end{tikzpicture}

          Now, $\texttt{i} \neq i'$ but $\texttt{j} = j'$.

    \item From $\texttt{i} \neq i'$ but $\texttt{j} = j'$ the next step is to increment \texttt{j}:

          \begin{tikzpicture}
              \matrix[matrix of nodes,nodes={draw=gray, anchor=center, minimum size=.6cm}, column sep=-\pgflinewidth, row sep=-\pgflinewidth] (A) {
                  $s_5$                  & \phantom{} & \phantom{} & 3                      & \textbf{3} & 11                             \\
                  $s_4$                  & \phantom{} & 3                      & 3                      & \textbf{2} & 7                              \\
                  $s_3$                  & 4                      & 2                      & \textbf{0} & \textbf{1} & 9                              \\
                  $s_2$                  & 3                      & \textbf{0} & 2                      & 3                      & 9                              \\
                  $s_1$                  & \textbf{0} & 3                      & 4                      & 6                      & \phantom{}         \\
                  \phantom{} & $p_1$                  & $p_2$                  & $p_3$                  & $p_4$                  & $p_5$ \\};
          \end{tikzpicture}

          As $\texttt{previous} = \texttt{current} = \texttt{j} \neq \texttt{ij}$, \texttt{runCount} is incremented to $2$.
\end{enumerate}

The algorithm terminates when the last element in $\mathcal{S}$ is reached. Note that this example does not show the case where \texttt{runCount} reaches \texttt{MaxRunCount}, but the idea is simple--when this occurs, the direction opposing the previously computed direction is taken.

Further, the output of the follower is triggered whenever $j'$ changes--this indicates that the follower deems that the position in the score has moved. The mapping that maps the specific $j'$ entry to the position in the score, previously obtained in the preprocessing step, is used to output the score position--this could be a timestamp.

%% file: 02_parts/03_impl/chapters/04b_dtw_impl.tex
\section{Implementation}
\label{impl_dtw_impl}

\subsection{Technology Stack}
\label{impl_dtw_impl_techstack}

Python was used, with similar reasons as those covered in \autoref{tb_quant_impl_details} and \autoref{impl_asm_impl_details}. Similar good practices in software engineering detailed in \autoref{tb_quant_bettereng} are used.

\subsection{Following Modes}

Two modes will be made available with different settings to the user: \textbf{Online Mode} and \textbf{Offline Mode}. The former runs with the whole performance sequence available to the follower (thus running the system as an aligner) and the latter otherwise.

\subsection{System Architecture}
\label{impl_dtw_impl_arch}

Here a specific focus is given to the \textbf{Online Mode}'s system architecture--the \textbf{Offline Mode}'s architecture is similar, save for the real-time streaming of performance audio to the system--the \textbf{Offline Mode} has full knowledge of the entire performance audio prior to beginning alignment.

The architecture of this follower is informed by the reasoning given in \autoref{tb_qual_impl_details_choice}--the interface of each component is clearly defined, and it is thus intended that individual components can be swapped out when needed. Three architecture diagrams each showing a different part of the system are presented here, before the full architecture of the system is shown.


\subsubsection{Preprocessor}

\begin{figure}[h]
    \centering
    \includegraphics[width=0.75\columnwidth]{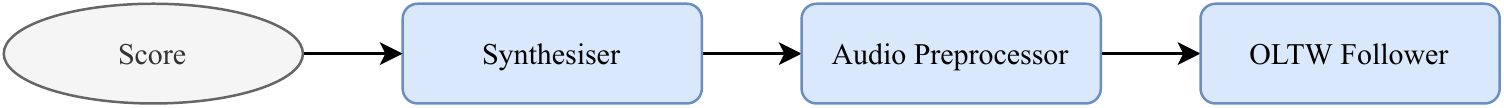}
    \caption{Architecture Diagram for the Preprocessor.}
    \label{fig:oltw_preprocessor}
\end{figure}

The architecture for the preprocessing step as covered in \autoref{impl_oltw_param} is given in \autoref{fig:oltw_preprocessor}.

Preprocessing is straightforward--here it is assumed that the score is MIDI data. Via the Synthesiser, the \texttt{pretty\_midi}\footnote{\url{https://github.com/craffel/pretty-midi}} and FluidSynth\footnote{\url{https://www.fluidsynth.org/}} libraries are used to synthesise score data into WAVE files. The \nameref{impl_dtw_sys_arch_audio_preproc} will be covered in its own architecture diagram in \autoref{fig:oltw_audiopreproc}--in short it produces $\mathcal{S}$ to be fed into the OLTW Follower. 

\subsubsection{Follower}

\begin{figure}[h]
    \centering
    \includegraphics[width=0.85\columnwidth]{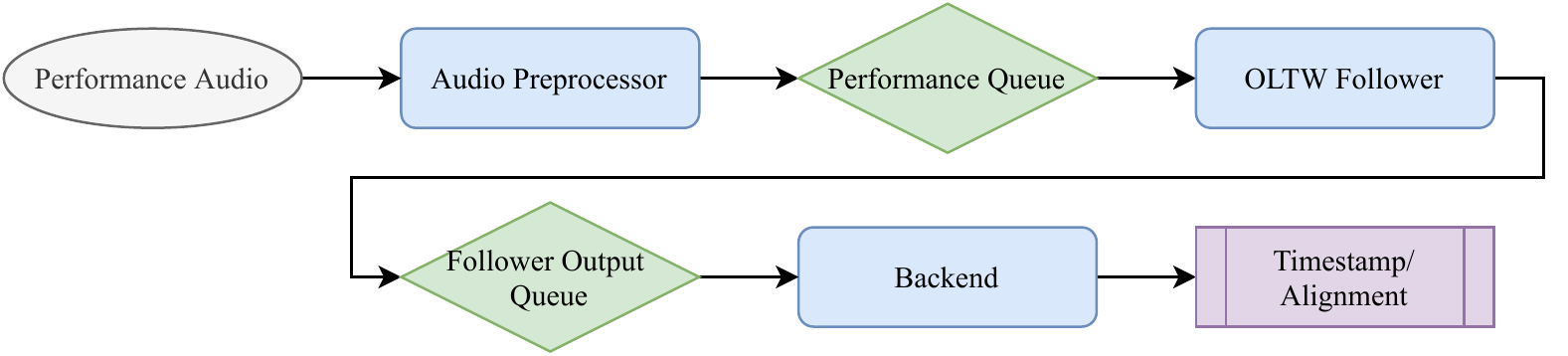}
    \caption{Architecture Diagram for the Following step.}
    \label{fig:oltw_step}
\end{figure}

\autoref{fig:oltw_step} shows the architecture diagram for the OLTW follower; this diagram describes the system when running the follower algorithm detailed in \autoref{impl_oltw_param}.

Performance Audio is streamed into the Audio Preprocessor, which converts the audio into pieces of $\mathcal{P}$ to be consumed by the OLTW Follower. The OLTW Follower then computes the best alignment and outputs it via a Backend module. The Backend module can be configured to output in either the \emph{timestamp} format (suitable for the qualitative testbench proposed in \autoref{tb_qual}) or the \emph{alignment} format (suitable for the quantitative testbench introduced in \autoref{tb_quant}).

The streaming of the performance audio can be configured with the following parameters:

\begin{itemize}
    \item \textbf{Simulate Performance}. When in online mode, this setting emulates the live-streaming of the performance audio file into the system. This can be turned off for testing purposes to save time at the expense of obtaining inaccurate real-time-sensitive metrics (such as those related to latency and offset). This is turned off by default.
    \item \textbf{Sleep Compensation}. The emulation of live-streaming the performance audio into the system is done by sleeping an appropriate amount of time between feeding two performance audio slices into the system. As most devices do not have real-time operating systems that allow exact sleep durations, the sleeping time is usually more than the ideal time. Sleep compensation allows the user to specify a small iota of time to reduce the sleep time between two slices to ensure that the sleep time more accurately emulates the streaming of the performance audio. This is set to $\si{0.0005\si{\second}}$ by default.
    \item \textbf{Play Performance Audio}. This setting, when turned on, plays the performance audio as it is streamed into the system. Listening to the performance audio is crucial when evaluating the system using the qualitative testbench as it allows a human user to judge the system by observing how accurate the alignment is based on the played audio. This setting is turned off by default.
\end{itemize}

Several tunable parameters are also available for the Backend:
\begin{itemize}
    \item \textbf{Backtracking}. With backtracking on, the follower is allowed to ``go back in score time'', essentially reporting aligned score events that precede the last aligned score event. This setting does not heavily affect the Backend in \emph{alignment} mode (as that mode only reports ``unseen'' notes). In the \emph{timestamp} mode however, turning off backtracking heavily stabilises the timestamps output. Backtracking is turned off by default.
    \item \textbf{Backend Compensation}. This setting only affects the \emph{timestamp} Backend mode. Because of the nature of streaming in slices, the follower essentially works on a slice of audio that is one frame length's amount of time behind, thus, the timestamps reported when this setting is on are each the sum of the original timestamp and a frame length's amount of time. Backend compensation is turned on by default.
\end{itemize}

As the follower runs time-sensitive real-time processes, parallelism is employed to reduce overall system latency. Here, two \textbf{multiprocessing queues}\footnote{\url{https://docs.python.org/3/library/multiprocessing.html}} are used to allow the Audio Preprocessor, OLTW Follower and Backend to run in parallel.
This architecture ensures that the three running components do not block each other--if the system is not designed in parallel, the preprocessor may lose slices of audio streamed real-time into the system, or the Backend may block the time-sensitive follower. Using parallelism also allows the system to take advantage of the increased computation power offered on modern machines, which are often multicore. Further, the OLTW follower is designed in a way to consume elements of $\mathcal{P}$ on demand and the blocking multiprocessing queue system fits well.

In terms of parallel processes run, as the OLTW Follower and Backend each run one process and the Audio Processor covered below runs two processes, a total of four parallel processes work together concurrently during following.

\subsubsection{Audio Preprocessor}
\label{impl_dtw_sys_arch_audio_preproc}

\begin{figure}[h]
    \centering
    \includegraphics[width=\columnwidth]{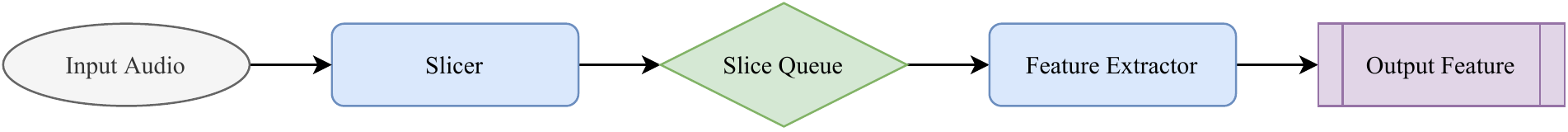}
    \caption{Architecture Diagram for the Audio Preprocessor.}
    \label{fig:oltw_audiopreproc}
\end{figure}

The architecture for the audio preprocessor is shown in \autoref{fig:oltw_audiopreproc}.

The slicer obtains slices of a specific length (detailed further in \autoref{cqt_impl}) of the input audio then writes it into the Slice Queue, another instance of a multiprocessing queue. This ensures that all input audio slices are obtained without latency or blocking issues, and implies that the preprocessor runs two processes in parallel. The Feature Extractor component then consumes slices of audio from the Slice Queue to be fed into the CQT implementation covered in~\autoref{cqt_impl}--the extracted CQT feature is then processed into the feature vector representations described in \autoref{impl_dtw_cqt_to_note}.

\subsubsection{Overall Architecture}

\begin{figure}[h]
    \centering
    \includegraphics[width=0.45\columnwidth]{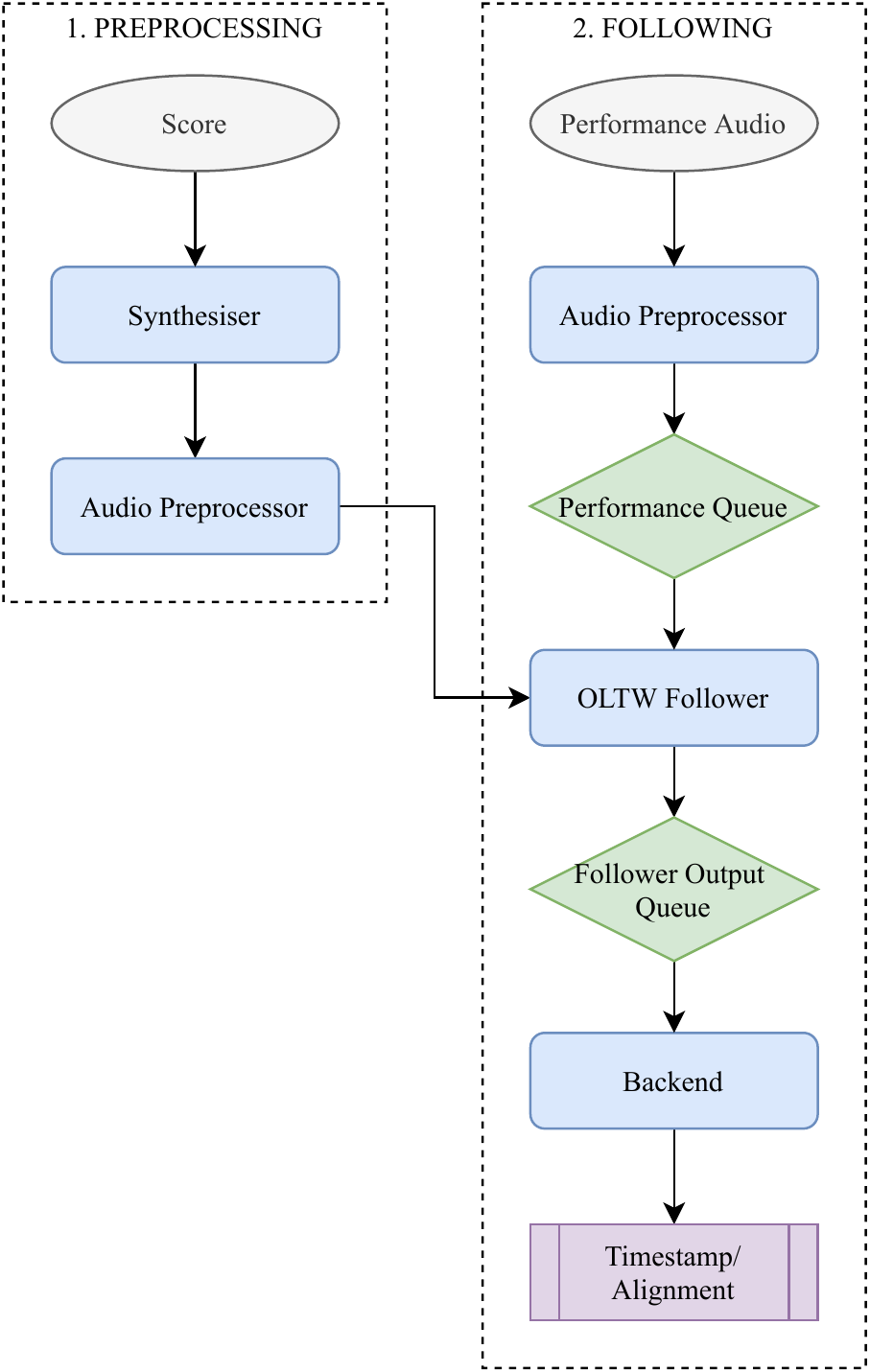}
    \caption{Overall Architecture.}
    \label{fig:oltw_archfull}
\end{figure}

The full architecture of the system is shown in \autoref{fig:oltw_archfull}. The system is segmented to delineate the two core steps of the system: preprocessing and following.

As discussed, in short, the preprocessing step--running two parallel processes--preprocesses score data prior to following. The following step then takes over, running four parallel processes to process streamed performance audio and align extracted features before outputting the timestamp or alignment output via the Backend component.

\subsection{CQT Implementation}
\label{cqt_impl}

The invertible real-time CQT algorithm--sliCQ--introduced by Dörfler, Holighaus, Grill and Velasco \cite{holighaus13,angelo11} is used. In particular, the Python implementation by Grill is used\footnote{\url{https://github.com/grrrr/nsgt}}. Further, some other available substitutes of sliCQ are detailed below.

\subsubsection{Offline Mode}
\label{impl_dtw_offline_mode_cqt}

The parameters used in Offline Mode are $f_s = 44100\si{\hertz}$ and $N_{bin}=12$. $f_{\min}$, $f_{\max}$, slice length and transition length can be customised via command line arguments, but otherwise default to the values defined below in \nameref{impl_dtw_online_mode_cqt}. 

In addition to the sliCQ algorithm, a reference CQT computation method using classical CQT as introduced in \cite{christiancqt}\footnote{\url{https://librosa.org/doc/main/generated/librosa.cqt.html}} can be selected.

\subsubsection{Online Mode}
\label{impl_dtw_online_mode_cqt}

In Online Mode, the default parameters are $f_s = 44100\si{\hertz}$, $N_{bin}=12$, $f_{\min} = 130.8 \si{\hertz}$ ($C3$) and $f_{\max} = 4186.0 \si{\hertz}$ ($C8$).
Further, the default frame length is $8192$ samples and the default hop length is $2048$ samples, which means that each hop is $\frac{2048}{44100} \approx 46\si{\milli\second}$ long. These frames are directly fed as slices into the sliCQ algorithm detailed in \autoref{impl_dtw_cqt_tech}; referring to \autoref{fig:tukeywins}, the default settings here imply $2N = 8192$ and $M = 2048$. In addition to sliCQ, ``real-time-capable'' CQT approximations provided in the \texttt{librosa} library, namely hybrid CQT\footnote{\url{https://librosa.org/doc/main/generated/librosa.hybrid_cqt.html}} and pseudo CQT\footnote{\url{https://librosa.org/doc/main/generated/librosa.pseudo_cqt.html}}, are available as options--these alternatives are, as expected, not as performant as sliCQ--see \autoref{impl_dtw_eval} for evaluation results.

For configuration purposes, $f_{\min}$, $f_{\max}$, slice length, transition length and other further parameters detailed in this system are available as configurable command line arguments--refer to the repository\footnote{\githubFlippy} for more information.

Multithreading for sliCQ computation is available as an option for performance, and, when required and if enabled, FFT calculations are sped up via using the FFTW library\footnote{\url{http://www.fftw.org/}}.

\subsection{DTW Implementation}
\label{impl_dtw_dtw}

\subsubsection{Offline Mode (Classical)}
\label{impl_dtw_dtw_offline}

In Offline Mode, users can choose between classical DTW (implementing \autoref{eqn:orig_dtw}) or the OLTW algorithm as will be covered below in \nameref{impl_dtw_dtw_online}. Command line settings are provided to switch between these two modes. This will allow comparison of classical offline DTW to OLTW. The weights for the local path constraint calculation (see \autoref{eqn:oltw_dtw}), $w_a$, $w_b$ and $w_c$, are all set to $1$ (as per standard configurations in DTW, see \autoref{eqn:orig_dtw}), but can be configured differently.

\subsubsection{Online Mode (OLTW)}
\label{impl_dtw_dtw_online}

In Online Mode, the OLTW algorithm as detailed in \autoref{impl_dtw_baseline_worked_example} is used. The two parameters listed in \autoref{impl_oltw_param}, \texttt{c} and \texttt{MaxRunCount}, are set to 500 and 3 respectively. With a hop length giving a hop time of approximately $46\si{\milli\second}$ long, the width of the search band is around $23\si{\second}$ long--relevant experiments showed that a search band of $10\si{\second}$ is sufficient \cite{Dixon2005MATCHAM, arzt08}. These parameters are available for customisation via the command line. The weights $w_a$, $w_b$ and $w_c$ are set to $0.5$, $1$ and $1$ respectively to favour full performances with minimal mistakes as in the evaluation runs discussed in \autoref{impl_dtw_eval}--the weights can be tuned differently to adjust the follower to work with a variety of performances.

%% file: 02_parts/03_impl/chapters/04c_dtw_eval.tex
\section{Evaluation}
\label{impl_dtw_eval}

All evaluation runs are performed on the system described in \autoref{appendix_benchmark_system}. \Cref{impl_dtw_prod_repro} contains instructions on how to reproduce these evaluations--note that real-time-sensitive metrics, such as latency and offset, depend heavily on the benchmarking system's hardware capabilities.


\Cref{impl_dtw_eval_quant} first covers results for quantitative evaluations run on the \nameref{impl_dtw_eval_bach10} and \nameref{impl_dtw_eval_bwv846} datasets. \Cref{impl_dtw_eval_qual} subsequently covers discusses results for qualitative evaluations run on the \textit{QualScofo} dataset to observe how well the system generalises to a variety of music.

\subsection{Quantitative Results}
\label{impl_dtw_eval_quant}

In the quantitative evaluations, the misalignment threshold $\theta_e$--introduced in \Cref{def:misalign}, is varied in the range $\si{50\milli\second}$ to $\si{3000\milli\second}$ in $\si{50\milli\second}$ increments for precision rate reports--other metrics are reported at the default $\theta_e = \si{300\milli\second}$ setting. The score note search bound (see \Cref{def:bound_ms}) is set to the default $\si{1\milli\second}$.

Four systems, two offline and two online, are put to the test. All the parameters used are, unless otherwise noted, the defaults stated in \autoref{impl_dtw_impl}. 

The offline systems use classical DTW as described in \autoref{impl_dtw_dtw} (\nameref{impl_dtw_dtw_offline}), and comprise:
\begin{itemize}
    \item \cqtOffline. This system extracts features from the audio using a classical CQT computation method as described in \cite{christiancqt}  and detailed further in \autoref{cqt_impl} (\nameref{impl_dtw_offline_mode_cqt}). This CQT computation method requires the full input signal and does not run in real time. This serves as a baseline system for offline alignment.
    \item \nsgtOffline. This system extracts features from the audio using the sliCQ method described in \cite{holighaus13} and detailed further in \autoref{cqt_impl} (\nameref{impl_dtw_online_mode_cqt}). While this method is real-time capable, it would be interesting to see how this CQT computation method performs in comparison to the method described in \cite{christiancqt}.
\end{itemize}

On the other hand, the online systems use OLTW, as described in \autoref{impl_dtw_dtw} (\nameref{impl_dtw_dtw_online}); they are:
\begin{itemize}
    \item \cqtOnline. This system uses FFT before binning the frequencies into imitate CQT features\footnote{Further information can be found at \url{https://librosa.org/doc/main/generated/librosa.pseudo_cqt.html}}. As seen in \autoref{fig:cqt_time}, this system is not real-time capable on the evaluation system and hence will cause latency and offset to build up; however, this will only affect real-time-sensitive metrics such as latency and offset. This feature extraction technique is similar to ones used by Dixon and Arzt \etal{} \cite{dixonOLTW,Dixon2005MATCHAM,arzt08,arzt16}, and serves as a baseline to study if using sliCQ features improves the score follower's performance.
    \item \nsgtOnline. This system extracts features from the audio using the sliCQ method described in \cite{holighaus13} and detailed further in \autoref{cqt_impl} (\nameref{impl_dtw_online_mode_cqt})--this setup is the ``flagship'' score follower of the project.
\end{itemize}

\subsubsection{Bach10}
\label{impl_dtw_eval_bach10}

The Bach10 dataset \cite{duan11bach10}, also used for evaluating the ASM score aligner in \autoref{impl_asm_eval_bach10}, comprises 10 short chorales, each with four voices played by four instruments (violin, clarinet, saxophone and bassoon).

The composite recording of all the instruments' audio tracks is used as the performance audio, and therefore this evaluation tests the ability of the system to extract features reliably from actual performance audio, discussed in \autoref{scofo_challenge_limitfe}. All pieces are highly polyphonic, and thus this test strongly evaluates the system's robustness against following polyphonic music, as detailed in \autoref{scofo_challenges_polyphonic}.

As the recordings are done by trained musicians in a recording setting, there are some, but not many, performance deviations; notably, the performance has significant tempo deviations as they emulate the phrasing of a choir performance (and thus contain appropriate pauses between lines where a choir may naturally pause and breathe). Thus, the system's reaction to performance deviations as detailed in \autoref{scofo_challenges_perfdev} is not so apparent in this evaluation suite. Further, the musical pieces do not contain many ornaments and underspecified performance directions, so the performance of the system against underspecified musical scores (see \autoref{scofo_challenges_underspec}) is not a priority here.

\begin{figure}[h]
    \centering
    \includegraphics[width=0.75\columnwidth]{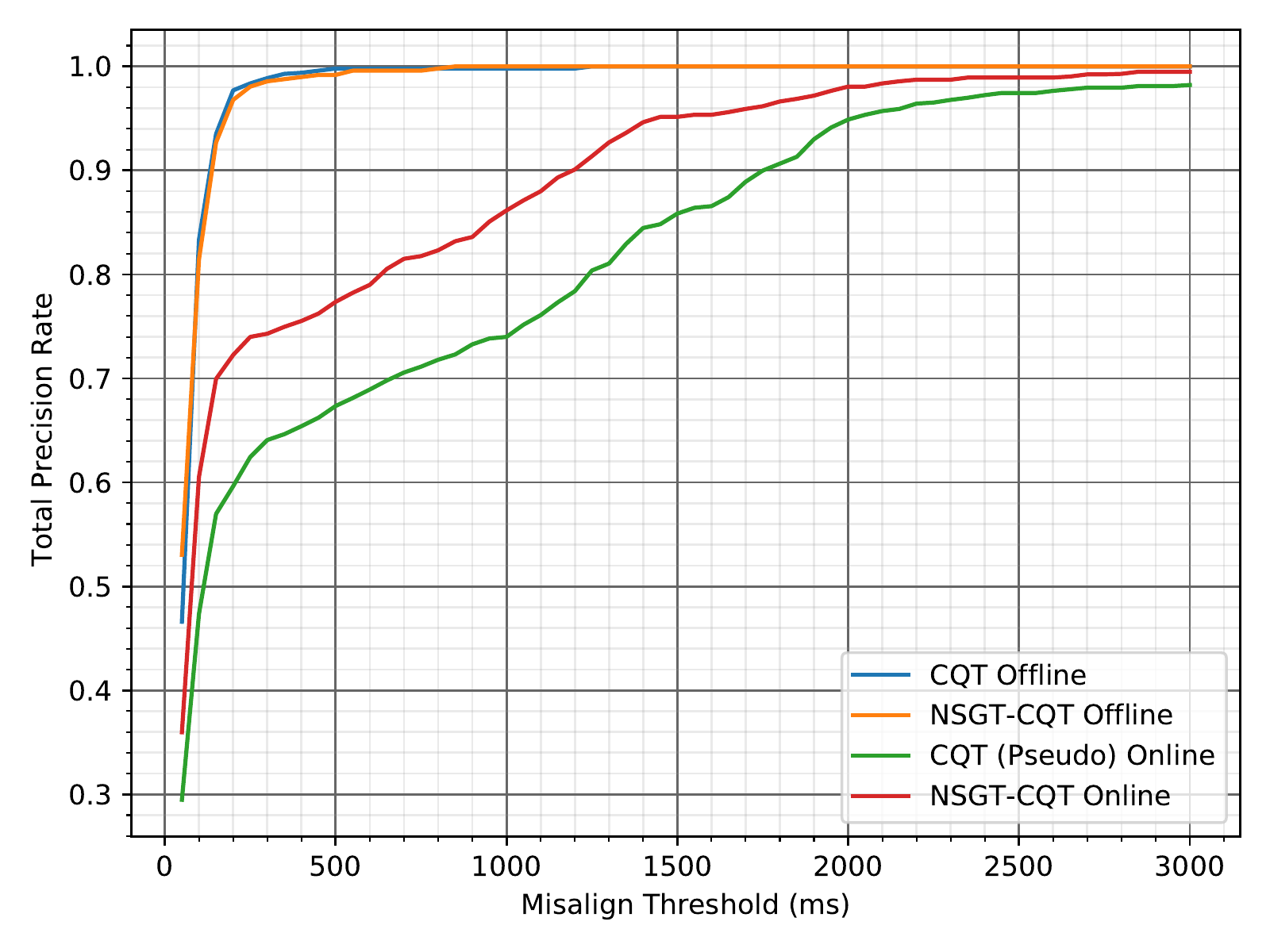}
    \caption{Total precision rate ($r_{pt}$) results of the four score following systems over varying misalign thresholds ($\theta_e$) on the Bach10 dataset.}
    \label{fig:bach10_plot_precision}
\end{figure}

\autoref{fig:bach10_plot_precision} shows the four systems' total precision rate ($r_{pt}$) performance on the Bach10 dataset. As expected, the \cqtOffline{} system performed very well. The \nsgtOffline{} system performs almost identically, showing sliCQ's robustness and reliability in extracting features from the audio.

More interestingly, the \cqtOnline{} system's performance consistently and significantly trails behind that of the \nsgtOnline{} system. This is exacerbated by the fact that the former system is not truly capable of running in real time on the test system (see \autoref{fig:cqt_time}). While this system was not evaluated against pieces in the MIREX evaluation dataset \cite{cont07}, the \nsgtOnline{}'s total precision rate for Bach10 at $\theta_e = \si{300\milli\second}$ outperforms all known pure audio-to-audio methods in the literature \cite{arzt08, suzuki10} when ran through the MIREX evaluation. Most notably, the fact that the \cqtOnline{} system is heavily based on Dixon's original solution \cite{Dixon2005MATCHAM} shows the robustness of sliCQ in extracting audio features over existing and commonly used FFT-based methods.

While it is apparent that some existing methods in the literature--a select few detailed in \autoref{scofo_recent_mirex}--(may) outperform this system at the MIREX and/or Bach10 dataset evaluations, this system, as mentioned, does not require complete score data as opposed to the aforementioned methods. Not requiring complete score data is an advantage as some musical pieces do not have full score data and users may have to resort to annotated score recordings \cite{arztconcert}. Nevertheless, most feature extraction techniques for complex high-performing systems rely on FFT-based methods to extract features, and sliCQ's robustness and reliability can potentially increase performance in the aforementioned systems. This system is also simple, flexible and easily extensible to be improved--\autoref{impl_dtw_dicsuss} will outline many areas in which this system can be improved.

For \nsgtOnline, the $95\%$ total precision rate at $\theta_e = \si{1400\milli\second}$ implies that about $95\%$ of notes in the correct dataset are correctly aligned within $\si{1400\milli\second}$ of the actual position. This further implies that this system is robust enough for APT, which does not require very high precision rates. At $\theta_e = \si{2650\milli\second}$, the system also reaches $99\%$ total precision rate, reaching performance very close to the offline approaches. This performance was not matched within $\theta_e \leq \si{3000\milli\second}$ by the \cqtOnline{} system. At the standard $\theta_e = \si{300\milli\second}$, the \nsgtOnline{} system's total precision rate is a respectable $74\%$, which is sufficient for performance analysis and APT.

Moving on, at the standard $\theta_e = \si{300\milli\second}$ setting, further piecewise (average across all pieces) metrics detailed in \autoref{table:eval_metrics_bach10} are studied to compare the four systems--\Cref{tb_quant} gives full details of these metrics. Online metrics, such as those related to latency and offset, are only reported for the online methods.

\begin{table}[ht]
    \caption{Metrics used to compare the approaches against the Bach10 dataset.}
    \centering 
    \begin{tabular}{c c} 
        \hline\hline 
        Metric                        & Symbol       \\ [0.5ex] 
        \hline 
        Miss Rate                     & $r_m$        \\
        Misalign Rate                 & $r_e$        \\
        Piece Completion              & $p_e$        \\
        Standard Deviation of Error   & $\sigma_e$   \\
        Mean Absolute Error           & $\text{MAE}$ \\
        Standard Deviation of Latency & $\sigma_l$   \\
        Mean Latency                  & $\mu_l$      \\
        Standard Deviation of Offset  & $\sigma_o$   \\
        Mean Absolute Offset          & $\text{MAO}$ \\
        Piecewise Precision Rate      & $r_{pp}$     \\
        Total Precision Rate          & $r_{pt}$     \\
        \hline 
    \end{tabular}
    \label{table:eval_metrics_bach10} 
\end{table}

\autoref{table:bach10_metrics} shows the piecewise results for the four systems. As discussed and seen in \autoref{fig:bach10_plot_precision}, the two offline systems performed almost identically, and have exceptional performance, reaching almost $99\%$ piecewise precision rate. On the other hand, between the two online systems, the \nsgtOnline{} system performs better across the board compared to the \cqtOnline{} system, save for the miss rate ($r_m$) metric where the latter system performed very slightly better. The \cqtOffline{} approach's non-real-time capability is demonstrated here with poor scores in mean latency ($\mu_l$) and mean absolute offset ($\text{MAO}$). As the pieces in the Bach10 dataset are not that long (the average piece length is under one minute), the build-up in lag is actually not as pronounced as that observed in the BWV846 evaluation presented later in \autoref{impl_dtw_eval_bwv846}.

Overall, the performance of the project's flagship online system,
\nsgtOnline{},
is excellent especially when compared to the FFT-based counterpart--\cqtOnline{}.
The performance of \nsgtOnline{} on the highly polyphonic music pieces with
each audio recording comprising four different instruments show how the follower system is reliable in extracting robust features from audio, dealing with polyphonic music and--to an extent--performance deviations.

\afterpage{%
    \clearpage
    \thispagestyle{empty}
    \begin{landscape}

        \begin{table}
            \caption{Piecewise results of the systems against the Bach10 dataset at $\theta_e = \si{300\milli\second}$. The arrows, where present, denote whether a metric should be minimised ($\downarrow$) or maximised ($\uparrow$). \textbf{Bolded} and \textit{italic} results denote the best score for the online and offline systems respectively in each metric.} 
            \centering 
            \begin{tabular}{l c c c c c c c c c c c} 
                \hline\hline 
                System         & $r_m\downarrow$ & $r_e\downarrow$ & $p_e\uparrow$  & $\sigma_e(\si{\milli\second})\downarrow$ & $\text{MAE}(\si{\milli\second})\downarrow$ & $\sigma_l(\si{\milli\second})\downarrow$ & $\mu_l(\si{\milli\second})\downarrow$ & $\sigma_o(\si{\milli\second})\downarrow$ & $\text{MAO}(\si{\milli\second})\downarrow$ & $r_{pp}\uparrow$ & $r_{pt}\uparrow$ \\ [0.5ex] 
                \hline 
                \cqtOffline{}  & 0.000           & \textit{0.012}  & 1.000          & \textit{48}                              & 63                                         & N/A                                      & N/A                                   & N/A                                      & N/A                                        & \textit{0.988}   & \textit{0.989}   \\
                \nsgtOffline{} & 0.000           & 0.013           & 1.000          & 72                                       & \textit{58}                                & N/A                                      & N/A                                   & N/A                                      & N/A                                        & 0.987            & 0.986            \\
                \hline 
                \cqtOnline{}   & \textbf{0.004}  & 0.349           & 0.983          & 88                                       & 76                                         & 962                                      & 4014                                  & 987                                      & 4051                                       & 0.647            & 0.641            \\
                \nsgtOnline{}  & 0.006           & \textbf{0.240}  & \textbf{0.996} & \textbf{66}                              & \textbf{64}                                & \textbf{883}                             & \textbf{1490}                         & \textbf{880}                             & \textbf{1529}                              & \textbf{0.754}   & \textbf{0.743}   \\
                \hline 
            \end{tabular}
            \label{table:bach10_metrics} 
        \end{table}

        \begin{table}
            \caption{Piecewise results of the systems against the BWV846 dataset at $\theta_e = \si{300\milli\second}$. The arrows, where present, denote whether a metric should be minimised ($\downarrow$) or maximised ($\uparrow$). \textbf{Bolded} and \textit{italic} results denote the best score for the online and offline systems respectively in each metric.} 
            \centering 
            \begin{tabular}{l c c c c c c c c c c c} 
                \hline\hline 
                System         & $r_m\downarrow$ & $r_e\downarrow$ & $p_e\uparrow$  & $\sigma_e(\si{\milli\second})\downarrow$ & $\text{MAE}(\si{\milli\second})\downarrow$ & $\sigma_l(\si{\milli\second})\downarrow$ & $\mu_l(\si{\milli\second})\downarrow$ & $\sigma_o(\si{\milli\second})\downarrow$ & $\text{MAO}(\si{\milli\second})\downarrow$ & $r_{pp}\uparrow$ & $r_{pt}\uparrow$ \\ [0.5ex] 
                \hline 
                \cqtOffline{}  & 0.000           & \textit{0.045}  & 0.979          & \textit{55}                              & \textit{53}                                & N/A                                      & N/A                                   & N/A                                      & N/A                                        & 0.957            & 0.957            \\
                \nsgtOffline{} & 0.000           & 0.025           & \textit{0.997} & 64                                       & 96                                         & N/A                                      & N/A                                   & N/A                                      & N/A                                        & \textit{0.978}   & \textit{0.976}   \\
                \hline 
                \cqtOnline{}   & 0.015           & 0.327           & 0.925          & 105                                      & 99                                         & 5149                                     & 9333                                  & 5095                                     & 9272                                       & 0.658            & 0.662            \\
                \nsgtOnline{}  & \textbf{0.013}  & \textbf{0.193}  & \textbf{0.940} & \textbf{88}                              & \textbf{72}                                & \textbf{307}                             & \textbf{1066}                         & \textbf{265}                             & \textbf{1030}                              & \textbf{0.794}   & \textbf{0.797}   \\
                \hline 
            \end{tabular}
            \label{table:bwv846_metrics} 
        \end{table}
    \end{landscape}
    \clearpage
}

\subsubsection{BWV846}
\label{impl_dtw_eval_bwv846}

The BWV846 dataset is produced by the ASM aligner as discussed in \autoref{impl_asm_eval_bwv846}. This evaluation suite demonstrates the last piece in the puzzle for the novel score following quantitative dataset procurement method detailed in \autoref{tb_quant_dataset_api}.

This dataset has key differences when compared to Bach10--most of which are addressed in \autoref{impl_asm_eval_bwv846}. Most importantly, the performances--especially the \textit{Fugue}'s--have more deviations from the score, which showcase the score follower's ability to handle performance deviations. Further, the \textit{Prelude} is highly monophonic, and the \textit{Fugue} otherwise, so some insight into the follower's ability to follow either monophonic and polyphonic music can be drawn.

\begin{figure}[h]
    \centering
    \includegraphics[width=0.75\columnwidth]{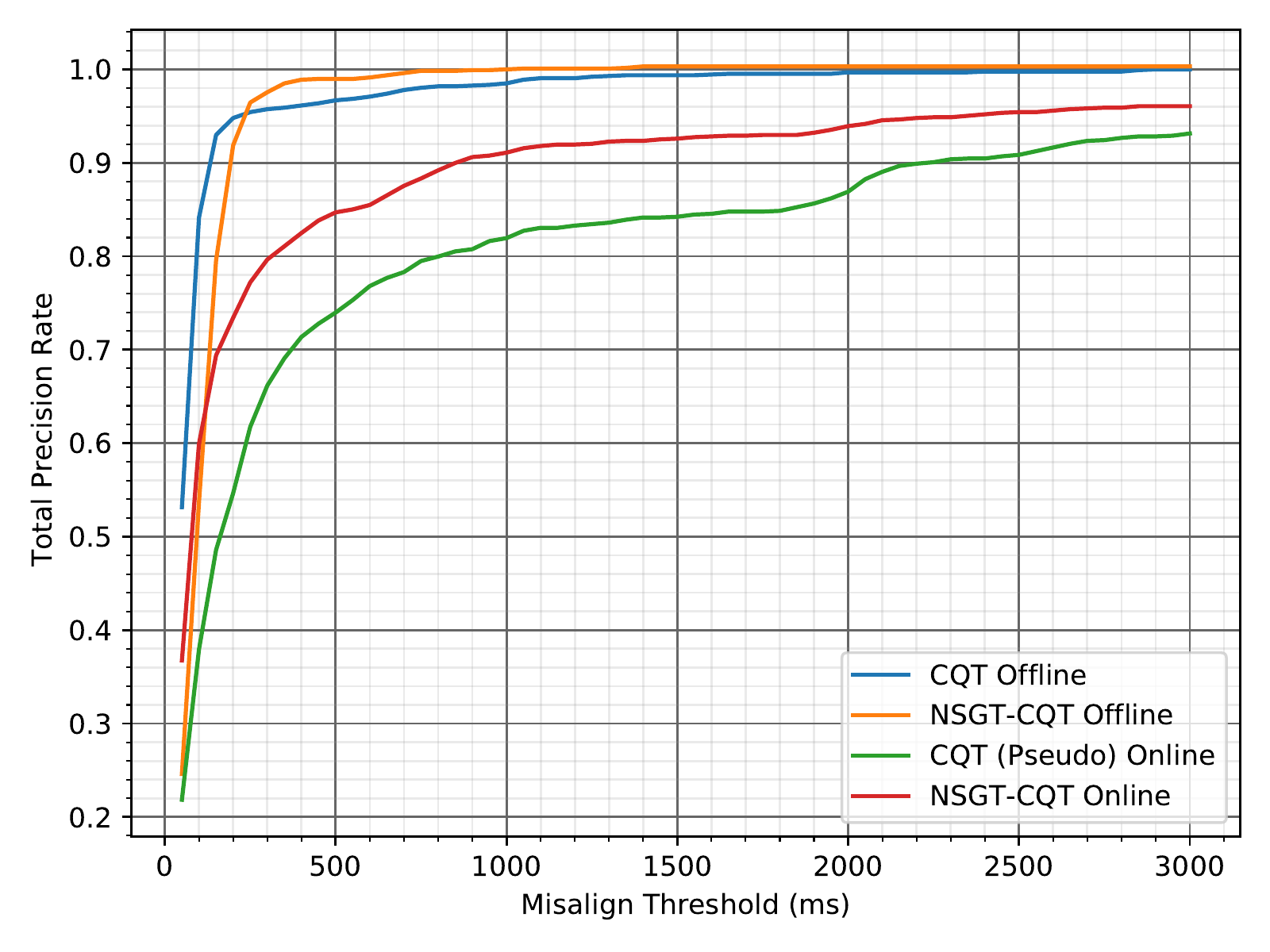}
    \caption{Total precision rate ($r_{pt}$) results of the four score following systems over varying misalign thresholds ($\theta_e$) on the BWV846 dataset.}
    \label{fig:bwv846_plot_precision}
\end{figure}

\autoref{fig:bwv846_plot_precision} shows the four systems' total precision rate ($r_{pt}$) performance on the Bach10 dataset. Generally, the performance of the four systems are similar to that seen in the quantitative evaluation on the Bach10 dataset.

For the two offline systems, the \cqtOffline{} system appeared to have an edge over the \nsgtOffline{} system at $\theta_e \leq \si{250\milli\second}$, after which the latter system consistently performs better before both systems reach almost perfect $r_{pt}$ approaching $\theta_e = \si{3000\milli\second}$.

Unsurprisingly, for the online systems, the \nsgtOnline{} system once again consistently outperforms the \cqtOnline{} system. Interestingly, the former system even surpassed the performance of the \nsgtOffline{} approach at $\theta_e \leq \si{150\milli\second}$--this was not expected. Though, unlike for the Bach10 evaluation suite, both online approaches did not approach $r_{pt} = 1$ as $\theta_e$ approached $\si{3000\milli\second}$. The \nsgtOnline{} system's performance plateaued at $r_{pt} \approx 0.96$ after $\theta_e \geq \si{2500\milli\second}$, suggesting that the gap to the almost-perfect-scoring offline systems stem from the system's inability to navigate around performance deviations which are more frequent in the BWV846 dataset than in the Bach10 dataset. Nevertheless, the performance is still impressive, considering the fact that for $\theta_e \leq \si{1000\milli\second}$, the performance of the online systems on the BWV846 dataset is significantly better than that of the BWV846 dataset as seen in \autoref{fig:bach10_plot_precision}--this may be attributable to the clearer note onsets produced by the piano in the BWV846 performance, which is in contrast to the less clear ones in the Bach10 recordings.

Piecewise metrics, identical to the ones presented in \autoref{table:eval_metrics_bach10}, are presented at the standard $\theta_e = \si{300\milli\second}$ for the systems in \autoref{table:bwv846_metrics}. The performances of the offline systems are very close to each other, with the \nsgtOffline{} approach slightly trumping over the \cqtOffline{} approach in the most important metrics: total precision rate ($r_{pt}$) and piecewise precision rate ($r_{pp}$). On the other hand, for the online systems, the performance advantage of the \nsgtOnline{} system over the \cqtOnline{} system is even more pronounced here than in the \nameref{impl_dtw_eval_bach10} evaluation. The \cqtOnline{}'s non-real-time property translated to very poor scores in the latency and offset metrics, and is worse here than in the \nameref{impl_dtw_eval_bach10} evaluation as both pieces in BWV846 are significantly longer (and thus the lag builds up). Importantly, the standard deviations of latency ($\sigma_l$) and offset ($\sigma_o$) of the \nsgtOnline{} systems are low as also seen in \autoref{table:bach10_metrics}--this suggests that there is a possible improvement area by letting the score follower report score events at a sensible $\mu_l$ ahead of the currently aligned position.

Exploiting the fact that the \textit{Prelude} is highly monophonic and the \textit{Fugue} otherwise, key metrics obtained via the two online systems at $\theta_e = \si{300\milli\second}$ for each piece are reported in \autoref{table:bwv846_eval_separate}. There is simply no contest between the two online approaches--the \nsgtOnline{} approach outperforms the \cqtOnline{} system in every metric measured. Perhaps unexpectedly, the systems fared slightly better (around $5\%$ improvement in $r_p$) in the highly polyphonic \textit{Fugue} than the highly monophonic \textit{Prelude}--it is however worth noting that this slight difference is not conclusive that the techniques used favour polyphonic pieces: the \textit{Prelude}, of which the first six bars can be seen in \autoref{fig:modern-staff-notation}, potentially poses another challenge--the majority of the \textit{Prelude} comprises broken \textit{arpeggios} repeated twice, and this repetition may mislead the system--this is in contrast to the well-structured \textit{Fugue} that has longer lines of melody and harmony--this can only be explained with qualitative analysis, which will be discussed in \autoref{impl_dtw_eval_qual}. Nevertheless, this comparison shows that the system performs robustly against highly monophonic and highly polyphonic piano pieces, and thus shows that the feature extraction method is reliable--it does not favour either category of music.

\begin{table}[ht]
    \caption{Results of the two online systems on the \textit{Prelude} and \textit{Fugue} of BWV846 with $\theta_e = \si{300\milli\second}$. The arrows, where present, denote whether a metric should be minimised ($\downarrow$) or maximised ($\uparrow$). \textbf{Bolded} results denote the best score for each piece in each metric.} 
    \centering 
    \begin{tabular}{c c c c c c c c} 
        \hline\hline 
        Piece                             & System        & $r_m\downarrow$ & $r_e\downarrow$ & $p_e\uparrow$  & $\sigma_e(\si{\milli\second})\downarrow$ & $\text{MAE}(\si{\milli\second})\downarrow$ & $r_{p}\uparrow$ \\ [0.5ex]
        \hline
        \multirow{2}{*}{\textit{Prelude}} & \cqtOnline{}  & 0.035           & 0.334           & 0.909          & 90                                       & 71                                         & 0.631           \\
                                          & \nsgtOnline{} & \textbf{0.031}  & \textbf{0.193}  & \textbf{0.914} & \textbf{80}                              & \textbf{60}                                & \textbf{0.776}  \\
        \hline
        \multirow{2}{*}{\textit{Fugue}}   & \cqtOnline    & 0.000           & 0.321           & 0.942          & 121                                      & 126                                        & 0.685           \\
                                          & \nsgtOnline{} & 0.000           & \textbf{0.193}  & \textbf{0.967} & \textbf{96}                              & \textbf{84}                                & \textbf{0.813}  \\
        \hline 
    \end{tabular}
    \label{table:bwv846_eval_separate} 
\end{table}

Overall, the performance of the flagship \nsgtOnline{} system is, as was observed in \autoref{impl_asm_eval_bach10}, exceptional especially when compared against the \cqtOnline{} system. The flagship system appears to generalise well to both monophonic and polyphonic pieces, and is able to perform well against performance deviations.

\subsection{Qualitative Evaluation Results}
\label{impl_dtw_eval_qual}

The quantitative evaluations performed in \autoref{impl_dtw_eval_quant} cover a variety of baroque pieces by Johann Sebastian Bach for piano and four voices--it contains some diversity in the music that measures the systems' performance against performance deviations and polyphonic music. The systems' robustness against limitations in feature extraction was also explored in the two evaluation suites. Nevertheless, the two suites do not reflect the performance of the systems in generalising to more diverse pieces performed under different circumstances; thus, qualitative analyses were performed for the flagship \nsgtOnline{} system (described in \autoref{impl_dtw_eval_quant}) against a diverse collection of music pieces. The system's performance against underspecified music scores will also be explored.

The dataset used for this suite of evaluations is the \textit{QualScofo} dataset presented in \autoref{tb_qual_prod_dataset}.

Qualitative evaluations are done via connecting this system to the qualitative testbench proposed in \autoref{tb_qual}. A human user judges the quality of following, mainly via how much the currently followed position deviates from the actual score position based on the performance audio. Should such deviations occur, how fast the follower corrects itself is also noted. Recordings of all the testbench runs are provided--see \autoref{impl_dtw_prod_demos} for more information. \Cref{impl_dtw_prod_repro} shows instructions on how to reproduce the evaluation.

Here, the performance of the follower on every piece of the \textit{QualScofo} dataset is discussed individually--overall comments are provided after these discussions. These pieces are organised by their respective groups. Readers are advised to watch the recordings provided in the repository (see \autoref{impl_dtw_prod_demos}) while perusing the analysis below.

\begin{itemize}
    \item{\texttt{cello}
          \begin{enumerate}
              \item{\texttt{suite1}

                    While this excerpt consists almost entirely of semiquavers, performers often introduce \textit{rubato} (temporal deviations)--this is in contrast to the structurally similar \textit{Prelude in C major, BWV 846}. In this performance, Yo-Yo Ma liberally holds many of the root notes present in the first and third beat of each bar. Further, as a consequence of the time-frequency uncertainty principle (\Cref{def:timefreq}), the lower register of the cello pose a challenge for feature extraction. This piece therefore challenges the score follower's ability to respond to temporal performance deviations and limitations in low-pitched feature extraction.

                    The score follower flawlessly handled this performance--the deviations were very minimal and mostly spanned  only about half a beat's length. The held root notes were easily identified and seamlessly followed.
                    }
          \end{enumerate}
          }
    \item \texttt{octet}
          \begin{enumerate}
              \item{\texttt{mendelssohn}

                    This octet excerpt consists of one melody line mainly driven by the first violin along with harmonic layers provided by the other instruments. The tempo is quite steady, and performance deviations are rare if any. This piece therefore tests the follower's ability to extract features from polyphonic audio arising from multiple instruments.

                    This excerpt was handled quite well by the follower with almost no deviations save for some instability in the first bar. The follower was able to follow the melodic notes of the violin--up to note-level--flawlessly.
                    }
          \end{enumerate}
    \item \texttt{orchestra}
          \begin{enumerate}
              \item{\texttt{eine}

                    This excerpt for a chamber string orchestra by Mozart contains some common ornaments in classical music, notably the \textit{trill} and \textit{acciaccatura}. The melody line is simple, with the lower register instruments (viola, cello and double bass) playing a more supportive role. This excerpt tests the follower's ability to follow performance deviations stemming from ornaments--for a chamber string orchestra performance.

                    This piece as a whole was handled quite well by the system, save for a bar's length of deviation between bars 8 and 11, which was quickly fixed as the next phrase started in bar 11. The mentioned ornaments posed no challenge to the score follower.
                    }
              \item{\texttt{peer}

                    This excerpt for a small romantic-era symphony orchestra starts by featuring several expressive solo melody lines--played interchangeably by the flute and the oboe--which contain a significant amount of artistic \textit{rubato}. The string section follows with a recapitulation of the melody lines, also flush with expressivity and \textit{rubato}--this piece therefore evaluates the follower's ability to perform against performance deviations stemming from both a solo instrument and a large ensemble.

                    The solo melodic lines were followed flawlessly and as a whole the excerpt was handled well. Some deviation--up to a beat's length--was found during the string unison section in bars 20-24; this deviation was rectified quickly after the phrase.
                    }
              \item{\texttt{1812-1}

                    This first excerpt from the \textit{1812 Overture} is the first part of the brass fanfare finale. The recording of this performance is not as clear despite it being a studio recording. This composition was also made for a large romantic-era symphony orchestra with a separate brass band and artillery (cannon). This excerpt does not contain the artillery section, but is still quite complex as it contains multiple layers of sound from the wide array of instruments. Further, there are running notes played by the strings between the very loud phrases of the fanfare that may cause a challenge in extracting reliable features from polyphonic audio produced by a large ensemble.

                    As expected, while the brass fanfare notes were well followed, the running notes of the strings were not followed well, with deviations of up to half a bar. These deviations were however quickly fixed as the next phrase of the brass fanfare began.
                    }
              \item{\texttt{1812-1}

                    This second excerpt from the \textit{1812 Overture} is the second part of the brass fanfare finale. This part of the finale contains the artillery (cannon) and bells section, and therefore strongly tests the follower's ability to generalise against non-standard instruments.

                    This excerpt was handled quite excellently (mostly minor deviations of up to two beats) save for the final twelve bars where a deviation of up to two bars occurred; nevertheless, this deviation was rectified as the piece progressed into the last two bars. The follower therefore was able to extract features from this complex excerpt, but the structure of the piece caused some confusion in the last few bars.
                    }
          \end{enumerate}
    \item \texttt{piano}
          \begin{enumerate}

              \item{\texttt{fugue}

                    Bach's polyphonic \textit{Fugue in C major, BWV 846} appears multiple times in this report--including in the quantitative analysis of this follower in \autoref{impl_dtw_eval_quant}--as a prime example of a polyphonic piano piece. The challenges from this piece arise not only from its polyphony, but also from its ornaments and the performer's artistic deviations. The \textit{mordent}s in the piece (see \autoref{fig:846fugue13} for an example) were played by the performer as a \textit{trill}, and an additional \textit{trill} was added in bar 23. Note that this performance, from the MAESTRO dataset \cite{MAESTRO}, is the same as that used in the quantitative analysis done in \autoref{impl_dtw_eval_quant}.

                    The $0.813$ total precision rate at $\theta_e = \si{300\milli\second}$ shown in \autoref{table:bwv846_eval_separate} was reflected well in this qualitative evaluation (note that $\si{300\milli\second}$ constitutes $30\%$ of a beat assuming a tempo of $60\text{BPM}$). Overall, the entirety of the piece was well handled, and the ornaments did not pose a challenge--most deviations were only up to half a beat's length and were quickly resolved. It could be shown from here that the follower is quite robust against polyphonic piano performances.
                    }
              \item{\texttt{prelude}

                    In contrast to the \textit{Fugue}, the \textit{Prelude in C major, BWV846} is primarily monophonic. It comprises repeating broken \textit{arpeggios}, with the top three notes repeated every iteration and the bottom two notes held over the whole length of the \textit{arpeggio}. Not many performance deviations are present in this piece, but the repeating nature of the top three notes in the broken \textit{arpeggios} may pose a challenge for score followers to align structurally similar melody lines. This performance, like the \textit{Fugue}, is taken from the MAESTRO dataset \cite{MAESTRO} and therefore is the same as that used in the quantitative analysis done in \autoref{impl_dtw_eval_quant}.

                    The $0.685$ total precision rate at $\theta_e = \si{300\milli\second}$ shown in \autoref{table:bwv846_eval_separate}--lower than the $0.815$ of the \textit{Fugue}--was reflected well in this qualitative evaluation. The repeating nature of the top three notes in the \textit{arpeggio}
                    caused significant misalignments of up to a beat in the first seven bars. Nevertheless, the piece was handled well beyond the deviations mentioned: after the first seven bars, the following stabilised. The performer's slowing down in the final few bars however caused noticeable deviations of up to a beat's length.
                    }
              \item{\texttt{turkish}

                    This simple classical-era piece by Mozart contains a simple melody line like many pieces of its era. The tempo of the performance is quite constant, and there are not many ornaments save for the added \textit{trill} by the performer in the second last bar in the excerpt.

                    Save for the follower's slight instability in the first bar, the performance was handled flawlessly by the score follower--the added \textit{trill} posed no challenge to the system.
                    }
              \item{\texttt{moonlight}

                    This popular piece by Beethoven comprises repeating triplets throughout the first movement with a melody line gradually developing. Slight temporal deviations are common in performances, including in this performed by the famous pianist Claudio Arrau. The repeating triplets are expected to pose a challenge to score followers as they may align the notes to the wrong iteration of the triplets.

                    The score follower started off with some significant instability in the beginning, which was resolved as the performance moved into bar 2. There were minimal deviations in the rest of the piece--deviations, if any, were only up to half a beat's length. Overall, the score follower's performance was excellent.
                    }
              \item{\texttt{unsospiro}

                    Like many of Liszt's pieces, \textit{Un sospiro} is challenging. It contains a very fast series of repeating broken \textit{arpeggios}, with a simple melody line layered of top. The \textit{tempo} of this romantic-era piece is also often varied artistically in performances, including in this live performance by Marc-André Hamelin in 1997.

                    The first two bars before the melodic line starts contained significant deviations and instability, but when the melodic line started, the following was generally stable towards the end of the excerpt, with some noticeable deviations (of up to half a beat's length) in the end of some melodic phrases.
                    }
              \item{\texttt{gnossienne}

                    This late 19th century piece by Erik Satie is unique in the sense that it is in \textit{free time}, i.e. it does not have any bar lines--this is common amongst many of Satie's ``unorthodox'' compositions. Nevertheless, there still exists a meter (4/4) in this excerpt. The \textit{free time} structure perhaps was given to emphasise the free-flowing long melodic phrases that should not be disconnected. Many performances of this piece, including this performance, contain significant \textit{rubato} to emphasise these melodic phrases. There also exists many \textit{acciaccatura} ornaments within the simple melodic lines.

                    This piece was handled reasonably well--the only obstacles were in following the relatively silent parts between melodic lines, where misalignments of up to two beats' length were observed. Nevertheless, these misalignments were resolved as the piece moves into the next melodic line. The \textit{acciaccatura} ornaments did not pose a challenge.
                    }
              \item{\texttt{entertainer}

                    This contemporary 20th century piece by Scott Joplin contains a playful melody line for the right hand supported by a simple accompaniment for the left hand. Notably, this performance was recorded on a \textit{pianola} (piano roll) and is not of studio quality. This piece serves as an indication on whether the follower can follow contemporary pieces, and also to observe whether the lower quality in recording affects the follower.

                    This piece was handled flawlessly, with little to no noticeable deviations.
                    }
              \item{\texttt{clair}

                    \textit{Clair de lune} starts off with a performance direction: ``\textit{Andante très expressif}'', which translates to ``moderatly slow, very expressive''. Bar 15 then has the direction ``\textit{Tempo rubato}''. This excerpt, performed by the famous pianist Lang Lang, is a romantic piece that showcases very rich \textit{rubato} and expressivity; thus, this piece contains many artistic performance deviations that are challenging to score followers.

                    This score follower handled the temporal deviations well, with deviations up to one beat in length, notably in bars 11-14. The deviations however were quickly fixed in bar 15.
                    }

          \end{enumerate}
    \item \texttt{violin}
          \begin{enumerate}
              \item{\texttt{chaconne-arp}

                    The \textit{Chaconne} was used extensively in \autoref{scofo_challenges} as an example of a challenging piece for score followers to follow. This excerpt features the \textit{arpeggios} discussed in \autoref{scofo_challenges_underspec}: the \textit{arpeggio} marking in bar 89 poses a challenge for score followers as performers may interpret this differently. Hilary Hahn's performance used here in fact contains two separate ways of playing the arpeggio. Hahn follows the suggested way from bars 89 to 100 before switching to another way until the end of the \textit{arpeggio} section. This may also be treated as \textit{improvisation} as discussed in \autoref{scofo_challenges_polyphonic}. The synthesised MIDI of the performance used by the follower merely contains the \textit{unbroken} chords as specified in the score; therefore, it is expected that significant deviations may occur during the \textit{arpeggio} section.

                    The section leading into the \textit{arpeggio} section--bars 84 to 88, despite containing some temporal deviations common to solo baroque performances, was handled flawlessly by the follower. The \textit{arpeggio} section was not handled as well, but was surprisingly still quite acceptable. The first part (bars 89 to 100) only contained deviations of up to two beats, but the other style of playing the \textit{arpeggios} employed by Hahn from bars 100 to 119 caused larger deviations of up to 1 bar (3 beats). This however was resolved in the second bar of the section succeeding the \textit{arpeggio} section (bar 121). All in all, the follower's robustness against such an extreme underspecification is remarkably impressive.
                    }
              \item{\texttt{chaconne-front}

                    This front excerpt of the \textit{Chaconne} comprises the polyphonic section shown in \autoref{fig:chaconnefirst10} which was described--in \autoref{scofo_challenges_perfdev_intentional}--as an excerpt that cannot be practically played on a violin. Hahn in this performance breaks down the chords in bars 1 to 8 into groups of two and does not hold the notes not in the top two of each chord. Hahn plays some chords in bars 9 to 16 in the ``rebound'' style, replaying the lower notes after executing the chord from bottom to top. These performance deviations are not encoded in the score, and therefore are expected to cause problems in score following.

                    This section, as expected, was not handled as well as other pieces in the dataset; nevertheless, the deviations, despite widespread, were mostly only up to a beat's length from the start to bar 15. From bar 16 to the end of the excerpt however, the deviation increased to up to a bar in length, and was not resolved by the end. It is however expected that the simpler structure beyond this excerpt would help the follower to resolve this deviation.
                    }
          \end{enumerate}

\end{itemize}

Overall, the score follower performed quite robustly against the wide variety of performances in the \textit{QualScofo} dataset. A common problem for most of the pieces is that there is some instability in the beginning of the following, but this soon goes away. Moreover, misalignments, if present and noticeable, are usually small (up to a bar's length in the worst case) and generally gets resolved further into the piece as the algorithm has more of the performance's information to work with. Only one piece--the \texttt{violin} group's \texttt{chaconne-front} piece--caused significant misalignments, resulting in a deviation of up to a bar that was not resolved within the excerpt.

Notably, the \textit{QualScofo} dataset contains pieces that cover not only the four main challenges targeted by the system discussed in \autoref{impl_dtw_req}, but also partially (via the \texttt{violin} group's \texttt{chaconne-arp} piece) the fifth (\nameref{scofo_challenge_music_improv}) given in \autoref{scofo_challenges}. The score follower is generally robust against these challenges especially the four targeted in \autoref{impl_dtw_req}, and therefore it can be said that the targeted requirements are fulfilled via the results of this qualitative analysis.

%% file: 02_parts/03_impl/chapters/04d_dtw_discussion.tex
\section{Discussion}
\label{impl_dtw_dicsuss}

Overall, the quantitative and qualitative evaluation results presented in \autoref{impl_dtw_eval} show that the follower fulfils all the requirements set out in \autoref{impl_dtw_req}.

Notably, it was shown by the quantitative analysis in \autoref{impl_dtw_eval_quant} that the sliCQ approach to obtain CQT coefficients from audio as features for score following consistently outperforms a commonly used traditional FFT-based approach.

Nevertheless, the OLTW algorithm as discussed in \autoref{impl_dtw_dtw_online} can be improved in many ways to obtain better performance--this, along with other possible areas of future work--will be discussed subsequently in \autoref{impl_dtw_future}. The component-based architecture described in \autoref{impl_dtw_impl_arch} means that extensions to the score follower would generally only require the implementation of an individual component that can be easily swapped into the system.

%% file: 02_parts/03_impl/chapters/04e_dtw_product.tex
\section{Resultant Product}

\subsection{Open-source Repository}

The score follower is released publicly under the GPLv3 Licence on GitHub\footnote{\githubFlippy}.

\subsection{Usage Guide}

The README\footnote{\url{https://github.com/flippy-fyp/flippy/blob/main/README.md}} of the repository shows a detailed usage guide of the score follower, including instructions on how to use it with the quantitative testbench (\autoref{tb_quant}) and qualitative testbench (\autoref{tb_qual}). A development guide is also provided.

\subsection{Demos}
\label{impl_dtw_prod_demos}

Video demonstrations of the system in action when used in conjunction with the qualitative testbench against all the pieces discussed in the qualitative evaluation analysis (\autoref{impl_dtw_eval_qual}) are provided in the repository under the \texttt{demos} directory\footnote{\url{https://github.com/flippy-fyp/flippy/tree/main/demos}}.

\subsection{Reproduction Suite}
\label{impl_dtw_prod_repro}

A reproduction suite is provided to reproduce results discussed in this chapter. This not only includes plots (Figures~\ref{fig:cqt_time}, \ref{fig:featuresbwv846}, \ref{fig:dtw_time}, \ref{fig:bach10_plot_precision} and \ref{fig:bwv846_plot_precision}) and metrics (Tables~\ref{table:bach10_metrics}, \ref{table:bwv846_metrics} and \ref{table:bwv846_eval_separate}), but also the qualitative analyses in \autoref{impl_dtw_eval_qual} producing the demos in \autoref{impl_dtw_prod_demos}.

The README of the repository contains further detailed information on the reproduction suite.

%% file: 02_parts/03_impl/chapters/04f_future.tex
\section{Future Work}
\label{impl_dtw_future}

While the sliCQ audio feature extraction technique performs remarkably well, there are many potential areas of improvement for this system. In this section, possible areas of future work are outlined and discussed.

\subsection{Improvements to OLTW}

The OLTW algorithm, first introduced by Dixon in 2005 \cite{Dixon2005MATCHAM}, has seen many iterations and improvements, some of which are detailed in \autoref{impl_dtw_tech} and \cite{arzt16}. It would be interesting to study the performance of this system after applying the improvements, which include, but are not limited to:

\begin{itemize}
    \item \textbf{Reconsidering past decisions}. A method to backtrack and reconsider past decisions in the OLTW algorithm was proposed by Arzt \etal{} in 2008 \cite{arzt08}; this work was performed to alleviate the OLTW algorithm's tendency to deviate significantly from the optimal path thus requiring a large amount of time to recover, as the algorithm only considers the current local information of the performance and score. This improvement provided two major improvements in following real-life performances with more significant performance deviations: it improved the algorithm's robustness against tempo changes and also its error tolerance.
    \item \textbf{Tempo models}. Score following can also be seen as a task of adapting the tempo of the score to the performance audio. Variations in tempo (known commonly as \textit{rubato}) is prevalent in most kinds of (classical) music (previously discussed in \autoref{scofo_challenges_perfdev}); therefore, feeding tempo information computed from the performance into the OLTW algorithm (and even learning tempo deviations from different performers before storing them) can significantly improve the score follower's performance. A version of this improvement was formalised and implemented by Arzt and Widmer in 2010 \cite{arzt10}. Chen and Jang in 2019 \cite{chen19} also incorporated a tempo model in their system.
    \item \textbf{Windowed Time Warping (WTW)}. WTW was first introduced as an improvement to OLTW that incorporates idea of the A* algorithm \cite{hart68} in 2010 by Macrae and Dixon \cite{macrae10dtw}. WTW was shown to be real-time capable and is in fact faster (by a factor of almost 100) than OLTW; however, WTW was only able to match the performance of OLTW in terms of alignment. Nevertheless, the usage of WTW would decrease the computation power required to compute the optimal path, which is important when using this system in more resource-constrained devices, such as mobile phones. Further, as some improvements may add to the latency and computational load of the following algorithm, the usage of WTW will open more doors to more novel and computationally heavy improvements.
\end{itemize}

As mentioned in \autoref{impl_dtw_impl_arch}, the component-based architecture of this project's score following system means that improvements detailed above (and beyond) can be easily created as another ``Follower'' component (as shown in \autoref{fig:oltw_step}) and plugged into the system.

\subsection{Online denoising and dereverberation of performance audio}

While this extension was briefly mentioned by numerous authors in the field \cite{henkel20}, no actual study and implementation was done--the closest to dealing with reverberation was in Henkel and Widmer's 2021 work \cite{henkel21} which used Impulse Responses (IRs) to augment synthesised performance audio in the MSMD dataset to allow their score follower to generalise better to performances in real-life settings.

Denoising and dereverberation techniques (which are more actively developed in the speech processing field \cite{naylor10,speechdenoising}, but are increasingly being researched for general audio \cite{engel2020ddsp}) may lead to significant improvements in the robustness of the score follower especially in end user environments. Users may use the score follower on devices with subpar microphones resulting in noisy audio, or attempt to use the system in different environments that cause the audio to be noisy or reverberant. 

\subsection{Using more Score Features}

While an advantage of this score following system is that an annotated audio representation of the score is sufficient, many high-performing systems in the MIREX evaluation (discussed in \autoref{scofo_recent_mirex}) exploit the availability of high-quality note onset information in the score to improve the follower's capability. This is used in implementations by Carabias-Orti \etal{} \cite{carabias12}, Rodriguez-Serrano \etal{} \cite{rodriguez16}, Bris-Peñalver \etal{} \cite{orti15,alonso16,alonso17} and Chen and Jang \cite{chen19}; all of which managed to produce high-quality score following evaluation results.

\subsection{Web Implementation}

Implementing this score follower on the Web (perhaps even to replace the current pair-device-based page turner implementation in \emph{TuneApp}\footnote{\url{https://tuneapp.co/pageturner}}) would allow the usage of the system on all Web-enabled devices easily without the hassle of manually installing a software artifact. Further, this system would be able to reach more users as a Web application, as opposed to only reaching a subset of users being a desktop- or mobile-exclusive application.

Significant challenges on porting this score follower to the Web exist; most notably, this score follower is written in Python (as mentioned in \autoref{impl_dtw_impl_techstack}). Python is (currently) not a language well supported by the Web. Viable strategies of implementing this include (but are not limited to):

\begin{itemize}
    \item \textbf{Rewriting in JavaScript or a JavaScript-supported language}. As mentioned in \autoref{tb_qual_impl_details_choice}, JavaScript is the \textit{lingua franca} of the Web. Thus, the follower can be rewritten in JavaScript (or a language that can transpile into JavaScript\footnote{\url{https://www.slant.co/topics/101/~best-languages-that-compile-to-javascript} covers a comprehensive list of languages that transpile into JavaScript.}, including but not limited to TypeScript\footnote{\url{https://www.typescriptlang.org/}}, Kotlin\footnote{\url{https://kotlinlang.org/docs/js-overview.html}} and F\#\footnote{\url{https://fable.io/}}). A problem with using JavaScript--or any language that transpiles into JavaScript--is that JavaScript is inherently single threaded, and would not fully support this system's multiprocessing architecture as detailed in \autoref{impl_dtw_impl}.
    \item \textbf{Rewriting in a WebAssembly-supported language}. WebAssembly\footnote{\url{https://webassembly.org/}} (abbreviated \textit{Wasm}) is a binary instruction format designed as a portable compilation target for various programming languages to be run on the Web. \textit{Wasm} is designed to be fast, safe and open. While there are fewer languages that provide full compilation support to \textit{Wasm}\footnote{\url{https://github.com/appcypher/awesome-wasm-langs} provides a comprehensive and updated list.} compared to the many languages that can transpile into JavaScript, major languages such as C++\footnote{\url{https://github.com/emscripten-core/emscripten}. Note that most languages that support emitting of LLVM (\url{https://llvm.org/ProjectsWithLLVM/}) are in fact supported by \textit{Emscripten}.} and Go\footnote{\url{https://golang.org/}} support compiling to \textit{Wasm}. Another advantage of \textit{Wasm} over (transpiling to) JavaScript is that \textit{Wasm} supports multithreading using \texttt{SharedArrayBuffer} in browsers\footnote{\url{https://emscripten.org/docs/porting/pthreads.html}}. Significant work however will be required to first rewrite the project into a \textit{Wasm}-supported language and then connect the emitted \textit{Wasm} output to the GUI.
    \item \textbf{Running Python on the Web}. This approach would theoretically require minimal effort as the project can remain in Python. There are two major classes of approaches that can enable Python on the browser: compiling into JavaScript (major implementations include Transcrypt\footnote{\url{https://www.transcrypt.org/}}, Brython\footnote{\url{https://brython.info/}} and Skulpt\footnote{\url{https://skulpt.org/}}) and incorporating a Python runtime in the browser (implementations include Pyodide\footnote{\url{https://pyodide.org/}}, which works by compiling the Python runtime into \textit{Wasm}). While there are numerous projects bringing Python to the browser, none of them currently fully support all the features required by this project (multiprocessing for instance), and thus block this approach. Pyodide (currently version 0.17.0 as of time of writing) shows great promise as it is actively developed, and incorporating the score follower into Pyodide would be an interesting project in the future when the features required for this score follower are implemented.
\end{itemize}

Note that porting the GUI part of the follower, which is mostly covered by the qualitative testbench introduced in \autoref{tb_qual} is comparatively trivial as it is written using Web technologies--see \autoref{tb_qual_impl_details_choice}.

\subsection{Real-life Testing}
\label{impl_dtw_future_reallife}

Unfortunately, in this work's time period, the COVID-19 pandemic prevented real-life tests from being carried out. It would be interesting to evaluate this project's system with musicians--of varying skill levels--in different real-life settings, be it music practice or performance. This would require augmenting the performance audio streaming component to take in streamed audio from a microphone.

\subsection{Score Following Applications}

The qualitative testbench introduced in \autoref{tb_qual} can be used in conjunction with this score follower as a means of carrying out performance analysis as discussed in \autoref{scofo_apps_perfanal}. The qualitative analyses--supported by the quantitative analyses--performed in \autoref{impl_asm_eval} strongly suggest that this system is accurate enough for this purpose. 

Moreover, the qualitative testbench can be used as an APT system, but more precisely an Automatic Page \emph{Scrolling} system. The results in the qualitative evaluations in \autoref{impl_dtw_eval_qual} show that this system is reliable and robust enough to be used to follow complete performances with appreciable performance deviations. Work remains to be done however to test this system on real performers--with varying skill levels--as per \autoref{impl_dtw_future_reallife}.

Score following has many applications beyond performance analysis and APT, some of which are detailed in \autoref{scofo_apps}. It would be interesting to explore the usage of this system in, for instance, computer-aided accompaniment and performance cues.

\subsection{Handling Improvised Performances}

This system was not tested thoroughly against improvised performances, which is a challenge of score following identified in \autoref{scofo_challenge_music_improv}. It would be interesting to see how well this system performs against improvised performances, and also to investigate the necessary adjustments and/or extensions required to allow the system to perform well on improvised performances.

%% file: 02_parts/01_conclusion.tex
\chapter{Conclusion}
\label{conclusion}

\epigraph{\itshape Those who have achieved all their aims probably set them too low.}{Herbert von Karajan}

Following the three-part structure of this report, the report is concluded by tackling the three parts one at a time.

\section*{\autoref{scofo}: \nameref{scofo}}
In this part a thorough review that covers score followers' applications (\autoref{scofo_apps}), basic preliminaries (\autoref{scofo_preliminaries}), challenges (\autoref{scofo_challenges}) and finally an extensive literature review (\autoref{scofo_litreview}) was presented. 

To the best of knowledge no such recent review was published in the literature, and new score following papers assume substantial prior score following knowledge. Moreover, it is common for new papers to only provide a literature review that covers a minimal subset of score followers related to their proposals.

Hence, it is with hope that this provides readers without any significant prior knowledge of score following to be able to get up to speed with score following in the last 40 years, making their study or research on score followers much easier. 

In the context of this report, this part gave extensive background on the ideas and concepts discussed and developed in the other two parts--the number of references to this part is testament to the significance of the review.

A challenge in this part is of course the fact that literature reviews are only up to date as of publishing, and the highly active nature of the field means that this review will be soon outdated. Where time allows, a future related effort is to instead make this review an open-source documentation webpage\footnote{Docusaurus (\url{https://docusaurus.io/}) is an easy-to-use and good option.}, where researchers can constantly update the page with more information procured from newer works, or even fill in gaps and add more detail on certain areas if there exists such a need. 

\section*{\autoref{testbench}: \nameref{testbench}}

This part contributes two testbenches by first attempting to identify difficulties of evaluating score followers in \autoref{tb_preliminaries}. The quantitative testbench was then presented in \autoref{tb_quant} before its qualitative counterpart was detailed in \autoref{tb_qual}.

Instead of attempting to come with a one-size-fits-all solution (which as mentioned in \autoref{tb_preliminaries} was quite impossible), the proposed pair of testbenches means that the score following research community could finally have open-source evaluation testbench solutions able to benchmark as many score followers as possible. The introduction of the novel way of procuring quantitative testbench data helps the community to more easily procure more quantitative datasets for benchmarking, and the \textit{QualScofo} dataset is a crucial component of the qualitative testbench. Adoption--and continuous development--of these testbenches will be a long-term challenge, but important to be surmounted to promote open and reproducible research. To this end, good engineering practices were introduced and upheld.

The usage of the testbenches for score follower implementations in \autoref{impl} also served as demonstrations of the proposed testbenches. In the future, it would be interesting to use the proposed new way of creating quantitative testbench data to produce more data for benchmarking, as well as to extend the proposed \textit{QualScofo} dataset to include more musical pieces spanning a larger variety of music and performer skill levels, or to incorporate different performers' interpretations of a piece of music.

\section*{\autoref{impl}: \nameref{impl}}

The beat-tracking-based \textit{TuneApp Conductor} was the first score follower implementation described in \autoref{impl_beat}. User feedback indicated that it is a successful interactive feature for \textit{TuneApp}. Most importantly, it showed a viable and entertaining application of score following. Moreover, \textit{TuneApp Conductor} answered the \textit{Radio-Baton} engineering sophistication question posed in \autoref{impl_beattrack_backstory}; now that sensors that could deliver the same information as the \textit{Radio-Baton} are ubiquitous on mobile devices, a score-following-based reimplementation of the \textit{Radio-Baton} that is immensely less sophisticated is indeed possible. As conducting definitely goes beyond just indicating beats, further work in the \textit{TuneApp Conductor} can incorporate computer vision approaches, notably pose detection \cite{posenet}, in which the user's body pose is taken into account in the music performance: controlling the tempo, volume, balance and perhaps even musical texture. It is also regrettable that the MusicXML-based conductor had to be dropped--however it was mentioned that it could one day be released after optimisation. (In addition, an application that works on facial features\footnote{Leonard Bernstein conducting with only his eyebrows: \url{https://www.youtube.com/watch?v=G7_6Z33eCaY}.} would be interesting.) 

Further, an ASM aligner was introduced in \autoref{impl_asm} to complement the quantitative testbench (by producing high-quality ground-truth data for evaluation) and also to provide a gentle introduction for the concepts in music alignment. Results showed that the ASM aligner was robust and reliable in producing ground-truth alignments from widely available music performance datasets. Moreover, the aligner's relation to a real-time score follower was discussed, which not only helped develop ideas used in the DTW implementation, but also showed why ASM-based score followers were quickly made obsolete by their HMM- and DTW-based counterparts. As music performance datasets containing performance MIDI information mostly are for piano currently, in the future it would be interesting to evaluate the ASM aligner's ability based on datasets involving other instruments.

Finally, a CQT-DTW score follower was implemented in \autoref{impl_dtw}. Evaluation results showed that the score follower performs well against four major challenges in score following detailed in \autoref{scofo_challenges}, namely \nameref{scofo_challenge_limitfe}, \nameref{scofo_challenges_perfdev}, \nameref{scofo_challenges_underspec} and \nameref{scofo_challenges_polyphonic}. Importantly, it was shown that the sliCQ CQT approach extracts significantly more reliable audio features for score following than a commonly used traditional FFT-based approach. The performance and design of the score follower were discussed at length in \autoref{impl_dtw_dicsuss}, and the potential future directions of the system were detailed in \autoref{impl_dtw_future}.

\section*{Closing Thoughts}

Score following is, no doubt, a challenging research topic that brings many useful applications in music practice, performance and production. The breadth and depth of knowledge required to overcome challenges to the problem are significant, encompassing the fields of music (theory and practical), mathematics (e.g. optimisation and stochastic methods), electrical engineering (e.g. signal processing, acoustical engineering and audio engineering) and computer science (e.g. computer vision, artificial intelligence and pattern recognition). These fields help bridge the gap between two infinitely complex human arts of music engraving and performance.

This project, while long and extensive, is only a small, one-year foray into this field. Nevertheless, substantial work in the form of code\footnote{See \autoref{appendix_project_code_metrics} for code metrics.}--mostly open source--were written to attain the practical goals of this project in demonstrating the contributions listed in \autoref{intro:contrib}. There are myriad extensions possible to all the work done here, and it is with hope that this report helps further push the boundaries of knowledge in score following.

%% file: 03_back_pages/appendix/appendix.tex
\appendix
\chapter{Extra notes to Figures}
\label{appendix_extranotestofigures}

\newenvironment{itemizefignotes}
{ \begin{itemize}
        \setlength{\itemsep}{0pt}
              \setlength{\parskip}{0pt}
              \setlength{\parsep}{0pt}     }
              { \end{itemize}                  }

\subsubsection{\autoref{fig:caa}}
\begin{itemizefignotes}
    \item Icons made by \textit{Freepik} from \url{https://www.flaticon.com}.
    \item Music excerpt from Beethoven's \textit{Violin Sonata No. 9 in A Minor, Op. 47}. Public Domain.
\end{itemizefignotes}

\subsubsection{\autoref{fig:generic-framework}}
\begin{itemizefignotes}
    \item Icons made by \textit{iconixar}, \textit{surang}, \textit{DinosoftLabs} and \textit{Eucalypt} from \url{https://www.flaticon.com}.
\end{itemizefignotes}

\subsubsection{\autoref{fig:mdr-conv}}
\begin{itemizefignotes}
    \item Icons made by \textit{Pixel perfect} and \textit{iconixar} from \url{https://www.flaticon.com}.
    \item MIDI logo from the \textit{MIDI Manufacturers Association}. Public Domain.
    \item MusicXML logo from \url{https://www.musicxml.com}.
\end{itemizefignotes}

\subsubsection{\autoref{fig:generic-framework}, \autoref{fig:audio-to-symbolic}, \autoref{fig:modern-staff-notation}, \autoref{fig:fuguebar7}, \autoref{fig:qualbwv846}, \autoref{fig:846fugue13}, \autoref{fig:prelude33}}
\begin{itemizefignotes}
    \item Music excerpts from Bach's \textit{Prelude and Fugue in C major, BWV 846}. Public Domain.
\end{itemizefignotes}

\subsubsection{\autoref{fig:chaconnefirst10}, \autoref{fig:chaconnelast2}, \autoref{fig:chaconnearp}}
\begin{itemizefignotes}
    \item Music excerpts from Bach's \textit{Chaconne from Partita II, BWV1004}.
    ©2019 by Hajo Dezelski. Creative Commons Attribution ShareAlike 3.0 (Unported) License - free to distribute, modify and perform.
\end{itemizefignotes}

\newpage
\chapter{Specifications of the System used for Evaluation and Benchmarks}
\label{appendix_benchmark_system}

\begin{table}[ht]
    \centering 
    \begin{tabular}{l c} 
        \hline\hline 
        Component & Model/Version                    \\ [0.5ex] 
        \hline 
        CPU       & Intel Core i7-8700 CPU (3.20GHz) \\
        RAM       & 16GB                             \\
        Storage   & 512GB SSD                        \\
        OS        & Ubuntu 20.04.2 LTS               \\
        \hline 
    \end{tabular}
    \label{table:systemspec} 
\end{table}

\chapter{Project Code Metrics}
\label{appendix_project_code_metrics}

Lines of Code (LoC) calculated using \textit{VS Code Counter}\footnote{\url{https://marketplace.visualstudio.com/items?itemName=uctakeoff.vscode-counter}}.

\begin{table}[ht]
    \centering 
    \begin{tabular}{l l r r} 
        \hline\hline 
        Repository                                                                                        & Language           & LoC  & Total \\ [0.5ex] 
        \hline 
        Quantitative Testbench (including ASM Aligner in \autoref{impl_asm})\footnote{\githubFlippyQuant} & Python             & 3193 & 3193  \\
        \hline
        Qualitative Testbench\footnote{\githubFlippyQual}                                                 & TypeScript (React) & 1304 &       \\
                                                                                                          & Python             & 71   & 1375  \\
        \hline
        \textit{TuneApp Conductor} (\Cref{impl_beat})                                                     & TypeScript (React) & 1471 &       \\
                                                                                                          & CSS                & 69   & 1540  \\
        \hline
        CQT-DTW Score Follower (\Cref{impl_dtw})\footnote{\githubFlippy}                                  & Python             & 3883 &       \\
                                                                                                          & Shell Script       & 154  & 4037  \\
        \hline
        \textbf{Total}                                                                                    &                    &      & \textbf{10145} \\
        \hline
    \end{tabular}
    \label{table:loc} 
\end{table}